\documentclass[11pt,fullpage,letterpaper]{article}

\usepackage[margin=1in]{geometry}
\usepackage{fancyhdr}
\usepackage[utf8]{inputenc}
\usepackage[T1]{fontenc}

\usepackage[english]{babel}
\usepackage[backend=biber,style=alphabetic,natbib=true,maxalphanames=4,maxnames=10]{biblatex}
\usepackage{titletoc}
\usepackage[toc, page, header]{appendix}
\newcommand{\stopmaintoc}{\setcounter{tocdepth}{0}}

\usepackage{url}
\usepackage{booktabs}
\usepackage{amsfonts}
\usepackage{nicefrac}
\usepackage{microtype}
\usepackage{xcolor}

\usepackage{parskip}

\title{The Power of Two Choices in Dynamic Pricing:\\
As Good As Full Feedback, Instance-Wise}

\author{Yan Dai\thanks{Operations Research Center, Massachusetts Institute of Technology. Email: \texttt{yandai20@mit.edu}.\\ Part of the research was done when Yan Dai was a research intern at Microsoft Research New England.} \and Tao Lin\thanks{School of Data Science, The Chinese University of Hong Kong, Shenzhen. Email: \texttt{lintao@cuhk.edu.cn}.\\ Part of the research was done when Tao Lin was a postdoctoral researcher at Microsoft Research New England.} \and Aleksandrs Slivkins\thanks{Microsoft Research NYC. Email: \texttt{slivkins@microsoft.com}.}}
\date{November 2026}

\usepackage{multirow}
\usepackage{colortbl}
\usepackage{arydshln}
\ADLnullwidehline
\newcommand{\hlcell}{\cellcolor{green!10}}
\usepackage{footnote}
\usepackage{enumitem}
\setitemize{leftmargin=*,nosep}
\setenumerate{leftmargin=*,nosep}

\usepackage{amsmath}
\usepackage{amsthm}
\usepackage{amssymb}
\usepackage{mathtools}
\usepackage{thmtools}
\usepackage{bbm}
\usepackage{bm}
\allowdisplaybreaks
\let\originalmiddle=\middle
\def\middle#1{\mathrel{}\originalmiddle#1\mathrel{}}

\usepackage{algorithm}
\usepackage[noEnd,commentColor=blue!70!white]{algpseudocodex}

\makeatletter
\newcommand{\algeq}[2][]{%
  \par\vspace{2pt}%
  \noindent
  \makebox[\columnwidth]{%
    \hfill #2%
    \if\relax\detokenize{#1}\relax
      \hfill
    \else
      \hfill (\refstepcounter{equation}\theequation)\label{#1}%
    \fi
  }%
  \par\vspace{2pt}%
}
\makeatother

\usepackage[colorlinks=true, linkcolor=blue!60!black, citecolor=blue!60!black,
urlcolor=black]{hyperref}
\usepackage[sort]{cleveref}
\makeatletter
\AddToHook{env/algorithmic/begin}{\def\@currentcounter{ALG@line}}
\makeatother
\crefalias{ALG@line}{line}
\AddToHook{cmd/appendix/before}{%
\crefalias{section}{appendix}%
\crefalias{subsection}{appendix}
}

\Crefname{assumption}{Assumption}{Assumptions}
\Crefname{reduction}{Reduction}{Reductions}
\Crefname{claim}{Claim}{Claims}
\Crefformat{equation}{Eq. #2(#1)#3}
\Crefrangeformat{equation}{Eqs. #3(#1)#4 to #5(#2)#6}
\Crefmultiformat{equation}{Eqs. #2(#1)#3}{ and #2(#1)#3}{, #2(#1)#3}{ and #2(#1)#3}
\Crefrangemultiformat{equation}{Eqs. #3(#1)#4 to #5(#2)#6}{ and #3(#1)#4 to #5(#2)#6}{, #3(#1)#4 to #5(#2)#6}{ and #3(#1)#4 to #5(#2)#6}

\newenvironment{subroutine}[1][!t]
  {\begin{algorithm}[#1]\floatname{algorithm}{Subroutine}\crefalias{algorithm}{subroutine}}
  {\end{algorithm}}
\Crefname{subroutine}{Subroutine}{Subroutines}

\newtheorem{theorem}{Theorem}[section]
\newtheorem{lemma}[theorem]{Lemma}

\newtheorem{corollary}[theorem]{Corollary}

\newtheorem{reduction}[theorem]{Reduction}

\theoremstyle{definition}
\newtheorem{definition}{Definition}

\newtheorem{example}[definition]{Example}
\newtheorem{remark}[definition]{Remark}
\AtBeginDocument{%
}

\usepackage{tikz}
\usetikzlibrary{positioning}
\usetikzlibrary{calc}
\usetikzlibrary{decorations.pathreplacing}
\usetikzlibrary{patterns}
\usepackage{pgfplots}
\pgfplotsset{compat=1.18}
\usepgfplotslibrary{groupplots}

\newcommand{\E}{\operatornamewithlimits{\mathbb{E}}}
\newcommand{\1}{\mathbbm{1}}
\renewcommand{\O}{\operatorname{\mathcal O}}
\newcommand{\Otil}{\operatorname{\tilde{\mathcal O}}}
\newcommand{\argmax}{\operatornamewithlimits{argmax}}
\newcommand{\diam}{\operatornamewithlimits{diam}}
\renewcommand{\tilde}{\widetilde}
\renewcommand{\hat}{\widehat}
\renewcommand{\bar}{\overline}

\newcommand{\mR}{\operatorname{Reg}}
\newcommand{\mF}{{\mathcal{F}}}
\newcommand{\mA}{{\mathcal{A}}}
\newcommand{\mC}{\mathcal{C}}
\newcommand{\mT}{{\mathcal{T}}}
\newcommand{\mI}{{\mathcal{I}}}
\newcommand{\mH}{{\mathcal{H}}}

\newcommand{\candidateSet}{\mathcal S}
\newcommand{\priceDist}{P}

\newcommand{\Sample}{\operatorname{\normalfont\textsc{Sample}}}
\newcommand{\Threshold}{\operatorname{\normalfont\textsc{Threshold}}}
\newcommand{\Interval}{\operatorname{\normalfont\textsc{Interval}}}
\newcommand{\AlgOL}{$\mathcal A^{\text{OL}}$\xspace}

\newcommand{\general}{{\mathrm{all}}}
\newcommand{\regular}{{\mathrm{regular}}}
\newcommand{\curv}{{\mathrm{curv}(\kappa)}}
\newcommand{\curvature}{\curv}

\newcommand{\MHR}{{\mathrm{MHR}}}
\newcommand{\CDF}{{\mathrm{CDF}}}

\renewcommand{\paragraph}[1]{\vspace{2pt} \noindent{\bf #1}}

\usepackage{xspace}
\newcommand{\ie}{{i.e.,~\xspace}}
\newcommand{\eg}{{e.g.,~\xspace}}

\newcommand{\rbr}[1]{\left(#1\right)}
\newcommand{\sbr}[1]{\left[#1\right]}
\newcommand{\cbr}[1]{\left\{#1\right\}}

\newcommand{\eps}{\epsilon}
\newcommand{\LDOTS}{\, ,\ \ldots\ ,}

\newcommand{\demand}{D}
\newcommand{\intervalP}{D}
\newcommand{\Reg}{\operatorname{Reg}}
\newcommand{\ADV}{\mathtt{ADV}}
\newcommand{\popt}{p^{\ast}}
\newcommand{\pseq}{\popt_{\mathtt{seq}}}

\begin{document}

\maketitle
\begin{abstract}

We study the dynamic pricing problem, where a firm sells a product to $T$ myopic customers and their values are independent samples from an unknown value distribution $F$. With posted prices, the firm only sees whether each customer buys, and the minimax regret is $\widetilde \Theta(T^{2/3})$ \citep{kleinberg2003value}. A \emph{menu} instead offers several options, each selling the product with some probability at some expected payment.  Infinitely many menu options allow the seller to fully reveal customer values (i.e., eliciting full feedback), reducing the regret to $\Theta(\sqrt T)$. How many options are needed?

Perhaps surprisingly, a second menu option suffices, achieving the same regret rates as infinite menus (up to $\log$ factors). This holds broadly: without knowing any property of $F$, our algorithm with two menu options attains $\widetilde{\mathcal O}(\sqrt T)$ regret in general, $\widetilde{\mathcal O}(T^{1/3})$ regret for monotone-hazard-rate distributions, and $\widetilde{\mathcal O}(T^{d/(d+1)})$ instance-dependent regret when $d\in(0,1)$ is a suitable notion of  ``dimension''. All are optimal even with infinite menus. Moreover, we argue, via a novel lower-bound construction, that our algorithm is near-optimal \emph{for each problem instance}.

Along the way, we study the query complexity of revenue maximization: An algorithm issues a \emph{query} to each customer, from some class of feasible queries, and learns something about their value. The goal is to learn the property of interest (either the best price or the CDF) up to some precision. Query complexity is the smallest number of queries to guarantee this, in the worst case over some class of value distributions. We find query complexity technically intertwined with dynamic pricing, with posted pricing and full feedback corresponding to, resp., threshold queries and value queries. Given these two extremes, how powerful are the ``intermediate'' query classes?

Our answer focuses on \emph{interval queries}, which reveal whether customer value lies in a given interval. We show that they suffice to achieve the same query complexity as value queries (up to $\log$ factors) for both properties (best price and CDF) and across several classes of value distributions. In particular, we provide a new interval-query algorithm for each property, which handles all distribution classes at once. The best-price-estimation algorithm is the key ingredient in our main dynamic pricing algorithm, and the CDF-estimation algorithm also underpins a simpler dynamic pricing algorithm which attains $\widetilde{\mathcal O}(\sqrt T)$ regret with three menu options.

\end{abstract}
\thispagestyle{empty}
\addtocounter{page}{-1}
\newpage
\tableofcontents
\setcounter{tocdepth}{2}
\thispagestyle{empty}
\addtocounter{page}{-1}
\newpage

\section{Introduction}\label{sec:intro}

Consider a firm introducing a new product to the market, whose values to customers -- the highest price they are willing to pay -- are randomized and unknown. Hence, at a price $p$, a customer buys with probability $\demand(p) := \Pr_{v\sim F}\{v\ge p\}$, where $F$ is a fixed but unknown distribution of values; the function $\demand(\cdot)$ is called the \emph{demand curve}. The firm learns this function (or equivalently, the value distribution $F$) via dynamic pricing and customers' buying decisions.

Concretely, consider the now-standard formulation of dynamic pricing \citep{kleinberg2003value}. A revenue-maximizing firm has one product for sale, with unlimited supply. In each round $t \in [T]$, a new customer arrives with value $v_t\in [0,1]$ drawn from $F$. The firm posts a price $p_t\in [0,1]$ based on the history, according to some algorithm. The customer buys the product if and only if $v_t\geq p_t$. The firm never sees $v_t$ but observes the buying decision $\1[v_t\ge p_t]$. If $F$ were known, the optimal algorithm would post the \emph{monopoly price} $p^\ast_F$ in every round: the price maximizing expected revenue $R(p):=p\cdot \demand(p)$ \citep{myerson1981optimal}. The firm aims to minimize its \emph{regret}, defined as $T\cdot R(p^\ast_F)$ minus its expected cumulative revenue in $T$ rounds. One can interpret regret as how much the firm would pay to know the demand curve in advance, e.g., via market research. This problem and many of its variants are well-studied; in particular, the minimax-optimal regret across all problem instances is of the order $\tilde \Theta(T^{2/3})$ \citep{kleinberg2003value}.\footnote{The \emph{minimax-optimal regret} is the optimal regret over all algorithms, in the worst case over a given class of problem instances, for a given time horizon $T$. A lower bound on this quantity is called a \emph{minimax lower bound}.} It is closely related to multi-armed bandits and particularly to Lipschitz bandits.\footnote{In multi-armed bandits \citep{LS19bandit-book,slivkins-MABbook}, in each round the algorithm chooses some ``arm'' and receives a reward for this arm (drawn from some fixed but unknown distribution). Lipschitz bandits \citep{kleinberg2019bandits,bubeck2011x} is a well-studied bandit problem where expected rewards obey a Lipschitz condition, such as $|R(p) - R(p')|\leq |p-p'|$, for all $p,p'\in [0,1]$. Dynamic pricing satisfies a one-sided version: whenever $p>p'$. See Related Work (\Cref{app:related}) for more background. \label{fn:LipMAB}}

However, firms commonly offer a \emph{menu} instead of a single price, \eg different subscription levels at different prices. Within our single-product framework, a menu is defined as a collection of options, where each option $(q,p)\in [0,1]^2$ indicates that the customer gets the item with probability $q$ and pays $p$ in expectation.\footnote{One could also interpret $q$ as quality, if varying qualities are possible and customer utility is linear in quality.} Option $(0,0)$, the ``outside option'' of not doing anything, is always implicitly included. Thus, a single posted price $p$ becomes a size-1 menu $\{(1,p)\}$. A customer with value $v$ chooses the option with the highest utility $qv-p$, or buys nothing if all utilities are negative. Such choices can potentially reveal information about the customer's value, allowing the seller to learn the unknown $F$ and optimize future prices. Our central question is:
\begin{center}
\textbf{\textit{How beneficial are menus for dynamic pricing?}}
\end{center}

Thus, we study \emph{dynamic pricing with menus}, where each customer is offered a menu designed by the seller's algorithm based on history, and the chosen option is observable. The optimal algorithm under the known value distribution $F$ remains the same: always post price $p^\ast_F$, defined as above, which holds even when infinitely many options (\emph{infinite menus}) are allowed \citep{myerson1981optimal}.

\paragraph{Warm-Up: Infinite Menus.} We achieve regret $\O(\sqrt{T})$ using infinite menus, with a complementary minimax lower bound. This is in stark contrast to the $\tilde \Theta(T^{2/3})$ regret rate without menus. While infinite menus are impractical, we treat them as a useful baseline and a stepping stone for finite-menu results. To elaborate on infinite menus as a baseline, we show that infinite menus are equivalent to \emph{full feedback}, where each customer's value $v_t$ is directly observed after each round.\footnote{Generically, \emph{full feedback} refers to observing the outcome of every possible action one could have taken.} That is, with either infinite menus or full feedback (or both), the seller achieves the same expected revenue in dynamic pricing; this holds for every realization of values $(v_t)_{t\in [T]}$.

\paragraph{Main Contributions.} Our main high-level result is that \emph{size-2 menus} are as powerful as \emph{infinite menus}: they achieve the same regret rates that are better than the rates without menus. We prove this in a very broad sense. First, we achieve $\tilde \Theta(\sqrt{T})$ regret for all problem instances. Second, we improve this to $\tilde \Theta(T^{1/3})$ for all instances with monotone hazard rate (MHR) or constant curvature (a similar property defined in \citep{kleinberg2003value}). Third, we obtain a continuous interpolation: $\tilde \Theta(T^{d/(d+1)})$ regret for all instances with \emph{interval dimension} $d\in [0,1]$, a concept similar to the zooming dimension in Lipschitz bandits \citep{kleinberg2019bandits,bubeck2011x}. In each scenario, we obtain an upper bound for \emph{size-2 menus}, and a matching minimax lower bound for \emph{infinite menus}. We summarize them in \Cref{table:regret}, along with the rates attainable without menus. All upper bounds are obtained by a single algorithm (\Cref{alg:size-2}) and via a unified analysis, which is our first key contribution from a technical perspective.

\begin{table}[!t]
\centering
\caption{Regret Bounds under Different Menu Sizes}
\label{table:regret}
\Crefname{theorem}{Thm}{Thms}
\Crefname{lemma}{Lm}{Lms}
\resizebox{\textwidth}{!}{%
\renewcommand{\arraystretch}{1.3}
\begin{tabular}{|c|c|c|c|} \hline
\begin{tabular}{@{}c@{}}Value\\Distribution\end{tabular} & \begin{tabular}{@{}c@{}}With Infinite Choices\\(Size-$\infty$ Menus)\end{tabular} & \begin{tabular}{@{}c@{}}With a Second Choice\\(Size-$2$ Menus)\end{tabular} & \begin{tabular}{@{}c@{}}Only Posted Price\\(Size-$1$ Menus)\end{tabular} \\\hline
General & $\Omega(\sqrt T)$ \textbf{(\Cref{thm:minimax lower}a)} & \hlcell $\Otil(\sqrt T)$ \textbf{(\Cref{thm:size-2}a)} & $\tilde \Theta(T^{2/3})$ \citep{kleinberg2003value} \\\hline
MHR & $\Omega(T^{1/3})$ \textbf{(\Cref{thm:minimax lower}b)} & \hlcell $\Otil(T^{1/3})$ \textbf{(\Cref{thm:size-2}b)} & $\tilde\Theta(\sqrt T)$ \citep{singla2024bandit,leme2023pricing} \\\hline
Constant curvature & $\Omega(T^{1/3})$ \textbf{(\Cref{thm:minimax lower}c)} & \hlcell $\Otil(T^{1/3})$ \textbf{(\Cref{thm:size-2}c)} & \begin{tabular}{@{}c@{}}$\Otil(T^{3/5})$ \citep{kleinberg2019bandits}\\$\Omega(T^{3/5})$ \textbf{(\Cref{lem:posted price curvature})}\end{tabular} \\\hline
\begin{tabular}{@{}c@{}}Interval Dimension\\$d\in (0,1)$\end{tabular} & $\Omega(T^{d/(d+1)})$ \textbf{(\Cref{thm:minimax lower}d)} & \hlcell $\Otil(T^{d/(d+1)})$ \textbf{(\Cref{thm:instance upper})} & \begin{tabular}{@{}c@{}}$\Otil(T^{(d+1)/(d+2)})$ \citep{kleinberg2019bandits}\\$\Omega(T^{(d+1)/(d+2)})$ \textbf{(\Cref{thm:fixed instance})}\end{tabular} \\\hline
\end{tabular}}
\end{table}

Interestingly, we obtain minimax regret rates $o(\sqrt{T})$, which is uncommon for bandit problems even under full feedback. This is where the special structure of dynamic pricing is essential, as opposed to, \eg the more generic structure of Lipschitz bandits. The latter incurs $\Omega(\sqrt T)$ minimax regret rate under full feedback, even when the interval dimension is $d=0$. In contrast, when only using posted prices, dynamic pricing yields the same minimax regret rates as Lipschitz bandits (see \Cref{sec:lipschitz}).

While standard minimax lower bounds are tailored to a particular $T$, we also provide \emph{anytime} lower bounds that apply to infinitely many $T$'s at once. They rule out any algorithm that beats every fixed-$T$ lower bound \emph{eventually}, after some instance-dependent initial time period. They handle problem instances with any fixed interval dimension $d\in (0,1)$, with the same rates as in \Cref{table:regret}.

Furthermore, we argue our algorithm is near-optimal \emph{for every value distribution $F$}, rather than (just) in the worst case over some class of problem instances. This is due to a more intricate lower bound we prove. The ``shape'' of such a lower bound must be non-trivial, as something must prevent the algorithm from always posting the monopoly price $p^\ast_F$. So, given $F$ and time horizon $T$, we construct two value distributions (called \emph{decoys}) with the following properties. First, letting $V$ be our algorithm's regret on $F$, no other algorithm, even with infinite menus, can have regret much better than $V$ simultaneously on $F$ and on both decoys. Second, the decoys are close to $F$: Their Hellinger distance to $F$ tends to $0$ as $T\to \infty$. Both properties can be made more concrete in various ways, under additional assumptions. In particular, the decoys can preserve the properties of $F$  in terms of interval dimension or MHR. These lower-bound constructions, unified by a common idea specifically tailored for dynamic pricing, jointly constitute our second key technical contribution.

\begin{table}[t]
\centering
\caption{Sample, Interval-Query, and Threshold-Query Complexities}
\label{table:complexity}
\Crefname{theorem}{Thm}{Thms}
\Crefname{lemma}{Lm}{Lms}
\begin{minipage}{\textwidth}
\resizebox{\textwidth}{!}{%
\begin{savenotes}
\renewcommand{\arraystretch}{1.3}
\begin{tabular}{|c|c|c|c|c|}\hline
Property & Value Distribution & Sample Complexity & Interval-Query Complexity & Threshold-Query Complexity \\\hline
\multirow{5}{*}{CDF} & \multirow{2}{*}{General} & $\O(\epsilon^{-2})$ \citep{dvoretzky1956asymptotic} & \hlcell & \multirow{2}{*}{$\tilde \Theta(\epsilon^{-3})$ \citep{leme2023pricing}} \\
& & $\Omega(\epsilon^{-2})$ \citep{huang2015making} & \hlcell & \\\cline{2-3}\cline{5-5}
& Regular & \multirow{2}{*}{$\Omega(\eps^{-2})$ \textbf{(\Cref{lem:sample lower bound MHR})}} & \hlcell & \multirow{2}{*}{\shortstack{$\Otil(\epsilon^{-2.5})$ \citep{leme2023description}\\$\Omega(\epsilon^{-2.5})$ \citep{leme2023pricing}\footnote{Their hard instance \citep[Theorem 3.7]{leme2023pricing} is also MHR, even though they do not state so explicitly.}}} \\\cline{2-2}
& MHR & & \hlcell & \\\cline{2-3}\cline{5-5}
& Constant Curvature & $\Omega(\eps^{-2})$ \textbf{(\Cref{lem:CDF sample lower bound curvature})} & \hlcell \multirow{-5}{*}{$\Otil(\epsilon^{-2})$ \textbf{(\Cref{lem:CDF interval query})}} & $\tilde \Theta(\epsilon^{-3})$ \textbf{(\Cref{lem:CDF threshold lower bound curvature})} \\\hline
\multirow{7}{*}{\shortstack{Monopoly\\Price}} & \multirow{2}{*}{General} & $\O(\epsilon^{-2})$ \citep{dvoretzky1956asymptotic} & \hlcell & \multirow{2}{*}{$\tilde \Theta(\epsilon^{-3})$ \citep{leme2023pricing}} \\
& & $\Omega(\epsilon^{-2})$ \citep{huang2015making} & \hlcell & \\\cline{2-3}\cline{5-5}
& Regular & $\Omega(\eps^{-2})$ \textbf{(\Cref{lem:lower regular})} & \hlcell \multirow{-3}{*}{$\Otil(\epsilon^{-2})$ \textbf{(\Cref{lem:CDF interval query})}} & \multirow{2}{*}{$\tilde \Theta(\epsilon^{-2})$ \citep{leme2023pricing}} \\\cline{2-4}
& MHR & $\tilde \Theta(\epsilon^{-1.5})$ \citep{huang2015making} & \hlcell & \\\cline{2-3}\cline{5-5}
& \multirow{2}{*}{Constant Curvature} & & \hlcell & \multirow{2}{*}{\shortstack{$\Otil(\epsilon^{-2.5})$ \citep{kleinberg2019bandits}\\$\Omega(\epsilon^{-2.5})$ \textbf{(\Cref{lem:threshold lower bound curvature})}}} \\
& & \multirow{-2}{*}{$\Omega(\epsilon^{-1.5})$ \textbf{(\Cref{lem:sample lower bound curvature})}} & \hlcell\multirow{-3}{*}{$\Otil(\epsilon^{-1.5})$ \textbf{(\Cref{thm:size-2})}} & \\\hline
\end{tabular}
\end{savenotes}}
\end{minipage}
\end{table}

\paragraph{Ancillary Contributions: Interval Query Complexity.} Dynamic pricing is closely related to query complexity of revenue maximization. Here, an algorithm faces a sequence of customers, with values as described above. In each round $t$, the algorithm issues a \emph{query} to the customer, from some class of feasible queries, and learns something about value $v_t$. The goal is to learn a property of interest: either the CDF of value distribution, or just the monopoly price $p^*_F$. The \emph{query complexity} is the smallest number of queries guaranteed to learn the property up to a given precision $\eps>0$.

Prior work studied \emph{threshold queries} that reveal $\1[v_t\ge p]$ for a given threshold $p$  \citep{leme2023description,leme2023pricing} and \emph{value queries} that directly reveal $v_t$ \citep{dvoretzky1956asymptotic,huang2015making}.\footnote{Value-query (a.k.a.\ \emph{sample}) complexity is widely studied in dynamic pricing \citep{cole2014sample,guo2019settling,guo2021generalizing,jin2024sample}.} These two query classes correspond to dynamic pricing with size-1 menus and full feedback, respectively, in terms of the information revealed after each round. Their query complexities are well-understood; see \Cref{table:complexity}.

While the threshold and value queries represent the two extremes, our concern is ``intermediate'' queries: ones that reveal more information than threshold queries but less than value queries. How powerful are they? This is poorly understood, much like dynamic pricing with finite menus.

We focus on \emph{interval queries} that reveal whether $v_t$ lies in a given interval $[a,b)$. This is, in some sense, the least informative query class above threshold queries.\footnote{If one can only perform $N$ threshold queries on a given customer, interval queries correspond to $N=2$.} Our high-level result is that \emph{interval queries are maximally powerful in terms of query complexity, \ie as powerful as value queries}. This again holds in a very broad sense: for both properties (CDF and monopoly price) and across several classes of value distributions (all distributions, regular distributions, MHR, and those of constant curvature). We provide a new interval-query algorithm for each property, which handles all distribution classes at once. See \Cref{table:complexity} for specific upper and lower bounds.

Most notably, we show that the minimax-optimal interval-query complexity is $\Otil(\eps^{-2})$ for both properties, matching the lower bounds from \citep{huang2015making}. We extend the $\tilde\Omega(\eps^{-2})$ lower bound to several classes of instances: MHR and constant-curvature (for CDF estimation) and regular value distributions (for both CDF and monopoly-price estimation). Moreover, we strengthen the monopoly-price \emph{upper} bound to $\Otil(\eps^{-1.5})$ for MHR and constant-curvature (and show that it is optimal).

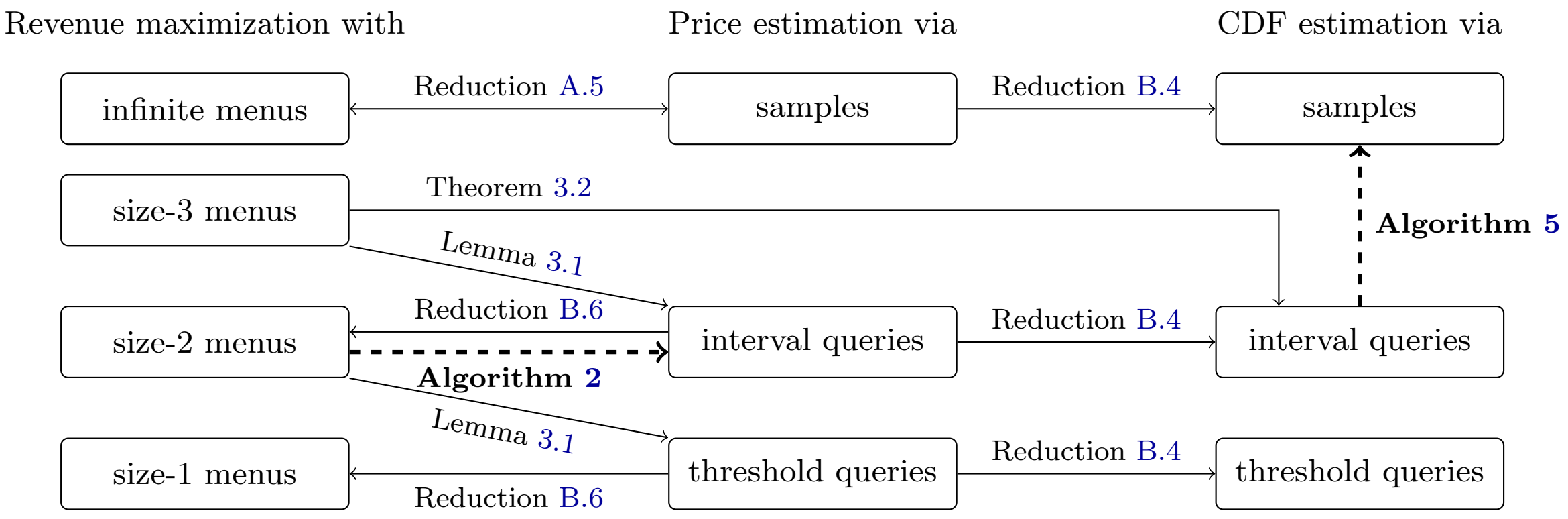
\begin{figure}[!t]
\centering
\begin{tikzpicture}[prob/.style={draw, rounded corners=2pt, align=center, text width=2.7cm, minimum height=0.7cm, inner sep=2pt, font=\footnotesize}, res/.style={font=\scriptsize}, new/.style={->, very thick, dashed}]
\node[font=\footnotesize] at (0,0.85) {Revenue maximization with};
\node[font=\footnotesize] at (6,0.85) {Price estimation via};
\node[font=\footnotesize] at (11.4,0.85) {CDF estimation via};
\node[prob] (minf) at (0,0) {infinite menus};
\node[prob] (m3) at (0,-1) {size-3 menus};
\node[prob] (m2) at (0,-2.3) {size-2 menus};
\node[prob] (m1) at (0,-3.6) {size-1 menus};
\node[prob] (ps) at (6,0) {samples};
\node[prob] (pi) at (6,-2.3) {interval queries};
\node[prob] (pt) at (6,-3.6) {threshold queries};
\node[prob] (cs) at (11.4,0) {samples};
\node[prob] (ci) at (11.4,-2.3) {interval queries};
\node[prob] (ct) at (11.4,-3.6) {threshold queries};
\draw[<->] (minf) -- node[res, above] {\Cref{reduction:infinite to samples}} (ps);
\draw[->] (m3) -| ([xshift=-8mm]ci.north);
\node[res, above] at (3,-1) {\Cref{thm:size-3 menus}};
\draw[->] ([yshift=1mm]pi.west) -- node[res, above] {\Cref{reduction:online to batch}} ([yshift=1mm]m2.east);
\draw[new] ([yshift=-1mm]m2.east) -- node[res, below] {\textbf{\Cref{alg:size-2}}} ([yshift=-1mm]pi.west);
\draw[->] (m3.south east) -- node[res, above, sloped] {\Cref{reduction:infinite to full-info}} (pi.north west);
\draw[->] (m2.south east) -- node[res, below, sloped] {\Cref{reduction:infinite to full-info}} (pt.north west);
\draw[->] (pt) -- node[res, below] {\Cref{reduction:online to batch}} (m1);
\draw[->] (ps) -- node[res, above] {\Cref{reduction:price to CDF}} (cs);
\draw[->] (pi) -- node[res, above] {\Cref{reduction:price to CDF}} (ci);
\draw[->] (pt) -- node[res, above] {\Cref{reduction:price to CDF}} (ct);
\draw[new] (ci) -- node[res, right] {\textbf{\Cref{alg:CDF interval query}}} (cs);
\end{tikzpicture}
\caption{Formal connections between query complexity and dynamic pricing with menus. Arrows stand for black-box reductions, while thick dashed arrows are our ad-hoc algorithms.}
\label{fig:reductions}
\end{figure}

Our results on query complexity are closely connected to those on dynamic pricing. We derive and leverage several formal connections between the two problems (see \Cref{fig:reductions}). Some of these connections are ``black-box reductions'', converting an arbitrary algorithm for one problem into an algorithm for another (and likewise for their performance guarantees). Some others are ``ad-hoc reductions'', converting a particular approach for one problem into an algorithm for another. In particular, our best-price-estimation algorithm is the key ingredient in our main algorithm for dynamic pricing using size-2 menus, and our CDF-estimation algorithm underpins a simple approach which attains $\widetilde{\mathcal O}(\sqrt T)$ minimax regret with three menu options.

\paragraph{Extension: Adversarial Dynamic Pricing with Menus.} We come back to dynamic pricing with menus and consider \emph{adversarial values}: Each value $v_t$ is chosen by an adversary, rather than drawn from a fixed distribution. The choice of $v_t$ can be randomized and adaptive to the history.

Using size-3 menus, we achieve $\Otil(\sqrt T)$ minimax regret against the best fixed price in hindsight, whose minimax optimum is $\tilde \Theta(T^{2/3})$ when only using posted prices \citep{kleinberg2003value}. This is the best possible even with infinite menus. Leveraging our machinery for i.i.d.\ values, we plug any suitable algorithm \AlgOL for \emph{online learning} (\ie adversarial bandits with full feedback) into our approach. Using Hedge \citep{freund1997decision}, we attain the aforementioned $\Otil(\sqrt T)$ bound; whereas using Fixed-Share \citep{herbster1998tracking}, we obtain $\Otil(\sqrt{ST})$ regret against any price sequence that switches the price at most $S\geq 1$ times.

\paragraph{Upper Bound Techniques.} Our dynamic pricing algorithms build on those for interval queries. With threshold queries, estimating a CDF up to $\epsilon$ takes $\tilde \Theta(\epsilon^{-3})$ queries \citep{leme2023pricing}: Within any of the $\epsilon^{-1}$ intervals of width $\epsilon$, one can move a mass of order $\epsilon$ from its right half to its left half, such that any threshold query outside it tells no difference. Hence one must spend $\Omega(\epsilon^{-2})$ threshold queries on each of these intervals, thus giving $\Omega(\epsilon^{-3})$. An interval query on the left half of it, in contrast, observes a coin whose mean, and hence variance, is $\O(\epsilon)$, so the same change is detectable in $\O(\epsilon^{-1})$ queries. In general, for an interval $I$ with probability mass $\intervalP(I)$, its outcome (which is Bernoulli) has variance no more than $\intervalP(I)$, and thus empirical Bernstein \citep{maurer2009empirical} estimates $\intervalP(I)$ up to $\epsilon$ using $\Otil(\frac{\intervalP(I)}{\epsilon^2}+\frac 1\epsilon)$ queries. We thus build a segment tree whose leaves are the $\epsilon^{-1}$ intervals of width $\epsilon$, and apply empirical Bernstein to estimate each tree interval up to $\epsilon/(\log_2 \epsilon^{-1})$. As the intervals of each depth are disjoint, their masses sum to $1$; since there are $\log_2 \epsilon^{-1}$ depths, this approach uses $\Otil(\epsilon^{-2})$ interval queries (\Cref{lem:CDF interval query}). Translating interval queries to menus directly, however, requires menus with 3 options, because we need an extra option to ``exploit'' the estimated monopoly price $\hat p^\ast$. It also cannot bypass the worst-case rate of $\Otil(\epsilon^{-2})$, due to our near-matching $\Omega(\epsilon^{-2})$ sample complexity lower bounds for regular, MHR, or curvature distributions (see \Cref{table:complexity}).

We then directly estimate the monopoly price $p^\ast$, which only requires estimating those near-optimal prices. Our starting point is the one-sided Lipschitzness of the revenue function $R$: For any $p<p'$,
\begin{equation}\label{eq:1-side Lip}
R(p')-R(p)=(p'-p)\Pr_{v\sim F}\{v\ge p'\}-p\Pr_{v\sim F}\{p\le v<p'\}\le p'-p.
\end{equation}
If $p<p'$ are both $\Delta$-optimal, i.e., $R(p),R(p')\ge R(p^\ast)-\Delta$, then $p\Pr_{v\sim F}\{p\le v<p'\}\le p'-p+\Delta$; therefore, \emph{close near-optimal prices also have little mass in between}. This allows a phased elimination approach (\Cref{alg:size-2}): For a halving $\Delta=\frac 12,\frac 14,\ldots$, we start with $4\Delta$-optimal candidates $p_0<\cdots<p_N$, estimate all revenue differences up to $\frac \Delta 4$ with a segment tree over candidates, refine our candidate set, and eliminate the sub-optimal ones. Using the one-sided Lipschitzness in \Cref{eq:1-side Lip}, the total queries spent on a $\Delta$ is $\Otil(\frac{p_N-p_0+\Delta}{\Delta^2})$ (\Cref{lem:size-2 rounds}). This not only gives $\Otil(\epsilon^{-2})$ in general (via $p_N-p_0\le 1$), but also gives $\Otil(\epsilon^{-1.5})$ for MHR or curvature distributions whose $\Delta$-optimal region can be no wider than $\sqrt \Delta$. Even better, our interval queries always have one $\Delta$-optimal endpoint, thus eliminating the need for an ``exploitation price'' and being \emph{implementable using size-2 menus}.

In general, for any value distribution $F$, let its covering number $N_F(\Delta)$ be the number of length-$\Delta$ intervals needed to cover the convex hull of all $\Delta$-optimal prices. \Cref{alg:size-2} spends $\Otil(N_F(4\Delta)/\Delta)$ rounds on $\Delta$ (since the $4\Delta$-optimal candidates satisfy $p_N-p_0\le 4\Delta\cdot N_F(4\Delta)$), each with regret $\O(\Delta)$. Therefore, its regret is $\Otil(V_T(F))$ with $V_T(F):=\sup_\Delta\min\{T\Delta,N_F(\Delta)\}$. When $N_F(\Delta)=\O(\Delta^{-d})$ for some $d\in [0,1]$ (when we say $F$ has interval dimension $d$), this becomes $\Otil(T^{d/(d+1)})$ (\Cref{thm:instance upper}).

\paragraph{Lower Bound Techniques.} Our lower bounds hold around \emph{every} $F$, which rules out the usual approach: Lower bounds for Lipschitz bandits plant a bump at an arbitrary location \citep[\S 5.1]{kleinberg2019bandits}, but in pricing, the revenues of all prices are determined by the same value distribution $F$, so making one price $\Delta$-better than another requires moving masses between them. A key fact is that \Cref{eq:1-side Lip} is tight in both directions: While two close $\Delta$-optimal prices have little mass in between (enabling our upper bounds), \emph{between two far apart ones}, say the smallest and largest $\Delta$-optimal prices,  \emph{the mass is at least $\Delta N_F(\Delta)$}. Hence scaling the density on a half of this interval by $1\pm\Theta(1/N_F(\Delta))$ makes either end $\Delta$-better than the other, and the resulting decoys $G_+,G_-$ have squared Hellinger distances from $F$ bounded by $\Delta/N_F(\Delta)$. No algorithm, even with infinite menus, tells them apart in $\Theta(N_F(\Delta)/\Delta)$ rounds, hence giving an $\Omega(\min\{T\Delta,N_F(\Delta)\})=\Omega(V_T(F))$ lower bound on $\{G_+,G_-\}$.

To claim instance-optimality of \Cref{alg:size-2}, we further require $V_T(G_\pm)$ not to be much larger than $V_T(F)$. But the previous move may lift a plateau of $R$, creating many more near-optimal prices. We again scale the density by a relative $\Theta(1/N_F(\Delta))$ within the new $\Delta$-optimal region, down on the left of the new $p_\pm^\ast$ and up on its right, so that the new $R_\pm$ decays at least linearly. This keeps $V_T(G_\pm)$ at most a constant factor larger than $V_T(F)$, giving \Cref{thm:instance full}. A more detailed proof sketch, as well as an illustration (\Cref{fig:hard instances}), can be found in \Cref{sec:constructions}. The refined versions, \Cref{thm:instance lower} for interval dimensions and \Cref{thm:lower MHR} for MHR, follow similar but more delicate constructions.

The decoy instances above depend on $T$, and we further develop anytime lower bounds using a similar idea. We give a (randomized) value distribution of interval dimension $d$ which almost surely ensures that any algorithm, even with infinite menus, suffers regret $\Omega(T^{d/(d+1)})$ for infinitely many $T$'s. Instead of the single-parameter family under posted prices \citep{kleinberg2003value}, which only gives $\Otil(1)$ regret under infinite menus, we construct a fractal-like value distribution $F$: We plant the two-point construction above at super-exponential scales $\Delta_1>\Delta_2>\cdots$ with $\Delta_{k+1}=\Delta_k^q$ for a constant $q>1$, each on the top of the previous one, with a fresh fair coin deciding which half is better. Then each scale $\Delta_k$ forces $\Omega(N(\Delta_k))=\Omega(T^{d/(d+1)})$ regret at $T\approx N(\Delta_k)/\Delta_k$ with probability $\frac 12$, and hence we derive an $\Omega(T^{d/(d+1)})$ lower bound that holds infinitely often almost surely (\Cref{thm:fixed instance}).

\section{Setup: Dynamic Pricing with Menus}\label{sec:setup}\label{sec:setup menus}

\paragraph{Notations.} For a positive integer $n$, let $[n]:=\{1,2,\ldots,n\}$. For a (possibly infinite) set $\mathcal X$, $\triangle\mathcal X$ denotes the set of distributions over $\mathcal X$. We denote the Dirac measure over $p$ by $\delta_p$. We use standard big-oh notations: $\O,\Omega,\Theta$ hide absolute constants, and $\Otil,\tilde \Omega,\tilde \Theta$ further hide logarithmic factors.

\paragraph{Problem Statement.} We have a $T$-round game between a seller and myopic customers. The seller has an infinite supply of a product. In each round $t\in [T]$, a customer with value $v_t\in[0,1]$ comes, where $v_t$ is independently sampled from a fixed but unknown \emph{value distribution} $F\in \triangle[0,1]$ (we extend to adversarial values in \Cref{sec:adv}). Neither the value $v_t$ nor the value distribution $F$ is known.

The seller offers a menu $M_t$ to this customer. A menu is defined as a collection of options $(q,p)\in [0,1]\times [0,1]$, where each option $(q,p)\in M_t$ indicates that the customer choosing it gets the item with probability $q$ and pays $p$ in expectation (i.e., whether or not they get the item). We call $|M_t|$ the \emph{size} of the menu, so that a single posted price $p_t$ is equivalent to a size-1 menu $M_t=\{(1,p_t)\}$.

The payoff of a customer with value $v$ under option $(q,p)$ is quasi-linear as $U(q,p;v):=q\times v-p$. Customer $t$ chooses an option $(q_t,p_t)\in M_t\cup \{(0,0)\}$ maximizing their utility $U(\,\cdot\,;v_t)$. The ``outside option'' $(0,0)$ for not doing anything is always implicitly included in the menu. When several options maximize utility, we assume the customer chooses the one maximizing allocation probability.

The seller observes the chosen option $(q_t,p_t)$ but not the value $v_t$. Thus, the observable history up to the end of round $t$ is $\mH_t = (M_s,q_s,p_s)_{s\in[t]}$. The menu $M_t$ can depend on the history $\mH_{t-1}$. We design algorithms for the seller to construct menus, so as to maximize the expected total revenue.

\paragraph{Monopoly Price and Regret.} The seller competes with the \emph{monopoly price}, defined as
\begin{equation}\label{eq:monopoly price}
p_F^\ast:=\argmax_{p\in [0,1]} R_F(p), \quad
\text{ where } ~~
\demand_F(p) := \Pr_{v\sim F}\{v\ge p\}
~~ \text{ and } ~~
R_F(p):=p\cdot \demand_F(p)
\end{equation}
are the \emph{demand curve} and the \emph{revenue function} of $F$, respectively. This price always exists since the demand curve is non-increasing and left-continuous in $p$; if there are several maximizers, $p_F^\ast$ is any fixed one of them. If $F$ were known, posting $p^\ast_F$ in all rounds is optimal across all algorithms, even if menus are allowed \citep{myerson1981optimal}. We define \emph{regret} w.r.t.\ the monopoly price:
\begin{equation*}
\textstyle \Reg_F(T) := T\cdot R_F(p_F^\ast)-\E\sbr{\sum_{t\in[T]} p_t},
\end{equation*}
mirroring the standard notion of pseudo-regret from (stochastic) multi-armed bandits. We omit the subscript $F$ when it is clear from the context. We are often interested in \emph{minimax-optimal regret}: the best possible $T$-round regret achievable by a dynamic pricing algorithm using menu size at most $m\in\mathbb{N}\cup\{\infty\}$, in the worst case over a given class $\mF$ of value distributions:
\begin{equation*}
\Reg_{\mF}^{(m)}(T):=\inf_{\text{algorithms using size-$m$ menus}}\; \sup_{F\in \mF}\; \Reg_F(T).
\end{equation*}

\paragraph{Classes of Interest.} We are particularly interested in the following classes of value distributions:
\begin{itemize}
\item General distributions ($\mF_{\general}$): all distributions over $[0,1]$, including those without densities.
\item Regular distributions \citep{myerson1981optimal} ($\mF_{\regular}$): value distributions with probability density function (PDF) $f$ such that the \emph{virtual value} function $\phi(v):=v-\frac{\demand_F(v)}{f(v)}$ is non-decreasing.
\item Monotone hazard rate (MHR) \citep{barlow1963properties} ($\mF_{\MHR}$): those with a non-decreasing \emph{hazard rate} function $h(v):=\frac{f(v)}{\demand_F(v)}$. Both regular and MHR are standard in economic theory with $\mF_\MHR\subseteq \mF_\regular$.
\item Distributions with \emph{curvature} $\kappa\in (0,2)$ \citep{kleinberg2003value} ($\mF_\curv$): value distributions which satisfy the property that $R(p^\ast)-R(p)\ge \kappa (p-p^\ast)^2$ for all prices $p\in [0,1]$.\footnote{$\mF_\curv=\varnothing$ for $\kappa>2$ and $F\in \mF_{\mathrm{curv}(2)}$ is trivial (\Cref{lem:curvature range}); we hence assume $\kappa\in (0,2)$. The original ``curvature and smooth'' assumption \citep[Theorem 1.2]{kleinberg2003value} assumes $R$ has a unique maximizer $p^\ast\in (0,1)$, $R\in \mathcal C^{(2)}$ near $p^\ast$, and $R''(p^\ast)<0$. Such value distributions lie in $\mF_\curv$ for some $\kappa>0$ \citep[Lemma 3.11]{kleinberg2003value}. Under their assumption, they achieve $\Otil(\sqrt T)$ regret for posted pricing, whereas we prove  $\Reg_\curv^{(1)}(T)\geq \Omega(T^{3/5})$ (\Cref{lem:posted price curvature}).\label{footnote:curvature assumption in KL}}

\end{itemize}

\paragraph{Other Notations.} For the ease of reading, we write $\mR^{(m)}_{\general}(T) :=\mR^{(m)}_{\mF_{\general}}(T)$, and similarly for the other classes. The \emph{price space} $[0,1]$ is the set of all feasible prices. A \emph{price interval} is an interval in the price space. Slightly abusing the $\demand_F(\cdot)$ notation, let $\intervalP_F(I) := \Pr_{v\sim F}\{v \in I\}$ be the probability of interval $I$ under $F$. A price $p$ is \emph{$\eps$-optimal} if $R(p)\ge R(p^\ast)-\eps$. A distribution $\priceDist\in \triangle[0,1]$ over prices is $\eps$-optimal if its expected revenue $\E_{p\sim\priceDist}[R(p)]\ge R(p^\ast)-\eps$.

\paragraph{Query Complexity.} For monopoly-price estimation, given a class $\mF$ of value distributions, we define the query complexities $\Sample_\mF(\eps)$, $\Threshold_\mF(\eps)$, and $\Interval_\mF(\eps)$ as, resp., the minimal number of value queries (samples), threshold queries, and interval queries needed to find an $\epsilon$-optimal price distribution. For CDF estimation, the goal is to find a value distribution $\hat F$ that is $\epsilon$-close to $F$ in L\'evy distance with probability $\frac 34$. The corresponding query complexities are denoted, resp., as $\Sample_\mF^\CDF(\eps)$, $\Interval_\mF^\CDF(\eps)$, and $\Threshold_\mF^\CDF(\eps)$. Formal definitions are in \Cref{app:complexity definitions}.

\section{Upper Bounds: Dynamic Pricing with Size-2 Menus}\label{sec:main}

In this section we present our main algorithmic results, as described in the Introduction. The main theorems are stated in \Cref{sec:size-2 regret}. The rest of the section builds towards these theorems. In \Cref{sec:menus}, we establish a key property that connects menus and interval queries. In \Cref{sec:size-3}, we design a new interval-query algorithm for CDF estimation, matching the query complexity of value queries. Invoking the reduction from \Cref{sec:menus}, we obtain a simple algorithm for dynamic pricing with size-3 menus which attains $\Otil(\sqrt T)$ regret. In \Cref{sec:price interval query}, we specify our main algorithm (\Cref{alg:size-2}) for dynamic pricing with size-2 menus, and prove a key technical lemma about its performance. Internally, this algorithm is based on monopoly-price estimation via interval queries (as CDF estimation is wasteful) and can be usefully repurposed for that problem (see \Cref{app:interval complexity}).

\subsection{Connecting Menus and Interval Queries}\label{sec:menus}

This subsection presents an interesting connection between menus and interval queries. Let $\priceDist$ be a \emph{price distribution} supported on $m$ points $0 \le p_1 \le \cdots \le p_m \le 1$. An interval query on two consecutive support points $[p_i, p_{i+1})$ asks whether the customer's value is in that interval. The key property enabling all our results is that any price distribution $\priceDist$ and an interval query on its support points can be ``implemented'' by a menu. This menu obtains the same expected revenue as a randomized posted price $p\sim\priceDist$ and (at least) the same information as the interval query.

\begin{lemma}\label{reduction:infinite to full-info}\label{lem:Myerson prices}
Fix a price distribution $\priceDist$ with discrete support $0\le p_1<\cdots<p_m\le 1$. For $i\in[m]$, let $w_i$ be the probability of $p_i$ under $\priceDist$. Construct a size-$m$ menu $M=\cbr{ \rbr{\sum_{j\le i}w_j,\;\sum_{j\le i} w_jp_j}\,\big \vert\, i\in [m] }$.

Given $M$, a customer with value $v\in [p_i,p_{i+1})$ chooses the $i$-th option, where $p_{m+1}:=\infty$, and one with $v<p_1$ chooses the outside option. Hence, the seller observes which of the $(m+1)$ intervals, $[0,p_1),[p_1,p_2),\ldots,[p_m,1]$, contains $v$; and the customer gets the item with probability $\Pr_{p\sim \priceDist}\{v\ge p\}$ and pays $\E_{p\sim \priceDist}\sbr{p\,\1[v\ge p]}$ in expectation, the same as posting a random price $p\sim \priceDist$.
\end{lemma}

Therefore, menus enable the seller to perform interval (instead of threshold) queries on $v_t$, while still collecting the same expected revenue as posting a random price. Specifically, to implement an interval query $[a,b)$ while exploiting a price $r$, such as the current estimate of $p^*$, we apply \Cref{lem:Myerson prices} to the price distribution $\priceDist=\gamma\delta_a+\gamma\delta_b+(1-2\gamma)\delta_r$ where $0<\gamma<1/2$ is arbitrarily small. The resulting menu has three options, reveals whether the customer’s value lies in $[a,b)$, and earns expected revenue at least $R(r)-2\gamma$. A size-2 menu can also ask an interval query $[a,b)$, but only exploits either $a$ or $b$. Hence dynamic pricing using size-3 menus reduces to estimating $p^\ast$ via interval queries, while size-2 menus additionally require every query to have a near-optimal endpoint.

\subsection{Size-3 Menus and Interval-Query Complexity of CDF Estimation}\label{sec:size-3}

Starting with CDF estimation, we provide a new interval-query algorithm, showing that interval queries are as powerful as value queries. This also enables dynamic pricing using size-3 menus.
\begin{theorem}\label{lem:CDF interval query}\label{thm:size-3 menus}\label{thm:size-3}
$\Interval_\general^\CDF(\epsilon)=\O(\epsilon^{-2}\log^4\frac 2\epsilon)$ and $\mR_\general^{(3)}(T)=\O(\sqrt T\log^3 T)$.
\end{theorem}

\begin{subroutine}[!t]
\caption{\textproc{EmpiricalBernstein}(interval $I$, precision $\epsilon$, failure probability $\delta$)}\label{alg:empirical Bernstein}
\begin{algorithmic}[1]
\For{$K=1,2,\ldots$}\label{line:empirical Bernstein}\Comment{This loop is adaptive and issues an indefinite number of queries}
\State Issue an interval query on $I$, observe a new feedback $X_K=\1[v\in I]$ with fresh $v\sim F$
\State Calculate $\hat \mu_K:=\frac 1K\sum_{k=1}^K X_k$ and the confidence radius $r_K$ according to \Cref{eq:confidence radius}
\algeq[eq:confidence radius]{$\displaystyle r_K:=\sqrt{\frac{2\iota}{K}\cdot \frac{1}{K(K-1)}\sum_{k<k'} (X_k-X_{k'})^2}+\frac{7\iota}{3(K-1)},\quad \iota=\log \frac 2\delta$}
\State Terminate if $K\ge 2$ and $r_K\le \epsilon$, in which case \Return estimation $\hat \intervalP(I):=\hat \mu_K$
\EndFor
\end{algorithmic}
\end{subroutine}

The key is \Cref{alg:empirical Bernstein} (subroutine is an algorithm that starts at some round $t_0$, inputs the history $\mH_{t_0}$, and controls the subsequent rounds until it exits). For an interval $I$ with $\intervalP(I)=\Pr_{v\sim F}\{v\in I\}$, the outcome of an interval query on $I$ has a variance at most $\intervalP(I)$. Therefore, by the empirical Bernstein inequality \citep{maurer2009empirical}, \Cref{alg:empirical Bernstein} estimates $\intervalP(I)$ up to $\epsilon$ using $\Otil(\intervalP(I)\epsilon^{-2}+\epsilon^{-1})$ queries.

We then estimate the CDF by building a segment tree over $[0,1)$: Let $N:=2^{\lceil \log_2 \epsilon^{-1}\rceil}$ so that $\frac 1N\le \epsilon$. In each depth $h=0,1,\ldots,\log_2 N$, the tree has the $2^h$ intervals $[\frac k{2^h},\frac{k+1}{2^h})$. Each non-leaf interval has its two halves as its children, decomposing any $[0,\frac iN)$ as a disjoint union of $\O(\log_2 N)$ tree intervals.

Thus, estimating the probability of every tree interval up to $\frac{\epsilon}{2\log_2(2N)}$ estimates every $\Pr_{v\sim F}\{v<\frac iN\}$ up to $\epsilon$, giving a value distribution $\hat F$ that is $\epsilon$-close to $F$ in L\'evy distance. As the tree intervals of the same depth are disjoint, their probabilities $\sum_{k=0}^{2^h-1} \intervalP([\frac k{2^h},\frac{k+1}{2^h}))$ sum to $1$. Hence each of the $\log_2 N$ depths takes $\Otil(\epsilon^{-2})$ queries in total, proving the first statement of \Cref{thm:size-3}. We defer the full pseudocode (\Cref{alg:CDF interval query}) and how to build segment trees (\Cref{alg:segment tree}) to \Cref{app:size-3}.

The second statement follows from \Cref{reduction:infinite to full-info}: By setting two $p$'s as the two endpoints of an interval query, each with a negligible probability $\gamma=T^{-2}$, and the other one as $\max\{\hat p^\ast-\epsilon,0\}$, where $\hat p^\ast$ is the monopoly price of the current estimated CDF $\hat F$, with a probability $1-2\gamma$, we reduce size-3 revenue maximization to CDF estimation with interval queries, thus giving $\mR_\general^{(3)}(T)=\Otil(\sqrt T)$.

\subsection{Size-2 Menus and Interval-Query Complexity of the Monopoly Price}\label{sec:price interval query}\label{sec:size-2}

Focusing on dynamic pricing with size-2 menus, we specify our main algorithm and prove a key technical guarantee which underpins our theorems. Internally, our algorithm is based on monopoly-price estimation via interval queries, as CDF estimation is wasteful for our purposes (recall \Cref{table:complexity}).

To estimate $p^\ast$, it suffices to consider the near-optimal regions of the price space. If two prices $p<p'$ are both $\Delta$-optimal, i.e., $R(p),R(p')\ge R(p^\ast)-\Delta$, the one-sided Lipschitzness of $R$ (\Cref{eq:1-side Lip}) gives
\begin{equation*}
\intervalP_F([p,p'))=\Pr_{v\sim F}\{p\le v<p'\}\le \frac{p'-p+\Delta}{p}.
\end{equation*}
Hence, two close $\Delta$-optimal prices must have little probability mass in between, so the interval query between them has a small variance. This recovers the key to the improvement in \Cref{lem:CDF interval query}.

Utilizing this, our \Cref{alg:size-2} runs in layers with halving $\Delta=\frac 12,\frac 14,\ldots$. At the beginning of each layer $\Delta$, it maintains a set $\candidateSet_{2\Delta}$ of candidate prices, which are all $4\Delta$-optimal and at least $\Delta$ apart from each other. We also ensure one of them lies in $[p^\ast-\Delta,p^\ast]$, i.e., $p^\ast\in \bigcup_{p\in \candidateSet_{2\Delta}}[p,p+\Delta]$. Then:
\begin{itemize}
\item \textbf{Estimate.} Write out the candidate prices $\candidateSet_{2\Delta}=\{p_0,p_1,\ldots,p_N\}$ in increasing order. As in \Cref{sec:size-3}, build a segment tree $\mT$ over $\candidateSet_{2\Delta}$ and estimate the probability mass of every tree interval via empirical Bernstein (precision is specified later). Also estimate the mass above $p_N$.
\item \textbf{Refine.} For every candidate $p_i\in \candidateSet_{2\Delta}$, add both $p_i$ and $p_i+\frac \Delta 2$ to $\candidateSet_\Delta$, and estimate the mass of $[p_i,p_i+\frac \Delta 2)$. Together with the tree interval estimates above, using a segment tree decomposition $\mI_\mT$, we estimate the revenue difference between any two candidate prices up to a precision $\frac \Delta 4$.
\item \textbf{Eliminate.} Remove those $p\in \candidateSet_\Delta$'s whose estimated revenue is $\Delta$-below the best candidate. The remaining prices are $2\Delta$-optimal, $\frac \Delta 2$ apart from each other, and contain a price in $[p^\ast-\frac \Delta 2,p^\ast]$.
\end{itemize}

This design not only improves the interval-query complexity (by ignoring sub-optimal prices), but also saves a menu option since all interval queries have their left endpoints in $\candidateSet_{2\Delta}$. Therefore, we eliminate the need for a dedicated exploitation price: When performing an interval query $[p_i,p_j)$ with $p_i<p_j$ and $p_i\in \candidateSet_{2\Delta}$, by \Cref{reduction:infinite to full-info}, the size-2 menu built from the price distribution $(1-\gamma)\delta_{p_i}+\gamma\delta_{p_j}$ with $\gamma=T^{-2}$ reveals which of $[0,p_i),[p_i,p_j),[p_j,1]$ contains $v_t$, and collects revenue $(1-\gamma)R(p_i)+\gamma R(p_j)\ge R(p^\ast)-4\Delta-\gamma$. Therefore, every round in layer $\Delta$ brings regret $\O(\Delta)$.

It remains to count the number of interval queries needed for each layer $\Delta$. In \Cref{eq:estimate revenue}, the first two terms come from the \textbf{Estimate} step, where $\mI_\mT[p_i,p_N)$ consists of $\O(\log N)$ segment tree intervals (\Cref{lem:segment tree}), and the last term comes from the \textbf{Refine} step. As the \textbf{Eliminate} step only compares two candidates $p\in \{p_i,p_i+\frac\Delta 2\}$ and $p'\in \{p_j,p_j+\frac\Delta 2\}$ with $p_i,p_j\in \candidateSet_{2\Delta}$, we only need to estimate all $\hat R(p')-\hat R(p)$ up to $\frac\Delta 4$. This leads to our choices of precisions when calling \textproc{EmpiricalBernstein} in \Cref{line:segT Bernstein,line:estimate init mass,line:non-segT Bernstein,line:discard}. They bound the number of rounds for each layer $\Delta$, in the following lemma.
\begin{lemma}[Number of Rounds]\label{lem:size-2 rounds}
For each $\Delta=2^{-k}$, \Cref{alg:size-2} spends $\Otil((p_N-p_0+\Delta)/\Delta^2)$ rounds where $p_0,p_N$ are both $4\Delta$-optimal prices, and incurs $\O(\Delta)$ regret in each round. Consequently, for any given $T$, the regret of \Cref{alg:size-2} is upper bounded as
\begin{equation}\label{eq:size-2 rounds regret}
\Otil\left (\sup_{\Delta=2^{-k},k\ge 0} \min\left \{T \Delta,\frac{\diam\{p\in [0,1]\mid R(p)\ge R(p^\ast)-\Delta\}}{\Delta}\right \}\right ),
\end{equation}
where $\diam\{A\}:=\sup\{A\}-\inf\{A\}$ is the diameter of a set $A$.
\end{lemma}

\begin{algorithm}[!t]
\caption{Revenue Maximization via Size-2 Menus}
\label{alg:size-2}
\begin{algorithmic}[1]
\State Let $\bar p\gets$ \Call{ScaleSearch}{} in \Cref{alg:scale search}. From now on, every price $p$ is offered as $\bar p\times p$, and $F$ now denotes the distribution of $\min\{\frac v{\bar p},1\}$ with $v\sim F$ \label{line:scale} \Comment{for an edge case in \Cref{footnote:ScaleSearch}}
\State Initialize $\candidateSet_1$ as the grid $\{0,\frac 12,1\}$, such that each $p\in \candidateSet_1$ is $1$-optimal and $p^\ast$ is $\frac 12$-close to $\candidateSet_1$
\For{$\Delta=\frac 12,\frac 14,\frac 18,\ldots$}
\State Write out $\candidateSet_{2\Delta}=\{p_0,p_1,\ldots,p_N\}$ in increasing order \Comment{all $p\in \candidateSet_{2\Delta}$'s are $4\Delta$-optimal}
\State Build a segment tree by $(\mT,\mI_\mT)\gets \Call{BuildSegTree}{p_0,p_1,\ldots,p_N}$ \Comment{see \Cref{alg:segment tree}}
\For{each tree interval $I=[p_i,p_j)\in \mT$}
\State Estimate $\hat \intervalP(I)\gets \Call{EmpiricalBernstein}{I,\frac{\Delta}{80(\log_2 N+2)(p_i+\Delta)},T^{-3}}$ \label{line:segT Bernstein} \Comment{see \Cref{alg:empirical Bernstein}}
\EndFor
\State Estimate $\hat \demand(p_N)\gets \Call{EmpiricalBernstein}{[p_N,1],\frac{\Delta}{8(p_N-p_0+\Delta)},T^{-3}}$\label{line:estimate init mass}
\State Initialize the new set of candidate prices $\candidateSet_\Delta:=\{p_0,p_0+\frac \Delta 2,p_1,p_1+\frac \Delta 2,\ldots,p_N,p_N+\frac \Delta 2\}\cap [0,1]$
\For{$i=0,1,\ldots,N-1$}
\State Estimate $\hat \intervalP([p_i,p_i+\frac \Delta 2))\gets \Call{EmpiricalBernstein}{[p_i,p_i+\frac \Delta 2),\frac{\Delta}{40(p_i+\Delta)},T^{-3}}$ \label{line:non-segT Bernstein}
\EndFor
\State Estimate $\hat \intervalP([p_N,p_N+\frac \Delta 2))\gets \Call{EmpiricalBernstein}{[p_N,p_N+\frac \Delta 2),\frac{\Delta}{40(p_N+\Delta)},T^{-3}}$, but terminate this call once $\hat \mu_K-r_K>\frac{2\Delta}{p_N}$ in \Cref{alg:empirical Bernstein} (then remove $p_N+\frac \Delta 2$ from $\candidateSet_\Delta$) \label{line:discard}
\For{each candidate price $p\in \candidateSet_\Delta$}
\State Estimate its revenue by \Cref{eq:estimate revenue} with $p_i:=\max\{p'\in \candidateSet_{2\Delta}\mid p'\le p\}$, where the three terms come from \Cref{line:estimate init mass}, \Cref{line:segT Bernstein}, \Cref{line:non-segT Bernstein}, and \Cref{line:discard} (with $\mI_\mT$ in \Cref{alg:segment tree})
\algeq[eq:estimate revenue]{$\displaystyle \hat R(p):=p\left(\hat \demand(p_N)+\sum_{I\in \mI_\mT[p_i,p_N)}\hat \intervalP(I)-\hat \intervalP([p_i,p))\right)$}
\EndFor
\State Eliminate all $p\in \candidateSet_\Delta$'s with $\max_{p'\in \candidateSet_\Delta} \hat R(p')-\hat R(p)\ge \Delta$ \Comment{all $p\in \candidateSet_\Delta$'s are $2\Delta$-optimal}
\EndFor
\end{algorithmic}
\end{algorithm}
\begin{proof}[Proof Sketch]
Given our choice of empirical Bernstein precisions, estimating each $[p_i,p_j)\in \mT$ only takes $\Otil(\frac{p_i\intervalP([p_i,p_j))}{\Delta^2}+\frac 1\Delta)$ queries. Since $p_i\intervalP([p_i,p_j))\le p_j-p_i+R(p_i)-R(p_j)$ from \Cref{eq:1-side Lip}, the first terms telescope to $p_N-p_0+R(p_0)-R(p_N)\le p_N-p_0+4\Delta$ over each depth (as $p_0,p_N\in \candidateSet_{2\Delta}$ are both $4\Delta$-optimal); and all those $\Delta^{-1}$ terms sum to $\Otil(\frac{p_N-p_0+\Delta}{\Delta^2})$ because all candidates in $\candidateSet_{2\Delta}$ are $\Delta$ apart. The same telescoping calculation also holds for those $[p_i,p_i+\frac\Delta 2)$ with $i<N$. The last interval $[p_N,p_N+\frac\Delta 2)$ cannot be absorbed in $(p_N-p_0)$ and thus requires extra care: In \Cref{line:discard}, if we are confident that $p_N+\frac\Delta 2$ earns less than $p_N$, we directly discard it. If this never happens, then $\intervalP([p_N,p_N+\frac\Delta 2))\le \frac{4\Delta}{p_N}$, and the number of queries is then $\Otil(\frac 1\Delta)$. Otherwise, after $\Otil(\frac{1}{\Delta})$ rounds, empirical Bernstein gives a multiplicatively close confidence bound, thus triggering the discard.
\end{proof}

\subsection{Upper Bounds for Dynamic Pricing with Size-2 Menus}\label{sec:size-2 regret}
Building upon \Cref{lem:size-2 rounds}, in \Cref{thm:size-2}, we first derive several minimax regret upper bounds for \Cref{alg:size-2}. We will further supply an instance-dependent upper bound in \Cref{thm:instance upper}.
\begin{theorem}[Upper Bounds]\label{thm:size-2}
For dynamic pricing with size-2 menus, \Cref{alg:size-2} ensures:
\begin{enumerate}[label=\textup{(\alph*)}]
    \item $\mR_\general^{(2)}(T)=\O(\sqrt T\log^4 T)$;
    \item $\mR_\MHR^{(2)}(T)=\O(T^{1/3}\log^4 T)$;
    \item $\mR_\curvature^{(2)}(T)=\O((\kappa^{-1/3}T^{1/3}+\log T)\log^4 T)$ for every curvature $\kappa\in (0,2)$.
\end{enumerate}
\end{theorem}

\Cref{thm:size-2} implies interval-query complexity $\Interval_\MHR(\epsilon)=\Otil(\epsilon^{-1.5})$ and $\Interval_\curvature(\epsilon)=\Otil(\kappa^{-1/2}\epsilon^{-1.5})$, matching the corresponding sample complexities (\Cref{thm:sample complexities}), whereas we know that $\Threshold_\MHR(\epsilon)=\tilde \Theta(\epsilon^{-2})$ \citep{leme2023pricing} and $\Threshold_\curvature(\epsilon)=\Omega(\epsilon^{-2.5})$ (\Cref{lem:threshold lower bound curvature}).

Looking at \Cref{eq:size-2 rounds regret}, for any $F$ and $\Delta\in (0,1]$, let $N_F(\Delta):=\lceil \diam \{p\in [0,1]\mid R(p)\ge R(p^\ast)-\Delta\}/\Delta\rceil$ be the $\Delta$-covering number of the convex hull of all $\Delta$-optimal prices of $F$. Inspired by the zooming dimension in Lipschitz bandits \citep{kleinberg2019bandits,bubeck2011x}, we define the interval dimension as follows:
\begin{definition}\label{def:zoom dim}
$F$ has interval dimension $d\in [0,1]$ with multiplier $C$ if $N_F(\Delta)\le C\Delta^{-d}$, $\forall \Delta\in (0,1]$.\footnote{$d=0$ for a point mass at $\frac 12$, where $R$ drops linearly; $d=\frac 12$ for MHR distributions (e.g., $\text{Unif}[0,1]$) and curvature distributions, where $R$ drops quadratically around $p^\ast$; and $d=1$ for the equal-revenue distribution $\Pr\{v\ge p\}=\frac 1{2p}$ on $[\frac 12,1]$, where $R$ does not drop at all. \Cref{app:examples} gives more economically natural examples, such as discrete values, kinked demands, Pareto distributions. In that section, we also give a family of value distributions with every $d\in [0,1)$.}
\end{definition}

Unlike Lipschitz bandits, $N_F(\Delta)$ covers the \emph{convex hull} of the $\Delta$-optimal prices rather than the prices themselves. The two concepts coincide when $R_F$ is unimodal (e.g., $F\in \mF_\regular$). But as we will see shortly in \Cref{footnote:convex hull}, this distinction is necessary rather than an artifact of our analysis. Plugging $N_F$ into \Cref{lem:size-2 rounds}, we derive the following instance-dependent regret upper bound.

\begin{theorem}[Instance-Dependent Regret Upper Bound]\label{thm:instance upper}
For any value distribution $F$ and any $T\ge 2$, the regret of \Cref{alg:size-2} is $\O(V_T(F)\log^5 T)$, where
\begin{equation}\label{eq:V_T}
V_T(F):=\sup_{\Delta\in (0,1]}\min\{T\Delta,N_F(\Delta)\},
\end{equation}
which implies regret $\O(CT^{d/(d+1)}\log^5 T)$ on any $F$ with interval dimension $d$ and multiplier $C$.
\end{theorem}

\begin{proof}[Proof Sketch of \Cref{thm:size-2,thm:instance upper}]
Since we always have $p_N-p_0\le 1$, the number of rounds spent on layer $\Delta$ is $\Otil(\Delta^{-2})$ by \Cref{lem:size-2 rounds}. As the per-round regret is $\O(\Delta)$, this gives $\Otil(\sqrt T)$ regret in the worst case. For MHR or curvature distributions, the near-optimal prices concentrate around $p^\ast$: With curvature $\kappa$, all $\Delta$-optimal prices are within $\sqrt{\Delta/\kappa}$ of $p^\ast$ by definition; for MHR distributions, they are also within $4\sqrt \Delta$ of $p^\ast$ when $\Delta\le \frac{R(p^\ast)}4$ (\Cref{lem:MHR shape}; a price-space analog of \citealt[Lemma 3.2]{huang2015making}). Hence $p_N-p_0=\O(\sqrt \Delta)$, and layer $\Delta$ spends $\Otil(\Delta^{-1.5})$ rounds, thus giving a refined regret upper bound of $\Otil(T^{1/3})$.\footnote{There is an edge case to take care of: Whenever $\Delta\ge R(p^\ast)$, every price is $\Delta$-optimal, so the early layers always refine the whole $[0,1]$ rather than one with width $\sqrt \Delta$, thus bringing an unnecessary term of order $\Delta^{-1}=1/R(p^\ast)$ (e.g., $\sqrt T$ for $\text{Unif}[0,T^{-1/2}]\in \mF_\MHR$). To avoid this, we first find a $\bar p\ge p^\ast$ by posting the prices $1,\frac 12,\frac 14,\ldots$ in turn (which we call \textsc{ScaleSearch}; \Cref{alg:scale search}), and then only zoom into the smaller interval $[0,\bar p]$. This gives a $\bar p<10p^\ast$ for MHR distributions while only incurring $\O(\log^2 T)$ regret, thus avoiding the unnecessary $1/R(p^\ast)$. See \Cref{app:scale search}.\label{footnote:ScaleSearch}} For any $F$, the $\O(V_T(F)\log^5 T)$ bound follows from \Cref{lem:size-2 rounds}, since $\diam\{p\in [0,1]\mid R(p)\ge R(p^\ast)-\Delta\}\le \Delta\cdot N_F(\Delta)$ by definition. This consequently proves \Cref{thm:instance upper}, as interval dimension $d$ with multiplier $C$ ensures $V_T(F)\le CT^{d/(d+1)}$: Every $\Delta\le T^{-1/(d+1)}$ has $T\Delta\le T^{d/(d+1)}$, and every other $\Delta$ has $N_F(\Delta)\le C\Delta^{-d}<CT^{d/(d+1)}$.
\end{proof}

\section{Lower Bounds: From Minimax to Instance-Optimality}\label{sec:lower bounds}

We introduce the minimax lower bounds in \Cref{sec:minimax lower}. Our main lower bounds are in \Cref{sec:instance}, showing the near-instance-optimality of \Cref{alg:size-2}: Around every value distribution $F$, no algorithm, even with infinite menus, can significantly beat \Cref{alg:size-2} on a few \emph{decoy instances} close to $F$. We overview how to construct these decoy instances in \Cref{sec:constructions}. We develop anytime lower bounds in \Cref{sec:fixed instance}, and compare dynamic pricing with Lipschitz bandits and experts in \Cref{sec:lipschitz}.

\subsection{Minimax Lower Bounds}\label{sec:minimax lower}
\begin{theorem}[Minimax Lower Bounds]\label{thm:minimax lower}
For dynamic pricing with infinite menus and full feedback:
\begin{enumerate}[label=\textup{(\alph*)}]
    \item $\mR_\general^{(\infty)}(T)=\Omega(\sqrt T)$;
    \item $\mR_\MHR^{(\infty)}(T)=\Omega(T^{1/3})$;
    \item $\mR_\curvature^{(\infty)}(T)=\Omega(\kappa^{-1/3}T^{1/3})$ for every curvature $\kappa\in (0,2)$;
    \item the minimax regret over value distributions with interval dimension $d\in (0,1)$ is $\Omega(T^{d/(d+1)})$.
\end{enumerate}
\end{theorem}

Hence, all rates in \Cref{thm:size-2} are minimax-optimal even with infinite menus. We defer the (mostly standard) proofs to \Cref{app:class lower}. A few remarks before moving on to our main lower bounds:
\begin{itemize}
\item Result (d) can also be extended to an \emph{anytime} lower bound, which we will detail in \Cref{sec:fixed instance}.
\item Our hard instances in (a) are regular, hence $\mR_\regular^{(\infty)}(T)=\Omega(\sqrt T)$, which is as hard as $\mF_\general$ and strictly harder than $\mF_\MHR$. This is in contrast to dynamic pricing with posted prices, where $\mR_\regular^{(1)}(T)=\tilde \Theta(\sqrt T)$ \citep{singla2024bandit,leme2023pricing}, thus strictly easier than $\mF_\general$ and as easy as $\mF_\MHR$.
\item Our hard instances in (a) are also \emph{quasi-MHR}, a broader-than-MHR class introduced in \citep{feng2025beyond}, thus answering their open question on learning reserves for quasi-MHR buyers (see \Cref{remark:quasi-MHR}).
\item As depicted in \Cref{table:regret}, \Cref{thm:size-2,thm:minimax lower} together show that size-2 menus are nearly as good as infinite menus (up to $\log$ factors) for all these classes of value distributions. On the other hand, algorithms using only posted prices are polynomially worse for all these classes.
\item All these results also imply lower bounds on sample complexities. Hence, for both CDF estimation and monopoly price estimation, interval queries are as powerful as samples (up to $\log$ factors), while being strictly stronger than threshold queries (see \Cref{table:complexity}; detailed in \Cref{app:complexities}).
\end{itemize}

\subsection{Near Instance-Optimality of \Cref{alg:size-2}}\label{sec:instance}
Are size-2 menus as good as infinite ones in general, or just for these specific classes in \Cref{thm:minimax lower}? We claim that, on \emph{every} possible value distribution $F$, size-2 menus near-match infinite menus.

To justify such an ``instance-optimality'' claim, ideally, one would want to lower bound any algorithm's regret on this $F$ alone. But this is impossible, because the algorithm always posting $p_F^\ast$ (somehow ``cheating'' by giving up on all other instances) has zero regret on $F$. A natural rescue, known as the Lai--Robbins-type \citep{lai1985asymptotically}, is to only consider algorithms that are good on all instances (e.g., with $o(T)$ regret). But it fails for our full-feedback problem since exploration is free: One can keep playing $p_F^\ast$ until the empirical CDF leaves the confidence ball of $F$ \citep{dvoretzky1956asymptotic}, and then switch to \Cref{alg:size-2}. This gives $\O(1)$ regret on $F$ while remaining $\Otil(\sqrt T)$ on every other value distribution. Instead, we construct a local-minimax-type lower bound \citep{cai2015framework,zhu2016local} in \Cref{thm:instance full}.
\begin{theorem}[Local Minimax Lower Bound]\label{thm:instance full}
Let $F$ be such that $R_F(p_F^\ast)>0$ and $T\ge 2$, and let $c_1:=\frac{R_F(p_F^\ast)^{12}}{2^{45}}$, $c_2:=\frac{325}{R_F(p_F^\ast)^3}$, and $c_3:=\frac{2^{20}}{R_F(p_F^\ast)^6}$. There are value distributions $G_+,G_-$ such that:
\begin{enumerate}[label=\textup{(\alph*)}]
    \item Any algorithm, even with infinite menus, suffers regret at least $c_1V_T(F)$ on $G_+$ or on $G_-$;
    \item $V_T(G_\pm)\le c_2V_T(F)$; and
    \item $H^2(F,G_\pm)\le \frac{c_3}{V_T(F)^2}$ where $H^2(F,G):=\frac 12\int(\sqrt{dF}-\sqrt{dG})^2$ is the squared Hellinger distance.
\end{enumerate}
\end{theorem}

Comparing with \Cref{thm:instance upper}, we conclude the \emph{near instance-optimality} of \Cref{alg:size-2}: \emph{\textbf{No algorithm, even using infinite menus, beats \Cref{alg:size-2} simultaneously on these converging-to-$F$ decoy instances}} beyond $\log$ factors. We summarize it as \Cref{cor:instance optimality}, which is much stronger than minimax lower bounds that only evaluate the algorithm on a few \emph{specific} hard instances.

\begin{corollary}[Near Instance-Optimality of \Cref{alg:size-2}]\label{cor:instance optimality}
Under the conditions of \Cref{thm:instance full},
\begin{enumerate}
\item \Cref{alg:size-2} has regret $\O(V_T(F)\log^5 T)$ on $F$ and $\O(R_F(p_F^\ast)^{-3}V_T(F)\log^5 T)$ on $G_+$ and on $G_-$.
\item Any algorithm, even with infinite menus, suffers $\Omega(R_F(p_F^\ast)^{12} V_T(F))$ regret on $G_+$ or on $G_-$.
\item If $F$ does not have interval dimension $0$, then $H^2(F,G_\pm)\to 0$ as $T\to \infty$ (see \Cref{lem:V_T infinite}). Otherwise, \Cref{alg:size-2} attains polylogarithmic regret on $F$, hence trivially near instance-optimal.
\end{enumerate}
\end{corollary}

\begin{remark}
\Cref{thm:instance full} implies that $N_F(\Delta)$ must cover the convex hull of all $\Delta$-optimal prices (as in \Cref{def:zoom dim}), not just these prices themselves (as in zooming dimension of Lipschitz bandits). Indeed, value distribution $F=\frac 12 \delta_{1/2}+\frac 12\delta_1$ has interval dimension 1 but zooming dimension 0. For all large $T$, there are two close-to-$F$ value distributions $G_\pm$, both also bimodal with zooming dimension 0, where any algorithm, even with infinite menus, must suffer $\Omega(V_T(F))=\Omega(\sqrt T)$ regret.\label{footnote:convex hull}
\end{remark}

While general, \Cref{cor:instance optimality} is a bit away from being perfect: The decoy instances $G_\pm$ do not include $F$, which does not exclude the possibility that $F$ is actually much easier; and $G_\pm$, while close to $F$ in Hellinger distances, no longer enjoy the good properties of $F$ (e.g., interval dimensions or MHR).

In \Cref{thm:instance lower}, we resolve them when $N_F(\Delta)=\Theta(\Delta^{-d})$ for all $\Delta$ small enough. For $T$ large enough, there now exists a decoy instance $G$ with interval dimension $\le d$ such that $\mR_{\{F,G\}}^{(\infty)}(T)=\Omega(T^{d/(d+1)})$. Since \Cref{alg:size-2} attains $\O(T^{d/(d+1)} \log^5 T)$ regret on both $F$ and $G$ (\Cref{thm:instance upper}), we can sharpen the near instance-optimality to: \emph{\textbf{No algorithm, even using infinite menus, beats \Cref{alg:size-2} by $\omega(\log^5 T)$ factors simultaneously on $F$ and a local $G$}} (which converges to $F$ as $T\to \infty$).

\begin{theorem}[Near Instance-Optimality via Interval Dimensions]\label{thm:instance lower}
If $F$ has interval dimension $d\in (0,1)$ with multiplier $C$, there is a constant $c>0$ only depending on $R_F(p_F^\ast),C$, and $d$. For any $T\ge \frac 1c$ satisfying $N_F(\Delta)\ge \frac{\Delta^{-d}}C$ with $\Delta=T^{-1/(d+1)}$, there is a value distribution $G$ that satisfies:
\begin{enumerate}[label=\textup{(\alph*)}]
    \item Any algorithm, even with infinite menus, suffers regret at least $cT^{d/(d+1)}$ on $F$ or on $G$;
    \item $G$ also has interval dimension $d$ with multiplier $C/c$; and
    \item $H^2(F,G)\le \frac 1{cT}$.
\end{enumerate}
\end{theorem}

We also refine \Cref{thm:instance full} to MHR distributions, motivated by the fact that interval (or zooming) dimensions are not the most suitable metric for MHR distributions: In dynamic pricing with posted prices, due to the $\frac 12$ zooming dimension, MHR distributions seem to bear a $\tilde \Theta(T^{3/5})$ rate \citep{kleinberg2019bandits}; but their minimax regret is actually $\Otil(\sqrt T)$ \citep{singla2024bandit}. With menus, while our \Cref{thm:instance lower} indeed suggests an $\Omega(T^{1/3})$ lower bound, the decoy instance $G$ may not be MHR. We resolve this in \Cref{thm:lower MHR}.
\begin{theorem}[Near Instance-Optimality for MHR Distributions]\label{thm:lower MHR}
Let $F$ be MHR. If on an open interval containing $p_F^\ast$, its hazard rate function $h(v)=\frac{f(v)}{\demand_F(v)}$ is differentiable with $h'$ bounded by some $[c,C]$ with $0<c<C$, then for every $T>0$, there is an MHR distribution $G$ with $H^2(F,G)\le \frac 1{64T}$, such that any algorithm, even with infinite menus, suffers regret $\mR_{\{F,G\}}^{(\infty)}(T)=\Omega(T^{1/3})$.
\end{theorem}

\subsection{Constructions of Decoy Instances}\label{sec:constructions}

We overview our constructions of the decoy instances in \Cref{thm:instance full}, and how to refine them in \Cref{thm:instance lower,thm:lower MHR}. We remark that while local minimax type of lower bounds exist in the statistics or learning literature \citep{cai2015framework,zhu2016local}, their techniques do not translate to our setting. Instead, our constructions are specifically tailored towards dynamic pricing and are highly non-trivial.

\begin{proof}[Proof Sketch of \Cref{thm:instance full}]
Throughout, all non-subscripted $p^\ast$, $R$, and $N$ stand for those of $F$, and all $\O$, $\Omega$, $\Theta$ notations treat $R(p^\ast)$ and $p^\ast$ as constants. While we still use the two-point method (standard for minimax bounds), we instead construct decoy instances locally around $F$.

\begin{figure}[!t]
\centering
\begin{tikzpicture}
\begin{groupplot}[group style={group size=3 by 2, horizontal sep=1.15cm, vertical sep=0.2cm, xticklabels at=edge bottom},
  scale only axis, width=4cm, scaled ticks=false, clip mode=individual, every axis plot/.append style={thick},
  tick label style={font=\scriptsize}, label style={font=\scriptsize}, title style={font=\footnotesize, align=center}, xticklabel style={align=center},
  yticklabel style={/pgf/number format/fixed, /pgf/number format/precision=3, text width=0.5cm, align=right}, xmajorgrids, ymajorgrids, grid style={black!10}, axis line style={black!70}, tick style={black!70}]
\nextgroupplot[title={(a) Initial $G_\pm$ (\Cref{thm:instance full})}, xmin=0.34, xmax=0.7, ymin=0.165, ymax=0.295, height=2.6cm, xtick={0.4,0.5,0.6,0.7}, ytick={0.17,0.21,0.25,0.29}, ylabel={Revenue}]
\fill[black, fill opacity=0.05] (axis cs:0.5,0.165) rectangle (axis cs:0.6,0.295);
\fill[black, fill opacity=0.1] (axis cs:0.34,0.24) rectangle (axis cs:0.7,0.25);
\node[right, inner sep=1.5pt, font=\scriptsize] at (axis cs:0.7,0.245) {$\Delta$};
\addplot[black, domain=0.34:0.7, samples=73] {x*(1-x)};
\addplot[blue!75!black, domain=0.5:0.6, samples=21] {x*(1-x)+0.8*x*(x-0.5)};
\addplot[blue!75!black, domain=0.6:0.7, samples=21] {1.2*x*(1-x)};
\addplot[red!75!black, densely dashed, domain=0.5:0.6, samples=21] {x*(1-x)-0.8*x*(x-0.5)};
\addplot[red!75!black, densely dashed, domain=0.6:0.7, samples=21] {0.8*x*(1-x)};
\addplot[only marks, mark=*, mark size=1.3pt, black] coordinates {(0.5,0.25)};
\addplot[only marks, mark=*, mark size=1.3pt, blue!75!black] coordinates {(0.6,0.288)};
\node[font=\scriptsize, text=black] at (axis cs:0.372,0.2196) {$R$};
\node[font=\scriptsize, text=blue!75!black] at (axis cs:0.655,0.2837) {$R_{G_+}$};
\node[font=\scriptsize, text=red!75!black] at (axis cs:0.62,0.1735) {$R_{G_-}$};
\nextgroupplot[title={(b) Sharpening (\Cref{thm:instance full})}, xmin=0.28, xmax=0.72, ymin=0.2, ymax=0.275, height=2.6cm, xtick={0.375,0.5,0.625}, ytick={0.21,0.23,0.25,0.27}]
\fill[black, fill opacity=0.05] (axis cs:0.375,0.2) rectangle (axis cs:0.625,0.275);
\fill[black, fill opacity=0.1] (axis cs:0.28,0.24) rectangle (axis cs:0.72,0.25);
\node[right, inner sep=1.5pt, font=\scriptsize] at (axis cs:0.72,0.245) {$\Delta$};
\addplot[black, domain=0.28:0.425, samples=25] {(0.255625-(x-0.5)*(x-0.5))};
\addplot[black, domain=0.425:0.575, samples=2] {0.25};
\addplot[black, domain=0.575:0.72, samples=25] {(0.255625-(x-0.5)*(x-0.5))};
\addplot[blue!75!black, domain=0.375:0.425, samples=9] {0.75*(0.255625-(x-0.5)*(x-0.5))+0.16*x};
\addplot[blue!75!black, domain=0.425:0.5, samples=13] {0.1875+0.16*x};
\addplot[blue!75!black, domain=0.5:0.575, samples=13] {1.07*(0.3125-0.125*x)};
\addplot[blue!75!black, domain=0.575:0.625, samples=9] {1.07*(1.25*(0.255625-(x-0.5)*(x-0.5))-0.125*x)};
\addplot[blue!75!black, domain=0.625:0.72, samples=17] {0.989193*(0.255625-(x-0.5)*(x-0.5))};
\addplot[only marks, mark=*, mark size=1.3pt, black] coordinates {(0.5,0.25)};
\addplot[only marks, mark=*, mark size=1.3pt, blue!75!black] coordinates {(0.5,0.2675)};
\node[font=\scriptsize, text=black] at (axis cs:0.56,0.244) {$R_{G\hphantom{'}}$};
\node[font=\scriptsize, text=blue!75!black] at (axis cs:0.56,0.2681) {$R_{G'}$};
\nextgroupplot[title={(c) Including $F$ (\Cref{thm:instance lower})}, xmin=0.36, xmax=0.66, ymin=0.215, ymax=0.258, height=2.6cm, xtick={0.4,0.45,0.5,0.55,0.6}, ytick={0.22,0.23,0.24,0.25}]
\fill[black, fill opacity=0.05] (axis cs:0.55,0.215) rectangle (axis cs:0.6,0.258);
\fill[black, fill opacity=0.1] (axis cs:0.36,0.24) rectangle (axis cs:0.66,0.25);
\fill[black, fill opacity=0.25] (axis cs:0.36,0.2475) rectangle (axis cs:0.66,0.25);
\node[right, inner sep=1.5pt, font=\scriptsize] at (axis cs:0.66,0.24375) {$\Delta$};
\node[right, inner sep=1.5pt, font=\scriptsize] at (axis cs:0.66,0.24875) {$\Delta/4$};
\addplot[black, domain=0.36:0.66, samples=73] {x*(1-x)};
\addplot[blue!75!black, domain=0.55:0.6, samples=13] {x*(1-x)+0.416667*x*(x-0.55)};
\addplot[blue!75!black, domain=0.6:0.66, samples=15] {1.052083*x*(1-x)};
\addplot[only marks, mark=*, mark size=1.3pt, black] coordinates {(0.5,0.25)};
\addplot[only marks, mark=*, mark size=1.3pt, blue!75!black] coordinates {(0.6,0.2525)};
\node[font=\scriptsize, text=black] at (axis cs:0.385,0.2321) {$R$};
\node[font=\scriptsize, text=blue!75!black] at (axis cs:0.625,0.2535) {$R_G$};
\nextgroupplot[xmin=0.34, xmax=0.7, ymin=0, ymax=2.5, height=1.5cm, xtick={0.4,0.5,0.6,0.7}, ytick={0,1,2}, xticklabels={{$0.4$\\$\ell$},{$0.5$\\$p^\ast$},{$0.6$\\$u$},{$0.7$}},ylabel={Density Ratio}]
\fill[black, fill opacity=0.05] (axis cs:0.5,0) rectangle (axis cs:0.6,2.5);
\draw[densely dashed, black!50] (axis cs:0.34,1) -- (axis cs:0.7,1);
\addplot[blue!75!black, const plot] coordinates {(0.34, 1) (0.5,1) (0.5,0.2) (0.6,1.2) (0.7,1.2)};
\addplot[red!75!black, densely dashed, const plot] coordinates {(0.34, 1) (0.5,1) (0.5,1.8) (0.6,0.8) (0.7,0.8)};
\node[font=\tiny,color=red!75!black] at (axis cs:0.55,2.05) {$\approx 1\mp\nicefrac{1}{N(\Delta)}$};
\node[font=\tiny,color=red!75!black] at (axis cs:0.65,0.455) {$\approx 1\pm\Delta$};
\nextgroupplot[xmin=0.28, xmax=0.72, ymin=0, ymax=1.8, height=1.5cm, xtick={0.375,0.5,0.625}, ytick={0,0.5,1,1.5}, xticklabels={{$0.375$\\$\ell'$},{$0.5$\\$p_G^\ast$},{$0.625$\\$u'$}}]
\fill[black, fill opacity=0.05] (axis cs:0.375,0) rectangle (axis cs:0.625,1.8);
\draw[densely dashed, black!50] (axis cs:0.28,1) -- (axis cs:0.72,1);
\addplot[blue!75!black, const plot] coordinates {(0.28, 1) (0.375,1) (0.375,0.75) (0.5,1.3375) (0.625,0.989193) (0.72,0.989193)};
\node[font=\tiny,color=blue!75!black] at (axis cs:0.4375,0.55) {$\approx 1-c$};
\node[font=\tiny,color=blue!75!black] at (axis cs:0.5625,1.5) {$\approx 1+c$};
\node[font=\tiny,color=blue!75!black] at (axis cs:0.67,0.8) {$\approx 1$};
\nextgroupplot[xmin=0.36, xmax=0.66, ymin=0, ymax=1.4, height=1.5cm, xtick={0.4,0.45,0.5,0.55,0.6}, ytick={0,0.5,1}, xticklabels={{$0.4$\\$\ell$},{$0.45$\\$a$},{$0.5$\\$p^\ast$},{$0.55$\\$b$},{$0.6$\\$u$}}]
\fill[black, fill opacity=0.05] (axis cs:0.55,0) rectangle (axis cs:0.6,1.4);
\draw[densely dashed, black!50] (axis cs:0.36,1) -- (axis cs:0.66,1);
\addplot[blue!75!black, const plot] coordinates {(0.36, 1) (0.55,1) (0.55,0.583333) (0.6,1.052083) (0.66,1.052083)};
\node[font=\tiny,color=blue!75!black] at (axis cs:0.575,0.45) {$\approx 1-\nicefrac{1}{N(\Delta)}$};
\node[font=\tiny,color=blue!75!black] at (axis cs:0.623,1.2) {$\approx 1+\Delta$};
\end{groupplot}
\end{tikzpicture}
\caption{Constructions in \Cref{thm:instance lower,thm:instance full} (under $F=\text{Unif}[0,1]$ and $\Delta=0.01$)}
\label{fig:hard instances}
\end{figure}
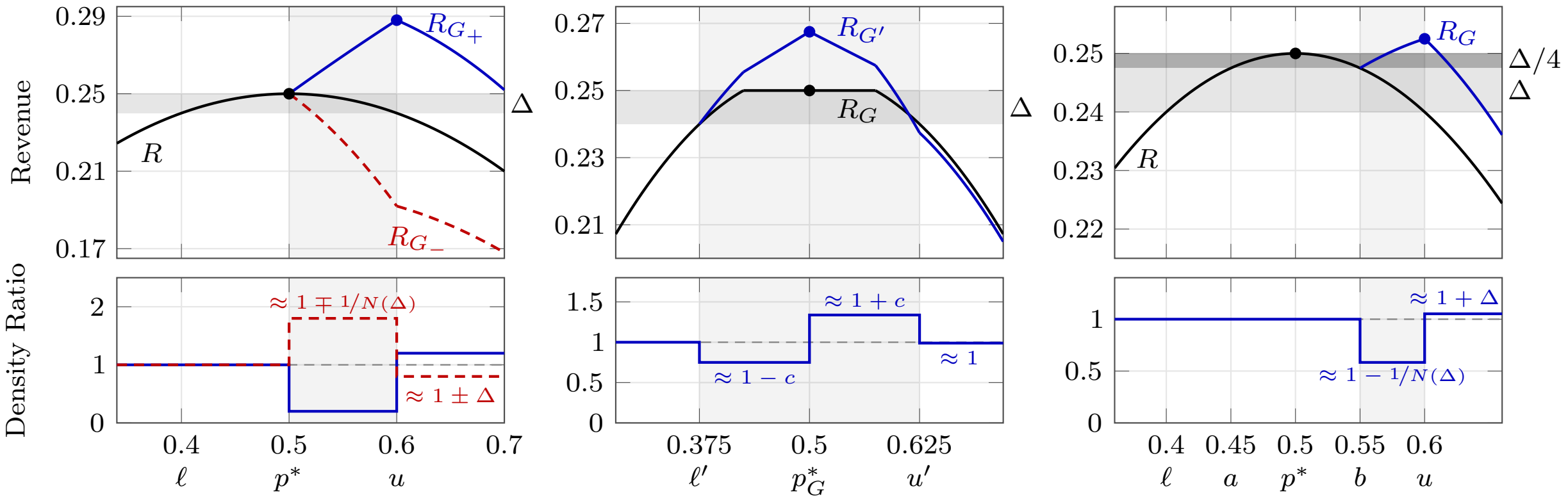

Let $\Delta$ be such that $\min\{T\Delta,N(\Delta)\}\ge \frac 12V_T(F)$ (\Cref{eq:V_T}), and assume $F$ has a probability density function (PDF) $f$ for the ease of presentation. Let $\ell\le u$ be the smallest and the largest $\Delta$-optimal prices of $F$, and call $R_G(p_G^\ast)-R_G(p)$ the \emph{gap} of a price $p$ under a value distribution $G$. It suffices to find two value distributions, both close to $F$, such that every price has a gap $\Omega(\Delta)$ under one of them. We do so by moving $\Theta(\Delta)$ probability mass from (a sub-interval of) $[\ell,u]$ to $[u,1]$, so that price $u$ gains $\Theta(\Delta)$ revenue relative to $\ell$ (or vice versa, giving two distributions $G_+$ and $G_-$). How to ensure $G_\pm$ are close to $F$? If this sub-interval has total probability mass $m$ in $F$, then this move rescales $f$ by $1\mp\Theta(\frac\Delta m)$. Hence $G_\pm$ are indistinguishable for $\Theta(\frac m{\Delta^2})$ rounds via the Hellinger distance, giving regret $\Omega(\min\{T\Delta,\frac m\Delta\})$. Therefore, finding an interval with $m=\Omega(\Delta \cdot N(\Delta))$ suffices. This is made possible by the other direction of \Cref{eq:1-side Lip}, namely far apart prices have large mass in between:
\begin{align*}
\intervalP_F([\ell,u))&\overset{(a)}{=}\frac{(u-\ell)\demand_F(u)-(R(u)-R(\ell))}{\ell}\ge \frac 1\ell \left (\frac{R(p^\ast)}2(u-\ell)-\Delta\right )\\
&\overset{(b)}{\ge} \frac 1\ell \left (\frac{R(p^\ast)}2(N(\Delta)-1)\Delta-\Delta\right )\overset{(c)}{\ge}\frac{R(p^\ast)}{4} \Delta \cdot N(\Delta)=\Omega(\Delta \cdot N(\Delta)),
\end{align*}
whenever $N(\Delta)\ge 2+\frac{4}{R(p^\ast)}=\Theta(1)$, where (a) uses \Cref{eq:1-side Lip}, (b) uses the definition that $(u-\ell)\ge (N(\Delta)-1)\Delta$, and (c) uses $N(\Delta)\ge 2+\frac{4}{R(p^\ast)}$ and $\ell\le 1$. Suppose that $[p^\ast,u)$ has at least half of this mass (the case of $[\ell,p^\ast)$ is similar). Construct $G_+$ and $G_-$, with PDFs $g_+$ and $g_-$, as follows:
\begin{equation*}
g_\pm:=\begin{cases}
f & \text{on }[0,p^\ast),\\
\left(1\mp\dfrac{4\Delta / p^\ast}{\intervalP_F([p^\ast,u))}\right)f & \text{on }[p^\ast,u),\\[2ex]
\left(1\pm\dfrac{4\Delta / p^\ast}{\demand_F(u)}\right)f & \text{on }[u,1].
\end{cases}
\end{equation*}
That is, $G_+$ moves a total mass $\theta:=4\Delta / p^\ast$ from $[p^\ast,u)$ to $[u,1]$ by rescaling $f$ on both intervals, and $G_-$ does the opposite; see \Cref{fig:hard instances} (a) for an illustration. They are indeed close to $F$:
\begin{equation*}
H^2(F,G_\pm)\le \int_0^1\frac{(g_\pm(v)-f(v))^2}{f(v)}dv=\frac{\theta^2}{\intervalP_F([p^\ast,u))}+\frac{\theta^2}{\demand_F(u)}=\O\left (\frac\Delta{N(\Delta)}+\Delta^2\right )=\O\left (\frac{1}{V_T(F)^2}\right ),
\end{equation*}
where we used $\theta=\Theta(\Delta)$, $\intervalP_F([p^\ast,u))=\Omega(\Delta \cdot N(\Delta))$, and $\demand_F(u)=\Theta(1)$. By direct calculation,
\begin{equation*}
R_{G_\pm}(p)=R(p)\pm\theta p\cdot\frac{\intervalP_F([p^\ast,p))}{\intervalP_F([p^\ast,u))},\quad\forall p\in [p^\ast,u].
\end{equation*}
Hence under $G_+$, the price $u$ gains $\theta u\ge 4\Delta$, whereas every $p\in [p^\ast,u]$ with $\intervalP_F([p^\ast,p))\le \frac 12\intervalP_F([p^\ast,u))$ gains at most $\frac{\theta u}2$ and thus has a gap at least $\frac{\theta u}2-\Delta\ge \Delta$. Under $G_-$, $p^\ast$ keeps its revenue, whereas every $p\in [p^\ast,u]$ with $\intervalP_F([p^\ast,p))\ge \frac 12\intervalP_F([p^\ast,u))$ loses at least $\frac{\theta p^\ast}2=2\Delta$ (hence a gap at least $\Delta$). Finally, no $p>u$ is $\Delta$-optimal under $F$ and hence under $G_-$; and no $p<p^\ast$ is $\Delta$-optimal under $G_+$. Hence, every $p\in [0,1]$ has a gap at least $\Delta$ under $G_+$ or under $G_-$, enabling the two-point method.

However, $V_T(G_\pm)$ can be much larger than $V_T(F)$: If $R_F$ decreases linearly at a suitable slope on the upper part of $[p^\ast,u]$, then $G_+$ exactly cancels this decrease, so that $R_{G_+}$ is maximized on a plateau of length $\Theta(\Delta \cdot N(\Delta))$ (see \Cref{fig:hard instances} (b) for an illustration). We thus perform a \emph{sharpening} step on each $G\in \{G_+,G_-\}$. Let $p_G^\ast$ be the monopoly price under $G$, and $\ell'\le p_G^\ast\le u'$ be the smallest and the largest $\Delta$-optimal prices under $G$. Construct value distribution $G'$ with PDF $g'$ as follows:
\begin{equation*}
g':=\begin{cases}
g & \text{on }[0,\ell'),\\
(1-c)\,g & \text{on }[\ell',p_G^\ast),\\
(1+c_1)(1+c)\,g & \text{on }[p_G^\ast,u'),\\
(1+c_1)(1-c_2)\,g & \text{on }[u',1],
\end{cases}
\end{equation*}
where $c:=\min_{\delta\in [\Delta,1]}\frac 1{N_F(\delta)}$, and $c_1:=c\cdot\frac{\intervalP_G([\ell',p_G^\ast))}{\demand_G(p_G^\ast)}$ and $c_2:=c\cdot\frac{\intervalP_G([p_G^\ast,u'))}{\demand_G(u')}$ keep the total mass unchanged (both $c_1$ and $c_2$ are $\O(\Delta)$). That is, by rescaling $g$ on different intervals, $G'$ moves a $c$ fraction of the mass on $[\ell',p_G^\ast)$ to $[p_G^\ast,1]$, and then scales up $[p_G^\ast,u')$ by $(1+c)$, taking the extra mass from $[u',1]$. Then $R_{G'}=R_G$ on $[0,\ell']$, $p_G^\ast$ remains optimal, no gap decreases, and
\begin{equation*}
R_{G'}(p_G^\ast)-R_{G'}(p)\ge R_G(p_G^\ast)-R_G(p)+c\,p_G^\ast\left\lvert \demand_G(p)-\demand_G(p_G^\ast)\right\rvert=\Omega\left(c\lvert p-p_G^\ast\rvert\right),\quad \forall p\in [\ell',u'],
\end{equation*}
where the last step applies \Cref{eq:1-side Lip} to $G$. Therefore, in both directions of $p_{G'}^\ast=p_G^\ast$, our new $R_{G'}$ is decreasing linearly fast throughout the $\Delta$-optimal region $[\ell',u']$. This gives $N_{G'}(\delta)=\O(\frac 1c)$ for all $\delta<\Delta$, where $\frac 1c=N_F(\delta')$ for some $\delta'\in [\Delta,1]$. For $\delta\ge \Delta$, as no gap decreases and $R_G=R_F+\O(\Delta)$, every $\delta$-optimal price under $G'$ is $\delta'$-optimal under $F$ for some $\delta'\ge \delta$, giving $N_{G'}(\delta)=\O(N_F(\delta'))$. Either way, for any $\delta\in (0,1]$, we have $N_{G'}(\delta)=\O(N_F(\delta'))$ for some $\delta'\ge \delta$. Therefore,
\begin{align*}
V_T(G')=\sup_{\delta\in (0,1]} \min\{T\delta,N_{G'}(\delta)\}&\le\sup_{\delta\in (0,1]} \sup_{\delta'\in [\delta,1]} \O\left (\min\{T\delta,N_F(\delta')\}\right )\\&\le \sup_{\delta'\in (0,1]} \O\left (\min\{T\delta',N_F(\delta')\}\right )=\O\left (V_T(F)\right ).
\end{align*}

As $g'=(1\pm\O(c))g$ on $[\ell',u')$ and $g'=(1\pm\O(\Delta))g$ on $[u',1]$, similar calculation gives $H^2(F,G')=\O(\frac\Delta{N(\Delta)})=\O(V_T(F)^{-2})$. Hence $G_+'$ and $G_-'$, sharpened from $G_+$ and $G_-$, are indistinguishable in $\Theta(N(\Delta)/\Delta)$ rounds. Consequently, by the two-point method (\Cref{lem:two-point}), any algorithm, even with infinite menus, must suffer regret $\Omega(\min\{T\Delta,N(\Delta)\})=\Omega(V_T(F))$ on one of them.
\end{proof}

\begin{proof}[Proof Sketch of \Cref{thm:instance lower}.]
Compared to the generic lower bound, we now additionally want $F$ itself to replace one of the two decoy instances. Let $\Delta:=T^{-1/(d+1)}$, so that $N(\Delta)\ge \frac{\Delta^{-d}}C$, and let $a\le b$ be the smallest and the largest $\Delta'$-optimal prices for $\Delta':=(4C^2)^{-1/(1-d)}\Delta$. Then
\begin{equation*}
b-a\le \Delta'N(\Delta')\le C\Delta'^{1-d}=\frac{\Delta^{1-d}}{4C}\le \frac 14\Delta \cdot N(\Delta).
\end{equation*}

Since $u-\ell=\Omega(\Delta\cdot N(\Delta))$, one of $[\ell,a)$ and $[b,u)$, say $[b,u)$, has length $\Omega(\Delta \cdot N(\Delta))$, and hence probability mass $\Omega(\Delta \cdot N(\Delta))$ by \Cref{eq:1-side Lip}. Similar to \Cref{thm:instance full}, let $G$ move a mass $\theta:=\frac{\Delta+\Delta'}u$ from $[b,u)$ to $[u,1]$, so $R_G=R$ on $[0,b]$, while $R_G(u)=R(u)+\theta u\ge R(p^\ast)+\Delta'$ (see \Cref{fig:hard instances} (c)). Thus every price in $[0,b]$ has gap at least $\Delta'$ under $G$, every price outside $[a,b]$ has gap more than $\Delta'$ under $F$, and $H^2(F,G)=\O(\frac\Delta{N(\Delta)})=\O(\Delta^{d+1})=\O(\frac 1T)$. Similarly sharpening $G$ gives a $G'$ such that $N_{G'}(\delta)=\O(N_F(\delta'))$ for some $\delta'\ge \delta$, which means $G'$ has interval dimension $d$. A two-point method (\Cref{lem:two-point}) on $F$ and $G'$ then yields the desired $\mR_{\{F,G'\}}^{(\infty)}(T)=\Omega(T^{d/(d+1)})$ lower bound.
\end{proof}

\begin{proof}[Proof Sketch of \Cref{thm:lower MHR}]
Recall that the demand of a value distribution $F$ is $\demand_F(p)=\Pr_{v\sim F}\{v\ge p\}$. By definition, the hazard rate $h=\frac{f}{\demand_F}=-(\log \demand_F)'$, and thus MHR is equivalent to having a concave $\log \demand_F$. In \Cref{thm:instance full}, we rescale the density $f$ by constants on intervals, which creates downward jumps in the hazard rate and breaks MHR. We instead directly work on the log-demand by constructing $\log \demand_G=\log \demand_F+\phi$ for some $\phi$. Since we want $G$ close to $F$, we only pick a small interval $[a,b]$ (that contains $p_F^\ast$) for $\phi$ to be non-constant. From the assumption that $h'\ge c$ around $p_F^\ast$, in order to make $\log \demand_F+\phi$ concave, we thus set $\phi(p)=\frac c2(\min\{p,b\}-a)_+^2$.

It remains to choose $a$ and $b$ such that every price has a large gap under $F$ or $G$, so the two-point method applies. Let $s=b-p_F^\ast$. Under $F$, every price above $p_F^\ast+\frac s2$ has a gap $\Omega(s^2)$ as $R_F$ drops quadratically around $p_F^\ast$; under $G$, since $\log R_G=\log p+\log \demand_G$ is concave, every price below $p_F^\ast+\frac s2$ has a gap $\Omega(s^2)$ once $\log R_G(b)-\log R_G(p_F^\ast+\frac s2)=\Omega(s^2)$. That is, the increase of $\phi$ from $p_F^\ast+\frac s2$ to $b$, which is of order $cs(b-a)$, must beat the drop of $\log R_F$ there by $\Omega(s^2)$. Since $h'\le C$, the drop is only $\O(s^2)$, and thus taking $a:=p_F^\ast-\Theta(s)$ with a large enough constant suffices.

The final step is verifying that $G$ is close to $F$: The density ratio $\frac{dG}{dF}=(1-\frac{\phi'}h)e^{\phi}$ deviates from $1$ by $\O(s)$ on $[a,b]$, an interval of length $\Theta(s)$, and by $\O(s^2)$ above $b$. Thus $H^2(F,G)=\O(s^3)$. Picking $s=\Theta(T^{-1/3})$, the two-point method (\Cref{lem:two-point}) gives $\Omega(Ts^2)=\Omega(T^{1/3})$ regret on $F$ or on $G$.
\end{proof}

\subsection{Anytime Lower Bounds}\label{sec:fixed instance}

The decoy instances in \Cref{thm:instance lower,thm:instance full,thm:lower MHR} are all chosen after seeing $T$ (in this aspect, similar to standard minimax lower bounds). It would be more favorable if a (randomized) instance is hard for infinitely many $T$'s almost surely, which we call \emph{anytime} lower bounds.

In dynamic pricing and the closely related Lipschitz bandits literature, \textit{i)} for curvature and smooth value distributions (see \Cref{footnote:curvature assumption in KL}), an $\Omega(\sqrt T)$ anytime lower bound is available \citep{kleinberg2003value}; and \textit{ii)} for Lipschitz bandits, $\Omega(T^{(d+1)/(d+2)})$ anytime lower bounds are also derived \citep[\S 5.1]{kleinberg2019bandits}. But they don't directly apply to dynamic pricing with posted prices or infinite menus. Utilizing our construction in \Cref{fig:hard instances}, we complement the literature by deriving an anytime lower bound for dynamic pricing: We construct a single randomized distribution $F$, whose interval dimension is a given $d\in (0,1)$, that forces $\Omega(T^{d/(d+1)})$ with infinite menus and $\Omega(T^{(d+1)/(d+2)})$ with posted prices.
\begin{theorem}[Anytime Lower Bounds]\label{thm:fixed instance}
For every $d\in (0,1)$, there is a constant $c>0$ and a probability distribution $\mu$ over value distributions with interval dimension $d$ and multiplier $11$, such that for every algorithm, the following holds with probability $1$ over the randomized $F\sim \mu$:
\begin{enumerate}
\item Even with infinite menus, its regret on $F$ is at least $cT^{d/(d+1)}$ for infinitely many $T$.
\item When using posted prices, its regret on $F$ is at least $cT^{(d+1)/(d+2)}$ for infinitely many $T$.
\end{enumerate}
\end{theorem}
\begin{proof}[Proof Sketch]
The construction of \citet{kleinberg2003value} uses a smooth one-parameter family, which can be learned at $\O(\log T)$ regret with infinite menus. We use the multi-scale needle-in-a-haystack idea for Lipschitz bandits \citep[\S 5.1]{kleinberg2019bandits}: Fix an infinite sequence of scales $\Delta_1>\Delta_2>\cdots$. Starting from a plateau $P_0$, each scale $k$ raises a part of the current plateau $P_{k-1}$ by $\Delta_k$. The top of it, whose half-width $w_k=\Delta_{k+1}^{1-d}$, becomes the next plateau $P_k$. To construct a single instance $F$ that forces infinite menus and posted prices at the same time, we alternate between two types of scales:
\begin{enumerate}
\item The first type is a \emph{needle} \citep[Theorem~5.7]{kleinberg2019bandits}, dividing $P_{k-1}$ into width-$\Delta_k$ intervals and raising a uniformly random one of them by $\Delta_k$. For posted pricing, this forces regret $\Omega(T_k \Delta_k)$ in $T_k=\Theta(\frac{w_{k-1}}{\Delta_k^3})$ rounds with probability at least $\frac 12$. Since $\Delta_k=w_{k-1}^{1/(1-d)}$, we have $T_k\Delta_k=\Theta(T_k^{(d+1)/(d+2)})$. Unlike in Lipschitz bandits, where the needle can be planted anywhere, the result must remain a revenue function, i.e., $\demand(p)=R(p)/p$ must stay non-increasing, or $\demand'=(R'-\demand)/p<0$. We ensure this by planting all scales on plateaus around $p=\frac 12$ such that $\demand(p)\ge 0.41$ and $R'(p)\le \frac 13$.
\item The second type is our new construction, which we call a \emph{tilt}. We divide $P_{k-1}$ into two halves, choose one using fresh randomness, and raise it by $\Delta_k$ by moving masses within it similar to \Cref{fig:hard instances} (a): the revenue rises linearly from both ends of this half, so that the middle part has a plateau $P_k$ that is $\Delta_k$ higher. These two halves are indistinguishable with infinite menus for $T_k=\Theta(\frac{w_{k-1}}{\Delta_k^2})$ rounds, thus forcing $T_k\Delta_k=\Omega(T_k^{d/(d+1)})$ regret (where again, $\Delta_k=w_{k-1}^{1/(1-d)}$).
\end{enumerate}

There is a unique challenge due to infinite menus. In posted prices, the finer scales are invisible to the seller unless a query falls into the small plateau $P_k$ that contains them, as the outcomes are otherwise identically distributed. With infinite menus, the seller sees every value (as if a threshold query is placed everywhere), so it could detect the finer scales within the first $T_k$ rounds. Because they sit inside $P_k$, this would reveal our choice at scale $k$. We therefore let the scales decay super-exponentially, $\Delta_{k+1}=\Delta_k^q$ for a large constant $q$, so that all finer scales together change the value distribution by only $\O(\Delta_{k+1})$ in squared Hellinger distance, which is invisible in the first $T_k$ rounds.

For algorithms using posted prices, due to needles, we have $\Omega(T^{(d+1)/(d+2)})$ regret for infinitely many $T$ almost surely, as each needle scale succeeds with probability at least $\frac 12$ using fresh randomness. For those using infinite menus, the needles are not obstacles: Full feedback reveals a needle within $\Otil(\Delta_k^{-1})$ rounds (as if all width-$\Delta_k$ intervals receive threshold queries simultaneously). Instead, those tilt scales force $\Omega(T^{d/(d+1)})$ regret for infinitely many $T$ almost surely by an identical argument.
\end{proof}

\subsection{Dynamic Pricing Compared to Lipschitz Bandits or Experts}\label{sec:lipschitz}

\begin{table}[!t]
\caption{Comparison between Dynamic Pricing and Lipschitz Bandits or Experts}\label{table:Lipschitz}
\centering
\renewcommand{\arraystretch}{1.2}
\begin{tabular}{|c|c|c|} \hline
Problem & Full Feedback / Infinite Menus & Bandit Feedback / Posted Prices \\ \hline
\multirow{2}{*}{\shortstack{Dynamic\\Pricing}} & \hlcell $\Otil(T^{d/(d+1)})$ using size-2 \textbf{(\Cref{thm:instance upper})} & $\Otil(T^{(d+1)/(d+2)})$ \citep{kleinberg2019bandits} \\
& \hlcell $\Omega(T^{d/(d+1)})$ \textbf{(\Cref{thm:instance lower,thm:instance full})} & $\Omega(T^{(d+1)/(d+2)})$ \textbf{(\Cref{thm:fixed instance})} \\\hline
\multirow{2}{*}{\shortstack{Lipschitz\\Bandits}} & $\Otil(\sqrt T)$ \citep{kleinberg2019bandits} & $\Otil(T^{(d+1)/(d+2)})$ \citep{kleinberg2019bandits} \\
& $\Omega(\sqrt T)$ even if $d=0$ \textbf{(\Cref{lem:experts lower bound})} & $\Omega(T^{(d+1)/(d+2)})$ \citep{contextualMAB-colt11} \\\hline
\end{tabular}
\end{table}

We conclude this section by comparing dynamic pricing to Lipschitz bandits or experts \citep{kleinberg2019bandits}. For Lipschitz bandits over $[0,1]$ (that is, each round's action is some $x\in [0,1]$, and each round's reward function $f_t$ is i.i.d.\ with a 1-Lipschitz expectation), the zooming algorithm attains $\Otil(T^{(d+1)/(d+2)})$ regret in terms of the zooming dimension $d$ \citep{kleinberg2019bandits}. This rate cannot be improved for $d\in (0,1)$, even \textit{i)} in terms of our interval dimension that is always no smaller, \textit{ii)} for dynamic pricing problems alone, and \textit{iii)} on a fixed value distribution $F$ independent of $T$ (see \Cref{thm:fixed instance}). So in terms of zooming or interval dimensions, dynamic pricing is one of the \emph{hardest} Lipschitz bandits.

With full feedback or menus, however, dynamic pricing is strictly easier than Lipschitz experts. Our $T^{d/(d+1)}$ is $o(\sqrt T)$ whenever $d<1$, whereas Lipschitz experts suffer $\Omega(\sqrt T)$ regret even if $d=0$ (all $\Delta$-optimal arms lie in an interval of length $4\Delta$; \Cref{lem:experts lower bound}). Indeed, comparing two arms whose expected payoffs differ by $\Delta$ takes about $\Delta^{-2}$ times their difference-in-payoff's second moment. Lipschitz experts only require the expected payoffs to be Lipschitz, so two very close arms can still have perfectly negatively correlated payoffs, making this second moment a constant. In pricing, the payoffs are determined by the same $v_t$, so for any two $\Delta$-optimal prices $p<p'$, \Cref{eq:1-side Lip} gives
\begin{equation*}
\E_{v\sim F}\Bigl[\bigl(p\1[v\ge p]-p'\1[v\ge p']\bigr)^2\Bigr]=p^2\Pr_{v\sim F}\{p\le v<p'\}+(p'-p)^2\Pr_{v\sim F}\{v\ge p'\}\le 3\Delta \cdot N(\Delta).
\end{equation*}
Hence comparing the $\Delta$-optimal prices only brings an $N(\Delta)$ regret overhead, rather than the $\Delta^{-1}$ in Lipschitz experts. We summarize the comparison in \Cref{table:Lipschitz}, where $d$ is the interval dimension in the first row (\Cref{def:zoom dim}) and the zooming dimension in the second row \citep{kleinberg2019bandits}.

\section{Extension: Adversarial Dynamic Pricing Using Size-3 Menus}\label{sec:adv}

We now consider \emph{adversarial dynamic pricing}, where an adaptive adversary chooses each value $v_t\in [0,1]$ based on the history. With a single posted price, the optimal regret against the best fixed price is $\tilde \Theta(T^{2/3})$ \citep{kleinberg2003value}. Should $v_t$ be revealed after each round (e.g., via infinite menus), one can achieve regret $\Otil(\sqrt T)$. Therefore, regret rates with size-1 menus and size-$\infty$ menus both match their i.i.d. counterparts, and we ask whether it is still possible to attain $\Otil(\sqrt T)$ regret using small menus. And indeed, we achieve this with size-3 menus, leveraging the approach from \Cref{sec:size-3}.

\paragraph{Preliminaries.} Adversarial dynamic pricing keeps most of the setup from the i.i.d. case: in each round $t$, the seller's algorithm offers a menu $M_t$ based on the history $\mH_{t-1}$, the customer with value $v_t\in[0,1]$ chooses an option in $M_t$ maximizing their utility, and the seller observes the chosen option. The revenue in round $t$, $\text{revenue}_t$, is the expected payment in the chosen menu option. The only difference is that each value $v_t$ is now chosen by an \emph{adaptive adversary}, \ie arbitrarily, possibly based on history $\mH_{t-1}$ and the previous values $v_1,v_2,\ldots,v_{t-1}$, possibly using its own randomness, but without observing the internal randomness of the algorithm in round $t$. Note that the adversary could tailor its choices to the algorithm's specification. The algorithm maximizes the total expected revenue in the worst case over all adversaries.

This is a special case of \emph{adversarial bandits} \citep{auer2002nonstochastic}, a standard bandit framework in which the arms' rewards in each round are chosen by an adaptive adversary (defined similarly). We adopt a (standard) notion of regret against a given price sequence $\pseq = \rbr{\popt_1 \LDOTS \popt_T}$ and adversary $\ADV$:
\begin{equation*}
\textstyle \Reg(T\mid \pseq,\ADV)
:= \E\sbr{\sum_{t\in[T]} \rbr{\popt_t\,\1[v_t\ge \popt_t] - \text{revenue}_t} \;\mid\; \ADV}.
\end{equation*}
We are interested in the worst case across all adversaries and all price sequences in a given class $\mC$.
\begin{equation*}
\Reg(T)
:= \sup_{\ADV,\; \pseq\in\mC} \quad \Reg(T\mid \pseq,\ADV).
\end{equation*}
The basic notion is against the best fixed price, \ie when $\mC$ consists of all fixed-price sequences. Competing against \emph{all} price sequences (a.k.a. \emph{dynamic regret}), results in $\Omega(T)$ regret even with full feedback \citep{zinkevich2003online}. We focus on the class $\mathcal C_S$ of sequences that ``switch'' (in the sense of $\popt_t\neq \popt_{t+1}$) at most $S$ times for some known integer $S\ge 0$; in particular, $\mathcal C_0$ consists of all fixed-price sequences.

We build on known results from adversarial bandits with full feedback, a.k.a. \emph{online learning}. For intuition on how they might help: one approach to achieve $\Otil(\sqrt{T})$ regret for adversarial dynamic pricing with full feedback is to take Hedge \citep{freund1997decision}, an algorithm for online learning, and run it over the price grid $\candidateSet:=\{0,\frac 1T,\frac 2T,\ldots,1\}$, \ie with prices in $\candidateSet$ as actions.

\subsection{Adversarial Dynamic Pricing Algorithm using Online Learning}\label{sec:adv size-3}
We follow the approach of \Cref{sec:size-3}, which has two key ingredients. First, by \Cref{reduction:infinite to full-info} (which is applicable to any sequence of realized values), a size-3 menu can exploit one price while asking an interval query via the other two. Second, a segment tree over the prices makes it suffice to estimate the probabilities of its tree intervals. The estimator in \Cref{sec:size-3}, however, fails against adversarial values. We instead estimate all tree intervals at once by importance weighting.

We run an online learning algorithm $\mA^{\text{OL}}$ over the grid $\candidateSet=\{0,\frac 1T,\frac 2T,\ldots,1\}$. In each round, we use \textit{i)} one menu option for the ``exploitation price'' suggested by $\mA^{\text{OL}}$; and \textit{ii)} two other menu options for an interval query on two ``exploration prices'', based on which we estimate the revenue of every price in $\candidateSet$. Using \Cref{reduction:infinite to full-info}, we put a probability mass $1-2\gamma$ (with $\gamma=1/T$) on the exploitation price, and $\gamma$ on each of the exploration prices, such that our algorithm's expected revenue is only $\O(1)$ different from always using prices suggested by $\mA^{\text{OL}}$. We start with our requirements for $\mA^{\text{OL}}$.

\paragraph{Online Learning Algorithm.} We allow any $\mA^{\text{OL}}$ with the following second-order guarantee:
\begin{definition}[Online Learning Algorithm]\label{def:online learning adv}
Let $\mathcal C\subseteq [K]^T$ be a class of comparator sequences. We say $\mA^{\text{OL}}$ is suitable for \Cref{alg:adv size-3}, if for some $\bm w_t\in \triangle[K]$ and $\bm q_t\in [0,1]^K$ that are both functions of $\bm \ell_1,\ldots,\bm \ell_{t-1}$, for every sequence of loss vectors $\bm \ell_1,\ldots,\bm \ell_T$ with $\ell_{t,i}\ge -1$ for all $t$ and $i$, it ensures
\begin{equation*}
\sum_{t=1}^T \left (\E_{i\sim \bm w_t}\left [\ell_{t,i}\right ]-\ell_{t,i_t^\ast}\right )\le \textsc{Penalty}(\mathcal C,T,K,\eta)+\eta\sum_{t=1}^T \E_{i\sim \bm w_t}\left [q_{t,i}\ell_{t,i}^2\right ]
\end{equation*}
for every $(i_1^\ast,i_2^\ast,\ldots,i_T^\ast)\in \mathcal C$, where \textsc{Penalty} is a function that only depends on $\mathcal C$, $T$, $K$, and $\eta$.
\end{definition}

For example, with $\eta\le 1$, Hedge with $w_{t,i}\propto \exp(-\eta\sum_{\tau<t}\ell_{\tau,i})$ \citep{freund1997decision} satisfies \Cref{def:online learning adv} with $\mathcal C=\mathcal C_0$ and $\textsc{Penalty}=\eta^{-1}\log K$. Fixed-Share \citep{herbster1998tracking}, which mixes each update of Hedge with the uniform distribution at rate $T^{-1}$, satisfies \Cref{def:online learning adv} with the class $\mathcal C_{S-1}$ of sequences that switch at most $(S-1)$ times and $\textsc{Penalty}=\eta^{-1}(S\log (KT)+1)$, for every $S\ge 1$. Both algorithms have $\bm q_t\equiv 1$, but this need not be the case for dynamic regret or small-loss algorithms. Equipped with $\mA^{\text{OL}}$ which takes care of exploitation, we then explain how to do exploration in \Cref{alg:adv size-3}.

\paragraph{Segment Tree Decomposition.} Similar to \Cref{sec:size-3} and \Cref{alg:size-2}, we build a segment tree $\mT$ over $\candidateSet$ such that all $[0,p)$'s are partitioned into $\O(\log T)$ tree intervals. In order to estimate the revenues of all $p\in \candidateSet$, it suffices to estimate $\1_t(I):=\1[v_t\in I]$ for every tree interval $I\in \mT$, because\footnote{We write $\1[v_t\ge p]=1-\sum_{I\in \mI_\mT[0,p)}\1_t(I)$ instead of $\1[v_t\ge p]=\sum_{I\in \mI_\mT[p,1]}\1_t(I)$ for a technical reason: Online learning algorithms like Hedge mostly require losses to be lower bounded by $-1$ (but allowed to be arbitrarily positive); see \Cref{def:online learning adv}. As we will discuss at the end of this section, this is also why \Cref{alg:adv size-3} fails to use size-2 menus.}
\begin{equation}\label{eq:adv decomposition}
p\1[v_t\ge p]=p\left(1-\sum_{I\in \mI_\mT[0,p)}\1_t(I)\right),\quad \forall p\in \candidateSet.
\end{equation}

\paragraph{Importance-Weighted Estimator.} We maintain an ``exploration distribution'' $\bm \mu_t$ over all tree intervals in $\mT$, whose construction is specified later, and explore (the two endpoints of) a random tree interval $I_t\sim \bm \mu_t$ in each round. Based on the interval query outcome $\1_t(I_t)$, from standard importance sampling properties, the estimator $\hat \1_t(I)=\1[I_t=I]\1_t(I_t) / \mu_t(I)$ (see \Cref{eq:importance weighting}) satisfies
\begin{equation*}
\E_t[\hat \1_t(I)]=\1_t(I),\quad \E_t[\hat \1_t^2(I)]=\frac{\1_t(I)}{\mu_t(I)},\quad \forall I\in \mT\text{ with }\mu_t(I)>0,
\end{equation*}
where $\E_t$ conditions on $v_t$ and the history before round $t$. Plugging $\hat \1_t$ into \Cref{eq:adv decomposition} then gives the unbiased estimator $\hat R_t(p)$ in \Cref{eq:income estimate}, which we feed into $\mA^{\text{OL}}$ as the revenue of every price $p\in \candidateSet$.
\begin{lemma}[Importance-Weighted Estimator]\label{lem:adv estimator}
For every round $t\in [T]$, for any price $p\in \candidateSet$ such that $w_t(p)>0$, we have $\E_t[\hat R_t(p)]=p\1[v_t\ge p]$. Meanwhile, $\hat R_t(p)\le 1$ is bounded from above.
\end{lemma}

\paragraph{Exploration Distribution.} It remains to choose the exploration distribution $\bm \mu_t$. We now apply \Cref{def:online learning adv}: With $-\hat R_t(p)$ as the loss for price $p\in \candidateSet$ in round $t\in [T]$, taking expectations,
\begin{equation}\label{eq:pathwise regret of Alg 3}
\E\left [\sum_{t=1}^T \left (\hat R_t(p_t^\ast)-\E_{p\sim \bm w_t} [\hat R_t(p)]\right )\right ]\le \textsc{Penalty}(\mathcal C,T,T+1,\eta)+\eta \E\left [\sum_{t=1}^T \sum_{p\in \candidateSet} w_t(p) q_{t}(p) \hat R_t(p)^2\right ].
\end{equation}

As $q_{t}(p)\le 1$ (recall \Cref{def:online learning adv}), for each $t\in [T]$, using the definition of $\hat R_t(p)$ and $\hat \1_t(I)$, we have
\begin{align*}
\E_t\left[\sum_{p\in \candidateSet}w_t(p)\hat R_t^2(p)\right]&\overset{(a)}{\le} \E_t\left [\sum_{p\in \candidateSet} w_t(p) p^2 \left (2+2\sum_{I\in \mI_\mT[0,p)} \hat \1_t(I)^2\right )\right ]\\
&\overset{(b)}{=}2 \sum_{p\in \candidateSet} w_t(p) p^2 + 2\sum_{p\in \candidateSet} w_t(p) p^2 \sum_{I\in \mI_\mT[0,p)} \frac{\E_t[\1[I_t=I]]\1_t(I)}{\mu_t(I)^2}\\
&\overset{(c)}{\le} 2\sum_{p\in \candidateSet}w_t(p)p^2+2\sum_{I\in \mT}\frac{\1_t(I)}{\mu_t(I)}\sum_{p:\mI_\mT[0,p)\ni I}w_t(p)p^2,
\end{align*}
where (a) uses the definition of $\hat R_t(p)$ in \Cref{eq:income estimate} and $(a+b)^2\le 2a^2+2b^2$, together with $\hat \1_t(I)=0$ for all $I\ne I_t$ (so that the cross terms vanish); (b) uses the definition of $\hat \1_t(I)$ in \Cref{eq:importance weighting} and $\1[\cdot]^2=\1[\cdot]$; (c) uses $\E_t[\1[I_t=I]]=\mu_t(I)$ and swaps the summations.

\begin{algorithm}[!t]
\caption{Adversarial Dynamic Pricing Using Size-3 Menus}\label{alg:adv size-3}
\begin{algorithmic}[1]
\Require{Time horizon $T$ (WLOG assumed to be a power of 2). Online learning algorithm $\mA^{\text{OL}}$}
\State Build a segment tree by $(\mT,\mI_\mT)\gets \Call{BuildSegTree}{0,\frac 1T,\frac 2T,\ldots,1}$ \Comment{see \Cref{alg:segment tree}}
\State Initialize an online learning algorithm $\mA^{\text{OL}}$ on the $T+1$ actions in $\candidateSet=\{0,\frac 1T,\frac 2T,\ldots,1\}$
\For{$t=1,2,\ldots,T$}
\State For any $p\in \candidateSet$, let $w_t(p)$ be the probability of $\mA^{\text{OL}}$ playing $p$
\State Calculate the exploration probability for all intervals $I\in \mT$ (uniform if $Z_t=0$):
\algeq[eq:exploration distribution]{$\displaystyle \mu_t(I):=\frac 1{Z_t}\sum_{p:\mI_\mT[0,p)\ni I} w_t(p) p^2,~\forall I\in \mT;\quad \text{where }Z_t:=\sum_{I\in \mT} \sum_{p:\mI_\mT[0,p)\ni I} w_t(p) p^2$}
\State Sample exploitation price $p_t\sim \bm w_t$ and exploration interval $I_t\sim \bm \mu_t$
\State Offer the size-3 menu built from putting $1-2\gamma$ mass on $p_t$ and $\gamma$ mass on each endpoint of $I_t$, where $\gamma=\frac 1T$. Observe the interval feedback $\1_t(I_t):=\1[v_t\in I_t]$
\State Construct the importance-weighted estimator for each tree interval $I\in \mT$:
\algeq[eq:importance weighting]{$\displaystyle \hat \1_t(I):=\frac{\1[I_t=I]\1_t(I_t)}{\mu_t(I)},\quad \forall I\in \mT.$}
\State For each candidate price $p\in \candidateSet$, feed $-\hat R_t(p)$ as their round-$t$ loss to $\mA^{\text{OL}}$, where
\algeq[eq:income estimate]{$\displaystyle \hat R_t(p):=p\left (1-\sum_{I\in \mI_\mT[0,p)} \hat \1_t(I)\right )\in (-\infty,1],\quad \forall p\in \candidateSet.$}
\EndFor
\end{algorithmic}
\end{algorithm}

In \Cref{eq:exploration distribution}, we thus set $\mu_t(I)$ proportional to $\sum_{p:\mI_\mT[0,p)\ni I}w_t(p)p^2$ so they cancel (only with a normalization $Z_t$ remaining). Then the last term becomes $2Z_t\sum_{I\in \mT}\1_t(I)$, and we further have \textit{i)} $Z_t\le 2(\log_2T+1)$ because each $\mI_\mT[0,p)$ contains at most two tree intervals of each depth; and \textit{ii)} $\sum_{I\in \mT}\1_t(I)\le \log_2T+1$ because $v_t$ lies in at most one tree interval of each depth. To summarize,
\begin{lemma}[Second Moment]\label{lem:adv second moment}
For every round $t$, we have $\E_t\bigl[\sum_{p\in \candidateSet} w_t(p)\hat R_t^2(p)\bigr]\le 2+4(\log_2T+1)^2$.
\end{lemma}

This consequently gives the following regret bound for adversarial dynamic pricing (for a comparator price $p$ with $w_t(p)=0$, we still have $\E_t[\hat R_t(p)]\ge p\1[v_t\ge p]$ by \Cref{lem:adv estimator full}).
\begin{theorem}[Adversarial Dynamic Pricing]\label{thm:adv}
For any comparator class $\mathcal C\subseteq \{0,\frac 1T,\ldots,1\}^T$, for any $\mA^{\text{OL}}$ satisfying \Cref{def:online learning adv} with $K=T+1$ and $\mathcal C$ translated to $[K]$, \Cref{alg:adv size-3} ensures
\begin{equation*}
\E\left[\sum_{t=1}^T p_t^\ast\1[v_t\ge p_t^\ast]\right]-\E\left[\sum_{t=1}^T\text{revenue}_t\right]\le \textsc{Penalty}(\mathcal C,T,T+1,\eta)+\O\bigl(\eta T\log^2 T+1\bigr),~\forall (p_1^\ast,\ldots,p_T^\ast)\in \mathcal C.
\end{equation*}
In particular, Hedge \citep{freund1997decision} with $\eta=(T\log T)^{-1/2}$ gives $\O\bigl(\sqrt{T\log^3 T}\bigr)$ regret against the best fixed price $\mathcal C_0$, and Fixed-Share \citep{herbster1998tracking} with $\eta=\sqrt{S/(T\log T)}$ gives $\O\Bigl(\sqrt{ST\log^3 T}\Bigr)$ against $\mathcal C_{S-1}$.
\end{theorem}

Hence size-3 menus attain $\Otil(\sqrt T)$ regret against the best fixed price, even for adaptive adversaries, which is optimal up to logarithmic factors even with i.i.d.\ values and infinite menus (\Cref{thm:minimax lower}).

It remains open whether size-2 menus can attain $\Otil(\sqrt T)$ regret against adversarial values. Unless it gives up the revenue of the round, a size-2 menu can only explore an interval with the exploitation price $p_t$ as an endpoint. How to estimate $\1[v_t\ge p]=\1[v_t\ge p_t]+\1[p\le v_t<p_t]$ for a $p<p_t$? If we use importance-weighting for both terms, the resulting estimators are unbounded from \emph{both} above and below (cf. \Cref{lem:adv estimator}), and \Cref{def:online learning adv} no longer applies. Such both-side unbounded losses have long been a challenge in online learning. While there are ad-hoc solutions for specific problems \citep[e.g.,][and the discussions therein]{dai2023refined}, we are unaware of any that applies to our \Cref{alg:adv size-3}.

We also briefly discuss why adversarial CDF estimation algorithms \citep{meister2021learning,okoroafor2023non}, where one estimates the empirical distribution $F_T$ of adversarial values $v_1,\ldots,v_T$ via threshold queries, cannot be used as a black-box. Suppose that, for every prefix $t\in [T]$, we are given an estimate $\hat F_t$ that's $\epsilon_t$-close to the current empirical distribution $F_t$. First, these existing works only give $\epsilon_t=\Otil(t^{-1/3})$ using threshold queries, and it is unclear whether we can get $\epsilon_t=\Otil(t^{-1/2})$ by interval queries (note that, our \Cref{alg:CDF interval query,alg:size-2} both rely on i.i.d.\ values). Even if we can, critically, an $\epsilon_t$ guarantee on the empirical distribution only translates to an $\epsilon_t t$ guarantee on estimating the \emph{cumulative} revenue function. However, online learning algorithms (e.g., Hedge) require estimating the \emph{per-round revenue}. To see why a cumulative guarantee is insufficient, consider a delayed online learning problem \citep{weinberger2002delayed}: Suppose that the true value $v_t$ is fully observed after $\epsilon_t t$ rounds, which also gives $\epsilon_t t$ close cumulative revenues (and stronger than the black-box guarantee), one must still suffer $\Omega\Bigl(\sqrt{\sum_{t=1}^T \epsilon_t t}\Bigr)$ regret. So even if $\epsilon_t=\Otil(t^{-1/2})$, a black-box algorithm likely only gives $\Otil(T^{3/4})$ regret. We finally remark that, while the estimator by \citet{meister2021learning} is unbiased, it has a second moment the same order as the number of candidate prices (namely $\lvert \candidateSet\rvert$), thus also only giving $\Otil(T^{2/3})$ regret.

\section{Further Discussion of Related Work}\label{app:related}

\paragraph{Dynamic Pricing.} The study of regret-minimizing dynamic pricing was initiated in \cite{kleinberg2003value} and attracted much attention afterwards. In all this work (with a notable exception of \cite{zuo2026learning}, discussed below), each customer is offered a single posted price. Minimax-optimal regret rates scale as $\tilde \Theta(T^{2/3})$ for all problem instances \citep{kleinberg2003value} and $\tilde \Theta(\sqrt T)$ for regular distributions \cite{singla2024bandit,leme2023pricing} or under a certain smoothness condition (\citep{kleinberg2003value}, see \Cref{footnote:curvature assumption in KL}). Subsequent work makes additional assumptions such as parametric structure \citep{broder2012dynamic,keskin2014dynamic} or finitely many values \citep{cesa2019dynamic}. Various extensions were considered, most notably: varying (\eg adversarial or drifting) value distributions  \citep{kleinberg2003value,contextualMAB-colt11,leme2021learning,Chara-OpRe23,AdvZooming-colt21-journal} and limited supply \citep{BZ09,babaioff2015dynamic,wang2014close,badanidiyuru2018bandits}.

Dynamic pricing was also studied \emph{beyond} the framework of one firm and one type of customers. For example, when multiple pricing agents interact with each other, competing \citep[\eg][]{bergemann1996learning,bergemann2000experimentation} and/or colluding \citep[\eg][]{calvano2020artificial,klein2021autonomous,hansen2021frontiers}; or when the chosen price(s) connect buyers and sellers, namely the bilateral trade \citep[\eg][]{cesa2024bilateral,cesa2023repeated,chen25tight}.

\paragraph{Lipschitz Bandits} (see \Cref{fn:LipMAB}) are closely related to dynamic pricing with a single posted price. Lipschitz bandits with real-valued actions, a.k.a.\ continuum-armed bandits \citep{agrawal-bandits-95,Bobby-nips04}, have the same minimax-optimal regret rate as dynamic pricing, $\tilde\Theta(T^{2/3})$, and this rate is obtained by the same approach: uniform discretization of the action space. \emph{Adaptive} discretization of the action space \citep{kleinberg2019bandits,bubeck2011x}, a major theme in Lipschitz bandits, seamlessly carries over to dynamic pricing as well. This approach leads to improved guarantees in terms of \emph{zooming dimension}, an instance-dependent quantity which (for $\mathbb R^d$-valued actions) is $d$ in the worst case and gets smaller if near-optimal actions comprise a low-dimensional region of the action space. Adaptive discretization and the concept of zooming dimension have been extended to contextual bandits \citep{contextualMAB-colt11,SmoothedRegret-colt19}, adversarial bandits \citep{AdvZooming-colt21-journal}, and reinforcement learning \citep{sinclair2019adaptive,cao2020provably}. More background on Lipschitz bandits can be found in \citep[Ch. 4]{slivkins-MABbook}.

Our notion of ``interval dimension'' is a variant of zooming dimension which is more stringent but retains instance-dependence. Our main algorithm performs adaptive discretization, albeit in a very different way. Interestingly, we need adaptive discretization even for our minimax result for size-2 menus on general distributions (even though uniform discretization suffices for the corresponding minimax result \emph{without} menus).

\paragraph{Bandits Lower Bounds.} Regret lower bounds in bandit problems, including those for dynamic pricing, must deal with the fundamental issue that they cannot consider a single problem instance (since they need to rule out an algorithm tailored to this problem instance). The most common approach, starting from \citep{auer2002nonstochastic}, fixes the time horizon $T$ and takes the worst case over some instance class.\footnote{Various lower bounds of this type were proved for dynamic pricing \citep{kleinberg2003value,babaioff2015dynamic} and Lipschitz bandits \citep[\eg][]{Bobby-nips04,contextualMAB-colt11}, as well as for other bandit structures such as linear bandits \citep[\eg][]{DaniHK-nips07,DaniHK-colt08}.} Some results for dynamic pricing and Lipschitz bandits refine this approach and construct problem instances that are difficult for infinitely many $T$'s at once, ruling out algorithms that attain a favorable regret rate after some initial time period \citep{kleinberg2003value,kleinberg2019bandits}.

A standard approach for \emph{per-instance} lower bounds, starting from \citep{lai1985asymptotically}, focuses on a particular problem instance but further requires non-trivial performance (\eg $o(T)$ regret) on several hard ``decoy'' instances. The quantity being lower-bounded is \emph{not} the dependence on $T$ (like in minimax lower bounds) but instead the ``leading constant'': the $C$ in a regret bound of the form $C\cdot f(T) + o(f(T))$. \cite{Combes-colt14} obtain such a result for Lipschitz bandits. Moreover, this approach fails under full-feedback problems, as detailed in \Cref{sec:instance}.

Our per-instance lower bounds, on the other hand, follow a different template (called \emph{local-minimax}) whereby the decoys are vanishingly close to the original problem instance and are subject to the same lower bound on regret. While this template was previously applied in \citep{cai2015framework,zhu2016local} to estimating convex functions from samples, their techniques do not apply to our setting.

\paragraph{Mechanism Design with Menus.} Closest to us, \citep{zuo2026learning} shows that for a buyer with a \emph{fixed but unknown value} (i.e., without randomness), menus avoid the $\Theta(\log\log T)$ regret of posted prices, and size-2 menus attain the optimal regret even if the buyer is strategic. While they also consider size-2 menus, their fixed value setup is very different from our dynamic pricing problem. \citet{han2024learning} also learn fixed type agents with menus: In online principal-agent problems, the agent's choice from a menu identifies the type in fewer rounds than a single strategy, \eg in a single round for pricing with an infinite menu. In other mechanism problems, menus can sometimes help elicit more revenue even if the agent's type distribution is known, \eg when selling multiple goods \citep{hart2019selling,BABAIOFF2022281} or in contract design \citep{guruganesh2023power,castiglioni2022designing,gan2022generalized}. In our single-item problem, however, no menu earns more than posting the monopoly price when $F$ is known \citep{myerson1981optimal}, and therefore menus are only helpful in our online dynamic pricing setup where types must be learned.

\paragraph{Sample and Query Complexity in Pricing.} Following the framework of \citet{cole2014sample}, \citet{huang2015making} characterize the sample complexity of pricing a single buyer, e.g., $\tilde \Theta(\epsilon^{-1.5})$ for MHR distributions. Later works develop refined multi-buyer bounds and extend to posted pricing \citep{guo2019settling,guo2021generalizing,jin2024sample}. \citet{leme2023pricing} introduce threshold (i.e., pricing) queries and separate them from samples, e.g., $\tilde \Theta(\epsilon^{-3})$ threshold queries versus $\O(\epsilon^{-2})$ samples for a general CDF; \citet{leme2023description} learn regular CDFs with $\Otil(\epsilon^{-2.5})$ threshold queries via a $\tilde \Theta(\epsilon^{-1/2})$-bit description; \citet{chen2026query} show that regularity no longer helps threshold queries for uniform pricing with multiple buyers; and \citet{tang2026pricing} study multiplicative approximations. With adversarial values, the median and the empirical CDF can be estimated with one threshold query per value \citep{meister2021learning,okoroafor2023non}; see the end of \Cref{sec:adv} for more discussions. We introduce interval queries, which lie between samples and threshold queries, and show that they are as good as samples for the CDF and for the monopoly price of all four classes.

\section{Conclusion and Future Research}\label{sec:conclusion}

We show that for dynamic pricing under unknown demands, a second offer is almost as good as infinite offers. This holds not only in the minimax sense over several distribution classes, but also locally around every value distribution. We also study the query-complexity version, where we show that between the extremes of threshold queries and value queries, the interval queries are almost as powerful as value queries. We also generalize to adversarial dynamic pricing, where our worst-case regret bound, attained by size-3 menus, is again as good as the one for infinite menus.

An immediate next-step question is whether \emph{size-2} menus are also as powerful as infinite menus in adversarial dynamic pricing (whose challenges are outlined at the end of \Cref{sec:adv}). Another direction is to extend beyond a single myopic customer per round: \eg to  multiple simultaneous customers with heterogeneous types, or to long-lived customers strategically optimizing their long-term payoffs.

\printbibliography

@inproceedings{kleinberg2003value,
  title={The Value of Knowing a Demand Curve: Bounds on Regret for Online Posted-Price Auctions},
  author={Kleinberg, Robert and Leighton, Tom},
  booktitle={Proceedings of the 44th Annual IEEE Symposium on Foundations of Computer Science},
  pages={594},
  year={2003}
}

@inproceedings{leme2023pricing,
  title={Pricing Query Complexity of Revenue Maximization},
  author={Leme, Renato Paes and Sivan, Balasubramanian and Teng, Yifeng and Worah, Pratik},
  booktitle={Proceedings of the 2023 Annual ACM-SIAM Symposium on Discrete Algorithms (SODA)},
  pages={399--415},
  year={2023},
  organization={SIAM}
}

@inproceedings{huang2015making,
  title={Making the Most of Your Samples},
  author={Huang, Zhiyi and Mansour, Yishay and Roughgarden, Tim},
  booktitle={Proceedings of the Sixteenth ACM Conference on Economics and Computation},
  pages={45--60},
  year={2015}
}

@article{dvoretzky1956asymptotic,
  title={Asymptotic Minimax Character of the Sample Distribution Function and of the Classical Multinomial Estimator},
  author={Dvoretzky, Aryeh and Kiefer, Jack and Wolfowitz, Jacob},
  journal={The Annals of Mathematical Statistics},
  pages={642--669},
  year={1956},
  publisher={JSTOR}
}

@article{broder2012dynamic,
  title={Dynamic Pricing under a General Parametric Choice Model},
  author={Broder, Josef and Rusmevichientong, Paat},
  journal={Operations Research},
  volume={60},
  number={4},
  pages={965--980},
  year={2012},
  publisher={INFORMS}
}

@article{keskin2014dynamic,
  title={Dynamic Pricing with an Unknown Demand Model: Asymptotically Optimal Semi-Myopic Policies},
  author={Keskin, N Bora and Zeevi, Assaf},
  journal={Operations Research},
  volume={62},
  number={5},
  pages={1142--1167},
  year={2014},
  publisher={INFORMS}
}

@inproceedings{cesa2019dynamic,
  title={Dynamic Pricing with Finitely Many Unknown Valuations},
  author={Cesa-Bianchi, Nicolo and Cesari, Tommaso and Perchet, Vianney},
  booktitle={Algorithmic Learning Theory},
  pages={247--273},
  year={2019},
  organization={PMLR}
}

@inproceedings{han2024learning,
  title={Learning in Online Principal-Agent Interactions: The Power of Menus},
  author={Han, Minbiao and Albert, Michael and Xu, Haifeng},
  booktitle={Proceedings of the AAAI Conference on Artificial Intelligence},
  volume={38},
  number={16},
  pages={17426--17434},
  year={2024}
}

@inproceedings{gan2022generalized,
  title={Generalized Principal-Agency: Contracts, Information, Games and Beyond},
  author={Gan, Jiarui and Han, Minbiao and Wu, Jibang and Xu, Haifeng},
  booktitle={Proceedings of the 20th Conference on Web and Internet Economics},
  year={2024}
}

@inproceedings{castiglioni2022designing,
  title={Designing Menus of Contracts Efficiently: The Power of Randomization},
  author={Castiglioni, Matteo and Marchesi, Alberto and Gatti, Nicola},
  booktitle={Proceedings of the 23rd ACM Conference on Economics and Computation},
  pages={705--735},
  year={2022}
}

@inproceedings{guruganesh2023power,
  title={The Power of Menus in Contract Design},
  author={Guruganesh, Guru and Schneider, Jon and Wang, Joshua and Zhao, Junyao},
  booktitle={Proceedings of the 24th ACM Conference on Economics and Computation},
  pages={818--848},
  year={2023}
}

@article{wang2014close,
  title={Close the Gaps: A Learning-While-Doing Algorithm for Single-Product Revenue Management Problems},
  author={Wang, Zizhuo and Deng, Shiming and Ye, Yinyu},
  journal={Operations Research},
  volume={62},
  number={2},
  pages={318--331},
  year={2014},
  publisher={INFORMS}
}

@article{badanidiyuru2018bandits,
  title={Bandits with Knapsacks},
  author={Badanidiyuru, Ashwinkumar and Kleinberg, Robert and Slivkins, Aleksandrs},
  journal={Journal of the ACM (JACM)},
  volume={65},
  number={3},
  pages={1--55},
  year={2018},
  publisher={ACM New York, NY, USA}
}

@inproceedings{feng2025beyond,
  title={Beyond Regularity: Simple versus Optimal Mechanisms, Revisited},
  author={Feng, Yiding and Jin, Yaonan},
  booktitle={2025 IEEE 66th Annual Symposium on Foundations of Computer Science (FOCS)},
  pages={303--342},
  year={2025},
  organization={IEEE}
}

@inproceedings{chen2026query,
  title={The Query Complexity of Uniform Pricing},
  author={Chen, Houshuang and Jin, Yaonan and Lu, Pinyan and Zhang, Chihao},
  booktitle={Proceedings of the ACM Web Conference 2026},
  pages={237--248},
  year={2026}
}

@inproceedings{leme2023description,
  title={Description Complexity of Regular Distributions},
  author={Leme, Renato Paes and Sivan, Balasubramanian and Teng, Yifeng and Worah, Pratik},
  booktitle={Proceedings of the 24th ACM Conference on Economics and Computation},
  pages={959},
  year={2023}
}

@inproceedings{maurer2009empirical,
  title={Empirical Bernstein Bounds and Sample-Variance Penalization},
  author={Maurer, Andreas and Pontil, Massimiliano},
  booktitle={Conference on Learning Theory},
  year={2009}
}

@inproceedings{dai2023refined,
  title={Refined Regret for Adversarial MDPs with Linear Function Approximation},
  author={Dai, Yan and Luo, Haipeng and Wei, Chen-Yu and Zimmert, Julian},
  booktitle={International Conference on Machine Learning},
  pages={6726--6759},
  year={2023},
  organization={PMLR}
}

@article{babaioff2015dynamic,
  title={Dynamic Pricing with Limited Supply},
  author={Babaioff, Moshe and Dughmi, Shaddin and Kleinberg, Robert and Slivkins, Aleksandrs},
  journal={ACM Transactions on Economics and Computation},
  volume={3},
  number={1},
  year={2015},
  publisher={ACM New York, NY, USA}
}

@inproceedings{singla2024bandit,
  title={Bandit Sequential Posted Pricing via Half-Concavity},
  author={Singla, Sahil and Wang, Yifan},
  booktitle={Proceedings of the 25th ACM Conference on Economics and Computation},
  pages={922--939},
  year={2024}
}

@article{myerson1981optimal,
  title={Optimal Auction Design},
  author={Myerson, Roger B},
  journal={Mathematics of Operations Research},
  volume={6},
  number={1},
  pages={58--73},
  year={1981},
  publisher={INFORMS}
}

@article{barlow1963properties,
  title={Properties of Probability Distributions with Monotone Hazard Rate},
  author={Barlow, Richard E and Marshall, Albert W and Proschan, Frank},
  journal={The Annals of Mathematical Statistics},
  pages={375--389},
  year={1963},
  publisher={JSTOR}
}

@article{kleinberg2019bandits,
  title={Bandits and Experts in Metric Spaces},
  author={Kleinberg, Robert and Slivkins, Aleksandrs and Upfal, Eli},
  journal={Journal of the ACM (JACM)},
  volume={66},
  number={4},
  pages={1--77},
  year={2019},
  publisher={ACM New York, NY, USA},
  note     = "Merged and revised version of conference papers
    in {\em ACM STOC 2008} and {\em ACM-SIAM SODA 2010}"
}

@article{bubeck2011x,
  title={X-Armed Bandits},
  author={Bubeck, S{\'e}bastien and Munos, R{\'e}mi and Stoltz, Gilles and Szepesv{\'a}ri, Csaba},
  journal={Journal of Machine Learning Research},
  volume={12},
  number={5},
  year={2011}
}

@article{cai2015framework,
  title={A Framework for Estimation of Convex Functions},
  author={Cai, T Tony and Low, Mark G},
  journal={Statistica Sinica},
  pages={423--456},
  year={2015},
  publisher={JSTOR}
}

@article{zhu2016local,
  title={Local Minimax Complexity of Stochastic Convex Optimization},
  author={Zhu, Yuancheng and Chatterjee, Sabyasachi and Duchi, John and Lafferty, John},
  journal={Advances in Neural Information Processing Systems},
  volume={29},
  year={2016}
}

@article{lai1985asymptotically,
  title={Asymptotically Efficient Adaptive Allocation Rules},
  author={Lai, Tze Leung and Robbins, Herbert},
  journal={Advances in Applied Mathematics},
  volume={6},
  number={1},
  pages={4--22},
  year={1985},
  publisher={Academic Press, Inc. Orlando, FL, USA}
}

@article{zuo2026learning,
  title={Learning Is Revelation in Disguise: Optimal Regret and Equivalence Results for Dynamic Pricing},
  author={Zuo, Shiliang},
  journal={arXiv Preprint arXiv:2604.24093},
  year={2026}
}

@article{tang2026pricing,
  title={Pricing Query Complexity of Multiplicative Revenue Approximation},
  author={Tang, Wei and Wang, Yifan and Zhang, Mengxiao},
  journal={arXiv Preprint arXiv:2602.10483},
  year={2026}
}

@InProceedings{chen25tight,
  author =  {Chen, Houshuang and Jin, Yaonan and Lu, Pinyan and Zhang, Chihao},
  title = {{Tight Regret Bounds for Fixed-Price Bilateral Trade}},
  booktitle = {53rd International Colloquium on Automata, Languages, and Programming (ICALP 2026)},
  pages = {57:1--57:21},
  year =  {2026},
  volume =  {374}
}

@article{jin2024sample,
  title={Sample Complexity of Posted Pricing for a Single Item},
  author={Jin, Billy and Kesselheim, Thomas and Ma, Will and Singla, Sahil},
  journal={Advances in Neural Information Processing Systems},
  volume={37},
  pages={82296--82317},
  year={2024}
}

@inproceedings{guo2021generalizing,
  title={Generalizing Complex Hypotheses on Product Distributions: Auctions, Prophet Inequalities, and Pandora's Problem},
  author={Guo, Chenghao and Huang, Zhiyi and Tang, Zhihao Gavin and Zhang, Xinzhi},
  booktitle={Conference on Learning Theory},
  pages={2248--2288},
  year={2021},
  organization={PMLR}
}

@inproceedings{leme2021learning,
  title={Learning to Price against a Moving Target},
  author={Leme, Renato Paes and Sivan, Balasubramanian and Teng, Yifeng and Worah, Pratik},
  booktitle={International Conference on Machine Learning},
  pages={6223--6232},
  year={2021},
  organization={PMLR}
}

@inproceedings{meister2021learning,
  title={Learning with Comparison Feedback: Online Estimation of Sample Statistics},
  author={Meister, Michela and Nietert, Sloan},
  booktitle={Algorithmic Learning Theory},
  pages={983--1001},
  year={2021},
  organization={PMLR}
}

@inproceedings{okoroafor2023non,
  title={Non-Stochastic CDF Estimation Using Threshold Queries},
  author={Okoroafor, Princewill and Gupta, Vaishnavi and Kleinberg, Robert and Goh, Eleanor},
  booktitle={Proceedings of the 2023 Annual ACM-SIAM Symposium on Discrete Algorithms (SODA)},
  pages={3551--3572},
  year={2023},
  organization={SIAM}
}

@inproceedings{cole2014sample,
  title={The Sample Complexity of Revenue Maximization},
  author={Cole, Richard and Roughgarden, Tim},
  booktitle={Proceedings of the Forty-Sixth Annual ACM Symposium on Theory of Computing},
  pages={243--252},
  year={2014}
}

@inproceedings{guo2019settling,
  title={Settling the Sample Complexity of Single-Parameter Revenue Maximization},
  author={Guo, Chenghao and Huang, Zhiyi and Zhang, Xinzhi},
  booktitle={Proceedings of the 51st Annual ACM SIGACT Symposium on Theory of Computing},
  pages={662--673},
  year={2019}
}

@article{hart2019selling,
  title={Selling Multiple Correlated Goods: Revenue Maximization and Menu-Size Complexity},
  author={Hart, Sergiu and Nisan, Noam},
  journal={Journal of Economic Theory},
  volume={183},
  pages={991--1029},
  year={2019},
  publisher={Elsevier}
}

@article{BABAIOFF2022281,
title = {The Menu-Size Complexity of Revenue Approximation},
journal = {Games and Economic Behavior},
volume = {134},
pages = {281--307},
year = {2022},
author = {Moshe Babaioff and Yannai A. Gonczarowski and Noam Nisan}
}

@article{freund1997decision,
  title={A Decision-Theoretic Generalization of On-Line Learning and an Application to Boosting},
  author={Freund, Yoav and Schapire, Robert E},
  journal={Journal of Computer and System Sciences},
  volume={55},
  number={1},
  pages={119--139},
  year={1997},
  publisher={Elsevier}
}

@article{herbster1998tracking,
  title={Tracking the Best Expert},
  author={Herbster, Mark and Warmuth, Manfred K},
  journal={Machine Learning},
  volume={32},
  number={2},
  pages={151--178},
  year={1998},
  publisher={Springer}
}

@article{auer2002nonstochastic,
  title={The Nonstochastic Multiarmed Bandit Problem},
  author={Auer, Peter and Cesa-Bianchi, Nicolo and Freund, Yoav and Schapire, Robert E},
  journal={SIAM Journal on Computing},
  volume={32},
  number={1},
  pages={48--77},
  year={2002},
  publisher={SIAM}
}

@inproceedings{zinkevich2003online,
  title={Online Convex Programming and Generalized Infinitesimal Gradient Ascent},
  author={Zinkevich, Martin},
  booktitle={Proceedings of the 20th International Conference on Machine Learning},
  pages={928--936},
  year={2003}
}

@article{weinberger2002delayed,
  title={On Delayed Prediction of Individual Sequences},
  author={Weinberger, Marcelo J and Ordentlich, Erik},
  journal={IEEE Transactions on Information Theory},
  volume={48},
  number={7},
  pages={1959--1976},
  year={2002},
  publisher={IEEE}
}

@article{sweezy1939demand,
  title={Demand under Conditions of Oligopoly},
  author={Sweezy, Paul M},
  journal={Journal of Political Economy},
  volume={47},
  number={4},
  pages={568--573},
  year={1939},
  publisher={The University of Chicago Press}
}

@article{hall1939price,
 author = {R. L. Hall and C. J. Hitch},
 journal = {Oxford Economic Papers},
 number = {2},
 pages = {12--45},
 publisher = {Oxford University Press},
 title = {Price Theory and Business Behaviour},
 year = {1939}
}

@article{bergemann1996learning,
  title={Learning and Strategic Pricing},
  author={Bergemann, Dirk and V{\"a}lim{\"a}ki, Juuso},
  journal={Econometrica},
  volume={64},
  number={5},
  pages={1125--1149},
  year={1996}
}

@article{bergemann2000experimentation,
  title={Experimentation in Markets},
  author={Bergemann, Dirk and V{\"a}lim{\"a}ki, Juuso},
  journal={The Review of Economic Studies},
  volume={67},
  number={2},
  pages={213--234},
  year={2000}
}

@article{calvano2020artificial,
  title={Artificial Intelligence, Algorithmic Pricing, and Collusion},
  author={Calvano, Emilio and Calzolari, Giacomo and Denicolo, Vincenzo and Pastorello, Sergio},
  journal={American Economic Review},
  volume={110},
  number={10},
  pages={3267--3297},
  year={2020},
  publisher={American Economic Association, 2014 Broadway, Suite 305, Nashville, TN 37203}
}

@article{klein2021autonomous,
  title={Autonomous Algorithmic Collusion: Q-Learning under Sequential Pricing},
  author={Klein, Timo},
  journal={The RAND Journal of Economics},
  volume={52},
  number={3},
  pages={538--558},
  year={2021},
  publisher={Wiley Online Library}
}

@article{hansen2021frontiers,
  title={Frontiers: Algorithmic Collusion: Supra-Competitive Prices via Independent Algorithms},
  author={Hansen, Karsten T. and Misra, Kanishka and Pai, Mallesh M.},
  journal={Marketing Science},
  volume={40},
  number={1},
  pages={1--12},
  year={2021},
  publisher={INFORMS}
}

@article{cesa2024bilateral,
  title={Bilateral Trade: A Regret Minimization Perspective},
  author={Cesa-Bianchi, Nicol{\`o} and Cesari, Tommaso and Colomboni, Roberto and Fusco, Federico and Leonardi, Stefano},
  journal={Mathematics of Operations Research},
  volume={49},
  number={1},
  pages={171--203},
  year={2024},
  publisher={INFORMS}
}

@inproceedings{cesa2023repeated,
  title={Repeated Bilateral Trade against a Smoothed Adversary},
  author={Cesa-Bianchi, Nicol{\`o} and Cesari, Tommaso R. and Colomboni, Roberto and Fusco, Federico and Leonardi, Stefano},
  booktitle={Proceedings of the Thirty Sixth Conference on Learning Theory},
  pages={1095--1130},
  year={2023},
  organization={PMLR}
}

@book{LS19bandit-book,
  title={Bandit Algorithms},
  author={Lattimore, Tor and Szepesv{\'a}ri, Csaba},
  year={2020},
  publisher={Cambridge University Press}
}

@article{slivkins-MABbook,
  title={Introduction to Multi-Armed Bandits},
  author={Slivkins, Aleksandrs},
  journal={Foundations and Trends$\circledR$ in Machine Learning},
  volume={12},
  number={1-2},
  pages={1--286},
  year={2019},
  note={Published with \emph{Now Publishers} (Boston, MA, USA). Also available at {\tt https://arxiv.org/abs/1904.07272}}
}

@article{contextualMAB-colt11,
  title={Contextual Bandits with Similarity Information},
  author={Slivkins, Aleksandrs},
  journal={Journal of Machine Learning Research},
  volume={15},
  number={1},
  pages={2533--2568},
  year={2014},
  note={Preliminary version in \emph{COLT 2011}}
}

@misc{AdvZooming-colt21-journal,
  title={Adaptive Discretization against an Adversary: Lipschitz Bandits, Dynamic Pricing, and Auction Tuning},
  author={Podimata, Chara and Slivkins, Aleksandrs},
  year={2025},
  note={Working paper},
  addendum={Preliminary version in \emph{COLT 2021}}
}

@article{BZ09,
  title={Dynamic Pricing without Knowing the Demand Function: Risk Bounds and Near-Optimal Algorithms},
  author={Besbes, Omar and Zeevi, Assaf},
  journal={Operations Research},
  volume={57},
  number={6},
  pages={1407--1420},
  year={2009},
  publisher={INFORMS}
}

@article{agrawal-bandits-95,
  title={The Continuum-Armed Bandit Problem},
  author={Agrawal, Rajeev},
  journal={SIAM Journal on Control and Optimization},
  volume={33},
  number={6},
  pages={1926--1951},
  year={1995},
  publisher={SIAM}
}

@article{Bobby-nips04,
  title={Nearly Tight Bounds for the Continuum-Armed Bandit Problem},
  author={Kleinberg, Robert},
  journal={Advances in Neural Information Processing Systems},
  volume={17},
  year={2004}
}

@article{DaniHK-nips07,
  title={The Price of Bandit Information for Online Optimization},
  author={Dani, Varsha and Hayes, Thomas P. and Kakade, Sham},
  journal={Advances in Neural Information Processing Systems},
  volume={20},
  year={2007}
}

@inproceedings{DaniHK-colt08,
  title={Stochastic Linear Optimization under Bandit Feedback},
  author={Dani, Varsha and Hayes, Thomas P. and Kakade, Sham},
  booktitle={Conference on Learning Theory},
  pages={355--366},
  year={2008}
}

@inproceedings{Combes-colt14,
  title={Lipschitz Bandits: Regret Lower Bound and Optimal Algorithms},
  author={Magureanu, Stefan and Combes, Richard and Proutiere, Alexandre},
  booktitle={Conference on Learning Theory},
  pages={975--999},
  year={2014},
  organization={PMLR}
}

@article{cao2020provably,
  title={Provably Adaptive Reinforcement Learning in Metric Spaces},
  author={Cao, Tongyi and Krishnamurthy, Akshay},
  journal={Advances in Neural Information Processing Systems},
  volume={33},
  pages={9736--9744},
  year={2020}
}

@article{sinclair2019adaptive,
  title={Adaptive Discretization for Episodic Reinforcement Learning in Metric Spaces},
  author={Sinclair, Sean R and Banerjee, Siddhartha and Yu, Christina Lee},
  journal={Proceedings of the ACM on Measurement and Analysis of Computing Systems},
  volume={3},
  number={3},
  pages={1--44},
  year={2019},
  publisher={ACM, New York, NY, USA}
}

@article{Chara-OpRe23,
  title={Contextual Search in the Presence of Adversarial Corruptions},
  author={Krishnamurthy, Akshay and Lykouris, Thodoris and Podimata, Chara and Schapire, Robert E.},
  journal={Operations Research},
  volume={71},
  number={4},
  pages={1120--1135},
  year={2023},
  publisher={INFORMS},
  note={Preliminary version in \emph{STOC 2021}}
}

@article{SmoothedRegret-colt19,
  title={Contextual Bandits with Continuous Actions: Smoothing, Zooming, and Adapting},
  author={Krishnamurthy, Akshay and Langford, John and Slivkins, Aleksandrs and Zhang, Chicheng},
  journal={Journal of Machine Learning Research},
  volume={21},
  number={137},
  pages={1--45},
  year={2020},
  note={Preliminary version at \emph{COLT 2019}}
}

\newpage
\appendix
\renewcommand{\appendixpagename}{\centering \LARGE Technical Appendices}
\appendixpage
\addtocontents{toc}{\protect\stopmaintoc}

\startcontents[section]
\printcontents[section]{l}{1}{\setcounter{tocdepth}{2}\let\stopmaintoc\relax}

\section{Warm-Up: Infinite Menus}\label{app:infinite}

This appendix proves our claims on infinite menus. The infinite menu built from a price distribution earns the same revenue as posting a random price from it, and reveals the value if this distribution has full support on $[0,1]$ (\Cref{lem:general Myerson}), which also proves \Cref{lem:Myerson prices} as a special case. Conversely, no menu earns more than a random posted price (\Cref{lem:menu revenue}). Hence infinite menus are as powerful as full feedback (\Cref{prop:infinite menus full feedback}), and their minimax regret is governed by the sample complexity (\Cref{reduction:infinite to samples}), which gives \Cref{thm:infinite menus}.

\begin{lemma}[Infinite Menus as Randomized Prices]\label{lem:general Myerson}
Let $\priceDist$ be a price distribution over $[0,1]$, and let $q(x):=\Pr_{p\sim \priceDist}\{p\le x\}$ for $x\in [0,1]$. Consider the following infinite menu
\begin{equation*}
M:=\left\{\left(q(x),\pi(x)\right)\middle| x\in [0,1]\right\},\quad \text{where }\pi(x):=q(x)\cdot x-\int_0^x q(y)~dy.
\end{equation*}
Then a customer with value $v$ chooses the option with $x=v$, which gives the highest payoff (``optimal''), and has (one of) the largest allocation probability among all optimal options. Hence:
\begin{itemize}
\item If $\priceDist$ has full support on $[0,1]$, the seller infers $v$ from the chosen option;
\item If $\priceDist$ has a finite support $p_1<p_2<\cdots<p_m$, the seller observes which of $[0,p_1),[p_1,p_2),\ldots,[p_m,1]$ contains $v$.
\item The customer gets the item with probability $q(v)=\Pr_{p\sim \priceDist}\{v\ge p\}$ and pays $\pi(v)=\E_{p\sim \priceDist}[p\1[v\ge p]]$, the same as posting a random price $p\sim \priceDist$.
\end{itemize}
\end{lemma}

\begin{proof}
We first show that the option with $x=v$ is optimal, then that it wins the tie-breaking, and finally compute what the seller observes and what the customer pays.

\textit{Optimality.} For any $x,v\in [0,1]$, the utility $U(q',\pi';v)=q'v-\pi'$ of an option $(q',\pi')$ satisfies
\begin{equation*}
U(q(v),\pi(v);v)-U(q(x),\pi(x);v)=\int_0^v q(y)~dy-\int_0^x q(y)~dy-q(x)(v-x)=\int_x^v \left(q(y)-q(x)\right)~dy\ge 0,
\end{equation*}
where the inequality holds as $q$ is non-decreasing, so $q(y)-q(x)$ is non-negative for $y>x$ and non-positive for $y<x$. Moreover, $U(q(v),\pi(v);v)=\int_0^v q(y)~dy\ge 0=U(0,0;v)$, so the option with $x=v$ is also at least as good as the outside option.

\textit{Tie-Breaking.} Suppose that the option with some $x$ is also optimal. If $q(x)=q(v)$, then $q$ is constant between $x$ and $v$, so $\pi(x)=\pi(v)$ and the two options are the same. Otherwise, the integral above is $0$, so $q$ is constant on the open interval between $x$ and $v$, which (by the right-continuity of $q$) requires $x<v$, so $q(x)<q(v)$. Hence the option with $x=v$ has the largest allocation probability among all optimal options, and the customer chooses it.

\textit{Observation and Payment.} If $\priceDist$ has full support on $[0,1]$, then $q$ is strictly increasing, so $q(v)$ determines $v$. If $\priceDist$ is supported on $p_1<p_2<\cdots<p_m$, then $q$ is constant on each of $[0,p_1),[p_1,p_2),\ldots,[p_m,1]$ and takes different values on different intervals, so $q(v)$ tells which of them contains $v$. Finally, $q(v)=\Pr_{p\sim \priceDist}\{v\ge p\}$ by definition, and by Fubini's theorem, $\int_0^v q(y)~dy=\E_{p\sim \priceDist}[(v-p)\1[v\ge p]]$, so $\pi(v)=q(v)\cdot v-\E_{p\sim \priceDist}[(v-p)\1[v\ge p]]=\E_{p\sim \priceDist}[p\1[v\ge p]]$.
\end{proof}

\begin{lemma}[Restatement of \Cref{lem:Myerson prices}]
Fix a price distribution $\priceDist$ with discrete support $0\le p_1<\cdots<p_m\le 1$. For $i\in[m]$, let $w_i$ be the probability of $p_i$ under $\priceDist$. Construct a size-$m$ menu $M=\cbr{ \rbr{\sum_{j\le i}w_j,\;\sum_{j\le i} w_jp_j}\,\big \vert\, i\in [m] }$.

Given $M$, a customer with value $v\in [p_i,p_{i+1})$ chooses the $i$-th option, where $p_{m+1}:=\infty$, and one with $v<p_1$ chooses the outside option. Hence, the seller observes which of the $(m+1)$ intervals, $[0,p_1),[p_1,p_2),\ldots,[p_m,1]$, contains $v$; and the customer gets the item with probability $\Pr_{p\sim \priceDist}\{v\ge p\}$ and pays $\E_{p\sim \priceDist}\sbr{p\,\1[v\ge p]}$ in expectation, the same as posting a random price $p\sim \priceDist$.
\end{lemma}
\begin{proof}
Apply \Cref{lem:general Myerson} to $\priceDist=\sum_{i=1}^m w_i\delta_{p_i}$, for which $q(x)=\sum_{j:p_j\le x}w_j$. By \Cref{lem:general Myerson}, $\pi(x)=\E_{p\sim \priceDist}[p\1[x\ge p]]$, so for $x\in [p_i,p_{i+1})$, the option indexed by $x$ is $(\sum_{j\le i}w_j,\sum_{j\le i}w_jp_j)$, i.e., the $i$-th option of $M$, and for $x<p_1$, it is the outside option $(0,0)$. So $M$ together with the outside option is the infinite menu in \Cref{lem:general Myerson}, and \Cref{lem:general Myerson} gives all claims, as $\priceDist$ is supported on $p_1<p_2<\cdots<p_m$.
\end{proof}

Conversely, no menu earns more than a random posted price, value by value:

\begin{lemma}[Menu Revenue]\label{lem:menu revenue}
For every menu $M$, there is a price distribution $\priceDist$ over $[0,1]$, which only depends on $M$, such that for every $v\in [0,1]$, a customer with value $v$ pays at most $\E_{p\sim \priceDist}[p\1[v\ge p]]$ under $M$, i.e., at most the expected payment when a random price $p\sim \priceDist$ is posted. Hence, for a value $v\sim G$ and a (possibly random) menu independent of $v$, with $\priceDist$ averaged over the randomness of the menu, the expected revenue is at most $\E_{p\sim \priceDist}[R_G(p)]$, and the regret of the menu is at least $\E_{p\sim \priceDist}[R_G(p_G^\ast)-R_G(p)]$.
\end{lemma}
\begin{proof}
It suffices to consider a deterministic menu $M$, as for a random menu, we can average $\priceDist$ over its randomness. We write the payment via the utility, and then compare it with the menu in \Cref{lem:general Myerson}.

Let $q(v)$ and $\pi(v)$ be the allocation probability and the payment of the option chosen by a customer with value $v$, and let $U(v):=q(v)v-\pi(v)$ be their utility. As $U(v)$ is the maximum of $q'v-\pi'$ over the options $(q',\pi')$ in $M$ and the outside option $(0,0)$, it is convex and $1$-Lipschitz, with $q(v)$ as a subgradient at $v$. Hence $q$ is non-decreasing, $U(v)=U(0)+\int_0^v q(y)\,dy$, and $U(0)\ge 0$. Let $\tilde q$ be the right-continuous version of $q$, and let $\priceDist$ be the price distribution over $[0,1]$ with $\Pr_{p\sim \priceDist}\{p\le x\}=\tilde q(x)+1-\tilde q(1)$ for every $x\in [0,1]$, which puts the remaining probability $1-\tilde q(1)$ on price $0$ and only depends on $M$. As $\tilde q\ge q$, and they differ on at most countably many points,
\begin{equation*}
\pi(v)=q(v)v-U(0)-\int_0^v q(y)\,dy\le \tilde q(v)v-\int_0^v \tilde q(y)\,dy=\E_{p\sim \priceDist}\left[p\1[v\ge p]\right],
\end{equation*}
where the equality is the payment $\pi(v)$ in \Cref{lem:general Myerson} for this $\priceDist$, in which the constant $1-\tilde q(1)$ cancels. Finally, by Fubini's theorem, $\E_{v\sim G}[\E_{p\sim \priceDist}[p\1[v\ge p]]]=\E_{p\sim \priceDist}[p\Pr_{v\sim G}\{v\ge p\}]=\E_{p\sim \priceDist}[R_G(p)]$, so the regret is at least $R_G(p_G^\ast)-\E_{p\sim \priceDist}[R_G(p)]=\E_{p\sim \priceDist}[R_G(p_G^\ast)-R_G(p)]$.
\end{proof}

\begin{theorem}[Infinite Menus, Full Feedback, and Both]\label{prop:infinite menus full feedback}
Consider three kinds of dynamic pricing algorithms: (a) those offering infinite menus, which observe the chosen options; (b) those posting a single random price in each round $t$ and observing $v_t$ afterwards, i.e., full feedback; and (c) those offering infinite menus and also observing full feedback. Let $\mA$ be an algorithm of one kind, and let $\gamma\in (0,1]$ be an arbitrarily small but fixed constant. Then each of the other two kinds has an algorithm $\mA'$ such that, for every $T$ and every sequence of realized values $v_1,v_2,\ldots,v_T\in [0,1]$ (not necessarily i.i.d.), the expected revenue of $\mA'$ in the first $T$ rounds is at least that of $\mA$ minus $\gamma$.
\end{theorem}

\begin{proof}
Algorithms of kinds (a) and (b) are also of kind (c), as an algorithm may ignore $v_t$, and a posted price is a size-1 menu. So it suffices to turn an algorithm $\mA$ of kind (c) into one of kind (b) without loss, and into one of kind (a) with loss at most $\gamma$.

Fix the values $v_1,v_2,\ldots$. In round $t$, let $M_t$ be the menu of $\mA$, which depends on $v_1,v_2,\ldots,v_{t-1}$ and the internal randomness of $\mA$, and let $\priceDist_t$ be the price distribution given by \Cref{lem:menu revenue} for $M_t$, so that the payment of a customer with value $v$ under $M_t$ is at most $\E_{p\sim \priceDist_t}[p\1[v\ge p]]$ for every $v$. Both algorithms below compute $M_t$ in round $t$, and after round $t$, they compute the option that $v_t$ chooses from $M_t$ and feed it to $\mA$ together with $v_t$. So $\mA$ receives the same feedback as when it runs by itself, and its menus have the same distribution.
\begin{itemize}
\item Kind (b): In round $t$, post a price drawn from $\priceDist_t$ using fresh randomness. Conditioning on $M_t$, its expected revenue in round $t$ is $\E_{p\sim \priceDist_t}[p\1[v_t\ge p]]$, which is at least that under $M_t$.
\item Kind (a): In round $t$, offer the menu in \Cref{lem:general Myerson} built from $(1-\gamma_t)\priceDist_t+\gamma_t\,\text{Unif}[0,1]$, where $\gamma_t:=\gamma 2^{-t}$. As this price distribution has full support on $[0,1]$, the chosen option reveals $v_t$. Its revenue in round $t$ is at least $(1-\gamma_t)\E_{p\sim \priceDist_t}[p\1[v_t\ge p]]$, which is at least the payment under $M_t$ minus $\gamma_t$.
\end{itemize}
Summing over the first $T$ rounds, the first algorithm loses nothing compared with $\mA$, and the second one loses at most $\sum_t\gamma_t\le \gamma$.
\end{proof}

As \Cref{prop:infinite menus full feedback} holds on every sequence of values, it also holds for i.i.d.\ values from every value distribution. Hence every regret bound for one kind, whether minimax over a class, local over a pair $\{G_+,G_-\}$, or on a fixed value distribution, also holds for the other two kinds up to an arbitrarily small additive constant. In particular, for every class $\mF$, the minimax regret $\mR_\mF^{(\infty)}(T)$ of infinite menus is also that of full feedback, with or without menus.

Consequently, the minimax regret of infinite menus is governed by the sample complexity (defined in \Cref{app:complexity definitions}), as we show in \Cref{reduction:infinite to samples}.

\begin{reduction}[Revenue Maximization with Infinite Menus $\leftrightarrow$ Sample Complexity]\label{reduction:infinite to samples}
Let $\mF$ be a class of value distributions.
\begin{enumerate}[label=\textup{(\alph*)}]
\item If for every $n\ge 1$, some estimator $\hat p$ from $n$ samples has $\E[R(p^\ast)-R(\hat p)]\le \epsilon_n$ under every $F\in \mF$, then $\mR_\mF^{(\infty)}(T)\le 2+\sum_{n=1}^{T-1}\epsilon_n$.
\item Conversely, if an algorithm with infinite menus has regret at most $\epsilon T$ in $T$ rounds under every $F\in \mF$, then $\Sample_\mF(\epsilon)\le T-1$.
\end{enumerate}
In particular, for $c\ge 1$, $\epsilon_n=\Otil(n^{-1/c})$ gives $\mR_\mF^{(\infty)}(T)=\Otil(T^{1-1/c})$, and $\mR_\mF^{(\infty)}(T)=\Otil(T^{1-1/c})$ gives $\Sample_\mF(\epsilon)=\Otil(\epsilon^{-c})$.
\end{reduction}

\begin{proof}
\textit{(a) Samples to Regret.} Consider the algorithm with full feedback that posts an arbitrary price in round $1$ and, in each round $t\ge 2$, posts the estimator $\hat p_t$ from $v_1,v_2,\ldots,v_{t-1}$. Its regret is at most $1+\sum_{t=2}^T\E[R(p^\ast)-R(\hat p_t)]\le 1+\sum_{n=1}^{T-1}\epsilon_n$, so \Cref{prop:infinite menus full feedback} with $\gamma=1$ gives an algorithm with infinite menus and regret at most $2+\sum_{n=1}^{T-1}\epsilon_n$.

\textit{(b) Regret to Samples.} Given $T-1$ samples, we run the algorithm for $T$ rounds, using the $t$-th sample as $v_t$ to compute the option chosen in each round $t\le T-1$. As the menu in each round $t\le T$ only depends on $v_1,v_2,\ldots,v_{t-1}$ and the internal randomness, this gives the menus of all $T$ rounds with the same distribution as in a real run. Draw $t\in [T]$ uniformly at random, and output $\hat p\sim \priceDist_t$, where $\priceDist_t$ is given by \Cref{lem:menu revenue} for the menu in round $t$. As this menu is independent of $v_t$, \Cref{lem:menu revenue} bounds the expected revenue in round $t$ by $\E[R(\hat p)\mid t]$. Averaging over $t$, $\E[R(\hat p)]$ is at least $\frac 1T$ times the expected revenue in the $T$ rounds, which is at least $R(p^\ast)-\epsilon$. Hence $\E[R(p^\ast)-R(\hat p)]\le \epsilon$, i.e., $\Sample_\mF(\epsilon)\le T-1$.

\textit{In particular.} We have $\sum_{n<T}\Otil(n^{-1/c})=\Otil(T^{1-1/c})$ for $c\ge 1$. If $\mR_\mF^{(\infty)}(T)=\Otil(T^{1-1/c})$, then some $T=\Otil(\epsilon^{-c})$ has $\mR_\mF^{(\infty)}(T)\le \epsilon T$, so (b) gives $\Sample_\mF(\epsilon)=\Otil(\epsilon^{-c})$.
\end{proof}

\begin{theorem}[Revenue Maximization with Infinite Menus]\label{thm:infinite menus}
With infinite menus, the minimax regret, the sample complexity of the monopoly price, and that of the CDF are as follows:
\begin{enumerate}[label=\textup{(\alph*)}]
\item General and regular distributions: $\mR_\general^{(\infty)}(T),\mR_\regular^{(\infty)}(T)=\Theta(\sqrt T)$, $\Sample_\general,\Sample_\regular=\Theta(\epsilon^{-2})$, and $\Sample_\general^\CDF,\Sample_\regular^\CDF=\Theta(\epsilon^{-2})$.
\item MHR distributions: $\mR_\MHR^{(\infty)}(T)=\tilde \Theta(T^{1/3})$, $\Sample_\MHR=\tilde \Theta(\epsilon^{-1.5})$, and $\Sample_\MHR^\CDF=\Theta(\epsilon^{-2})$.
\item Distributions with curvature $\kappa\in (0,2)$: $\mR_\curvature^{(\infty)}(T)=\tilde \Theta(\kappa^{-1/3}T^{1/3})$, $\Sample_\curvature=\tilde \Theta(\kappa^{-1/2}\epsilon^{-1.5})$, and $\Sample_\curvature^\CDF=\Theta(\epsilon^{-2})$.
\end{enumerate}
\end{theorem}

\begin{proof}
All the sample complexities are in \Cref{thm:sample complexities}, so it remains to bound the regret.

\textit{Upper Bounds.} Those for $\mF_\general$ also hold for $\mF_\regular$. By \Cref{lem:empirical revenue}, some estimator from $n$ samples has $\E[R(p^\ast)-R(\hat p)]=\O(n^{-1/2})$ under every value distribution, so \Cref{reduction:infinite to samples}(a) gives $\mR_\general^{(\infty)}(T)=\O(\sqrt T)$. For MHR and curvature distributions, as infinite menus include size-2 menus, $\mR_\mF^{(\infty)}(T)\le \mR_\mF^{(2)}(T)$, so \Cref{thm:size-2} gives the upper bounds.

\textit{Lower Bounds.} Those for $\mF_\regular$ also hold for $\mF_\general$. They are in \Cref{thm:minimax lower}, where \Cref{lem:regret lower regular} gives the one for $\mF_\regular$.
\end{proof}

\section{Query Complexities}\label{app:complexities}\label{sec:setup complexities}\label{app:setup}

This appendix collects all our results on query complexities. We define the complexities in \Cref{table:complexity} (\Cref{app:complexity definitions}) and relate them to each other and to the regret (\Cref{app:complexity relations}). We then state the two lower-bound methods, used throughout this section when proving lower bounds (\Cref{app:lower methods}). We then prove the bounds in \Cref{table:complexity} for samples (\Cref{app:sample complexity}), interval queries (\Cref{app:interval complexity}), and threshold queries (\Cref{app:threshold complexity}).

\subsection{Definitions: Sample, Interval-Query, and Threshold-Query Complexities}\label{app:complexity definitions}

We define query complexity in a unified way. In each round $t$, a new customer arrives, with value $v_t\sim F$ like in dynamic pricing. The algorithm issues a \emph{query} to the customer, chosen from some class of feasible queries, and learns something about $v_t$. The goal is to learn a property of interest: either the CDF of value distribution $F$, or just the monopoly price $p^*_F$. The \emph{query complexity} is the smallest number of queries guaranteed to learn the property within a given distance $\eps>0$ (\emph{$\eps$-learn} it) whenever $F$ belongs to a given class $\mF$ of value distributions.

The precise meaning of ``$\eps$-learn" depends on the property. For monopoly-price estimation, it means to find a (randomized) price $\hat p$ whose expected revenue is $\epsilon$-optimal:\footnote{Similar to \citet{huang2015making}, we define the complexities of the monopoly price via a randomized price whose expected revenue is near-optimal (where the expectation is over both the samples, or queries, and the randomness of $\hat p$). By Markov's inequality, \Cref{eq:near-optimal monopoly price} implies $\Pr\{R(p^\ast)-R(\hat p)\le 4\epsilon\}\ge \frac 34$, a guarantee with constant success probability as the version of \citet{leme2023pricing}. This is for the ease of reduction to regret.}
\begin{equation}\label{eq:near-optimal monopoly price}
R(p^\ast)-\E[R(\hat p)]\le \epsilon,\quad \text{where $p^\ast$ is the monopoly price in \Cref{eq:monopoly price}},
\end{equation}
and for CDF estimation, it means to construct a value distribution $\hat F$ that is, with probability $\frac 34$, $\epsilon$-close to $F$ in L\'evy distance, i.e.,
\begin{equation}\label{eq:Levy}
\demand_F(x+\epsilon)-\epsilon\le \demand_{\hat F}(x)\le \demand_F(x-\epsilon)+\epsilon,\quad \forall x\in[0,1].
\end{equation}

We consider three query classes: \emph{threshold queries} that reveal $\1[v_t\ge p]$ for a given threshold $p$, \emph{value queries} that directly reveal $v_t$, and an intermediate class of \emph{interval queries} that reveal whether $v_t$ lies in a given interval $[a,b)$. The corresponding query complexity is denoted as, resp., $\Threshold_\mF(\eps)$, $\Sample_\mF(\eps)$, $\Interval_\mF(\eps)$ for monopoly-price estimation, and $\Threshold_\mF^\CDF(\eps)$, $\Sample_\mF^\CDF(\eps)$, $\Interval_\mF^\CDF(\eps)$ for CDF estimation.\footnote{Recall that \emph{sample complexity} refers to query complexity with value queries.}

Finally, we present the following fact justifying the range $\kappa\in (0,2)$ in \Cref{sec:setup}:

\begin{lemma}[Range of Curvature]\label{lem:curvature range}
The class $\mF_\curvature$ is empty for $\kappa>2$, and every $F\in \mF_{\mathrm{curv}(2)}$ has $p^\ast=\frac 12$, so $\Sample_\curvature(\epsilon)=0$ for $\kappa=2$.
\end{lemma}
\begin{proof}
Let $F\in \mF_\curvature$. Taking $p\in \{0,1\}$ in the definition of $\mF_\curvature$ gives $\kappa(p^\ast)^2\le R(p^\ast)$ and $\kappa(1-p^\ast)^2\le R(p^\ast)-R(1)\le R(p^\ast)$, where $R(p^\ast)\le p^\ast$ (so $p^\ast>0$). Hence $\kappa\le \min\{\frac 1{p^\ast},\frac{p^\ast}{(1-p^\ast)^2}\}$, which is less than $2$ unless $p^\ast=\frac 12$, as the first term is decreasing and the second is increasing in $p^\ast$, and both are $2$ at $p^\ast=\frac 12$. So $\mF_\curvature$ is empty for $\kappa>2$, and every $F\in \mF_{\mathrm{curv}(2)}$ has $p^\ast=\frac 12$, in which case outputting $\hat p=\frac 12$ needs no sample.
\end{proof}

\subsection{Relations and Reductions Between Complexities}\label{app:complexity relations}

Samples reveal more than interval queries, which reveal more than threshold queries; an accurate CDF gives a near-optimal price; an algorithm with small regret gives a near-optimal price; and conversely, an estimator of the monopoly price or of the CDF gives an algorithm with small regret. These together give the black-box reductions in \Cref{fig:reductions}, formalized as follows.

\begin{theorem}[Relations Between Complexities]\label{thm:complexity relations}
For every class $\mF$ and $\epsilon\in (0,1]$:
\begin{enumerate}[label=\textup{(\alph*)}]
\item $\Sample_\mF(\epsilon)\le \Interval_\mF(\epsilon)\le \Threshold_\mF(\epsilon)$, and $\Sample_\mF^\CDF(\epsilon)\le \Interval_\mF^\CDF(\epsilon)\le \Threshold_\mF^\CDF(\epsilon)$.
\item $\Sample_\mF(6\epsilon)\le 11\log\frac 2\epsilon\cdot \Sample_\mF^\CDF(\epsilon)$, and the same holds for interval and threshold queries.
\item If an algorithm using size-$m$ menus has $T$-round regret at most $\frac{\epsilon T}2$ under every $F\in \mF$, where $T\ge \log\frac 2\epsilon$, then the interval-query complexity $\Interval_\mF(\epsilon)\le 4(m+1)T$. Moreover, if an algorithm using posted prices has $T$-round regret at most $\epsilon T$ under every $F\in \mF$, then the threshold-query complexity $\Threshold_\mF(\epsilon)\le T$.
\item If $\Interval_\mF(\epsilon)\le n(\epsilon)$ for every $\epsilon\in (0,1]$, where $n$ is a non-increasing function, then $\mR_\mF^{(3)}(T)\le 2+2\sum_{j\ge 1}2^{-j}\min\{n(2^{-j}),T\}$. The same holds for $\Threshold_\mF$ and $\mR_\mF^{(2)}(T)$ in place of $\Interval_\mF$ and $\mR_\mF^{(3)}(T)$.
\item If $\Interval_\mF^\CDF(\epsilon)\le n(\epsilon)$ for every $\epsilon\in (0,1]$, where $n$ is a non-increasing function, then $\mR_\mF^{(3)}(T)\le 2+2\sum_{j\ge 1}2^{-j}\min\{11\log(12\cdot 2^j)\cdot n(2^{-j}/6),T\}$.
\end{enumerate}
\end{theorem}
\begin{proof}
(a) is \Cref{lem:query hierarchy}, (b) is \Cref{reduction:price to CDF}, (c) is \Cref{reduction:online to batch}, and (d) is \Cref{lem:complexity to regret}. For (e), (b) gives $\Interval_\mF(\epsilon)\le 11\log\frac{12}\epsilon\cdot n(\frac\epsilon 6)$ for every $\epsilon\in (0,1]$, which is non-increasing in $\epsilon$, so (d) applies.
\end{proof}

\begin{lemma}[Samples, Interval Queries, and Threshold Queries]\label{lem:query hierarchy}
For every class $\mF$ and $\epsilon\in (0,1]$, $\Sample_\mF(\epsilon)\le \Interval_\mF(\epsilon)\le \Threshold_\mF(\epsilon)$, and $\Sample_\mF^\CDF(\epsilon)\le \Interval_\mF^\CDF(\epsilon)\le \Threshold_\mF^\CDF(\epsilon)$.
\end{lemma}
\begin{proof}
A threshold query at $p$ is answered by the interval query on $[0,p)$, as $\1[v\ge p]=1-\1[v\in [0,p)]$, and an interval query on $[a,b)$ is answered by a sample $v$, as $\1[v\in [a,b)]$ is a function of $v$. So an algorithm using $n$ threshold queries can be run with $n$ interval queries, and one using $n$ interval queries can be run with $n$ samples, without changing the distribution of its output.
\end{proof}

\begin{reduction}[Monopoly Price $\to$ CDF]\label{reduction:price to CDF}
Suppose that $\hat F$ is $\epsilon$-close to $F$ in L\'evy distance (\Cref{eq:Levy}), and let $\hat p:=\max\{\hat p^\ast-\epsilon,0\}$, where $\hat p^\ast$ maximizes $\hat R(p):=p\Pr_{v\sim \hat F}\{v\ge p\}$. Then $R(p^\ast)-R(\hat p)\le 5\epsilon$. Consequently, $\Sample_\mF(6\epsilon)\le 11\log\frac 2\epsilon\cdot \Sample_\mF^\CDF(\epsilon)$ for every class $\mF$ and $\epsilon\in (0,1]$, and the same holds for interval and threshold queries.
\end{reduction}

\begin{proof}
We show that the estimated revenue $\hat R(\hat p^\ast)$ is at least $R(p^\ast)-3\epsilon$, and that lowering $\hat p^\ast$ by $\epsilon$ keeps at least $\hat R(\hat p^\ast)-2\epsilon$ of the true revenue. Here $\hat p^\ast$ exists, as $\hat R$ is upper semi-continuous.

If $p^\ast<2\epsilon$, then $R(p^\ast)\le p^\ast<2\epsilon$ and the claim is trivial, so we assume $p^\ast\ge 2\epsilon$. Using L\'evy distance, $\demand_{\hat F}(p^\ast-2\epsilon)\ge \demand_F(p^\ast-\epsilon)-\epsilon\ge \demand_F(p^\ast)-\epsilon$, so
\begin{equation*}
\hat R(\hat p^\ast)\ge \hat R(p^\ast-2\epsilon)=(p^\ast-2\epsilon)\demand_{\hat F}(p^\ast-2\epsilon)\ge (p^\ast-2\epsilon)\left(\demand_F(p^\ast)-\epsilon\right)\ge R(p^\ast)-3\epsilon.
\end{equation*}
If $\hat p^\ast<\epsilon$, then $R(p^\ast)\le \hat R(\hat p^\ast)+3\epsilon\le \hat p^\ast+3\epsilon<4\epsilon$ and the claim is again trivial, so we assume $\hat p^\ast\ge \epsilon$, which gives $\hat p=\hat p^\ast-\epsilon$. Again using L\'evy distance, $\demand_F(\hat p)=\demand_F(\hat p^\ast-\epsilon)\ge \demand_{\hat F}(\hat p^\ast)-\epsilon$, so
\begin{equation*}
R(\hat p)\ge \hat p\left(\demand_{\hat F}(\hat p^\ast)-\epsilon\right)\ge \hat p^\ast\demand_{\hat F}(\hat p^\ast)-2\epsilon=\hat R(\hat p^\ast)-2\epsilon.
\end{equation*}
Therefore, $R(\hat p)\ge R(p^\ast)-5\epsilon$.

To complete the proof, we use \Cref{lem:CDF boosting} stated below to ``boost'' a CDF estimator (which only succeeds with probability $\frac 34$). Specifically, let $k$ be the smallest odd integer at least $8\log\frac 2\epsilon$, so that $k\le 11\log\frac 2\epsilon$ and $e^{-k/8}\le \frac\epsilon 2$. Run the CDF estimator $k$ times on fresh samples or queries, and output the above $\hat p$ for the median $\tilde F$ of its $k$ outputs, defined in \Cref{lem:CDF boosting}. \Cref{lem:CDF boosting} then guarantees that $\tilde F$ is $\epsilon$-close to $F$ with probability $1-\frac\epsilon 2$. We hence get $\E[R(p^\ast)-R(\hat p)]\le 5\epsilon+\frac\epsilon 2\le 6\epsilon$.
\end{proof}

\begin{lemma}[Boosting CDF Estimates]\label{lem:CDF boosting}
Let $k$ be odd, and let $\hat F_1,\hat F_2,\ldots,\hat F_k$ be independent value distributions, each $\epsilon$-close to $F$ in L\'evy distance with probability at least $\frac 34$. Then their median $\tilde F$, the value distribution with $\demand_{\tilde F}(x):=\mathrm{median}\{\demand_{\hat F_1}(x),\demand_{\hat F_2}(x),\ldots,\demand_{\hat F_k}(x)\}$, is $\epsilon$-close to $F$ with probability at least $1-e^{-k/8}$.
\end{lemma}
\begin{proof}
By Hoeffding's inequality, more than $\frac k2$ of the $\hat F_i$'s are $\epsilon$-close to $F$ with probability at least $1-e^{-2k(1/4)^2}=1-e^{-k/8}$. On this event, for every $x\in [0,1]$, more than half of $\demand_{\hat F_1}(x),\ldots,\demand_{\hat F_k}(x)$ lie in $[\demand_F(x+\epsilon)-\epsilon,\demand_F(x-\epsilon)+\epsilon]$, and hence so does their median. Finally, $\tilde F$ is a value distribution, i.e., $\demand_{\tilde F}$ is non-increasing and left-continuous with $\demand_{\tilde F}(0)=1$, as the median of $k$ numbers is non-decreasing and continuous in each of them.
\end{proof}

\begin{reduction}[Interval-Query and Threshold-Query Complexities $\to$ Revenue Maximization with Finite Menus]\label{reduction:online to batch}
Suppose that an algorithm using size-$m$ menus has $T$-round regret at most $\frac{\epsilon T}2$ under every $F\in \mF$, where $m$ is a positive integer and $T\ge \log\frac 2\epsilon$. Then the interval-query complexity $\Interval_\mF(\epsilon)\le 4(m+1)T$. Moreover, if an algorithm using posted prices ($m=1$) has $T$-round regret at most $\epsilon T$ under every $F\in \mF$, then the threshold-query complexity $\Threshold_\mF(\epsilon)\le T$.
\end{reduction}
\begin{proof}
We first simulate the algorithm for $T$ rounds by queries, as detailed at the end of this proof. Then we draw $t\in [T]$ uniformly at random and output $\hat p\sim \priceDist_t$, where $\priceDist_t$ is given by \Cref{lem:menu revenue} for the menu in round $t$. As this menu is independent of $v_t$, \Cref{lem:menu revenue} bounds the expected revenue in round $t$ by $\E[R(\hat p)\mid t]$, so averaging over $t$, $\E[R(p^\ast)-R(\hat p)]$ is at most the regret divided by $T$ if all $T$ rounds are simulated. It remains to simulate the option chosen in each round by a few queries.

\textit{Options as Intervals.} As the allocation probability of the chosen option is non-decreasing in the value, and ties are broken towards larger allocation probabilities, the values choosing each of the at most $m+1$ options of a menu (including the outside option) form an interval $[a,b)$ or $[a,1]$.

\textit{Rejection Sampling.} If $m=1$, these are $[0,a)$ and $[a,1]$, so a threshold query at $a$ reveals the chosen option. Then the algorithm takes exactly $T$ threshold queries, and $\E[R(p^\ast)-R(\hat p)]\le \epsilon$. In general, we simulate each round by rejection sampling: We repeatedly pick one of these $m+1$ intervals uniformly at random and query it on a fresh sample until the answer is $1$, where querying $[a,1]$ means negating the query on $[0,a)$, and then feed the option of the last queried interval to the algorithm. Each attempt stops at an interval $I$ with probability $\frac{\intervalP_F(I)}{m+1}$, so the fed option has the same distribution as the option chosen by a fresh value, and each attempt succeeds with probability exactly $\frac 1{m+1}$, independently of the others. Hence the simulation does not change the distribution of the algorithm.

\textit{Stopping.} We stop after $4(m+1)T$ queries, and output an arbitrary price if the $T$ rounds are not all simulated by then. This only happens if fewer than $T$ of the $4(m+1)T$ attempts succeed, which by the Chernoff bound has probability at most $e^{-(3/4)^2\cdot 4T/2}\le e^{-T}\le \frac\epsilon 2$, as $T\ge \log\frac 2\epsilon$. Therefore, $\E[R(p^\ast)-R(\hat p)]\le \frac\epsilon 2+\frac\epsilon 2=\epsilon$.
\end{proof}

Conversely, an estimator using interval (or threshold) queries gives an algorithm using size-3 (or size-2) menus, as \Cref{lem:Myerson prices} lets each menu ask a query while exploiting the previous estimate:

\begin{lemma}[Revenue Maximization $\to$ Interval-Query Complexity]\label{lem:complexity to regret}
Let $\mF$ be a class, and let $n:(0,1]\to [0,\infty)$ be non-increasing. If $\Interval_\mF(\epsilon)\le n(\epsilon)$ for every $\epsilon\in (0,1]$, then for every $T\ge 1$,
\begin{equation*}
\mR_\mF^{(3)}(T)\le 2+2\sum_{j\ge 1}2^{-j}\min\left\{n(2^{-j}),T\right\}.
\end{equation*}
If $\Threshold_\mF(\epsilon)\le n(\epsilon)$ for every $\epsilon\in (0,1]$, then the same bound holds for $\mR_\mF^{(2)}(T)$.
\end{lemma}
\begin{proof}
For $j=1,2,\ldots$, let $\epsilon_j:=2^{-j}$, and fix an algorithm that uses at most $n(\epsilon_j)$ interval queries and outputs a $\hat p_j$ with $\E[R(p^\ast)-R(\hat p_j)]\le \epsilon_j$ under every $F\in \mF$. Consider the algorithm using size-3 menus that runs these algorithms one after another, in phases $j=1,2,\ldots$ until the $T$ rounds end, and answers each interval query $[a,b)$ of phase $j$ in one round by the menu built from the price distribution $\gamma\delta_a+\gamma\delta_b+(1-2\gamma)\delta_{\hat p_{j-1}}$, where $\gamma:=\frac 1T$ and $\hat p_0:=0$. By \Cref{lem:Myerson prices}, this menu reveals whether the value lies in $[a,b)$, so each phase runs its algorithm on fresh values, and the menu earns at least $(1-2\gamma)R(\hat p_{j-1})\ge R(\hat p_{j-1})-2\gamma$ in expectation. As $\hat p_{j-1}$ only depends on the earlier phases, each round of phase $j$ has expected regret at most $\epsilon_{j-1}+2\gamma=2\epsilon_j+2\gamma$, where $\epsilon_0:=1$ bounds the gap of $\hat p_0$. As phase $j$ has at most $\min\{n(\epsilon_j),T\}$ rounds, the regret is at most $2\gamma T+\sum_j2\epsilon_j\min\{n(\epsilon_j),T\}$, which is the claimed bound. For threshold queries, the same argument applies to the size-2 menu built from $\gamma\delta_a+(1-\gamma)\delta_{\hat p_{j-1}}$ for a threshold query at $a$.
\end{proof}

\subsection{Two-Point and Needle-in-a-Haystack Methods}\label{app:lower methods}

Throughout, for a value distribution $G$, its demand is $\demand_G(p):=\Pr_{v\sim G}\{v\ge p\}$, and $R_G(p_G^\ast)-R_G(p)$ is the \emph{gap} of a price $p$ under $G$. For $G\ll G'$ (i.e., $G$'s support is a subset of that of $G'$), let $\chi^2(G\|G'):=\int(\frac{dG}{dG'}-1)^2\,dG'$, so that $H^2(G,G')\le \frac 12\chi^2(G\|G')$. As $1-H^2$ is multiplicative over independent samples, $n$ samples have total variation distance at most $\sqrt{2nH^2(G,G')}$ between $G$ and $G'$.

A two-point lower bound takes two value distributions that are close but share no near-optimal price:

\begin{lemma}[Two-Point Method for Complexities]\label{lem:two-point complexity}
Let $G_+$ and $G_-$ be two value distributions in a class $\mF$. If every price has gap at least $4\epsilon$ either under $G_+$ or under $G_-$, that is,
\begin{equation*}
\max\left \{R_{G_+}(p_{G_+}^\ast)-R_{G_+}(p),R_{G_-}(p_{G_-}^\ast)-R_{G_-}(p)\right \}\ge 4\epsilon,\quad \forall p\in [0,1],
\end{equation*}
then $\Sample_\mF(\epsilon)\ge \frac 1{8H^2(G_+,G_-)}$. If no value distribution is $\epsilon$-close to both $G_+$ and $G_-$ in L\'evy distance, then $\Sample_\mF^\CDF(\epsilon)\ge \frac 1{8H^2(G_+,G_-)}$.
\end{lemma}
\begin{proof}
Prove by contradiction. Suppose that $n<\frac 1{8H^2(G_+,G_-)}$ samples suffice, so that these samples have total variation distance at most $\sqrt{2nH^2(G_+,G_-)}<\frac 12$ between $G_+$ and $G_-$. Let $\hat p$ be the randomized price found in \Cref{eq:near-optimal monopoly price}, and let $A$ be the event that $\hat p$ has gap less than $4\epsilon$ under $G_+$. By Markov's inequality, \Cref{eq:near-optimal monopoly price} gives $\Pr_{G_+}\{A\}\ge \frac 34$, and as every price in $A$ has gap at least $4\epsilon$ under $G_-$, it also gives $\Pr_{G_-}\{A\}\le \frac 14$. Then $\Pr_{G_+}\{A\}-\Pr_{G_-}\{A\}\ge \frac 12$, a contradiction. Similarly, for CDF estimation, let $\hat F$ be the value distribution found in \Cref{eq:Levy}, and let $A$ be the event that $\hat F$ is $\epsilon$-close to $G_+$. Then $A$ has probability at least $\frac 34$ under $G_+$ and at most $\frac 14$ under $G_-$, giving the same contradiction.
\end{proof}

We now introduce the following needle-in-a-haystack method \citep{kleinberg2019bandits}, which says that if one perturbs a distribution $G$ into $J$ decoys, each differing from $G$ only on a small interval (disjoint across decoys), it is hard for any algorithm using few threshold queries to tell $G$ apart from most of these decoys.

\begin{lemma}[Needle-in-a-Haystack Method]\label{lem:haystack}
Let $G,F_1,F_2,\ldots,F_J$ be value distributions and $I_1,I_2,\ldots,I_J$ be disjoint sets, such that for every $j$, $\demand_{F_j}=\demand_G$ outside $I_j$, whereas $\lvert \demand_{F_j}-\demand_G\rvert\le \eta$ and $\demand_{F_j},\demand_G\in [\sigma,1-\sigma]$ on $I_j$ for some $\sigma\in (0,\frac 12]$. If the observations of an algorithm using $n$ threshold queries have total variation distance at least $\frac 13$ between $G$ and $F_j$ for at least $\frac{3J}4$ indices $j$, then $n\ge \frac{\sigma J}{36\eta^2}$.
\end{lemma}
\begin{proof}
A threshold query at $p$ observes a Bernoulli variable with mean $\demand_G(p)$ or $\demand_{F_j}(p)$, whose KL divergence is $0$ if $p\notin I_j$, and at most $\frac{(\demand_G(p)-\demand_{F_j}(p))^2}{\demand_{F_j}(p)(1-\demand_{F_j}(p))}\le \frac{2\eta^2}\sigma$ otherwise. By the chain rule, the observations have KL divergence at most $\frac{2\eta^2}\sigma n_j$ between $G$ and $F_j$, where $n_j$ is the expected number of queries in $I_j$ under $G$. As $\sum_jn_j\le n$, at most $\frac J4$ indices have $n_j>\frac{4n}J$, so some $j$ in the assumption has $n_j\le \frac{4n}J$, and Pinsker's inequality gives $\frac 13\le \sqrt{\frac{\eta^2n_j}\sigma}\le \sqrt{\frac{4\eta^2n}{\sigma J}}$.
\end{proof}

\subsection{Results on Sample Complexities}\label{app:sample complexity}

Several constructions below use the uniform distribution over $[\frac 12,1]$, denoted by $U:=\text{Unif}[\frac 12,1]$. Its demand is $1$ on $[0,\frac 12]$ and $2-2p$ on $[\frac 12,1]$, so that its revenue is $R_U(p)=\min\{p,2p(1-p)\}$, maximized at $\frac 12$ with
\begin{equation}\label{eq:uniform half}
\frac 12-R_U(p)\ge 2\left(p-\frac 12\right)^2\qquad\text{for all }p\in [0,1].
\end{equation}
Hence $U\in \mF_{\mathrm{curv}(2)}\subseteq \mF_\curvature$ for every $\kappa\in(0,2]$, and so is every distribution $F$ whose revenue function $R_F$ satisfies $R_F(\frac 12)=R_U(\frac 12)$, and $R_F(p)\le R_U(p)$ for all $p\in [0,1]$.

\begin{theorem}[Sample Complexities]\label{thm:sample complexities}
For every $\kappa\in (0,2)$, as $\epsilon\to 0$:
\begin{enumerate}[label=\textup{(\alph*)}]
\item $\Sample_\mF^\CDF(\epsilon)=\Theta(\epsilon^{-2})$ for each $\mF\in \{\mF_\general,\mF_\regular,\mF_\MHR,\mF_\curvature\}$;
\item $\Sample_\general(\epsilon)=\Theta(\epsilon^{-2})$ and $\Sample_\regular(\epsilon)=\Theta(\epsilon^{-2})$;
\item $\Sample_\MHR(\epsilon)=\tilde\Theta(\epsilon^{-1.5})$;
\item $\Sample_\curvature(\epsilon)=\tilde\Theta(\kappa^{-1/2}\epsilon^{-1.5})$.
\end{enumerate}
\end{theorem}
\begin{proof}
\textit{Upper Bounds.} Those for $\mF_\general$ also hold for its subclasses. \Cref{lem:empirical CDF} gives (a), and \Cref{lem:empirical revenue} gives (b). (c) is due to \citet{huang2015making}, and (d) is \Cref{lem:sample upper bound curvature}.

\textit{Lower Bounds.} Those for $\mF_\regular$ also hold for $\mF_\general$. \Cref{lem:lower regular} gives (b). For (a), \Cref{lem:sample lower bound MHR} gives the bound for MHR distributions, and hence for regular and general ones, as $\mF_\MHR\subseteq \mF_\regular\subseteq \mF_\general$, and \Cref{lem:CDF sample lower bound curvature} gives the bound for curvature distributions. (c) is due to \citet{huang2015making}, and (d) is \Cref{lem:sample lower bound curvature}.
\end{proof}

\begin{lemma}[Empirical CDF]\label{lem:empirical CDF}
Let $\hat F_n$ be the empirical distribution of $n$ samples from $F$, and $\mathrm{err}_n:=\sup_x\lvert \Pr_{v\sim \hat F_n}\{v\le x\}-\Pr_{v\sim F}\{v\le x\}\rvert$. Then $\hat F_n$ is $\mathrm{err}_n$-close to $F$ in L\'evy distance, and $\Pr\{\mathrm{err}_n>x\}=\O(e^{-2nx^2})$ for every $x>0$. In particular, $\Sample_\mF^\CDF(\epsilon)=\O(\epsilon^{-2})$ for every class $\mF$.
\end{lemma}
\begin{proof}
For every $x\in [0,1]$, taking left limits gives $\lvert \demand_{\hat F_n}(x)-\demand_F(x)\rvert\le \mathrm{err}_n$, so $\demand_F(x+\mathrm{err}_n)-\mathrm{err}_n\le \demand_F(x)-\mathrm{err}_n\le \demand_{\hat F_n}(x)\le \demand_F(x)+\mathrm{err}_n\le \demand_F(x-\mathrm{err}_n)+\mathrm{err}_n$, so $\hat F_n$ is $\mathrm{err}_n$-close to $F$. The tail bound is the Dvoretzky--Kiefer--Wolfowitz inequality \citep{dvoretzky1956asymptotic}, so some $n=\O(\epsilon^{-2})$ has $\Pr\{\mathrm{err}_n>\epsilon\}\le \frac 14$.
\end{proof}

\begin{lemma}[Empirical Revenue]\label{lem:empirical revenue}
Let $\hat p$ maximize the empirical revenue $\hat R(p):=p\Pr_{v\sim \hat F_n}\{v\ge p\}$ of $n$ samples from $F$. Then $R(p^\ast)-R(\hat p)\le 2\mathrm{err}_n$, where $\mathrm{err}_n$ is as in \Cref{lem:empirical CDF}, so $\E[R(p^\ast)-R(\hat p)]=\O(n^{-1/2})$. In particular, $\Sample_\general(\epsilon)=\O(\epsilon^{-2})$.
\end{lemma}
\begin{proof}
For every $p\in [0,1]$, $\lvert \hat R(p)-R(p)\rvert\le p\lvert \Pr_{v\sim \hat F_n}\{v<p\}-\Pr_{v\sim F}\{v<p\}\rvert\le \mathrm{err}_n$, as $\mathrm{err}_n$ also bounds the difference of the left limits. Hence $R(p^\ast)-R(\hat p)\le \hat R(p^\ast)-\hat R(\hat p)+2\mathrm{err}_n\le 2\mathrm{err}_n$. By \Cref{lem:empirical CDF}, $\E[\mathrm{err}_n]=\int_0^\infty\Pr\{\mathrm{err}_n>x\}\,dx=\O(n^{-1/2})$, so some $n=\O(\epsilon^{-2})$ has $\E[R(p^\ast)-R(\hat p)]\le \epsilon$.
\end{proof}

\begin{lemma}[Monopoly Price of Curvature Distributions via Samples]\label{lem:sample upper bound curvature}
For every $\kappa\in (0,2)$, $\Sample_\curvature(\epsilon)=\Otil(\kappa^{-1/2}\epsilon^{-1.5})$.
\end{lemma}
\begin{proof}
By \Cref{thm:size-2}(c), \Cref{alg:size-2}, which uses size-2 and hence infinite menus, has regret at most $c(\kappa^{-1/3}T^{1/3}+\log T)\log^4T$ under every $F\in \mF_\curvature$ for an absolute constant $c$. As $\kappa\le 2$, this is at most $\epsilon T$ for some $T=\Otil(\kappa^{-1/2}\epsilon^{-1.5})$, so \Cref{reduction:infinite to samples}(b) gives $\Sample_\curvature(\epsilon)\le T-1$.
\end{proof}

\begin{lemma}[Sample Lower Bound for Regular Distributions]\label{lem:lower regular}
For every $\epsilon\in (0,\frac 18]$, $\Sample_\regular(\frac\epsilon{25})\ge \frac 1{32\epsilon^2}$, even for the class of value distributions that are both regular and quasi-MHR.
\end{lemma}
\begin{proof}
By \Cref{lem:regular pair} below, the two value distributions $F_\pm$ there are regular and quasi-MHR, every price has gap at least $\frac\epsilon 6>4\cdot\frac\epsilon{25}$ under $F_+$ or under $F_-$, and $H^2(F_+,F_-)\le 4\epsilon^2$. So \Cref{lem:two-point complexity} gives $\Sample_\regular(\frac\epsilon{25})\ge \frac 1{32\epsilon^2}$, even for the smaller class.
\end{proof}

The pair tilts the revenue plateau of a regular distribution to the left or to the right, which is also used for the regret (\Cref{lem:regret lower regular}):

\begin{lemma}[A Regular Pair]\label{lem:regular pair}
For $\epsilon\in (0,\frac 18]$, let $F_\pm$ be the value distributions with
\begin{equation*}
\demand_{F_\pm}(p)=1\text{ on }[0,\tfrac 12],\quad \demand_{F_\pm}(p)=\frac 1{2p}\left(1\pm\epsilon\left(\frac 1p-2\right)\right)\text{ on }[\tfrac 12,\tfrac 34],\quad \demand_{F_\pm}(p)=\frac 83\left(1\mp\frac{2\epsilon}3\right)(1-p)\text{ on }[\tfrac 34,1].
\end{equation*}
Then $F_+$ and $F_-$ are regular and quasi-MHR, $R_{F_\pm}(p_{F_\pm}^\ast)\ge \frac 12$, every price has gap at least $\frac\epsilon 6$ under $F_+$ or under $F_-$, and $H^2(F_+,F_-)\le 4\epsilon^2$.
\end{lemma}
\begin{proof}
With $\epsilon=0$, this is a regular distribution $F_0$ whose revenue is $\frac 12$ on the plateau $[\frac 12,\frac 34]$, and $F_\pm$ tilt this plateau in opposite directions, as $R_{F_\pm}=\frac 12\pm\frac\epsilon 2(\frac 1p-2)$ on $[\frac 12,\frac 34]$. Their densities are $f_\pm=f_0(1\pm\epsilon g)$, where $f_0=\frac 1{2p^2}$ and $g=\frac 2p-2\in [\frac 23,2]$ on $(\frac 12,\frac 34)$, and $f_0=\frac 83$ and $g=-\frac 23$ on $(\frac 34,1)$. We check the class, the gaps, and the distance.
\begin{itemize}
\item \emph{Regular.} The virtual values of $F_\pm$ are $\frac{\pm\epsilon}{1\pm\epsilon g}$ on $(\frac 12,\frac 34)$, which are non-decreasing as $g$ is decreasing, and at most $2\epsilon\le \frac 14$ in absolute value, while they are $2p-1\ge \frac 12$ on $(\frac 34,1)$.
\item \emph{Quasi-MHR}, i.e., $\frac 1p\ln\frac 1{\demand_{F_\pm}(p)}$ is non-decreasing \citep{feng2025beyond}: It is $0$ on $(0,\frac 12]$ and continuous, and on $(\frac 12,1)$, its derivative has the sign of $\frac{pf_\pm}{\demand_{F_\pm}}-\ln\frac 1{\demand_{F_\pm}}$, which is positive, as $\frac{pf_\pm}{\demand_{F_\pm}}\ge \frac{1-2\epsilon}{1+2\epsilon/3}>\frac 23>\ln\frac{18}{11}\ge \ln\frac 1{\demand_{F_\pm}}$ on $(\frac 12,\frac 34)$, and $\frac{pf_\pm}{\demand_{F_\pm}}=\frac p{1-p}\ge \ln\frac 1{1-p}>\ln\frac 1{\demand_{F_\pm}}$ on $(\frac 34,1)$.
\item \emph{Gaps.} On $[\frac 12,\frac 34]$, $R_{F_+}$ decreases from its maximum $\frac 12$ and $R_{F_-}$ increases to its maximum $\frac 12+\frac\epsilon 3$, so as $R_{F_+}+R_{F_-}=1$ there, the two gaps of every price there sum to $\frac\epsilon 3$. Below $\frac 12$, $R_{F_-}=p<\frac 12$, and above $\frac 34$, $R_{F_+}=(1-\frac{2\epsilon}3)R_{F_0}\le \frac 12-\frac\epsilon 3$. Hence every price has gap at least $\frac\epsilon 6$ under $F_+$ or under $F_-$, and both maxima are at least $\frac 12$.
\item \emph{Distance.} $H^2(F_+,F_-)\le \epsilon^2\int f_0g^2\le 4\epsilon^2$.\qedhere
\end{itemize}
\end{proof}

\begin{lemma}[CDF Sample Lower Bound for MHR Distributions]\label{lem:sample lower bound MHR}
For every $\epsilon\in (0,\frac 1{22}]$, $\Sample_\MHR^\CDF(\epsilon)\ge \frac 1{968\epsilon^2}$.
\end{lemma}
\begin{proof}
We compare $\text{Unif}[0,1]$ with a slightly tilted distribution. Let $\eta:=22\epsilon\in (0,1]$, and let $F_1$ and $F_2$ be the value distributions with demands $\demand_{F_1}(p)=1-p$ and $\demand_{F_2}(p)=1-p^{1+\eta}$ on $[0,1]$.
\begin{itemize}
\item \emph{MHR.} Both are MHR, as $\frac{(1+\eta)x^\eta}{1-x^{1+\eta}}$ is non-decreasing.
\item \emph{L\'evy separation.} At $x=e^{-1}$, we have $\demand_{F_2}(x)-\demand_{F_1}(x)=e^{-1}(1-e^{-\eta})\ge \frac\eta{2e}>4\epsilon$, so $\demand_{F_2}(x)>\demand_{F_1}(x-2\epsilon)+2\epsilon$, i.e., $F_1$ and $F_2$ are not $2\epsilon$-close in L\'evy distance, and no value distribution is $\epsilon$-close to both.
\item \emph{Distance.} As $\frac{1+\eta}{(1+\eta/2)^2}=1-\frac{\eta^2}{(2+\eta)^2}$, we have $H^2(F_1,F_2)=1-\int_0^1\sqrt{(1+\eta)x^\eta}\,dx=1-\frac{\sqrt{1+\eta}}{1+\eta/2}\le \frac{\eta^2}4$.
\end{itemize}
So \Cref{lem:two-point complexity} gives $\Sample_\MHR^\CDF(\epsilon)\ge \frac 1{2\eta^2}=\frac 1{968\epsilon^2}$.
\end{proof}

\begin{lemma}[CDF Sample Lower Bound for Curvature Distributions]\label{lem:CDF sample lower bound curvature}
For every $\kappa\in (0,2]$ and $\epsilon\in (0,\frac 1{33}]$, $\Sample_\curvature^\CDF(\epsilon)\ge \frac 1{2200\epsilon^2}$.
\end{lemma}
\begin{proof}
We map the pair in \Cref{lem:sample lower bound MHR} into $[\frac 12,1]$. Let $\eta:=33\epsilon\in (0,1]$, let $F_1$ and $F_2$ be the value distributions with demands $1-p$ and $1-p^{1+\eta}$ on $[0,1]$, and let $G_i$ be the distribution of $1-\frac v2$ for $v\sim F_i$.
\begin{itemize}
\item \emph{Curvature.} $G_1=U$, and $\demand_{G_2}(p)=\min\{1,(2-2p)^{1+\eta}\}\le \demand_{G_1}(p)$ with equality at $\frac 12$, so $R_{G_2}\le R_U$ with equality at $\frac 12$. Hence $G_1,G_2\in \mF_{\mathrm{curv}(2)}\subseteq\mF_\curvature$ by \Cref{eq:uniform half}.
\item \emph{L\'evy separation.} At $x:=1-\frac 1{2e}$, we have $\demand_{G_1}(x)-\demand_{G_2}(x)=e^{-1}(1-e^{-\eta})>6\epsilon$, whereas $\demand_{G_1}(x+2\epsilon)=\demand_{G_1}(x)-4\epsilon$, so $\demand_{G_2}(x)<\demand_{G_1}(x+2\epsilon)-2\epsilon$, and no value distribution is $\epsilon$-close to both.
\item \emph{Distance.} As the map $v\mapsto 1-\frac v2$ is injective, $H^2(G_1,G_2)=H^2(F_1,F_2)=1-\frac{\sqrt{1+\eta}}{1+\eta/2}\le \frac{\eta^2}4$.
\end{itemize}
So \Cref{lem:two-point complexity} gives $\Sample_\curvature^\CDF(\epsilon)\ge \frac 1{2\eta^2}\ge \frac 1{2200\epsilon^2}$.
\end{proof}

\begin{lemma}[Sample Lower Bound for Curvature Distributions]\label{lem:sample lower bound curvature}
For every $\kappa\in (0,2)$ and $\epsilon\le \frac 1{1024}\min\{\kappa,(2-\kappa)^2\}$, we have $\Sample_\curvature(\epsilon)\ge \frac 1{1.1\cdot 10^5\sqrt\kappa\,\epsilon^{1.5}}$.
\end{lemma}
\begin{proof}
We use the two pairs in \Cref{lem:curvature pair small,lem:curvature pair large} below, depending on $\kappa$.
\begin{itemize}
\item If $\kappa\le \frac 18$, let $s:=\sqrt{4\epsilon/\kappa}$, which is at most $\frac 1{16}$ as $\epsilon\le \frac\kappa{1024}$. Under the pair $G_\pm$ in \Cref{lem:curvature pair small}, every price has gap at least $2\kappa s^2=8\epsilon>4\epsilon$ under $G_+$ or under $G_-$, and $H^2(G_+,G_-)\le 1600\kappa^2s^3$. So \Cref{lem:two-point complexity} gives $\Sample_\curvature(\epsilon)\ge \frac 1{12800\kappa^2s^3}\ge \frac 1{1.1\cdot 10^5\sqrt\kappa\,\epsilon^{1.5}}$, as $\kappa^2s^3=8\kappa^{1/2}\epsilon^{1.5}$.
\item If $\kappa>\frac 18$, let $s:=\sqrt{10\epsilon}$, which is at most $\frac{2-\kappa}8$ as $\epsilon\le \frac{(2-\kappa)^2}{640}$. Under the pair $U,G_3$ in \Cref{lem:curvature pair large}, every price has gap at least $\frac{s^2}2=5\epsilon>4\epsilon$ under $U$ or under $G_3$, and $H^2(U,G_3)\le 24s^3$. So \Cref{lem:two-point complexity} gives $\Sample_\curvature(\epsilon)\ge \frac 1{192s^3}\ge \frac 1{6100\epsilon^{1.5}}$, which is at least $\frac 1{1.1\cdot 10^5\sqrt\kappa\,\epsilon^{1.5}}$ as $\kappa>\frac 18$.\qedhere
\end{itemize}
\end{proof}

\begin{lemma}[A Pair with Small Curvature]\label{lem:curvature pair small}
Let $\kappa\in (0,\frac 18]$ and $s\in (0,\frac 1{16}]$. Let $Q(p):=\frac 14-2\kappa(p-\frac 12)^2$, and let $\psi$ be the odd function that equals $x$ on $[0,s]$, $2s-x$ on $[s,2s]$, and $0$ on $[2s,\infty)$. Then there are value distributions $G_\pm$ with revenue functions $R_{G_\pm}(p)=\min\{p,Q(p)\pm 4\kappa s\psi(p-\frac 12)\}$, such that $G_\pm\in \mF_\curvature$ with monopoly prices $\frac 12\pm s$, every price has gap at least $2\kappa s^2$ under $G_+$ or under $G_-$, and $H^2(G_+,G_-)\le 1600\kappa^2s^3$.
\end{lemma}
\begin{proof}
These are valid revenue functions, i.e., $\demand_{G_\pm}=\frac{R_{G_\pm}}p$ is non-increasing, as $\frac Qp$ is decreasing where $Q<p$, and on $J:=(\frac 12-2s,\frac 12+2s)\subseteq (\frac 38,\frac 58)$, the derivative of $Q\pm 4\kappa s\psi(p-\frac 12)$ is at most $8\kappa s+4\kappa s\le \frac 3{32}<\frac{R_{G_\pm}}p$.
\begin{itemize}
\item \emph{Curvature.} For $x:=p-\frac 12$, $R_{G_+}(\frac 12+s)-R_{G_+}(p)\ge 2\kappa(s^2+x^2-2s\psi(x))\ge \kappa(x-s)^2$, where the last step is checked on $\lvert x\rvert\le s$, $s<\lvert x\rvert\le 2s$, and $\lvert x\rvert>2s$ separately. So $G_+\in \mF_\curvature$ with monopoly price $\frac 12+s$, and symmetrically, $G_-\in \mF_\curvature$ with monopoly price $\frac 12-s$.
\item \emph{Gaps.} Both maxima are $\frac 14+2\kappa s^2$, whereas $R_{G_+}+R_{G_-}=2Q\le \frac 12$ on $J$ and $R_{G_\pm}\le \frac 14$ elsewhere, so every price has gap at least $2\kappa s^2$ under $G_+$ or under $G_-$.
\item \emph{Distance.} $G_\pm$ only differ on $J$, where their densities are $f_0\mp 4\kappa s(\frac{\psi(p-1/2)}p)'$ with $f_0:=\frac Q{p^2}-\frac{Q'}p\ge \frac 25$ and $\lvert (\frac{\psi(p-1/2)}p)'\rvert\le \frac 1p+\frac s{p^2}\le \frac{28}9$. As $(\sqrt a-\sqrt b)^2\le \frac{(a-b)^2}{a+b}$, we have $H^2(G_+,G_-)\le \frac 12\cdot 4s\cdot \frac{(8\kappa s\cdot 28/9)^2}{4/5}\le 1600\kappa^2s^3$.\qedhere
\end{itemize}
\end{proof}

For larger curvature, we shift the kink of $R_U$ to the right, as in \citet[Theorem 4.12]{huang2015making}:

\begin{lemma}[A Pair with Large Curvature]\label{lem:curvature pair large}
Let $\kappa\in (\frac 18,2)$ and $s\in (0,\frac{2-\kappa}8]$. Let $G_3$ multiply the density of $U$ by $1-4s$ on $[\frac 12,\frac 12+s)$ and by $1+\theta$ on $[\frac 12+s,1]$, where $\theta:=\frac{4s^2}{1/2-s}\le 16s^2$ keeps the total mass $1$. Then $U,G_3\in \mF_\curvature$, every price has gap at least $\frac{s^2}2$ under $U$ or under $G_3$, and $H^2(U,G_3)\le 24s^3$.
\end{lemma}
\begin{proof}
$R_{G_3}$ is $p$ on $[0,\frac 12]$, $p(1-2(1-4s)(p-\frac 12))$ on $[\frac 12,\frac 12+s]$, and $2(1+\theta)p(1-p)$ on $[\frac 12+s,1]$.
\begin{itemize}
\item \emph{Curvature.} $R_{G_3}$ increases on $[0,\frac 12+s]$, as its left derivative at $\frac 12+s$ is $16s^2>0$, and it is quadratic with second derivative at most $-4(1-4s)$ on both sides of $\frac 12+s$, so $R_{G_3}(\frac 12+s)-R_{G_3}(p)\ge 2(1-4s)(p-\frac 12-s)^2\ge \kappa(p-\frac 12-s)^2$ for $p\ge \frac 12$. For $p\le \frac 12$, the same inequality holds as $R_{G_3}(\frac 12+s)-p-\kappa(\frac 12+s-p)^2$ is concave in $p$, and non-negative at $p\in \{0,\frac 12\}$ as $\kappa\le 2(1-4s)\le \frac 2{(1+2s)^2}$. So $G_3\in \mF_\curvature$, and so is $U$ by \Cref{eq:uniform half}.
\item \emph{Gaps.} Every $p\ge \frac 12+\frac s2$ has gap at least $\frac{s^2}2$ under $U$, and as $R_{G_3}$ increases on $[0,\frac 12+s]$, every $p<\frac 12+\frac s2$ has gap at least $R_{G_3}(\frac 12+s)-R_{G_3}(\frac 12+\frac s2)\ge \frac{s^2}2$ under $G_3$.
\item \emph{Distance.} $H^2(U,G_3)=s(1-\sqrt{1-4s})^2+(\frac 12-s)(\sqrt{1+\theta}-1)^2\le 16s^3+32s^4\le 24s^3$.\qedhere
\end{itemize}
\end{proof}

\subsection{Results on Interval-Query Complexities}\label{app:interval complexity}

\begin{theorem}[Interval-Query Complexities]\label{thm:interval complexities}
For every $\kappa\in (0,2)$, as $\epsilon\to 0$:
\begin{enumerate}[label=\textup{(\alph*)}]
\item $\Interval_\mF^\CDF(\epsilon)=\tilde\Theta(\epsilon^{-2})$ for each $\mF\in \{\mF_\general,\mF_\regular,\mF_\MHR,\mF_\curvature\}$;
\item $\Interval_\general(\epsilon)=\tilde\Theta(\epsilon^{-2})$ and $\Interval_\regular(\epsilon)=\tilde\Theta(\epsilon^{-2})$;
\item $\Interval_\MHR(\epsilon)=\tilde\Theta(\epsilon^{-1.5})$;
\item $\Interval_\curvature(\epsilon)=\tilde\Theta(\kappa^{-1/2}\epsilon^{-1.5})$.
\end{enumerate}
That is, interval queries are as powerful as samples up to logarithmic factors (\Cref{thm:sample complexities}).
\end{theorem}
\begin{proof}
By \Cref{thm:complexity relations}(a), every interval-query complexity is at least the corresponding sample complexity, so the lower bounds follow from \Cref{thm:sample complexities}. For the upper bounds, \Cref{lem:CDF interval complexity,lem:interval upper general} give (a) and (b) for $\mF_\general$, and hence for its subclasses, and \Cref{lem:interval upper MHR,lem:interval upper curvature} give (c) and (d).
\end{proof}

\begin{lemma}[CDF Estimation via Interval Queries]\label{lem:CDF interval complexity}
For every $\epsilon\in (0,1]$, $\Interval_\general^\CDF(\epsilon)=\O(\epsilon^{-2}\log^4\frac 2\epsilon)$.
\end{lemma}
\begin{proof}
Run \Cref{alg:CDF interval query} with $\delta:=\epsilon^4$, and stop it after $160H^3\iota\epsilon^{-2}=\O(\epsilon^{-2}\log^4\frac 2\epsilon)$ queries (so that the number of queries won't exceed our budget), where $H$ and $\iota$ are as in \Cref{lem:CDF segment tree}. By \Cref{lem:CDF segment tree}, with probability at least $1-\O(\epsilon\log^3\frac 2\epsilon)$, it is not stopped, and it outputs an $\hat F$ that is $\epsilon$-close to $F$ in L\'evy distance. This probability is at least $\frac 34$ when $\epsilon$ is below an absolute constant, so the bound holds for all $\epsilon\in(0,1]$, as $\Interval_\general^\CDF(\epsilon)$ is non-increasing in $\epsilon$.
\end{proof}

The same algorithm also estimates the monopoly price in expectation, as its failure probability can be made small:

\begin{lemma}[Monopoly Price of General Distributions via Interval Queries]\label{lem:interval upper general}
For every $\epsilon\in (0,1]$, $\Interval_\general(\epsilon)=\O(\epsilon^{-2}\log^4\frac 2\epsilon)$.
\end{lemma}
\begin{proof}
Run \Cref{alg:CDF interval query} with accuracy $\frac\epsilon 6$ and $\delta:=\epsilon^5$, and stop it after $160H^3\iota(\frac\epsilon 6)^{-2}=\O(\epsilon^{-2}\log^4\frac 2\epsilon)$ queries, where $H$ and $\iota$ are as in \Cref{lem:CDF segment tree} for accuracy $\frac\epsilon 6$. Output the price $\hat p$ given by \Cref{reduction:price to CDF} for its output, or an arbitrary price if stopped. By \Cref{lem:CDF segment tree}, with probability at least $1-\O(\epsilon^2\log^3\frac 2\epsilon)$, it is not stopped and its output is $\frac\epsilon 6$-close to $F$, in which case $R(p^\ast)-R(\hat p)\le \frac{5\epsilon}6$. As every gap is at most $1$, $\E[R(p^\ast)-R(\hat p)]\le \frac{5\epsilon}6+\O(\epsilon^2\log^3\frac 2\epsilon)\le \epsilon$ when $\epsilon$ is below an absolute constant, and the bound holds for all $\epsilon\in (0,1]$ as $\Interval_\general(\epsilon)$ is non-increasing in $\epsilon$.
\end{proof}

\begin{lemma}[Monopoly Price of MHR Distributions via Interval Queries]\label{lem:interval upper MHR}
$\Interval_\MHR(\epsilon)=\Otil(\epsilon^{-1.5})$.
\end{lemma}
\begin{proof}
By \Cref{thm:size-2}(b), \Cref{alg:size-2} uses size-2 menus and has regret at most $cT^{1/3}\log^4T$ under every $F\in \mF_\MHR$ for an absolute constant $c$. This is at most $\frac{\epsilon T}2$ for some $T=\Otil(\epsilon^{-1.5})$ with $T\ge \log\frac 2\epsilon$, so \Cref{reduction:online to batch} with $m=2$ gives $\Interval_\MHR(\epsilon)\le 12T=\Otil(\epsilon^{-1.5})$.
\end{proof}

\begin{lemma}[Monopoly Price of Curvature Distributions via Interval Queries]\label{lem:interval upper curvature}
For every $\kappa\in (0,2)$, $\Interval_\curvature(\epsilon)=\Otil(\kappa^{-1/2}\epsilon^{-1.5})$.
\end{lemma}
\begin{proof}
By \Cref{thm:size-2}(c), \Cref{alg:size-2} uses size-2 menus and has regret at most $c(\kappa^{-1/3}T^{1/3}+\log T)\log^4T$ under every $F\in \mF_\curvature$ for an absolute constant $c$. As $\kappa\le 2$, this is at most $\frac{\epsilon T}2$ for some $T=\Otil(\kappa^{-1/2}\epsilon^{-1.5})$ with $T\ge \log\frac 2\epsilon$, so \Cref{reduction:online to batch} with $m=2$ gives $\Interval_\curvature(\epsilon)\le 12T=\Otil(\kappa^{-1/2}\epsilon^{-1.5})$.
\end{proof}

\subsection{Results on Threshold-Query Complexities}\label{app:threshold complexity}

For threshold queries, we prove extra lower bounds for the curvature distributions. All remaining upper and lower bounds in \Cref{table:complexity} are from \citep{leme2023description,leme2023pricing,kleinberg2019bandits}.

\begin{theorem}[Threshold-Query Lower Bounds for Curvature Distributions]\label{thm:threshold complexities}
For every $\kappa\in (0,2)$, as $\epsilon\to 0$ and $T\to \infty$:
\begin{enumerate}[label=\textup{(\alph*)}]
\item $\Threshold_\curvature^\CDF(\epsilon)=\Omega(\epsilon^{-3})$;
\item $\Threshold_\curvature(\epsilon)=\Omega((2-\kappa)^{3/2}\epsilon^{-2.5})$;
\item $\mR_\curvature^{(1)}(T)=\Omega((2-\kappa)^{3/5}T^{3/5})$, i.e., posted prices suffer this regret.
\end{enumerate}
Compared with \Cref{thm:interval complexities}, interval queries are thus polynomially more powerful than threshold queries for curvature distributions, for both properties.
\end{theorem}
\begin{proof}
(a) is \Cref{lem:CDF threshold lower bound curvature}, (b) is \Cref{lem:threshold lower bound curvature}, and (c) is \Cref{lem:posted price curvature}.
\end{proof}

\begin{lemma}[CDF Threshold-Query Lower Bound for Curvature Distributions]\label{lem:CDF threshold lower bound curvature}
For every $\kappa\in (0,2]$ and $\epsilon\in (0,\frac 1{480}]$, $\Threshold_\curvature^\CDF(\epsilon)\ge \frac 1{6\cdot 10^6\epsilon^3}$.
\end{lemma}
\begin{proof}
Using needle-in-a-haystack, we hide a needle at one of $\Theta(\epsilon^{-1})$ locations of $U$, each of which a threshold query can only probe separately. Let $\eta:=8\epsilon$ and $J:=\lfloor \frac 1{30\eta}\rfloor\ge \frac 1{60\eta}$. For $j\in [J]$, let $p_j:=\frac 35+6\eta j-3\eta$, $I_j:=(p_j-3\eta,p_j+3\eta)$, and $\phi_j(p):=\max\{0,1-\frac{\lvert p-p_j\rvert}{3\eta}\}$, so that $I_1,I_2,\ldots,I_J\subseteq (\frac 35,\frac 45)$ are disjoint. Let $F_j$ have $\demand_{F_j}:=\demand_U-\eta\phi_j$, i.e., $F_j$ moves mass $\eta$ from the right half of $I_j$ to its left half.
\begin{itemize}
\item \emph{Curvature.} As $\demand_{F_j}\le \demand_U$ with equality at $\frac 12$, we have $R_{F_j}\le R_U$ with equality at $\frac 12$, so $F_j\in \mF_{\mathrm{curv}(2)}\subseteq \mF_\curvature$ by \Cref{eq:uniform half}.
\item \emph{L\'evy separation.} An $\hat F$ satisfying \Cref{eq:Levy} for $F_j$ has $\demand_{\hat F}(p_j)\le \demand_{F_j}(p_j-\epsilon)+\epsilon=2-2p_j-\eta+\frac{10\epsilon}3$, whereas one for $U$ has $\demand_{\hat F}(p_j)\ge \demand_U(p_j+\epsilon)-\epsilon=2-2p_j-3\epsilon$, so no $\hat F$ satisfies both.
\end{itemize}
Hence, if an algorithm outputs an $\hat F$ satisfying \Cref{eq:Levy} with probability at least $\frac 34$ under $U$ and under every $F_j$, then its observations have total variation distance at least $\frac 34-\frac 14=\frac 12$ between $U$ and every $F_j$. As $\demand_U,\demand_{F_j}\in [\frac 15,\frac 45]$ on $I_j$, \Cref{lem:haystack} with $\sigma=\frac 15$ gives $n\ge \frac J{180\eta^2}\ge \frac 1{6\cdot 10^6\epsilon^3}$.
\end{proof}

\begin{lemma}[Threshold-Query Lower Bound for Curvature Distributions]\label{lem:threshold lower bound curvature}
For every $\kappa\in (0,2)$ and all sufficiently small $\epsilon>0$ (depending on $\kappa$), $\Threshold_\curvature(\epsilon)\ge \frac{(2-\kappa)^{3/2}}{2\cdot 10^7\epsilon^{2.5}}$.
\end{lemma}
\begin{proof}
We hide a needle near the smooth peak of a uniform distribution, at one of $\Theta(\epsilon^{-1/2})$ locations within distance $\Theta(\sqrt\epsilon)$ of the peak. Let $c_0:=\frac{6+\kappa}4\in (\kappa,2)$, $w:=\frac{2-c_0}4=\frac{2-\kappa}{16}$, and $\delta:=\sqrt{(2-\kappa)/8}$. Let $F_0:=\text{Unif}[1-\frac 1{c_0},1]$, so that $R_{F_0}(p)=c_0p(1-p)$ for $p\ge 1-\frac 1{c_0}$ and $R_{F_0}(p)=p$ otherwise, and hence $\frac{c_0}4-R_{F_0}(p)\ge c_0(p-\frac 12)^2$ for all $p\in [0,1]$. Unlike $U$, whose revenue has a kink at its monopoly price, $R_{F_0}$ has a smooth peak at $\frac 12$ with curvature $c_0>\kappa$. Let $\eta:=8\epsilon$ and $J:=\lfloor \frac\delta{3\sqrt\eta}\rfloor$, and for $j\in [J]$, let $p_j:=\frac 12-\delta\sqrt\eta+6\eta j-3\eta$, $I_j:=(p_j-3\eta,p_j+3\eta)$, and $\phi_j(p):=\max\{0,1-\frac{\lvert p-p_j\rvert}{3\eta}\}$. Let $F_j$ have $R_{F_j}:=R_{F_0}+\eta\phi_j$. For all sufficiently small $\epsilon$, $\demand_{F_j}=\frac{R_{F_j}}p$ is decreasing, and:
\begin{itemize}
\item \emph{$F_j\in \mF_\curvature$ with monopoly price $p_j$.} As $\lvert p_j-\frac 12\rvert\le \delta\sqrt\eta$, every $p\notin I_j$ has gap at least $(1-c_0\delta^2)\eta+c_0(p-\frac 12)^2\ge \kappa(p-p_j)^2$, where the last step uses $\kappa(p-p_j)^2\le c_0(p-\frac 12)^2+\frac{c_0\kappa}{c_0-\kappa}\delta^2\eta$ and $\frac{c_0^2\delta^2}{c_0-\kappa}=\frac{(6+\kappa)^2}{96}<1$. Every $p\in I_j$ has gap $c_0(p-p_j)(p+p_j-1)+\frac{\lvert p-p_j\rvert}3\ge \frac{\lvert p-p_j\rvert}4\ge \kappa(p-p_j)^2$.
\item \emph{Gaps.} Every price outside $I_j$ has gap at least $(1-c_0\delta^2)\eta\ge \frac\eta 2=4\epsilon$ under $F_j$, so by Markov's inequality, \Cref{eq:near-optimal monopoly price} under $F_j$ gives $\hat p\in I_j$ with probability at least $\frac 34$.
\item \emph{Needles.} On $I_j$, $\lvert \demand_{F_j}-\demand_{F_0}\rvert\le \frac\eta p\le 3\eta$, and $\demand_{F_0},\demand_{F_j}\in [\frac 12,1-w]$, as $\demand_{F_0}(\frac 12)=\frac{c_0}2=1-2w$.
\end{itemize}
As $\sum_j\Pr_{F_0}\{\hat p\in I_j\}\le 1$, at least $\frac{3J}4$ indices $j$ have $\Pr_{F_0}\{\hat p\in I_j\}\le \frac 4J\le \frac 13$, and hence total variation distance at least $\frac 34-\frac 13>\frac 13$ between $F_0$ and $F_j$. So \Cref{lem:haystack} with $\sigma=w$ and $3\eta$ in place of $\eta$ gives $n\ge \frac{wJ}{324\eta^2}\ge \frac{(2-\kappa)^{3/2}}{2\cdot 10^7\epsilon^{2.5}}$.
\end{proof}

By the online-to-batch conversion, this bound also forces posted prices to suffer more regret than menus (\Cref{thm:minimax lower}):

\begin{lemma}[Posted Prices for Curvature Distributions]\label{lem:posted price curvature}
For every $\kappa\in (0,2)$ and all sufficiently large $T$ (depending on $\kappa$), $\mR_\curvature^{(1)}(T)=\Omega((2-\kappa)^{3/5}T^{3/5})$.
\end{lemma}
\begin{proof}
If an algorithm using posted prices had regret at most $\gamma(2-\kappa)^{3/5}T^{3/5}$ under every $F\in \mF_\curvature$, then \Cref{reduction:online to batch} with $m=1$ would give $\Threshold_\curvature(\gamma(2-\kappa)^{3/5}T^{-2/5})\le T$, which contradicts \Cref{lem:threshold lower bound curvature} for a small enough absolute constant $\gamma>0$ and large enough $T$.
\end{proof}

\section{\texorpdfstring{Omitted Proofs in \Cref{sec:main} (Upper Bounds)}{Omitted Proofs: Upper Bounds}}\label{app:size-3}\label{app:upper}

We first state the two tools shared by all our algorithms (\Cref{app:tools}), and then prove \Cref{thm:size-3} (\Cref{app:CDF interval query}), \Cref{lem:size-2 rounds} (\Cref{app:size-2 rounds}), \textsc{ScaleSearch} and the properties of MHR distributions (\Cref{app:scale search}), and \Cref{thm:size-2,thm:instance upper} (\Cref{app:size-2}). \Cref{app:examples} lists examples of interval dimensions and characterizes when $V_T(F)$ tends to infinity. The resulting query complexities are collected in \Cref{app:complexities}.

\subsection{Empirical Bernstein and Segment Trees}\label{app:tools}

All our algorithms estimate the probability $\intervalP(I)=\Pr_{v\sim F}\{v\in I\}$ of interval $I$ by \Cref{alg:empirical Bernstein}, and combine these estimates via segment trees. This subsection states the properties of these two tools.

\paragraph{Empirical Bernstein.} \Cref{alg:empirical Bernstein} enters our analysis through two facts: its confidence radius $r_K$ is valid, and it shrinks as $\sqrt{\intervalP(I)/K}$ (even without knowing $\intervalP(I)$) rather than $\sqrt{1/K}$.

\begin{lemma}[Confidence Radius Shrinks with the Mean]\label{lem:empirical Bernstein}
Let $X_1,X_2,\ldots$ be i.i.d.\ Bernoulli random variables with mean $\mu$, and for $K\ge 2$, let $\hat\mu_K$ and $r_K$ be their empirical mean and the confidence radius in \Cref{eq:confidence radius} after the first $K$ of them, where $\iota=\log\frac 2\delta$ for some $\delta\in(0,\frac 12]$. For any $K_0\ge 2$, with probability at least $1-3K_0\delta$, every $K\in\{2,3,\ldots,K_0\}$ satisfies
\begin{equation}\label{eq:EB radius}
\lvert\hat\mu_K-\mu\rvert\le r_K\le \sqrt{\frac{2\iota\mu}K}+\frac{15\iota}{2K}.
\end{equation}
In particular, this applies to a call of \Cref{alg:empirical Bernstein} on an interval $I$ with $\mu=\intervalP(I)$, if its queries use fresh i.i.d.\ values $v\sim F$ given the history before the call.
\end{lemma}
\begin{proof}
We apply two bounds of \citet{maurer2009empirical}, and then bound the sample variance by the mean. By \Cref{eq:confidence radius}, $r_K=\sqrt{2\iota V_K/K}+\frac{7\iota}{3(K-1)}$, where $V_K:=\frac 1{K(K-1)}\sum_{k<k'\le K}(X_k-X_{k'})^2$ is the sample variance. For each $K\in\{2,3,\ldots,K_0\}$:
\begin{itemize}
\item $\lvert\hat\mu_K-\mu\rvert\le r_K$ with probability at least $1-2\delta$, by \citep[Theorem 4]{maurer2009empirical} applied to both $\{X_k\}$ and $\{1-X_k\}$, which have the same $V_K$;
\item $\sqrt{V_K}\le\sqrt{\mu(1-\mu)}+\sqrt{\frac{2\iota}{K-1}}$ with probability at least $1-\delta$, by \citep[Theorem 10]{maurer2009empirical}.
\end{itemize}
By a union bound, both hold for all such $K$ with probability at least $1-3K_0\delta$. Plugging the second one into $r_K$,
\begin{equation*}
r_K\overset{(a)}{\le}\sqrt{\frac{2\iota\mu}K}+\sqrt{\frac{2\iota}K\cdot\frac{4\iota}K}+\frac{14\iota}{3K}\overset{(b)}{\le}\sqrt{\frac{2\iota\mu}K}+\frac{15\iota}{2K},
\end{equation*}
where (a) uses $\mu(1-\mu)\le\mu$ and $K-1\ge\frac K2$, and (b) uses $\sqrt 8+\frac{14}3\le\frac{15}2$. This proves \Cref{eq:EB radius}. Finally, a call of \Cref{alg:empirical Bernstein} on $I$ observes a prefix of such a sequence with $\mu=\intervalP(I)$, as each query uses a fresh value.
\end{proof}

Both terms in \Cref{eq:EB radius} are at most $\frac\epsilon 2$ once $K\ge 8\iota\mu\epsilon^{-2}+15\iota\epsilon^{-1}$. Thus, intervals with smaller probabilities need fewer queries:

\begin{lemma}[Fewer Queries for Smaller
Probability Intervals] \label{lem:empirical Bernstein stop} Consider a call of \Cref{alg:empirical Bernstein} on an interval $I$ with precision $\epsilon\in(0,1]$ and failure probability $\delta\in(0,\frac 12]$, whose queries use fresh i.i.d.\ values $v\sim F$ given the history before the call, and let $\iota:=\log\frac 2\delta$. With probability at least $1-78\iota\epsilon^{-2}\delta$, it stops within $8\iota\intervalP(I)\epsilon^{-2}+15\iota\epsilon^{-1}+2$ queries and outputs an estimate $\hat\intervalP(I)$ with $\lvert\hat\intervalP(I)-\intervalP(I)\rvert\le\epsilon$.
\end{lemma}
\begin{proof}
Apply \Cref{lem:empirical Bernstein} with $K_0:=\lceil 25\iota\epsilon^{-2}\rceil$, which is at most $26\iota\epsilon^{-2}$ as $\iota\epsilon^{-2}\ge 1$, so it fails with probability at most $3K_0\delta\le 78\iota\epsilon^{-2}\delta$. On its event, $r_K\le\epsilon$ for every $K\ge K^\ast:=\max\{2,\lceil 8\iota\intervalP(I)\epsilon^{-2}+15\iota\epsilon^{-1}\rceil\}$, where $K^\ast\le 8\iota\intervalP(I)\epsilon^{-2}+15\iota\epsilon^{-1}+2\le 25\iota\epsilon^{-2}\le K_0$ as $\intervalP(I),\epsilon\le 1\le\iota$. So the call stops at some $K\le K^\ast$, and its output $\hat\mu_K$ satisfies $\lvert\hat\mu_K-\intervalP(I)\rvert\le r_K\le\epsilon$.
\end{proof}

\paragraph{Segment Trees.} Given endpoints $p_0<p_1<\cdots<p_N$, \Cref{alg:segment tree} builds a segment tree $\mT$ by recursively halving $[p_0,p_N)$ at its middle endpoint, and decomposes every $[p_i,p_j)$ into tree intervals $\mI_\mT[p_i,p_j)$. We use the properties in \Cref{lem:segment tree}.

\begin{subroutine}[!t]
\caption{\textproc{BuildSegTree}(endpoints $p_0<p_1<\cdots<p_N$)}\label{alg:segment tree}
\begin{algorithmic}[1]
\State Initialize the tree $\mT$ as a single root interval, namely $\mT:=\{[p_0,p_N)\}$
\For{tree interval $I=[p_i,p_j)\in \mT$}
\If{$j-i\ge 2$}\Comment{Otherwise, $j=i+1$ and then $[p_i,p_j)$ is a leaf}
\State Add its two children $[p_i,p_k)$ and $[p_k,p_j)$ to $\mT$, where $k=\lfloor \frac{i+j}{2}\rfloor$
\EndIf
\EndFor
\State Initialize the tree decomposition for the empty interval as $\mI_\mT(\varnothing)=\varnothing$
\For{all $0\le i<j\le N$ in the increasing order of $(j-i)$}
\If{$[p_i,p_j)$ belongs to $\mT$, i.e., it is a tree interval}
\State The tree decomposition $\mI_\mT[p_i,p_j):=\{[p_i,p_j)\}$ is itself
\Else
\State Let $I\in \mT$ be the smallest tree interval containing $[p_i,p_j)$. Let its two children be $I_L$ and $I_R$, both belonging to $\mT$. Then $\mI_\mT[p_i,p_j):=\mI_\mT([p_i,p_j)\cap I_L) \cup \mI_\mT([p_i,p_j)\cap I_R)$
\EndIf
\EndFor
\State \Return segment tree $\mT$ and the associated tree decomposition $\mI_\mT$
\end{algorithmic}
\end{subroutine}

\begin{lemma}[Segment Tree Decomposition]\label{lem:segment tree}
The segment tree $\mT$ built by \Cref{alg:segment tree} on $p_0<p_1<\cdots<p_N$ has $2N-1$ tree intervals in at most $\lceil \log_2 N\rceil+1$ depths, and tree intervals of the same depth are disjoint. Moreover, for any $0\le i<j\le N$, $\mI_\mT[p_i,p_j)$ partitions $[p_i,p_j)$ into tree intervals, at most two of each depth.
\end{lemma}
\begin{proof}
Each non-leaf tree interval is the disjoint union of its two children. Hence any two tree intervals are either nested or disjoint, tree intervals of the same depth are disjoint, and as the leaves are the $N$ intervals $[p_{i-1},p_i)$, there are $2N-1$ tree intervals. As each child of $[p_i,p_j)$ contains at most $\lceil \frac{j-i}2\rceil$ leaves, a tree interval of depth $h$ contains at most $\lceil N/2^h\rceil$ leaves. A tree interval with children contains at least $2$ leaves, so its depth is less than $\log_2 N$, and hence all depths are at most $\lceil \log_2 N\rceil$.

For decomposition, if $[p_i,p_j)$ is not a tree interval, let $I$ be the smallest tree interval containing it, with children $I_L$ and $I_R$. By the minimality of $I$, neither child contains $[p_i,p_j)$, so $[p_i,p_j)\cap I_L$ and $[p_i,p_j)\cap I_R$ are non-empty and partition $[p_i,p_j)$. By induction on $j-i$, $\mI_\mT[p_i,p_j)$ thus partitions $[p_i,p_j)$ into tree intervals. To count them, note that $[p_i,p_j)\cap I_L$ shares the right endpoint of $I_L$, and $[p_i,p_j)\cap I_R$ shares the left endpoint of $I_R$. Hence it suffices to show the following claim, as each of the two parts then contributes at most one tree interval of each depth:

\textit{If $S\subseteq J$ shares the right endpoint of a tree interval $J$, then $\mI_\mT(S)$ contains at most one tree interval of each depth, all of which are deeper than $J$ unless $S=J$. The same holds if $S$ shares the left endpoint of $J$.}

We prove the claim by induction on the number of leaves in $S$ (the case of left endpoints is symmetric). If $S$ is a tree interval, then $\mI_\mT(S)=\{S\}$, and $S$ is deeper than $J$ unless $S=J$. Otherwise, let $I'$ be the smallest tree interval containing $S$, so that $S\subsetneq I'\subseteq J$ and $S$ shares the right endpoint of $I'$, and let $I'_L$ and $I'_R$ be the children of $I'$. As $I'_R\not\supseteq S$ and both end at the same point, we have $S\supsetneq I'_R$, so $\mI_\mT(S)=\mI_\mT(S\cap I'_L)\cup \{I'_R\}$, where $S\cap I'_L\ne I'_L$ shares the right endpoint of $I'_L$. By induction, $\mI_\mT(S\cap I'_L)$ contains at most one tree interval of each depth, all deeper than $I'_L$ and hence than its sibling $I'_R$. So $\mI_\mT(S)$ contains at most one tree interval of each depth, all at least as deep as $I'_R$, which is deeper than $J$ as $I'\subseteq J$. This proves the claim and hence the lemma.
\end{proof}

\subsection{\texorpdfstring{Proof of \Cref{lem:CDF interval query} (Revenue Maximization with Size-3 Menus)}{Proof: Revenue Maximization with Size-3 Menus}}\label{app:CDF interval query}

\begin{algorithm}[!t]
\caption{CDF Estimation via Interval Queries}\label{alg:CDF interval query}
\begin{algorithmic}[1]
\Require{accuracy $\epsilon\in (0,1]$, failure probability $\delta\in (0,\frac 12]$}
\State Let $N=2^{\lceil \log_2 \epsilon^{-1}\rceil}$. Let $(\mT,\mI_\mT)\gets \Call{BuildSegTree}{0,\frac 1N,\frac 2N,\ldots,1}$ \Comment{see \Cref{alg:segment tree}}
\For{each interval $I$ in the segment tree $\mT$}
\State Estimate $\intervalP(I)$ by $\hat \intervalP(I)\gets \Call{EmpiricalBernstein}{I,\frac{\epsilon}{2\log_2(2N)},\delta}$ \Comment{see \Cref{alg:empirical Bernstein}}
\EndFor
\State For $i=0,1,\ldots,N$, let $\hat d_i:=1-\sum_{I\in \mI_\mT[0,\frac iN)}\hat \intervalP(I)$, which estimates $\demand(\frac iN)$
\State \Return the value distribution $\hat F$ with
\algeq[eq:CDF estimate]{$\displaystyle \demand_{\hat F}(x):=\max\left\{0,\min_{0\le i\le \lceil Nx\rceil}\hat d_i\right\},\quad \forall x\in [0,1].$}
\end{algorithmic}
\end{algorithm}

\Cref{alg:CDF interval query} is the CDF estimation algorithm sketched in \Cref{sec:size-3}. With endpoints $0,\frac 1N,\frac 2N,\ldots,1$, \Cref{alg:segment tree} builds the segment tree $\mT$ described there, whose tree intervals of depth $h$ are $[\frac k{2^h},\frac{k+1}{2^h})$ (for $k=0,1,\ldots,2^h-1$), and $\mI_\mT[0,\frac iN)$ decomposes $[0,\frac iN)$ into $\O(\log_2N)$ tree intervals (\Cref{lem:segment tree}). As the estimates $\hat d_i$ may not be monotone in $i$ (and thus fail to induce a CDF), \Cref{eq:CDF estimate} takes their running minimum and truncates it at $0$, so that $\demand_{\hat F}$ is non-increasing and left-continuous with $\demand_{\hat F}(0)=\hat d_0=1$.

\begin{theorem}[Restatement of \Cref{thm:size-3}; CDF Estimation via Interval Queries and Size-3 Menus]
$\Interval_\general^\CDF(\epsilon)=\O(\epsilon^{-2}\log^4\frac 2\epsilon)$ and $\mR_\general^{(3)}(T)=\O(\sqrt T\log^3 T)$.
\end{theorem}
\begin{proof}
The first claim is \Cref{lem:CDF interval complexity}, which runs \Cref{alg:CDF interval query} with $\delta:=\epsilon^4$ and applies \Cref{lem:CDF segment tree} below. The second claim is \Cref{lem:size-3 regret} below, which runs \Cref{alg:CDF interval query} in phases using size-3 menus.
\end{proof}

\begin{lemma}[CDF Estimation via a Segment Tree]\label{lem:CDF segment tree}
For any $F\in \mF_\general$, $\epsilon\in(0,1]$, and $\delta\in(0,\frac 12]$, let $H:=\lceil \log_2 \epsilon^{-1}\rceil+1$ and $\iota:=\log \frac 2\delta$. With probability at least $1-\O(H^2\iota\epsilon^{-3}\delta)$, \Cref{alg:CDF interval query} uses at most $160H^3\iota\epsilon^{-2}$ interval queries, and its output $\hat F$ in \Cref{eq:CDF estimate} is $\epsilon$-close to $F$ in L\'evy distance.
\end{lemma}
\begin{proof}
We show that all tree intervals are estimated accurately, then count the queries, and finally bound the L\'evy distance. As $N=2^{H-1}$, \Cref{alg:CDF interval query} calls \Cref{alg:empirical Bernstein} on each of the $2N-1$ tree intervals with precision $\eta:=\frac\epsilon{2H}$.

\textit{Accurate Estimates.} Each query uses a fresh value $v\sim F$, so \Cref{lem:empirical Bernstein stop} applies to each call given the history before it. As $2N-1<4\epsilon^{-1}$, a union bound shows that, with probability at least $1-4\epsilon^{-1}\cdot 78\iota\eta^{-2}\delta=1-\O(H^2\iota\epsilon^{-3}\delta)$, every tree interval $I$ satisfies $\lvert\hat\intervalP(I)-\intervalP(I)\rvert\le\eta$ and takes at most $8\iota\intervalP(I)\eta^{-2}+15\iota\eta^{-1}+2$ queries. We assume this event from now on.

\textit{Number of Queries.} By \Cref{lem:segment tree}, the tree intervals of each of the $H$ depths are disjoint, so their probabilities sum to at most $1$, and there are at most $2N\le 4\epsilon^{-1}$ tree intervals. Hence the number of queries is at most
\begin{equation*}
8H\iota\eta^{-2}+4\epsilon^{-1}\left(15\iota\eta^{-1}+2\right)=32H^3\iota\epsilon^{-2}+120H\iota\epsilon^{-2}+8\epsilon^{-1}\le 160H^3\iota\epsilon^{-2}.
\end{equation*}

\textit{L\'evy Distance.} For each $i\in[N]$, \Cref{lem:segment tree} shows that $\mI_\mT[0,\frac iN)$ partitions $[0,\frac iN)$ into tree intervals, at most two of each depth, so $\hat d_i$ in \Cref{eq:CDF estimate} satisfies
\begin{equation*}
\left\lvert \hat d_i-\demand_F\left(\tfrac iN\right)\right\rvert\le\sum_{I\in\mI_\mT[0,\frac iN)}\left\lvert\hat\intervalP(I)-\intervalP(I)\right\rvert\le 2H\eta=\epsilon.
\end{equation*}
This also holds for $i=0$, as $\hat d_0=1=\demand_F(0)$. As $\demand_F$ is non-increasing, the minimum over $j\le i$ and the truncation at $0$ in \Cref{eq:CDF estimate} keep these bounds, so $\lvert\demand_{\hat F}(\frac iN)-\demand_F(\frac iN)\rvert\le\epsilon$. For any $x\in(\frac{i-1}N,\frac iN]$, we have $\demand_{\hat F}(x)=\demand_{\hat F}(\frac iN)$ and $\demand_F(x+\epsilon)\le\demand_F(\frac iN)\le\demand_F(x-\epsilon)$ as $\frac 1N\le\epsilon$; for $x=0$, we have $\demand_{\hat F}(0)=1$. Therefore, $\demand_F(x+\epsilon)-\epsilon\le\demand_{\hat F}(x)\le\demand_F(x-\epsilon)+\epsilon$ for all $x\in[0,1]$, i.e., $\hat F$ is $\epsilon$-close to $F$ in L\'evy distance.
\end{proof}

The second claim of \Cref{thm:size-3} runs \Cref{alg:CDF interval query} in phases of halving accuracies, exploiting in each phase the price estimated in the previous one (\Cref{alg:size-3 phased}):

\begin{algorithm}[!t]
\caption{Revenue Maximization using Size-3 Menus}\label{alg:size-3 phased}
\begin{algorithmic}[1]
\Require{Time horizon $T\ge 4$}
\State Let $\gamma:=T^{-2}$ and $\hat p_0:=0$
\For{phase $j=1,2,\ldots$ until the $T$ rounds end}
\State Run \Cref{alg:CDF interval query} with $\epsilon_j=2^{-j}$ and $\delta=T^{-5}$, answering each of its interval queries $[a,b)$ by the size-3 menu built from the price distribution $\gamma\delta_a+\gamma\delta_b+(1-2\gamma)\delta_{\hat p_{j-1}}$ via \Cref{lem:Myerson prices}
\State Let $\hat p_j$ be the price given by \Cref{reduction:price to CDF} for the output of \Cref{alg:CDF interval query} in this phase
\EndFor
\end{algorithmic}
\end{algorithm}

\begin{lemma}[Size-3 Menus via Phased CDF Estimation]\label{lem:size-3 regret}Let $T\ge 4$. The regret of \Cref{alg:size-3 phased} is $\O(\sqrt T\log^3T)$ under every $F\in\mF_\general$.
\end{lemma}
\begin{proof}
By \Cref{lem:Myerson prices}, each menu reveals whether $v_t\in[a,b)$, so every phase runs \Cref{alg:CDF interval query} faithfully, and it earns the expected revenue of a random price drawn from the price distribution, which is at least $(1-2\gamma)R(\hat p_{j-1})$ in phase $j$. We first show that every phase succeeds with high probability, and then bound the regret of each phase.

\textit{Number of Phases.} Phase $i$ estimates $2^{i+1}-1$ tree intervals, each with at least $2$ queries, so it takes at least $2^{i+1}$ rounds. Hence, if phase $j$ is reached, the earlier phases took at least $\sum_{i<j}2^{i+1}=2^{j+1}-4$ rounds, so $2^{j+1}-4\le T$. As $T\ge 4$, at most $\log_2 T$ phases are reached, and all of them have $\epsilon_j\ge\frac 1T$.

\textit{Good Event.} Each phase uses fresh values, so \Cref{lem:CDF segment tree} applies to each phase given the history before it, with $H_j:=j+1\le\log_2T+1$ and $\iota:=\log(2T^5)$. By a union bound over the at most $\log_2 T$ phases, with probability at least $1-\O(\log^4T/T^2)$, every completed phase $j$ outputs an $\hat F$ (\Cref{eq:CDF estimate}) that is $\epsilon_j$-close to $F$, and every phase $j$ lasts at most $160H_j^3\iota\epsilon_j^{-2}$ rounds. Its complement contributes at most $T\cdot\O(\log^4T/T^2)=\O(1)$ to the regret.

\textit{Regret of Each Phase.} On this event, \Cref{reduction:price to CDF} gives $R(p^\ast)-R(\hat p_j)\le 5\epsilon_j$ for every completed phase $j$, which also holds for $j=0$ as $\epsilon_0=1$. So each round of phase $j$ has regret at most $5\epsilon_{j-1}+2\gamma=10\epsilon_j+2T^{-2}$, and phase $j$ has regret at most
\begin{equation*}
10\epsilon_j\min\left\{T,160H_j^3\iota\epsilon_j^{-2}\right\}+2\le 10\sqrt{160H_j^3\iota T}+2,
\end{equation*}
where we used $\min\{a,b\}\le\sqrt{ab}$. Summing over the at most $\log_2T$ phases, the regret is $\O(\sqrt T\log^3T)$.
\end{proof}

\subsection{\texorpdfstring{\textsc{ScaleSearch} for MHR Distributions (\Cref{line:scale} in \Cref{alg:size-2})}{ScaleSearch for MHR Distributions}}\label{app:scale search}

As explained in \Cref{footnote:ScaleSearch}, running the layers of \Cref{alg:size-2} directly on $F$ is wasteful whenever $\Delta\ge R_F(p_F^\ast)$, at which every price is $\Delta$-optimal. We resolve this by rescaling the prices: \Cref{alg:size-2} first calls \textsc{ScaleSearch} (\Cref{line:scale}) to find a scale $\bar p\ge p_F^\ast$, and then runs on the rescaled distribution
\begin{equation}\label{eq:rescaled G}
G:=\text{the distribution of }\min\left\{\frac v{\bar p},1\right\}\text{ for }v\sim F,\quad\text{i.e., }\demand_G(p)=\demand_F(\bar pp)\text{ for all }p\in [0,1].
\end{equation}
Whenever \Cref{alg:size-2} posts $p$ to $G$, we instead post $\bar pp$ to $F$. As $R_G(p)=\frac{R_F(\bar pp)}{\bar p}$, the monopoly price $p_G^\ast=\frac{p_F^\ast}{\bar p}$ of $G$ lies in $[0,1]$, and every regret under $G$ is multiplied by $\bar p\le 1$ under $F$. In this way, instead of suffering the $R_F(p_F^\ast)^{-1}$ dependency in \Cref{footnote:ScaleSearch}, we only suffer $R_G(p_G^\ast)^{-1} \bar p = \frac{\bar p^2}{R_F(p_F^\ast)}$, as $R_G(p_G^\ast)=\frac{R_F(p_F^\ast)}{\bar p}$. In this section, we prove that for MHR distributions, \textsc{ScaleSearch} ensures $\bar p<10p_F^\ast$ (\Cref{lem:scale search}), and $R_F(p_F^\ast)\ge \frac{p_F^\ast}e$ (\Cref{lem:MHR tail}), so $R_G(p_G^\ast)^{-1}=\frac{\bar p}{R_F(p_F^\ast)}\le 10e$.

\textsc{ScaleSearch} searches the geometric grid $x_j=2^{-j}$, which must contain a price $x_{j^\ast}\in (\frac{p_F^\ast}2,p_F^\ast]$ earning at least $\frac{R_F(p_F^\ast)}2$ (recall the one-sided Lipschitzness \Cref{eq:1-side Lip}). It must never rule out $x_{j^\ast}$, but must also rule out the prices far above $p_F^\ast$, so it combines pessimism and optimism:
\begin{itemize}
\item \emph{Pessimism.} The largest lower confidence bound among this grid, namely $L$ in \Cref{alg:scale search}, implies that $R_F(p_F^\ast)\ge L$ with high probability.
\item \emph{Optimism.} A grid price is kept if its upper confidence bound is at least $\frac L2$. Because $x_{j^\ast}$ earns at least $\frac{R_F(p_F^\ast)}2\ge \frac L2$, it is always kept, so twice the largest kept price is at least $p_F^\ast$.
\end{itemize}

\textsc{ScaleSearch} thus runs empirical Bernstein on all these grid prices simultaneously (i.e., in each round, ask all prices once). It waits until all confidence radii drop below $\frac L{16}$, so every kept price indeed earns a constant fraction of $R_F(p_F^\ast)$. For MHR distributions, the revenue decays exponentially above $p_F^\ast$, so all kept prices are $\O(p_F^\ast)$. This also does not bring a significant regret overhead: Posting $x_j$ gives a revenue sample with variance $x_jR_F(x_j)\le R_F(p_F^\ast)$, so $\Otil(R_F(p_F^\ast)^{-1})$ rounds per grid price suffice, each with regret at most $R_F(p_F^\ast)$, and the total regret is $\O(\log^2T)$.

This intuition is summarized in \Cref{lem:scale search} below. Further utilizing the tail and shape of MHR distributions in \Cref{lem:MHR tail,lem:MHR shape} respectively, we then derive the MHR bound in \Cref{thm:size-2}(b).

\begin{subroutine}[!t]
\caption{\textproc{ScaleSearch}(time horizon $T$)}\label{alg:scale search}
\begin{algorithmic}[1]
\State Let $J:=\lceil \log_2 T\rceil$ and $x_j:=2^{-j}$ for $0\le j\le J$
\For{$K=1,2,\ldots$}
\State Post each of $x_0,x_1,\ldots,x_J$ once using the size-1 menu $\{(1,x_j)\}$
\State For each $j$, let $\hat R(x_j)$ be the average of $x_j\1[v\ge x_j]$ over the $K$ rounds posting $x_j$
\State For each $j$, let $r_{j,K}$ be the confidence radius $r_K$ in \Cref{eq:confidence radius} for the outcomes $\1[v\ge x_j]$ of these $K$ rounds, with $\delta=T^{-3}$
\State Let $L:=\max_j(\hat R(x_j)-x_jr_{j,K})$. If $K\ge 2$ and $\max_jx_jr_{j,K}\le \frac L{16}$, \Return $\bar p:=\min\{1,2\max\{x_j:\hat R(x_j)+x_jr_{j,K}\ge \frac L2\}\}$
\EndFor
\end{algorithmic}
\end{subroutine}

\begin{lemma}[Scale Search]\label{lem:scale search}
Let $J:=\lceil \log_2 T\rceil$. With probability at least $1-3(J+1)T^{-2}$:
\begin{enumerate}[label=\textup{(\alph*)}]
\item \Cref{alg:scale search} has regret $\O(\log^2 T)$.
\item If $p_F^\ast\ge 2^{-J}$, its output $\bar p$ satisfies $p_F^\ast\le \bar p\le 1$.
\item Additionally, if $F$ is MHR, then the output $\bar p<10p_F^\ast$.
\end{enumerate}
\end{lemma}
\begin{proof}
\textsc{ScaleSearch} posts $1,\frac 12,\ldots,2^{-J}$ in turn until all revenue estimates are accurate compared with the best lower confidence bound $L$, and returns twice the largest price that may earn at least $\frac L2$. Let $\iota:=\log (2T^3)$. The outcomes $\1[v\ge x_j]$ are i.i.d.\ Bernoulli with mean $\frac{R_F(x_j)}{x_j}$. So by \Cref{lem:empirical Bernstein} with $K_0:=T$ for each $j$ and a union bound, with probability at least $1-3(J+1)T^{-2}$, every $j$ and every $K\ge 2$ satisfy
\begin{equation*}
\lvert \hat R(x_j)-R_F(x_j)\rvert\le x_jr_{j,K}\le\sqrt{\frac{2\iota x_jR_F(x_j)}K}+\frac{15\iota x_j}{2K}.
\end{equation*}
We assume this good event from now on. If $p_F^\ast<2^{-J}$, then $R_F(p_F^\ast)<\frac 1T$ and the regret is less than $1$, so assume $p_F^\ast\ge 2^{-J}$, and let $x_{j^\ast}\in (\frac{p_F^\ast}2,p_F^\ast]$, so that $R_F(x_{j^\ast})\ge x_{j^\ast}\demand_F(p_F^\ast)\ge \frac{R_F(p_F^\ast)}2$. We always have $L\le \max_jR_F(x_j)\le R_F(p_F^\ast)$.
\begin{itemize}
\item \emph{Regret.} As $x_jR_F(x_j)\le R_F(p_F^\ast)$ and $x_j\le 1$, once every $x_j$ has been posted $K\ge 10908\iota R_F(p_F^\ast)^{-1}$ times, every $x_jr_{j,K}$ is at most $\frac{R_F(p_F^\ast)}{36}$. Then $L\ge R_F(x_{j^\ast})-2x_{j^\ast}r_{j^\ast,K}\ge \frac 49R_F(p_F^\ast)\ge 16\max_jx_jr_{j,K}$, so \textsc{ScaleSearch} stops. Hence it takes $\O((J+1)\iota R_F(p_F^\ast)^{-1})$ rounds, each with regret at most $R_F(p_F^\ast)$, so its regret is $\O(\log^2 T)$.
\item \emph{Output.} When \textsc{ScaleSearch} stops, $L\ge R_F(x_{j^\ast})-2x_{j^\ast}r_{j^\ast,K}\ge \frac{R_F(p_F^\ast)}2-\frac L8$, so $L\ge \frac 49R_F(p_F^\ast)$. As $\hat R(x_{j^\ast})+x_{j^\ast}r_{j^\ast,K}\ge R_F(x_{j^\ast})\ge \frac{R_F(p_F^\ast)}2\ge \frac L2$, we have $\bar p\ge \min\{1,2x_{j^\ast}\}\ge p_F^\ast$. Moreover, every $j$ with $\hat R(x_j)+x_jr_{j,K}\ge \frac L2$ has $R_F(x_j)\ge \frac L2-2x_jr_{j,K}\ge \frac{3L}8\ge \frac{R_F(p_F^\ast)}6$. If $F$ is MHR, \Cref{lem:MHR tail} gives $R_F(p_F^\ast+u)\le R_F(p_F^\ast)(1+x)e^{-x}$ with $x:=\frac u{p_F^\ast}$, which is less than $\frac{R_F(p_F^\ast)}6$ if $x\ge 4$, as $5e^{-4}<\frac 16$. So all such $x_j$ are less than $5p_F^\ast$, and $\bar p<10p_F^\ast$.\qedhere
\end{itemize}
\end{proof}

\begin{lemma}[MHR Tails]\label{lem:MHR tail}
Let $F$ be MHR, and let $q^\ast:=\demand_F(p_F^\ast)$. Then $q^\ast\ge e^{-1}$, so $R_F(p_F^\ast)\ge \frac{p_F^\ast}e$, and every $u\ge 0$ satisfies $R_F(p_F^\ast+u)\le R_F(p_F^\ast)(1+x)e^{-x}$ with $x:=\frac u{p_F^\ast}$.
\end{lemma}
\begin{proof}
Let $h:=\frac f{\demand_F}$ be the hazard rate, so that $R_F'(p)=\demand_F(p)(1-ph(p))$ for almost every $p$. As $h$ is non-decreasing and $p_F^\ast$ maximizes $R_F$, we have $h(p)\le \frac 1{p_F^\ast}$ for $p<p_F^\ast$ and $h(p)\ge \frac 1{p_F^\ast}$ for $p>p_F^\ast$. Hence $q^\ast=\exp(-\int_0^{p_F^\ast}h)\ge e^{-1}$, and $\demand_F(p_F^\ast+u)=q^\ast\exp(-\int_{p_F^\ast}^{p_F^\ast+u}h)\le q^\ast e^{-x}$, so $R_F(p_F^\ast+u)\le (p_F^\ast+u)q^\ast e^{-x}=R_F(p_F^\ast)(1+x)e^{-x}$.
\end{proof}

\begin{lemma}[MHR Shape]\label{lem:MHR shape}
Let $F$ be MHR and $\Delta\le \frac{R_F(p_F^\ast)}4$. Then every $\Delta$-optimal price $p$ satisfies $\lvert p-p_F^\ast\rvert\le 4\sqrt \Delta$.
\end{lemma}
\begin{proof}
We lower bound the gap $R_F(p_F^\ast)-R_F(p)$ on both sides of $p_F^\ast$, with $h$ and $q^\ast$ as in \Cref{lem:MHR tail}.
\begin{itemize}
\item \emph{Left side.} For $s\in [0,p_F^\ast]$, as $h\le \frac 1{p_F^\ast}$ below $p_F^\ast$, $R_F'(p_F^\ast-s)\ge q^\ast(1-\frac{p_F^\ast-s}{p_F^\ast})=\frac{q^\ast s}{p_F^\ast}$. Integrating, $R_F(p_F^\ast)-R_F(p_F^\ast-u)\ge \frac{q^\ast u^2}{2p_F^\ast}\ge \frac{u^2}{2e}$, which exceeds $\Delta$ if $u>\sqrt{2e\Delta}$.
\item \emph{Right side.} By \Cref{lem:MHR tail}, $R_F(p_F^\ast)-R_F(p_F^\ast+u)\ge R_F(p_F^\ast)(1-(1+x)e^{-x})$ with $x:=\frac u{p_F^\ast}$. Since $1-(1+x)e^{-x}=\int_0^x ye^{-y}\,dy$, it is at least $\frac{x^2}{2e}$ for $x\le 1$ and at least $1-\frac 2e>\frac 14$ for $x\ge 1$. So $p_F^\ast+u$ is not $\Delta$-optimal if $x\ge 1$; otherwise, its gap is at least $\frac{R_F(p_F^\ast)u^2}{2e(p_F^\ast)^2}=\frac{q^\ast u^2}{2ep_F^\ast}\ge \frac{u^2}{2e^2}$, which exceeds $\Delta$ if $u>e\sqrt{2\Delta}$.
\end{itemize}
As $\sqrt{2e}<e\sqrt 2<4$, every $\Delta$-optimal price is within $4\sqrt\Delta$ of $p_F^\ast$.
\end{proof}

\subsection{\texorpdfstring{Proof of \Cref{lem:size-2 rounds} (Number of Rounds)}{Proof: Number of Rounds}}\label{app:size-2 rounds}

Throughout, we analyze \Cref{alg:size-2} after \Cref{line:scale}, i.e., on the rescaled distribution $G$ in \Cref{eq:rescaled G}, so that $R_G$, $p_G^\ast$, and $\demand_G$ below are the revenue, monopoly price, and demand of $G$. The candidates $\candidateSet_\Delta$ of layer $\Delta$ are initialized in \Cref{alg:size-2} and then shrink in its Eliminate step, and layer $\frac\Delta 2$ starts from the survivors. We use the one-sided Lipschitzness \Cref{eq:1-side Lip} in the following form: For any $p<p'$,
\begin{equation}\label{eq:1-side Lip mass}
p\intervalP_G([p,p'))=R_G(p)-R_G(p')+(p'-p)\demand_G(p')\le R_G(p)-R_G(p')+(p'-p).
\end{equation}

\begin{lemma}[Restatement of \Cref{lem:size-2 rounds}; Number of Rounds]\begin{enumerate}[label=\textup{(\alph*)}]
\item For each $\Delta=2^{-k}$, \Cref{alg:size-2} spends $\Otil((p_N-p_0+\Delta)/\Delta^2)$ rounds where $p_0,p_N$ are both $4\Delta$-optimal prices.
\item For each $\Delta=2^{-k}$, \Cref{alg:size-2} incurs $\O(\Delta)$ regret in each of these rounds.
\item Consequently, for any given $T$, the regret of \Cref{alg:size-2} is upper bounded as
\begin{equation*}
\Otil\left (\sup_{\Delta=2^{-k},k\ge 0} \min\left \{T \Delta,\frac{\diam\{p\in [0,1]\mid R(p)\ge R(p^\ast)-\Delta\}}{\Delta}\right \}\right ),
\end{equation*}
where $\diam\{A\}:=\sup\{A\}-\inf\{A\}$ is the diameter of a set $A$.
\end{enumerate}
\end{lemma}
\begin{proof}
By \Cref{lem:size-2 good event}, all estimates are accurate except with probability $3T^{-1}$, which contributes $\O(1)$ regret, so we assume this event $\mathcal E$. Fix a layer $\Delta$ with candidates $\candidateSet_{2\Delta}=\{p_0<p_1<\cdots<p_N\}$, which are all $4\Delta$-optimal by \Cref{lem:approx price} below for the previous layer (or trivially for $\Delta=\frac 12$).

\textit{Claim (b): Regret per Round.} Every menu in layer $\Delta$ puts probability at least $1-T^{-2}$ on a price in $\candidateSet_{2\Delta}$, and \Cref{line:estimate init mass} posts $p_N\in\candidateSet_{2\Delta}$. As these prices are $4\Delta$-optimal, every round has regret $4\Delta+T^{-2}$.

\textit{Claim (a): Number of Rounds.} By \Cref{lem:size-2 good event}, a call estimating $\intervalP_G(I)$ with precision $\epsilon$ takes at most $8\iota\intervalP_G(I)\epsilon^{-2}+15\iota\epsilon^{-1}+2$ rounds, where $\iota:=\log(2T^3)$. We bound the first term by telescoping \Cref{eq:1-side Lip mass} over disjoint intervals, and the other two by the number of calls. If $N=0$, layer $\Delta$ only makes the calls in \Cref{line:estimate init mass,line:discard}, which take $\O(\iota\Delta^{-1})$ rounds by the last two bullets below, so we assume $N\ge 1$ and let $c_N:=80(\log_2N+2)$.
\begin{itemize}
\item \emph{One interval.} Every interval in \Cref{line:segT Bernstein,line:non-segT Bernstein,line:discard} is some $I=[a,b)$ with $a\in\candidateSet_{2\Delta}$, estimated with precision $\epsilon=\frac\Delta{c(a+\Delta)}$, where $c=c_N$ for tree intervals and $c=40$ otherwise. As $(a+\Delta)^2\le 2(a+\Delta^2)$ and $a+\Delta\le 2$, this call takes at most $16\iota c^2\Delta^{-2}(a\intervalP_G(I)+\Delta^2\intervalP_G(I))+30\iota c\Delta^{-1}+2$ rounds.
\item \emph{Telescoping across depth.} The tree intervals of each of the at most $\log_2N+2$ depths are disjoint subintervals of $[p_0,p_N)$ (\Cref{lem:segment tree}), and so are the intervals $[p_i,p_i+\frac\Delta 2)$ with $i<N$, as the candidates are $\Delta$ apart. For each such family of disjoint intervals $[a,b)$, summing \Cref{eq:1-side Lip mass} over a partition of $[p_0,p_N)$ that contains them, the revenues telescope; as $p_0$ and $p_N$ are $4\Delta$-optimal and the probabilities sum to at most $1$,
\begin{equation*}
\sum_{[a,b)}a\,\intervalP_G([a,b))\le p_N-p_0+R_G(p_0)-R_G(p_N)\le p_N-p_0+4\Delta\quad\text{and}\quad\sum_{[a,b)}\Delta^2\intervalP_G([a,b))\le\Delta.
\end{equation*}
As $\mT$ has $2N-1$ tree intervals (\Cref{lem:segment tree}), the tree intervals take at most $16\iota c_N^2\Delta^{-2}(\log_2N+2)(p_N-p_0+5\Delta)+(2N-1)(30\iota c_N\Delta^{-1}+2)$ rounds.
\item \emph{All $[p_i,p_i+\frac \Delta 2)$ except for $i=N$.} The calculation in the two bullets above carries over, giving that all $[p_i,p_i+\frac \Delta 2)$ with $i<N$ take at most $16\cdot 40^2\iota\Delta^{-2}(p_N-p_0+5\Delta)+N(1200\iota\Delta^{-1}+2)$ rounds.
\item \emph{The last} $I=[p_N,p_N+\frac \Delta 2)$, which is not inside $[p_0,p_N)$. If $\intervalP_G(I)\ge\frac{4\Delta}{p_N}$, the lower confidence bound $\hat\mu_K-r_K\ge\intervalP_G(I)-2r_K$ exceeds $\frac{2\Delta}{p_N}$ once $r_K\le\frac{\intervalP_G(I)}5$, so by \Cref{eq:EB radius}, \Cref{line:discard} stops this call within $\O(\iota\intervalP_G(I)^{-1})=\O(\iota\Delta^{-1})$ rounds. Otherwise, $p_N\intervalP_G(I)+\Delta^2\intervalP_G(I)\le 5\Delta$, so it takes $\O(\iota\Delta^{-1})$ rounds by the first bullet.
\item \emph{Rounds when posting $p_N$.} \Cref{line:estimate init mass} estimates $\demand_G(p_N)\le 1$ up to $\frac{\Delta}{8(p_N-p_0+\Delta)}$, which takes $\O(\iota(p_N-p_0+\Delta)^2\Delta^{-2})=\O(\iota(p_N-p_0+\Delta)\Delta^{-2})$ rounds, as $p_N-p_0+\Delta\le 2$.
\end{itemize}
As the candidates in $\candidateSet_{2\Delta}$ are $\Delta$ apart, $N\le \frac{p_N-p_0}\Delta$, and as every call takes at least two rounds, $N\le T$. Hence layer $\Delta$ takes $\O(\iota(p_N-p_0+\Delta)\Delta^{-2}(\log_2N+2)^3)=\O((p_N-p_0+\Delta)\Delta^{-2}\log^4 T)$ rounds, which proves claim (a).

\textit{Claim (c).} Let $S_\Delta$ be the set of $\Delta$-optimal prices of $F$, let $s(\Delta):=\frac{\diam S_\Delta}\Delta$, and let
\begin{equation}\label{eq:size-2 S}
S:=\sup_{k\ge 0}\min\left\{T2^{-k},s(2^{-k})\right\}=\sup_{k\ge 0}\min\left\{T2^{-k},\frac{\diam S_{2^{-k}}}{2^{-k}}\right\}
\end{equation}
be the supremum in claim (c), so that $S\ge \min\{T,s(1)\}=1$ as every price is $1$-optimal. The failure event in \Cref{lem:scale search} contributes $\O(1)$ regret, and if $p_F^\ast<2^{-J}$ with $J:=\lceil\log_2T\rceil$, then $R_F(p_F^\ast)<\frac 1T$ and the regret is less than $1$. Otherwise, \textsc{ScaleSearch} has regret $\O(\log^2T)$ and outputs $\bar p\in [p_F^\ast,1]$.

For each $\Delta$, as $R_G(p)=\frac{R_F(\bar pp)}{\bar p}$ by \Cref{eq:rescaled G}, $G$'s $\Delta$-optimal prices are $\frac 1{\bar p}S_{\bar p\Delta}$. So in layer $\Delta$, the candidates satisfy $p_N-p_0\le \frac 1{\bar p}\diam S_{4\bar p\Delta}$, and by (a) and (b), the layer has regret $\O((p_N-p_0+\Delta)\Delta^{-1}\log^4T)$ under $G$. As every regret under $G$ is multiplied by $\bar p\le 1$ under $F$, the layer has regret $\O((s(4\bar p\Delta)+1)\log^4T)$ under $F$, and each of its rounds has regret at most $\bar p(4\Delta+T^{-2})$.

For $i\ge 0$ and $\Delta\in (2^{-i-1},2^{-i}]$, as $S_\Delta\subseteq S_{2^{-i}}$, we have $s(\Delta)\le 2s(2^{-i})$ and $T\Delta\le T2^{-i}$, so $\min\{T\Delta,s(\Delta)\}\le 2S$. For $\Delta>1$, $s(\Delta)\le \frac 1\Delta<S$.

Let $\Delta_j:=4\bar p2^{-j}$. For every $k\ge 0$, by the bound for one layer above, the layers $2^{-j}$ with $1\le j\le k$ have regret $\O(\log^4T\sum_{j=1}^k(s(\Delta_j)+1))$ in total, and all later rounds have regret $\O(T\Delta_k+1)$ in total. Let $k_T:=\lceil \log_2T\rceil+2$, so that $T\Delta_{k_T}\le 1$. If $s(\Delta_j)\le 2S$ for all $1\le j\le k_T$, then $k=k_T$ gives regret $\O(S\log^5T)$. Otherwise, let $k^\ast\le k_T$ be the smallest $j\ge 1$ with $s(\Delta_j)>2S$. By the previous paragraph, $T\Delta_{k^\ast}\le 2S$, so $k=k^\ast-1$ gives regret $\O(\log^4T\cdot k^\ast(2S+1)+2T\Delta_{k^\ast}+1)=\O(S\log^5T)$.

Adding the $\O(\log^2T)$ regret of \textsc{ScaleSearch}, the regret remains $\O(S\log^5T)$, which proves (c).
\end{proof}

\begin{lemma}[All Estimates Are Accurate]\label{lem:size-2 good event}
Let $\iota:=\log(2T^3)$. With probability at least $1-3T^{-1}$, every call of \Cref{alg:empirical Bernstein} in \Cref{alg:size-2}, say on an interval $I$, satisfies \Cref{eq:EB radius} with $\mu=\intervalP_G(I)$ for every number $K\ge 2$ of queries it makes. On this event, which we denote by $\mathcal E$, every call with precision $\epsilon$ outputs $\hat\intervalP(I)$ with $\lvert\hat\intervalP(I)-\intervalP_G(I)\rvert\le\epsilon$, and it stops within $8\iota\intervalP_G(I)\epsilon^{-2}+15\iota\epsilon^{-1}+2$ queries unless the $T$ rounds end or \Cref{line:discard} stops it earlier.
\end{lemma}
\begin{proof}
Each call uses fresh values and makes at most $T$ queries, so by \Cref{lem:empirical Bernstein} with $K_0:=T$ and $\delta=T^{-3}$, it violates \Cref{eq:EB radius} with probability at most $3T^{-2}$. As there are at most $T$ calls, a union bound proves the first claim. On $\mathcal E$, a call stops once $r_K\le\epsilon$, which holds for every $K\ge 8\iota\intervalP_G(I)\epsilon^{-2}+15\iota\epsilon^{-1}$ by \Cref{eq:EB radius}, and then its output $\hat\mu_K$ satisfies $\lvert\hat\mu_K-\intervalP_G(I)\rvert\le r_K\le\epsilon$.
\end{proof}

\begin{lemma}[Revenue Estimation Error]\label{lem:rev est err}
On $\mathcal E$, for every layer $\Delta$ and every two candidates $p<p'$ in $\candidateSet_\Delta$ before the Eliminate step, the estimates in \Cref{eq:estimate revenue} satisfy
\begin{equation*}
\left \lvert \left (\hat R(p')-\hat R(p)\right )-\left (R_G(p')-R_G(p)\right )\right \rvert\le \frac \Delta 4.
\end{equation*}
\end{lemma}
\begin{proof}
The key is that the error of $\hat\demand(p_N)$ enters $\hat R(p)$ multiplied by $p$, so it only contributes $(p'-p)$ times this error to the difference. Formally, fix a candidate $p$, and let $p_i:=\max\{p''\in \candidateSet_{2\Delta}\mid p''\le p\}$ as in \Cref{eq:estimate revenue}, so that $p\in\{p_i,p_i+\frac\Delta 2\}$. As $\mI_\mT[p_i,p_N)$ partitions $[p_i,p_N)$ (\Cref{lem:segment tree}),
\begin{equation*}
R_G(p)=p\demand_G(p)=p\left(\demand_G(p_N)+\sum_{I\in \mI_\mT[p_i,p_N)}\intervalP_G(I)-\intervalP_G([p_i,p))\right),
\end{equation*}
where $\intervalP_G([p_i,p_i)):=0$. Comparing with \Cref{eq:estimate revenue}, $\hat R(p)-R_G(p)=p(e_{\mathrm{tail}}+e_p)$, where $e_{\mathrm{tail}}$ is the error of $\hat \demand(p_N)$ in \Cref{line:estimate init mass}, and $e_p$ collects the errors of the tree intervals in $\mI_\mT[p_i,p_N)$ (\Cref{line:segT Bernstein}) and of $\hat \intervalP([p_i,p))$ (\Cref{line:non-segT Bernstein,line:discard}).
\begin{itemize}
\item \emph{The term $pe_p$.} $\mI_\mT[p_i,p_N)$ is empty if $N=0$, and otherwise has at most $2(\log_2N+2)$ tree intervals (\Cref{lem:segment tree}), each starting at a price at least $p_i$, so on $\mathcal E$ each of their errors is at most $\frac{\Delta}{80(\log_2N+2)(p_i+\Delta)}$, and the error of $\hat\intervalP([p_i,p))$ is at most $\frac{\Delta}{40(p_i+\Delta)}$. As $p\le p_i+\Delta$, we get $\lvert pe_p\rvert\le\frac\Delta{40}+\frac\Delta{40}=\frac\Delta{20}$.
\item \emph{The term $pe_{\mathrm{tail}}$.} On $\mathcal E$, $\lvert e_{\mathrm{tail}}\rvert\le\frac{\Delta}{8(p_N-p_0+\Delta)}$. As $p'-p\le p_N+\frac\Delta 2-p_0$, we get $(p'-p)\lvert e_{\mathrm{tail}}\rvert\le\frac\Delta 8$.
\end{itemize}
The error of $\hat R(p')-\hat R(p)$ is $(p'-p)e_{\mathrm{tail}}+p'e_{p'}-pe_p$, whose absolute value is thus at most $\frac\Delta 8+\frac\Delta{20}+\frac\Delta{20}<\frac\Delta 4$.
\end{proof}

\begin{lemma}[Candidates Always Approximate the Monopoly Price]\label{lem:approx price}
On $\mathcal E$, after the Eliminate step of every layer $\Delta$, the candidates $\candidateSet_\Delta$ contain a $\hat p$ with $p_G^\ast-\frac \Delta2\le \hat p\le p_G^\ast$ and $R_G(\hat p)\ge R_G(p_G^\ast)-\frac \Delta 2$, and every $p\in \candidateSet_\Delta$ satisfies $R_G(p)>R_G(p_G^\ast)-2\Delta$.
\end{lemma}
\begin{proof}
We use induction on the layers. For $\candidateSet_1=\{0,\frac 12,1\}$, some $\hat p\in \candidateSet_1$ has $p_G^\ast-\frac 12\le \hat p\le p_G^\ast$. Now consider layer $\Delta$, and suppose that some $p'\in \candidateSet_{2\Delta}$ has $p_G^\ast-\Delta\le p'\le p_G^\ast$.
\begin{itemize}
\item \emph{A candidate close to $p_G^\ast$.} The Refine step adds $p'$ to $\candidateSet_\Delta$, and also $p'+\frac \Delta 2$ if $p'+\frac \Delta 2\le 1$. One of them, say $\hat p$, satisfies $p_G^\ast-\frac \Delta 2\le \hat p\le p_G^\ast$: namely $p'+\frac \Delta 2$ if $p'+\frac \Delta 2\le p_G^\ast$ (which is then added, as $p_G^\ast\le 1$), and $p'$ otherwise. By \Cref{eq:1-side Lip}, $R_G(\hat p)\ge R_G(p_G^\ast)-(p_G^\ast-\hat p)\ge R_G(p_G^\ast)-\frac\Delta 2$.
\item \emph{It survives \Cref{line:discard}.} This line only removes $p_N+\frac \Delta 2$, so it only matters if $\hat p=p_N+\frac \Delta 2$, i.e., $p'=p_N$ and $p_N+\frac\Delta 2\le p_G^\ast$. Then \Cref{eq:1-side Lip mass} and $R_G(p_N)\le R_G(p_G^\ast)$ give $p_N\intervalP_G([p_N,p_G^\ast))\le p_G^\ast-p_N\le \Delta$. So $\intervalP_G([p_N,p_N+\frac \Delta 2))\le\frac\Delta{p_N}$, and on $\mathcal E$, its lower confidence bound never exceeds $\frac{2\Delta}{p_N}$.
\item \emph{It survives the Eliminate step.} Let $\hat p^\ast:=\argmax_{p\in \candidateSet_\Delta}\hat R(p)$. By \Cref{lem:rev est err}, $\hat R(\hat p^\ast)-\hat R(\hat p)\le R_G(\hat p^\ast)-R_G(\hat p)+\frac\Delta 4\le R_G(p_G^\ast)-R_G(\hat p)+\frac\Delta 4\le \frac 34\Delta<\Delta$.
\end{itemize}
This proves the first claim for $\candidateSet_\Delta$, which is also the induction hypothesis for the next layer. For the second claim, every survivor $p$ has $\hat R(\hat p)-\hat R(p)\le \hat R(\hat p^\ast)-\hat R(p)<\Delta$, so by \Cref{lem:rev est err} again,
\begin{equation*}
R_G(p_G^\ast)-R_G(p)\le \left(R_G(p_G^\ast)-R_G(\hat p)\right)+\left(\hat R(\hat p)-\hat R(p)\right)+\frac \Delta 4<\frac \Delta 2+\Delta+\frac \Delta 4<2\Delta.\qedhere
\end{equation*}
\end{proof}

\subsection{\texorpdfstring{Proofs of \Cref{thm:size-2,thm:instance upper} (Upper Bounds with Size-2 Menus)}{Proofs: Upper Bounds with Size-2 Menus}}\label{app:size-2}

We first reduce \Cref{thm:size-2} to the covering numbers of the rescaled distribution $G$ in \Cref{eq:rescaled G}:
\begin{lemma}[Reduction to Covering Numbers]\label{lem:size-2 reduction}
Let $F$ be a value distribution with $R_F(p_F^\ast)>0$. Except on events that contribute $\O(1)$ to the regret, either the regret of \Cref{alg:size-2} on $F$ is $\O(\log^2T)$, or \textsc{ScaleSearch} outputs some $\bar p\in [p_F^\ast,1]$, and the regret is at most
\begin{equation}\label{eq:size-2 regret via W}
\O(\log^2T)+\O\left(\bar pW_G\log^4T\right),\qquad\text{where }W_G:=\min_{k\ge 0}\left\{T2^{-k}+\sum_{j=0}^kN_G(2^{-j})\right\},
\end{equation}
and $N_G$ is the covering number of the rescaled distribution $G$ in \Cref{eq:rescaled G}, which satisfies
\begin{equation}\label{eq:N_G}
N_G(\Delta)\le N_F(\bar p\Delta)\quad\text{and}\quad N_G(\Delta)\le\frac 2\Delta\qquad\text{for all }\Delta\in(0,1].
\end{equation}
\end{lemma}
\begin{proof}
Let $J:=\lceil \log_2T\rceil$. The failure events of \Cref{lem:size-2 good event,lem:scale search} contribute $\O(1)$ regret, and if $p_F^\ast<2^{-J}\le\frac 1T$, the regret is at most $TR_F(p_F^\ast)<1$. Otherwise, by \Cref{lem:scale search}, \textsc{ScaleSearch} has regret $\O(\log^2T)$, which is the whole regret if the $T$ rounds end before it stops; otherwise, it outputs $\bar p\in[p_F^\ast,1]$. The rest of \Cref{alg:size-2} runs on $G$ in \Cref{eq:rescaled G}: posting $p\in[0,1]$ to $G$ corresponds to posting $\bar pp$ to $F$, and $R_G(p)=\frac{R_F(\bar pp)}{\bar p}$ is maximized at $\frac{p_F^\ast}{\bar p}\in[0,1]$, so every regret under $G$ is multiplied by $\bar p$ under $F$. Hence by \Cref{lem:regret widths} below for $G$, the total regret is at most \Cref{eq:size-2 regret via W}. Finally, if $p$ is $\Delta$-optimal for $G$, then $\bar pp$ is $\bar p\Delta$-optimal for $F$, so \Cref{lem:covering} below for $G$ gives \Cref{eq:N_G}.
\end{proof}

Compared with $V_T(F)$ in \Cref{thm:instance upper} and with the supremum in \Cref{lem:size-2 rounds} (which has $\frac{\diam}\Delta$ in place of $N$ and only dyadic $\Delta$), $W_G$ differs in two ways. First, $W_G$ sums over the scales, whereas the other two only take the largest scale; the sum is at most $\O(\log T)$ times the largest one (claim (c) of \Cref{lem:size-2 rounds}), which brings the extra $\log T$ in \Cref{thm:instance upper}, while summing directly keeps $\log^4T$ in \Cref{thm:size-2}. Second, $W_G$ is for the rescaled $G$ rather than $F$, which matters for MHR distributions: $\bar pW_G=\O(T^{1/3})$ for every MHR distribution (see the proof below), whereas $V_T(F)$ itself can be $\Theta(\sqrt T)$ (\Cref{footnote:ScaleSearch}).

\begin{theorem}[Restatement of \Cref{thm:size-2}; Upper Bounds]
For dynamic pricing with size-2 menus, \Cref{alg:size-2} ensures:
\begin{enumerate}[label=\textup{(\alph*)}]
\item $\mR_\general^{(2)}(T)=\O(\sqrt T\log^4 T)$;
\item $\mR_\MHR^{(2)}(T)=\O(T^{1/3}\log^4 T)$;
\item $\mR_\curvature^{(2)}(T)=\O((\kappa^{-1/3}T^{1/3}+\log T)\log^4 T)$ for every curvature $\kappa\in (0,2)$.
\end{enumerate}
\end{theorem}
\begin{proof}
If $R_F(p_F^\ast)=0$, then $v=0$ a.s.\ and every algorithm has regret $0$. Otherwise, by \Cref{lem:size-2 reduction} above, it suffices to bound $\bar pW_G$ in \Cref{eq:size-2 regret via W} for each class, recalling $\bar p\in [p_F^\ast,1]$, the rescaled $G$ in \Cref{eq:rescaled G}, and \Cref{eq:N_G}.
\begin{enumerate}[label=\textup{(\alph*)}]
\item \emph{General distributions.} Taking $k$ with $2^k\in[\sqrt T,2\sqrt T)$, \Cref{eq:N_G} gives $\bar pW_G\le T2^{-k}+\sum_{j\le k}2^{j+1}\le \sqrt T+2^{k+2}\le 9\sqrt T$.
\item \emph{MHR distributions.} By \Cref{lem:scale search}, $\bar p<10p_F^\ast$, and by \Cref{lem:MHR tail}, $R_F(p_F^\ast)\ge\frac{p_F^\ast}e$, so $\frac{R_F(p_F^\ast)}{4\bar p}\ge\frac 1{40e}$. As \Cref{lem:MHR shape} gives $N_F(\Delta)\le 1+8\Delta^{-1/2}$ for $\Delta\le \frac{R_F(p_F^\ast)}4$, \Cref{eq:N_G} gives $N_G(\Delta)\le 1+8(\bar p\Delta)^{-1/2}$ for $\Delta\le \frac 1{40e}$. If $\bar pT^2<1$, then $R_F(p_F^\ast)\le\bar p<T^{-2}$ and the regret is less than $1$. Otherwise, take $k$ with $2^k\in[\bar p^{1/3}T^{2/3},2\bar p^{1/3}T^{2/3})$. In $W_G$, the terms with $2^{-j}>\frac 1{40e}$ sum to at most $\sum_{j:2^j<40e}2^{j+1}\le 160e$ by \Cref{eq:N_G}, and the other terms sum to at most $\sum_{j\le k}(1+8(\bar p2^{-j})^{-1/2})\le k+1+28\bar p^{-1/2}2^{k/2}$. Hence $\bar pW_G\le\bar p(T2^{-k}+160e+k+1+28\bar p^{-1/2}2^{k/2})=\O(\bar p^{2/3}T^{1/3}+\log T)=\O(T^{1/3})$.
\item \emph{Distributions with curvature $\kappa$.} $G$ has curvature $\kappa\bar p$, as $R_G(\frac{p_F^\ast}{\bar p})-R_G(p)=\frac{R_F(p_F^\ast)-R_F(\bar pp)}{\bar p}\ge \kappa\bar p(p-\frac{p_F^\ast}{\bar p})^2$. So every $\Delta$-optimal price of $G$ is within $(\Delta/(\kappa\bar p))^{1/2}$ of $\frac{p_F^\ast}{\bar p}$, and $N_G(\Delta)\le 1+2(\kappa\bar p\Delta)^{-1/2}$. If $\kappa\bar pT^2<1$, the regret is at most $TR_F(p_F^\ast)\le T\bar p<\frac 1{\kappa T}$, which is at most $\kappa^{-1/3}T^{1/3}$ if $\kappa\ge T^{-2}$, while $\kappa^{-1/3}T^{1/3}>T$ otherwise. Otherwise, take $k$ with $2^k\in[(\kappa\bar p)^{1/3}T^{2/3},2(\kappa\bar p)^{1/3}T^{2/3})$, so that $k=\O(\log T)$ as $\kappa\le 2$ (\Cref{lem:curvature range}), and $\bar pW_G\le \bar p(T2^{-k}+k+1+7(\kappa\bar p)^{-1/2}2^{k/2})=\O(\kappa^{-1/3}T^{1/3}+\log T)$.
\end{enumerate}
Plugging these bounds into \Cref{eq:size-2 regret via W} proves the theorem.
\end{proof}

\begin{theorem}[Restatement of \Cref{thm:instance upper}; Instance-Dependent Regret Upper Bound]
For any value distribution $F$ and any $T\ge 2$, the regret of \Cref{alg:size-2} is $\O(V_T(F)\log^5 T)$, where
\begin{equation*}
V_T(F):=\sup_{\Delta\in (0,1]}\min\{T\Delta,N_F(\Delta)\},
\end{equation*}
which implies regret $\O(CT^{d/(d+1)}\log^5 T)$ on any $F$ with interval dimension $d$ and multiplier $C$.
\end{theorem}
\begin{proof}
By claim (c) of \Cref{lem:size-2 rounds}, where $\Otil$ hides $\log^5T$, it suffices to show that its supremum in \Cref{eq:size-2 S} satisfies
\begin{equation*}
S=\sup_{k\ge 0}\min\left\{T2^{-k},\frac{\diam S_{2^{-k}}}{2^{-k}}\right\}\le V_T(F),
\end{equation*}
where $S_\Delta$ is the set of $\Delta$-optimal prices of the original value distribution $F$. Indeed, for every $\Delta=2^{-k}$, \Cref{lem:covering} gives $\frac{\diam S_\Delta}\Delta=\frac{u_\Delta-\ell_\Delta}\Delta\le N_F(\Delta)$, so $\min\{T\Delta,\frac{\diam S_\Delta}\Delta\}\le \min\{T\Delta,N_F(\Delta)\}\le V_T(F)$.

For the second claim, it suffices to show $V_T(F)\le CT^{d/(d+1)}$. Indeed, $C\ge N_F(1)=1$, and every $\Delta\in (0,1]$ has either $\Delta\le T^{-1/(d+1)}$, so $T\Delta\le T^{d/(d+1)}$, or $\Delta>T^{-1/(d+1)}$, so $N_F(\Delta)\le C\Delta^{-d}<CT^{d/(d+1)}$.
\end{proof}

We now prove the lemmas used above, in order of first use.

\begin{lemma}[Covering Numbers]\label{lem:covering}
For any value distribution $F$ and $\Delta>0$, the $\Delta$-optimal prices have a smallest one $\ell_\Delta$ and a largest one $u_\Delta$, where $u_\Delta>\ell_\Delta$, and the covering number $N_F(\Delta)=\lceil \frac{u_\Delta-\ell_\Delta}\Delta\rceil$ (defined in \Cref{sec:size-2 regret} for $\Delta\le 1$, and by the same formula for $\Delta>1$) satisfies
\begin{equation*}
\max\left\{1,\frac{u_\Delta-\ell_\Delta}\Delta\right\}\le N_F(\Delta)\le 1+\frac{u_\Delta-\ell_\Delta}\Delta.
\end{equation*}
In particular, $N_F(\Delta)=1$ for $\Delta\ge 1$, and $N_F(\Delta)\le\frac 2\Delta$ for $\Delta\in(0,1]$.
\end{lemma}
\begin{proof}
As $\demand_F(p)$ is non-increasing and left-continuous in $p$, $R_F$ is upper semi-continuous, so the set of $\Delta$-optimal prices is closed, and it has a smallest and a largest element. They differ: If $\Delta\ge R_F(p_F^\ast)$, then every price is $\Delta$-optimal as $R_F\ge 0$; otherwise, $p_F^\ast\ge R_F(p_F^\ast)>\Delta$, and every $p\in [p_F^\ast-\Delta,p_F^\ast]$ is $\Delta$-optimal by \Cref{eq:1-side Lip}. As $u_\Delta-\ell_\Delta>0$, the display follows from the definition of $N_F(\Delta)$. Finally, every price is $1$-optimal as $R_F\le 1$, so $u_\Delta-\ell_\Delta\le 1$ for all $\Delta$, which gives the last two claims.
\end{proof}

\begin{lemma}[Regret of the Layers via Covering Numbers]\label{lem:regret widths}
Run \Cref{alg:size-2} without \Cref{line:scale}, i.e., with $\bar p=1$, on a value distribution $G$ with covering number $N_G$. Its regret is $\O(W_G\log^4 T)$, where $W_G:=\min_{k\ge 0}\{T2^{-k}+\sum_{j=0}^k N_G(2^{-j})\}$.
\end{lemma}
\begin{proof}
We bound the regret of each layer by its covering number, and the regret of all rounds after a given layer by its scale. The failure event in \Cref{lem:size-2 good event} contributes $\O(1)$ regret. Otherwise, fix a layer $\Delta$ and write $\candidateSet_{2\Delta}=\{p_0<p_1<\cdots<p_N\}$. By \Cref{lem:approx price}, all of $p_0,\ldots,p_N$ are $4\Delta$-optimal, so each round in this layer has regret at most $4\Delta+T^{-2}$, and $p_N-p_0\le u_{4\Delta}-\ell_{4\Delta}\le 4\Delta N_G(4\Delta)$ by \Cref{lem:covering}. As $N_G(4\Delta)\ge 1$, \Cref{lem:size-2 rounds} shows that this layer takes $\O(N_G(4\Delta)\Delta^{-1}\log^4 T)$ rounds, so its regret is $\O(N_G(4\Delta)\log^4 T)$. Now fix any $k\ge 0$. The layers $\Delta=2^{-i}$ with $1\le i\le k+2$ have $4\Delta=2^{-(i-2)}$, so their total regret is $\O(\log^4 T\cdot (N_G(2)+\sum_{j=0}^k N_G(2^{-j})))$, where $N_G(2)=1$. Each round in a later layer has regret at most $4\cdot 2^{-k-3}+T^{-2}\le 2^{-k}+T^{-2}$. Hence the regret is $\O(\log^4 T\cdot (T2^{-k}+\sum_{j=0}^k N_G(2^{-j})))$ for every $k$, and minimizing over $k$ proves the lemma.
\end{proof}

\subsection{Examples of Interval Dimensions}\label{app:examples}

We list value distributions with each interval dimension $d$, on which \Cref{alg:size-2} beats the worst-case $\Otil(\sqrt T)$ regret whenever $d<1$. Most of them follow from the following lemma.

\begin{lemma}[Polynomial Drops Give Interval Dimensions]\label{lem:drop dimension}
Let $\alpha\ge 1$ and $c>0$. If a value distribution $F$ satisfies $R(p^\ast)-R(p)\ge c\lvert p-p^\ast\rvert^\alpha$ for all $p\in [0,1]$, then $F$ has interval dimension $1-\frac 1\alpha$ with multiplier $2c^{-1/\alpha}+1$.
\end{lemma}
\begin{proof}
Let $\Delta\in (0,1]$. Every $\Delta$-optimal $p$ has $c\lvert p-p^\ast\rvert^\alpha\le \Delta$, i.e., $\lvert p-p^\ast\rvert\le (\frac\Delta c)^{1/\alpha}$, so $u_\Delta-\ell_\Delta\le 2(\frac\Delta c)^{1/\alpha}$. By \Cref{lem:covering}, $N(\Delta)\le 1+2c^{-1/\alpha}\Delta^{1/\alpha-1}\le (2c^{-1/\alpha}+1)\Delta^{-(1-1/\alpha)}$, as $\Delta^{1/\alpha-1}\ge 1$.
\end{proof}

\begin{example}[Interval Dimension $0$]\label{ex:d=0}
In each of the following, $R$ has a unique maximizer $p^\ast$ and drops at least linearly around it, i.e., $R(p^\ast)-R(p)\ge c\lvert p-p^\ast\rvert$ for some $c>0$ and all $p\in [0,1]$. So by \Cref{lem:drop dimension} with $\alpha=1$, $N(\Delta)\le \frac 2c+1$ for all $\Delta\in (0,1]$, i.e., $F$ has interval dimension $0$.
\begin{itemize}
\item \emph{Finitely many values} \citep{cesa2019dynamic}, with a unique optimal price $p^\ast$. Below the smallest value and between two consecutive values, the demand $\demand(p)$ is constant, so $R$ increases linearly there; above the largest value, $R=0$. Hence $p^\ast>0$ is one of the values, $R$ is linear with slope $\demand(p^\ast)>0$ on $(v',p^\ast]$, where $v'$ is the previous value (or $0$), and the other local maxima of $R$ are at the finitely many other values, so $R\le R(p^\ast)-g$ outside $(v',p^\ast]$ for some $g>0$. As $\lvert p-p^\ast\rvert\le 1$, $c:=\min\{\demand(p^\ast),g\}$ works.
\item \emph{Kinked demands}, i.e., at a unique maximizer $p^\ast$, $R$ has a positive left derivative and a negative right derivative. This models the kinked-demand theory of oligopoly \citep{sweezy1939demand,hall1939price}: The rivals of a firm match its price cuts but not its price increases, so its demand barely grows when it cuts its price below the prevailing price $p^\ast$ (i.e., the price elasticity $\varepsilon(p):=\frac{pf(p)}{\demand(p)}$ is less than $1$, so $R'(p)=\demand(p)(1-\varepsilon(p))>0$), but quickly drops when it raises its price above $p^\ast$ (i.e., $\varepsilon(p)>1$, so $R'(p)<0$). The one-sided derivatives give some $c',r>0$ with $R(p^\ast)-R(p)\ge c'\lvert p-p^\ast\rvert$ for $\lvert p-p^\ast\rvert<r$. As $R$ is upper semi-continuous and $p^\ast$ is its unique maximizer, $R\le R(p^\ast)-g$ for some $g>0$ on the compact set $\{p\in [0,1]\mid \lvert p-p^\ast\rvert\ge r\}$. So $c:=\min\{c',g\}$ works.
\item \emph{Pareto distributions} with shape $\beta>1$ and scale $p_0<1$, with values above $1$ rounded down to $1$, i.e., $\demand(p)=\min\{1,(\frac{p_0}p)^\beta\}$. Then $R(p)=p$ below $p_0$, and $R(p)=p_0^\beta p^{1-\beta}$ above $p_0$, which is convex and decreasing. So $p^\ast=p_0$, $R(p_0)-R(p)=p_0-p$ for $p<p_0$, and by convexity, $R(p_0)-R(p)\ge \frac{R(p_0)-R(1)}{1-p_0}(p-p_0)$ for $p>p_0$. So $c:=\min\{1,\frac{p_0-p_0^\beta}{1-p_0}\}$ works.
\item \emph{Uniform distributions} $\text{Unif}[a,b]$ with $\frac b2<a<b\le 1$, another case where the monopoly price is the lowest value. Here $R(p)=p$ below $a$, $R(p)=\frac{p(b-p)}{b-a}$ on $[a,b]$, which is concave with right derivative $\frac{b-2a}{b-a}<0$ at $a$, and $R=0$ above $b$. So $p^\ast=a$, and $R(a)-R(p)\ge \frac{2a-b}{b-a}(p-a)$ for $p\in [a,b]$ by concavity. As $R(a)-R(p)=a$ for $p>b$, $c:=\min\{a,\frac{2a-b}{b-a}\}$ works.
\end{itemize}
\end{example}

\begin{example}[Interval Dimension $\frac 12$]\label{ex:d=1/2}
Each of the following has interval dimension $\frac 12$.
\begin{itemize}
\item \emph{MHR distributions} with $R(p^\ast)>0$. By \Cref{lem:MHR shape}, $N(\Delta)\le 1+8\Delta^{-1/2}\le 9\Delta^{-1/2}$ for $\Delta\le \frac{R(p^\ast)}4$, and by \Cref{lem:covering}, $N(\Delta)\le \frac 2\Delta<4R(p^\ast)^{-1/2}\Delta^{-1/2}$ for $\Delta>\frac{R(p^\ast)}4$. So $F$ has interval dimension $\frac 12$ with multiplier $\max\{9,4R(p^\ast)^{-1/2}\}$.
\item \emph{Distributions with curvature $\kappa$}, i.e., $F\in \mF_\curvature$, which satisfies $R(p^\ast)-R(p)\ge \kappa(p-p^\ast)^2$ for all $p$ by definition. So \Cref{lem:drop dimension} with $\alpha=2$ gives interval dimension $\frac 12$ with multiplier $2\kappa^{-1/2}+1$.
\item \emph{Smooth around a unique maximizer.} If $F$ has a density that is continuously differentiable around a unique maximizer $p^\ast\in (0,1)$ of $R$ with $R''(p^\ast)<0$, which satisfies the ``curvature and smooth'' assumption of \citet[Theorem 1.2]{kleinberg2003value} (see \Cref{footnote:curvature assumption in KL}), then $F\in \mF_\curvature$ for some $\kappa\in (0,2)$ by \citet[Lemma 3.11]{kleinberg2003value}, so the previous bullet applies. Examples include the uniform, exponential, logistic (i.e., logit demands), and normal distributions, rescaled or truncated to $[0,1]$ with an interior maximizer.
\end{itemize}
\end{example}

\begin{example}[Interval Dimension $1$]\label{ex:d=1}
Every value distribution always has interval dimension $1$ with multiplier $2$, because $N(\Delta)\le \frac 2\Delta$ (\Cref{lem:covering}). In each of the following, we verify that $N(\Delta)\ge \frac{c'}\Delta$ for some $c'>0$ and all $\Delta\in (0,1]$, so it has no interval dimension $d<1$.
\begin{itemize}
\item \emph{The equal-revenue distribution} $\demand(p)=\min\{1,\frac 1{2p}\}$, i.e., unit-elastic demand. Its revenue is $R(p)=\frac 12$ for all $p\in [\frac 12,1]$, so all these prices are optimal, and $N(\Delta)\ge \frac 1{2\Delta}$.
\item \emph{Two optimal prices} where $R$ has another maximizer $p'\ne p^\ast$, e.g., finitely many values with two optimal prices. As $p'$ and $p^\ast$ are $\Delta$-optimal for every $\Delta$, $N(\Delta)\ge \frac{\lvert p'-p^\ast\rvert}\Delta$.
\end{itemize}
\end{example}

\begin{example}[Every Interval Dimension $d\in [0,1)$]\label{ex:F alpha}
We now give a family of distributions that can have any interval dimension $d\in [0,1)$, which will be useful for the construction of our lower bounds.
\begin{itemize}
\item \emph{The family $F_\alpha$.} For $\alpha\ge 1$, let $F_\alpha$ be given by $\demand_{F_\alpha}(p)=\frac{1-(2p-1)^\alpha}{2p}$ for $p\in [\frac 12,1]$ and $\demand_{F_\alpha}(p)=1$ for $p<\frac 12$. It is non-increasing, as its revenue function $R(p)=\frac{1-(2p-1)^\alpha}2$ is non-increasing on $[\frac 12,1]$, and $R(p)=p$ below $\frac 12$. Its $\Delta$-optimal prices form $[\frac 12-\Delta,\frac 12+\frac 12(2\Delta)^{1/\alpha}]$ for $\Delta\le \frac 12$, so it has interval dimension $1-\frac 1\alpha$ with multiplier $3$. It interpolates from $F_1$ with $\demand_{F_1}(p)=\frac 1p-1$ ($d=0$), through $F_2=\text{Unif}[\frac 12,1]$ ($d=\frac 12$), to the equal-revenue distribution as $\alpha\to\infty$ ($d\to 1$).
\item \emph{Random distributions.} The random value distributions in \Cref{thm:fixed instance} also have every interval dimension $d\in (0,1)$.
\end{itemize}
\end{example}

\textit{When $V_T(F)$ Tends to Infinity.} We also discuss when $V_T(F)$, the instance-dependent quantity defined in \Cref{eq:V_T}, tends to infinity. We claim that it stays bounded only for value distributions with interval dimension $0$, which are exactly those whose revenue drops at least linearly as in \Cref{ex:d=0}:
\begin{lemma}[When $V_T(F)$ Tends to Infinity]\label{lem:V_T infinite}
Let $F$ be a value distribution. Then $V_T(F)$ is non-decreasing in $T$, and $\lim_{T\to\infty}V_T(F)=\sup_{\Delta\in (0,1]}N_F(\Delta)$. Moreover, the following are equivalent:
\begin{enumerate}
\item $V_T(F)$ is bounded, i.e., $\sup_{T\ge 1}V_T(F)<\infty$.
\item $F$ has interval dimension $0$.
\item There is $c>0$ such that $R(p^\ast)-R(p)\ge c\lvert p-p^\ast\rvert$ for all $p\in [0,1]$.
\end{enumerate}
In particular, $V_T(F)\to\infty$ as $T\to\infty$ if $R$ has another maximizer $p'\ne p^\ast$; or if $p^\ast\in (0,1)$ and $R$ is differentiable at $p^\ast$, e.g., if $F$ has a density around $p^\ast$ that is continuous at $p^\ast$.
\end{lemma}
\begin{proof}
Write $R^\ast:=R(p^\ast)$, $N:=N_F$, and $N^\ast:=\sup_{\Delta\in (0,1]}N(\Delta)\in [0,\infty]$. For $T\le T'$ and every $\Delta$, we have $\min\{T\Delta,N(\Delta)\}\le \min\{T'\Delta,N(\Delta)\}\le N(\Delta)$, so $V_T(F)$ is non-decreasing in $T$ and at most $N^\ast$. Conversely, for every $\Delta\in (0,1]$ and $T\ge \frac{N(\Delta)}\Delta$, we have $V_T(F)\ge \min\{T\Delta,N(\Delta)\}=N(\Delta)$. Hence $\lim_{T\to\infty}V_T(F)=N^\ast$, so (1) holds if and only if $N^\ast<\infty$. This is equivalent to (2): If $N^\ast<\infty$, then $F$ has interval dimension $0$ with multiplier $N^\ast$, and if $F$ has interval dimension $0$ with multiplier $C$, then $N^\ast\le C$.

For (2) $\Rightarrow$ (3), let $N(\Delta)\le C$ for all $\Delta\in (0,1]$. Fix $p\in [0,1]$, and let $\Delta:=R^\ast-R(p)\in [0,1]$. For every $\Delta'\in (0,1]$ with $\Delta'\ge \Delta$, both $p$ and $p^\ast$ are $\Delta'$-optimal, so $\lvert p-p^\ast\rvert\le u_{\Delta'}-\ell_{\Delta'}\le \Delta'N(\Delta')\le C\Delta'$ by \Cref{lem:covering}. If $\Delta>0$, then taking $\Delta'=\Delta$ gives $R^\ast-R(p)\ge \frac 1C\lvert p-p^\ast\rvert$. If $\Delta=0$, then letting $\Delta'\to 0$ gives $p=p^\ast$. Hence (3) holds with $c=\frac 1C$.

For (3) $\Rightarrow$ (2), every $\Delta$-optimal price $p$ satisfies $\lvert p-p^\ast\rvert\le \frac{R^\ast-R(p)}c\le \frac\Delta c$, so $u_\Delta-\ell_\Delta\le \frac{2\Delta}c$ and $N(\Delta)\le \lceil \frac 2c\rceil$ for all $\Delta\in (0,1]$. That is, $F$ has interval dimension $0$ with multiplier $\lceil \frac 2c\rceil$, and $V_T(F)\le \lceil \frac 2c\rceil$ for all $T$.

Finally, (3) fails in both cases of the last claim. If $p'\ne p^\ast$ is another maximizer, then $R^\ast-R(p')=0<c\lvert p'-p^\ast\rvert$ for every $c>0$. Indeed, as $p'$ and $p^\ast$ are $\Delta$-optimal for every $\Delta$, we have $N(\Delta)\ge \frac{\lvert p'-p^\ast\rvert}\Delta$, so taking $\Delta=\sqrt{\lvert p'-p^\ast\rvert/T}$ gives $V_T(F)\ge \sqrt{\lvert p'-p^\ast\rvert T}$ for $T\ge \lvert p'-p^\ast\rvert$. If $p^\ast\in (0,1)$ and $R$ is differentiable at $p^\ast$, then $R'(p^\ast)=0$ as $p^\ast$ maximizes $R$, so $\frac{R^\ast-R(p)}{\lvert p-p^\ast\rvert}\to 0$ as $p\to p^\ast$. Moreover, if $F$ has a density $f$ around $p^\ast$ that is continuous at $p^\ast$, then $\demand(p)$ is differentiable at $p^\ast$ with derivative $-f(p^\ast)$, so $R(p)=p\demand(p)$ is differentiable at $p^\ast$ as well.
\end{proof}

\section{\texorpdfstring{Omitted Proofs in \Cref{sec:lower bounds} (Lower Bounds)}{Omitted Proofs: Lower Bounds}}\label{app:class lower}

All our regret lower bounds compare two close value distributions, which we state as the two-point method for regret in \Cref{app:two-point}. We then prove the minimax lower bounds (\Cref{app:minimax lower}), the local minimax lower bounds (\Cref{app:instance full,app:instance lower,app:lower MHR}), the anytime lower bounds (\Cref{app:fixed instance}), and a lower bound for Lipschitz experts (\Cref{app:experts}).

\subsection{Two-Point Method for Regret}\label{app:two-point}

As in \Cref{app:lower methods}, for a value distribution $G$, its demand at a price $p$ is $\demand_G(p):=\Pr_{v\sim G}\{v\ge p\}$, the probability of an interval $I$ is $\intervalP_G(I):=\Pr_{v\sim G}\{v\in I\}$, and $R_G(p_G^\ast)-R_G(p)$ is the \emph{gap} of a price $p$ under $G$. For $G\ll G'$ (i.e., $G$'s support is a subset of that of $G'$), let $\chi^2(G\|G'):=\int(\frac{dG}{dG'}-1)^2\,dG'$, so that $H^2(G,G')\le \frac 12\chi^2(G\|G')$. As $1-H^2$ is multiplicative over independent samples, $n$ samples have total variation distance at most $\sqrt{2nH^2(G,G')}$ between $G$ and $G'$. As in \Cref{lem:two-point complexity}, a two-point lower bound takes two value distributions that are close but share no near-optimal price.

\begin{lemma}[Two-Point Method for Regret]\label{lem:two-point}
Let $\Delta>0$, and let $G_+,G_-$ be two value distributions such that every price has gap at least $\Delta$ under $G_+$ or under $G_-$. If for some $T'\le T$, the first $T'-1$ values have total variation distance at most $\frac 12$ between $G_+$ and $G_-$, e.g., if $H^2(G_+,G_-)\le \frac 1{8T'}$, then any algorithm, even with infinite menus and full feedback, suffers regret at least $\frac{\Delta T'}4$ on $G_+$ or on $G_-$.
\end{lemma}
\begin{proof}
We show that, in each round $t\le T'$, the regrets under $G_+$ and under $G_-$ sum to at least $\frac\Delta 2$. Fix such a round $t$, and let $\priceDist$ be the price distribution in \Cref{lem:menu revenue} for the menu in round $t$, which only depends on the first $t-1$ values and the internal randomness. Let $B_\pm$ be the set of prices with gap at least $\Delta$ under $G_\pm$, and $\Phi_\pm:=\Pr_{p\sim \priceDist}\{p\in B_\pm\}$. As $B_+\cup B_-=[0,1]$, we have $\Phi_++\Phi_-\ge 1$. By \Cref{lem:menu revenue}, the round-$t$ regret under $G_\pm$ is at least $\Delta\E_{G_\pm}[\Phi_\pm]$. As the first $t-1$ values have total variation distance at most $\frac 12$, the sum of the two is at least $\Delta(\E_{G_+}[\Phi_+]+\E_{G_+}[\Phi_-]-\frac 12)\ge \frac \Delta 2$. Summing over $t\le T'$, the regrets in the first $T'$ rounds under $G_+$ and under $G_-$ sum to at least $\frac{\Delta T'}2$, so one of them is at least $\frac{\Delta T'}4$. As every round has non-negative regret by \Cref{lem:menu revenue}, the same holds for the $T$-round regret.
\end{proof}

\subsection{\texorpdfstring{Proof of \Cref{thm:minimax lower} (Minimax Lower Bounds)}{Proof: Minimax Lower Bounds}}\label{app:minimax lower}

\begin{theorem}[Restatement of \Cref{thm:minimax lower}; Minimax Lower Bounds]
For dynamic pricing with infinite menus and full feedback:
\begin{enumerate}[label=\textup{(\alph*)}]
\item $\mR_\general^{(\infty)}(T)=\Omega(\sqrt T)$;
\item $\mR_\MHR^{(\infty)}(T)=\Omega(T^{1/3})$;
\item $\mR_\curvature^{(\infty)}(T)=\Omega(\kappa^{-1/3}T^{1/3})$ for every curvature $\kappa\in (0,2)$;
\item the minimax regret over value distributions with interval dimension $d\in (0,1)$ is $\Omega(T^{d/(d+1)})$.
\end{enumerate}
\end{theorem}
\begin{proof}
Each bound takes two value distributions in the class that are close in Hellinger distance but share no near-optimal price, and applies \Cref{lem:two-point}:
\begin{itemize}
\item (a) General and regular distributions: \Cref{lem:regret lower regular} below uses the regular pair in \Cref{lem:regular pair}, which tilts the revenue plateau of a regular distribution to the left or to the right.
\item (b) MHR distributions: \Cref{thm:lower MHR} applies to $F=\text{Unif}[0,1]$, which is MHR with monopoly price $\frac 12$, and whose hazard rate $\frac 1{1-v}$ has derivative in $[1,16]$ on $(0,\frac 34)$.
\item (c) Curvature $\kappa$: \Cref{lem:regret lower curvature} below uses the pairs in \Cref{lem:curvature pair small,lem:curvature pair large}, which shift a quadratic revenue peak or the kink of $R_U$.
\item (d) Interval dimension $d\in (0,1)$: Let $F$ be $F_\alpha$ in \Cref{app:examples} with $\alpha=\frac 1{1-d}$, which has interval dimension $d$ with multiplier $3$, and $N(\Delta)\ge \frac{\Delta^{-d}}2$ for $\Delta\le \frac 12$, as its $\Delta$-optimal prices form $[\frac 12-\Delta,\frac 12+\frac 12(2\Delta)^{1/\alpha}]$. Using \Cref{thm:instance lower}, there exists a constant $c>0$ and, for every $T\ge \frac 1c$, a $G$ such that $F$ and $G$ both have interval dimension $d$ with multiplier $\frac 3c$, and any algorithm suffers regret at least $cT^{d/(d+1)}$ on one of them.\qedhere
\end{itemize}
\end{proof}

\begin{lemma}[Regret Lower Bound for Regular Distributions]\label{lem:regret lower regular}
For every $T\ge 2$, $\mR_\regular^{(\infty)}(T)\ge \frac{\sqrt T}{136}$, even if the seller observes $v_t$ in each round, and even for the class of value distributions that are both regular and quasi-MHR.
\end{lemma}
\begin{proof}
Let $\epsilon:=(32T)^{-1/2}\le \frac 18$, and let $F_\pm$ be the pair in \Cref{lem:regular pair}, which are regular and quasi-MHR. Then $H^2(F_+,F_-)\le 4\epsilon^2=\frac 1{8T}$, and every price has gap at least $\frac\epsilon 6$ under $F_+$ or under $F_-$. So \Cref{lem:two-point} with $\Delta=\frac\epsilon 6$ and $T'=T$ gives $\mR_{\{F_+,F_-\}}^{(\infty)}(T)\ge \frac{\epsilon T}{24}\ge \frac{\sqrt T}{136}$.
\end{proof}

\begin{remark}[Learning a Uniform Reserve for Quasi-MHR Buyers]\label{remark:quasi-MHR}
\citet{feng2025beyond} introduce quasi-MHR distributions, i.e., those with a non-decreasing \emph{conditional expected hazard rate} $\frac 1v\ln\frac 1{\demand_F(v)}$, which generalize MHR distributions. Among many problems, they study the sample complexity of the \emph{empirical uniform reserve}, i.e., learning from samples a uniform reserve price whose revenue is a $(1-\epsilon)$-approximation of the optimal one. The tight bound for quasi-MHR buyers is left open, which they show to be between $\tilde\Omega(\epsilon^{-1.5})$ and $\Otil(\epsilon^{-2})$. We close this by showing their $\Otil(\epsilon^{-2})$ upper bound is tight up to logarithmic factors: Even for a single buyer, for whom a uniform reserve is a posted price, the two value distributions in \Cref{lem:regular pair} are regular and quasi-MHR, with optimal revenues at least $\frac 12$, and any estimator needs $\Omega(\epsilon^{-2})$ samples (\Cref{lem:lower regular}). The main reason is that, while quasi-MHR distributions keep the tail bounds of MHR distributions, the faster rates for MHR distributions (\Cref{thm:size-2} and \citep{huang2015making}) rely on the quadratic drop of $R$ around $p^\ast$ (\Cref{lem:MHR shape}), which is lost for quasi-MHR distributions: The revenues of the pair above are almost flat on $[\frac 12,\frac 34]$.
\end{remark}

\begin{lemma}[Regret Lower Bound for Curvature Distributions]\label{lem:regret lower curvature}
For every $\kappa\in (0,2)$ and $T\ge 3\max\{\kappa^{-2},(2-\kappa)^{-3}\}$, $\mR_\curvature^{(\infty)}(T)\ge \frac{\kappa^{-1/3}T^{1/3}}{1100}$, even if the seller observes $v_t$ in each round.
\end{lemma}
\begin{proof}
We use the two pairs in \Cref{lem:curvature pair small,lem:curvature pair large}, depending on $\kappa$.
\begin{itemize}
\item If $\kappa\le \frac 18$, let $s:=(12800\kappa^2T)^{-1/3}$, which is at most $\frac 1{16}$ as $T\ge \kappa^{-2}$, and let $G_\pm$ be the pair in \Cref{lem:curvature pair small}. Then $H^2(G_+,G_-)\le 1600\kappa^2s^3=\frac 1{8T}$, so \Cref{lem:two-point} with $\Delta=2\kappa s^2$ and $T'=T$ gives $\mR_{\{G_+,G_-\}}^{(\infty)}(T)\ge \frac{\kappa s^2T}2\ge \frac{\kappa^{-1/3}T^{1/3}}{1100}$.
\item If $\kappa>\frac 18$, let $s:=(192T)^{-1/3}$, which is at most $\frac{2-\kappa}8$ as $T\ge 3(2-\kappa)^{-3}$, and let $U,G_3$ be the pair in \Cref{lem:curvature pair large}. Then $H^2(U,G_3)\le 24s^3=\frac 1{8T}$, so \Cref{lem:two-point} with $\Delta=\frac{s^2}2$ and $T'=T$ gives $\mR_{\{U,G_3\}}^{(\infty)}(T)\ge \frac{s^2T}8\ge \frac{T^{1/3}}{270}$, which is at least $\frac{\kappa^{-1/3}T^{1/3}}{540}$ as $\kappa>\frac 18$.\qedhere
\end{itemize}
\end{proof}

\subsection{\texorpdfstring{Proofs of \Cref{thm:instance full} and \Cref{cor:instance optimality} (Local Minimax Lower Bound)}{Proofs: Local Minimax Lower Bound}}\label{app:instance full}

Throughout \Cref{app:instance full,app:instance lower}, we fix an $F$ with $R^\ast:=R(p^\ast)>0$, and let the smallest and the largest $\Delta$-optimal prices of $F$ be $\ell_\Delta<u_\Delta$. Let $m(\Delta):=\intervalP_F([\ell_\Delta,u_\Delta))$ be the probability between them. As $N(\Delta)$ may not be monotone in $\Delta$, we use its non-increasing envelope $\bar N(\Delta):=\sup_{\Delta'\ge \Delta}N(\Delta')$, so that $1\le N\le \bar N$. Note that $1\le V_T(F)\le \sqrt{2T}$, as $N(1)=1$ and $N(\Delta)\le \frac 2\Delta$. Furthermore, this envelope keeps $V_T$:
\begin{lemma}[Envelope Keeps $V_T$]\label{lem:envelope}
If a value distribution $G$ satisfies $N_G(\delta)\le K\bar N(\delta)$ for all $\delta\in (0,1]$ with some $K\ge 1$, then $V_T(G)\le KV_T(F)$.
\end{lemma}
\begin{proof}
Fix $\delta\in (0,1]$. As $N$ is integer-valued and at most $\frac 2\delta$ on $[\delta,1]$, and $N=1$ on $[1,\infty)$ by \Cref{lem:covering}, the supremum in $\bar N(\delta)$ is attained at some $\delta'\in [\delta,1]$. So $\min\{T\delta,\bar N(\delta)\}\le \min\{T\delta',N(\delta')\}\le V_T(F)$, and hence $\min\{T\delta,N_G(\delta)\}\le K\min\{T\delta,\bar N(\delta)\}\le KV_T(F)$. Taking the supremum over $\delta$ proves the lemma.
\end{proof}

By the two-point method (\Cref{lem:two-point}), it suffices to find two decoy instances that are close to $F$, are not much harder than $F$, and share no price with a small gap:

\begin{definition}[Decoy Pair]\label{def:decoy pair}
Fix a value distribution $F$ and a horizon $T$. For $\Delta>0$, two value distributions $G_+$ and $G_-$ form a \emph{$\Delta$-decoy pair} if:
\begin{enumerate}
\item (Close) $H^2(F,G_\pm)\le \frac 1{64T}$;
\item (Not harder) $N_{G_\pm}(\delta)\le \frac{325}{(R^\ast)^3}\bar N(\delta)$ for all $\delta\in (0,1]$;
\item (Separating) every price has gap at least $\Delta$ under $G_+$ or under $G_-$.
\end{enumerate}
\end{definition}

As sketched in \Cref{fig:hard instances}, we construct decoy pairs in two steps: \Cref{lem:moving mass} moves mass between $\ell_\Delta$ and $u_\Delta$, which brings little Hellinger distance when $m(\Delta)\approx \Delta \cdot N(\Delta)$ is large, and \Cref{lem:sharpening} then sharpens the new revenue peaks, so that $G_\pm$ are not much harder than $F$. The proof below applies this at a scale $\Delta^\ast$ where $V_T(F)$ is nearly attained (in the special case where $V_T(F)$ is a constant, use $\Delta=\Theta(\frac 1T)$ instead).

\textit{Perturbing by a Density Ratio.} Both moving-mass lemmas (\Cref{lem:moving mass} here and \Cref{lem:one-sided moving mass} in \Cref{app:instance lower}) perturb $F$ as follows. For a function $h$ with $\int h\,dF=0$ and $\lvert \theta h\rvert\le \frac 12$, let
\begin{equation*}
\frac{dG}{dF}:=1+\theta h,\quad \Psi(p):=\int h(v)\1[v\ge p]\,dF(v).
\end{equation*}
Then
\begin{equation}\label{eq:perturbation}
R_G(p)=R(p)+p\cdot \theta \Psi(p)\quad\text{and}\quad\chi^2(G\|F)=\theta^2\int h^2\,dF.
\end{equation}
That is, $G$ moves mass from where $h<0$ to where $h>0$, and $\theta\Psi(p)$ is the net mass moved to above $p$, which changes the demand at $p$ by $\theta\Psi(p)$. We always take $h=\pm(-\frac{\1_I}{\intervalP_F(I)}+\frac{\1_{J}}{\intervalP_F(J)})$ for two disjoint intervals $I,J$, which moves mass $\theta$ between $I$ and $J$ and gives $\chi^2(G\|F)=\theta^2(\frac 1{\intervalP_F(I)}+\frac 1{\intervalP_F(J)})$.

Moving mass between $\ell_\Delta$ and $u_\Delta$ gives two value distributions close to $F$ that share no price with gap less than $\Delta$, with a squared Hellinger distance to $F$ inversely proportional to $m(\Delta)$:

\begin{lemma}[Moving Mass]\label{lem:moving mass}
Let $\Delta\le \frac{(R^\ast)^2}{10}$ and $m:=m(\Delta)\ge \frac{20\Delta}{R^\ast}$. There are two value distributions $G_+$ and $G_-$ with $\frac{dG_\pm}{dF}\in [\frac 12,\frac 32]$, $H^2(F,G_\pm)\le \frac{38\Delta^2}{(R^\ast)^3m}$, and $\lvert R_{G_\pm}(p)-R(p)\rvert\le \frac{5\Delta}{R^\ast}$ for all $p$, such that every price has gap at least $\Delta$ under $G_+$ or under $G_-$.
\end{lemma}
\begin{proof}
Write $\ell=\ell_\Delta$ and $u=u_\Delta$, so that $\ell\le p^\ast\le u$, and let $G_\pm$ be given by \Cref{eq:perturbation} with $\pm\theta$. As $m=\intervalP_F([\ell,p^\ast))+\intervalP_F([p^\ast,u))$, one of the following two cases holds.

\textit{Case 1: $\intervalP_F([p^\ast,u))\ge \frac m2$.} This is the case in \Cref{fig:hard instances} (a). We move mass $\theta$ between $I:=[p^\ast,u)$ and $[u,1]$: $G_+$ lifts $u$ above $p^\ast$, and $G_-$ lowers the upper part of $I$. Formally, let $\beta:=\demand_F(u)\ge R(u)\ge 0.9R^\ast$, $h:=-\frac{\1_I}{\intervalP_F(I)}+\frac{\1_{[u,1]}}\beta$, and $\theta:=\frac{4\Delta}{p^\ast}\le \frac{4\Delta}{R^\ast}$. Then $\Psi=0$ on $[0,p^\ast]$, $\Psi(p)=\frac{\intervalP_F([p^\ast,p))}{\intervalP_F(I)}\in [0,1]$ on $[p^\ast,u]$, and $\Psi(p)=\frac{\demand_F(p)}\beta\in [0,1]$ on $[u,1]$. As $\frac\theta{\intervalP_F(I)}\le \frac{8\Delta}{R^\ast m}\le 0.4$ and $\frac\theta\beta\le 0.45$, we have $\lvert \theta h\rvert\le \frac 12$.
\begin{itemize}
\item Under $G_+$, $R_{G_+}(u)\ge R^\ast-\Delta+\theta u$. Every $p\le p^\ast$ has $R_{G_+}(p)=R(p)\le R^\ast$, so its gap is at least $\theta u-\Delta\ge 3\Delta$; every $p\in (p^\ast,u)$ with $\Psi(p)\le \frac 12$ has $R_{G_+}(p)\le R^\ast+\frac{\theta u}2$, so its gap is at least $\frac{\theta u}2-\Delta\ge \Delta$.
\item Under $G_-$, $R_{G_-}(p^\ast)=R^\ast$. Every $p\in [p^\ast,u]$ with $\Psi(p)\ge \frac 12$ has $R_{G_-}(p)\le R^\ast-\frac{\theta p^\ast}2=R^\ast-2\Delta$; every $p>u$ has $R_{G_-}(p)\le R(p)<R^\ast-\Delta$.
\item $\chi^2(G_\pm\|F)=\theta^2(\frac 1{\intervalP_F(I)}+\frac 1\beta)\le \frac{16\Delta^2}{(R^\ast)^2}(\frac 2m+\frac{1.12}{R^\ast})\le \frac{50\Delta^2}{(R^\ast)^3m}$.
\end{itemize}

\textit{Case 2: $\intervalP_F([\ell,p^\ast))\ge \frac m2$.} The difference from Case 1 is that the probability mass to be moved now lies below $p^\ast$, so we move mass across $p^\ast$ rather than across $u$. We include the full construction for completeness. We move mass $\theta$ between $I:=[\ell,p^\ast)$ and $[p^\ast,1]$: $G_+$ lifts $p^\ast$ above $\ell$, and $G_-$ lowers the upper part of $I$ and all prices above $p^\ast$. Formally, let $\beta:=\demand_F(p^\ast)\ge R^\ast$, $h:=-\frac{\1_I}{\intervalP_F(I)}+\frac{\1_{[p^\ast,1]}}\beta$, and $\theta:=\frac{5\Delta}{R^\ast}$. Then $\Psi=0$ on $[0,\ell]$, $\Psi(p)=\frac{\intervalP_F([\ell,p))}{\intervalP_F(I)}\in [0,1]$ on $[\ell,p^\ast]$, and $\Psi(p)=\frac{\demand_F(p)}\beta\in [0,1]$ on $[p^\ast,1]$. As $\frac\theta{\intervalP_F(I)}\le \frac{10\Delta}{R^\ast m}\le \frac 12$ and $\frac\theta\beta\le \frac{5\Delta}{(R^\ast)^2}\le \frac 12$, we have $\lvert \theta h\rvert\le \frac 12$. Also, $\ell\ge R(\ell)\ge 0.9R^\ast$.
\begin{itemize}
\item Under $G_+$, $R_{G_+}(p^\ast)=R^\ast+\theta p^\ast$. Every $p\le \ell$ has gap at least $\theta p^\ast\ge 5\Delta$; every $p\in (\ell,p^\ast)$ with $\Psi(p)\le \frac 12$ has gap at least $\frac{\theta p^\ast}2\ge 2.5\Delta$.
\item Under $G_-$, $R_{G_-}(\ell)=R(\ell)\ge R^\ast-\Delta$. Every $p\in [\ell,p^\ast]$ with $\Psi(p)\ge \frac 12$ has $R_{G_-}(p)\le R^\ast-\frac{\theta\ell}2\le R^\ast-2.25\Delta$; every $p>p^\ast$ has $R_{G_-}(p)=R(p)(1-\frac\theta\beta)\le R(p)(1-\theta)$, which is at most $R^\ast-2.5\Delta$ if $R(p)\ge \frac{R^\ast}2$ and less than $\frac{R^\ast}2\le R^\ast-5\Delta$ otherwise. So these prices have gap at least $1.25\Delta$.
\item $\chi^2(G_\pm\|F)\le \frac{25\Delta^2}{(R^\ast)^2}(\frac 2m+\frac 1{R^\ast})\le \frac{75\Delta^2}{(R^\ast)^3m}$.
\end{itemize}

In both cases, the prices with gap at least $\Delta$ under $G_+$ and under $G_-$ together cover $[0,1]$. Finally, $H^2(F,G_\pm)\le \frac 12\chi^2(G_\pm\|F)$, and $\lvert R_{G_\pm}(p)-R(p)\rvert=\theta p\lvert \Psi(p)\rvert\le \theta\le \frac{5\Delta}{R^\ast}$ by \Cref{eq:perturbation}.
\end{proof}

Moving mass may lift a plateau of the revenue, creating many near-optimal prices. We therefore need to ``sharpen'' the new revenue peak as follows:

\begin{lemma}[Sharpening]\label{lem:sharpening}
Let $K\ge 1$, $\Delta\in (0,\frac{R^\ast}{8K}]$, and $G$ be a value distribution with $\lvert R_G(p)-R(p)\rvert\le K\Delta$ for all $p$. Then there is $G'$ with $\frac{dG'}{dG}\in [\frac 12,\frac 32]$ and $\chi^2(G'\|G)\le \frac{102K\Delta}{(R^\ast)^3\bar N(\Delta)}$ such that
\begin{enumerate}
\item no price has a smaller gap under $G'$ than under $G$;
\item $N_{G'}(\delta)\le \frac{65K}{(R^\ast)^2}\bar N(\delta)$ for all $\delta\in (0,1]$.
\end{enumerate}
\end{lemma}
\begin{proof}
\textit{Plan.} As sketched in \Cref{sec:constructions}, the first step moves a small fraction of the mass on the $\Delta$-optimal region below $p_G^\ast$ to above $p_G^\ast$, and the second step scales up the $\Delta$-optimal region above $p_G^\ast$, taking the extra mass from the prices above this region. Then $R_{G'}$ decreases linearly fast in both directions of $p_G^\ast$ throughout the $\Delta$-optimal region, making only those prices close to $p_G^\ast$ near-optimal.

\textit{Setup.} Let $\Gamma(p):=R_G(p_G^\ast)-R_G(p)$ be the gap of $p$ under $G$, and $\ell'\le p_G^\ast\le u'$ be the smallest and the largest prices with $\Gamma\le \Delta$. As $R_G(p_G^\ast)\ge R^\ast-K\Delta\ge \frac 78R^\ast$, every $p$ with $\Gamma(p)\le \Delta$ has $\min\{p,\demand_G(p)\}\ge R_G(p)\ge \frac 34R^\ast$. Let $c:=\min\{\frac{R^\ast}{24},\frac 1{\bar N(\Delta)}\}$.

\textit{Construction.} We construct $G'$ in two steps. Each step keeps $p_G^\ast$ a maximizer and does not decrease the gap of any price.
\begin{itemize}
\item \emph{Left.} Let $\frac{dG_1}{dG}:=1-c$ on $[\ell',p_G^\ast)$, $1+c_1$ on $[p_G^\ast,1]$, and $1$ elsewhere, where $c_1:=\frac{c\intervalP_G([\ell',p_G^\ast))}{\demand_G(p_G^\ast)}$. Then $R_{G_1}=(1+c_1)R_G$ on $[p_G^\ast,1]$, $R_{G_1}=R_G$ on $[0,\ell')$, and $R_{G_1}(p)=R_G(p)+cp\intervalP_G([\ell',p))$ on $[\ell',p_G^\ast)$. So no gap decreases, and for $p\in [\ell',p_G^\ast)$, by \Cref{eq:1-side Lip} for $G$, the gap of $p$ under $G_1$ is at least
\begin{equation*}
\Gamma(p)+cp_G^\ast \intervalP_G([p,p_G^\ast))\ge \Gamma(p)+c\left((p_G^\ast-p)\demand_G(p_G^\ast)-\Gamma(p)\right)\ge \frac 34cR^\ast(p_G^\ast-p).
\end{equation*}
\item \emph{Right.} Let $\frac{dG'}{dG_1}:=1+c$ on $[p_G^\ast,u')$, $1-c_2$ on $[u',1]$, and $1$ elsewhere, where $c_2:=\frac{c\intervalP_{G_1}([p_G^\ast,u'))}{\intervalP_{G_1}([u',1])}$. Then $R_{G'}=R_{G_1}$ on $[0,p_G^\ast]$, $R_{G'}=(1-c_2)R_{G_1}$ on $(u',1]$, and $R_{G'}(p)=R_{G_1}(p)-cp\intervalP_{G_1}([p_G^\ast,p))$ on $[p_G^\ast,u']$. So no gap decreases, and for $p\in (p_G^\ast,u']$, by \Cref{eq:1-side Lip} for $G_1$, $p_G^\ast \intervalP_{G_1}([p_G^\ast,p))\ge (p-p_G^\ast)\intervalP_{G_1}([p,1])\ge (p-p_G^\ast)\demand_G(p)$, so the gap of $p$ under $G'$ is at least $c(p-p_G^\ast)\demand_G(p)$, which is at least $\frac 34cR^\ast(p-p_G^\ast)$ if $\Gamma(p)\le \Delta$.
\end{itemize}
The first claim therefore follows.

\textit{Covering Numbers.} For the second claim, let $p$ have gap at most $\delta$ under $G'$, so $\Gamma(p)\le \delta$.
\begin{itemize}
\item If $\delta\ge \Delta$, then $R(p)\ge R_G(p)-K\Delta\ge R^\ast-(1+2K)\delta$, so $p\in [\ell_{(1+2K)\delta},u_{(1+2K)\delta}]$, an interval of length at most $(1+2K)\delta N((1+2K)\delta)\le 3K\delta\bar N(\delta)$. Hence $N_{G'}(\delta)\le 1+3K\bar N(\delta)\le 4K\bar N(\delta)$.
\item If $\delta<\Delta$, then $p\in [\ell',u']$, so the two steps give $\lvert p-p_G^\ast\rvert\le \frac{4\delta}{3cR^\ast}$. Hence $N_{G'}(\delta)\le 1+\frac 8{3cR^\ast}$, which is at most $1+\frac 8{3R^\ast}\bar N(\Delta)\le \frac{11}{3R^\ast}\bar N(\delta)$ if $c=\frac 1{\bar N(\Delta)}$ (as $\delta<\Delta$), and at most $1+\frac{64}{(R^\ast)^2}\le \frac{65}{(R^\ast)^2}\bar N(\delta)$ otherwise.
\end{itemize}

\textit{Distance.} Every $p$ with $\Gamma(p)\le\Delta$ also has $R(p)\ge R_G(p)-K\Delta\ge R^\ast-(1+2K)\Delta$, so
\begin{equation*}
u'-\ell'\le u_{(1+2K)\Delta}-\ell_{(1+2K)\Delta}\le (1+2K)\Delta N((1+2K)\Delta)\le 3K\Delta\bar N(\Delta).
\end{equation*}
By \Cref{eq:1-side Lip} for $G$, $\ell'\intervalP_G([\ell',u'))\le u'-\ell'+\Delta$, so $m':=\intervalP_G([\ell',u'))\le \frac{16K}{3R^\ast}\Delta\bar N(\Delta)$ as $\bar N(\Delta)\ge 1$. Finally, as $\demand_G(p_G^\ast)\ge \frac 78R^\ast$ and $\demand_G(u')\ge \frac 34R^\ast$, we have $c_1\le \frac{8c}{7R^\ast}m'$ and $c_2=\frac{c\intervalP_G([p_G^\ast,u'))}{\demand_G(u')}\le \frac{4c}{3R^\ast}m'$. As $c\le \frac{R^\ast}{24}$ and $m'\le 1$, all of $c,c_1,c_2$ are at most $\frac 1{12}$, so $\frac{dG'}{dG}\in [\frac 12,\frac 32]$. Moreover, $\lvert \frac{dG'}{dG}-1\rvert$ is $c$ on $[\ell',p_G^\ast)$, at most $c+\frac{13}{12}c_1\le \frac{3c}{R^\ast}$ on $[p_G^\ast,u')$, at most $c_1+c_2\le \frac{3c}{R^\ast}m'$ on $[u',1]$, and $0$ elsewhere. Hence
\begin{equation*}
\chi^2(G'\|G)\le c^2m'+\frac{9c^2}{(R^\ast)^2}m'+\frac{9c^2}{(R^\ast)^2}m'^2\le \frac{19c^2m'}{(R^\ast)^2}\le \frac{102Kc^2\Delta\bar N(\Delta)}{(R^\ast)^3}\le \frac{102K\Delta}{(R^\ast)^3\bar N(\Delta)}.\qedhere
\end{equation*}
\end{proof}

We now prove \Cref{thm:instance full} from \Cref{lem:moving mass,lem:sharpening}.

\begin{theorem}[Restatement of \Cref{thm:instance full}; Local Minimax Lower Bound]
Let $F$ be such that $R_F(p_F^\ast)>0$ and $T\ge 2$, and let $c_1:=\frac{R_F(p_F^\ast)^{12}}{2^{45}}$, $c_2:=\frac{325}{R_F(p_F^\ast)^3}$, and $c_3:=\frac{2^{20}}{R_F(p_F^\ast)^6}$. There are value distributions $G_+,G_-$ such that:
\begin{enumerate}[label=\textup{(\alph*)}]
\item Any algorithm, even with infinite menus, suffers regret at least $c_1V_T(F)$ on $G_+$ or on $G_-$;
\item $V_T(G_\pm)\le c_2V_T(F)$; and
\item $H^2(F,G_\pm)\le \frac{c_3}{V_T(F)^2}$ where $H^2(F,G):=\frac 12\int(\sqrt{dF}-\sqrt{dG})^2$ is the squared Hellinger distance.
\end{enumerate}
\end{theorem}
\begin{proof}
Write $V:=V_T(F)$ and $c_0:=\frac{(R^\ast)^6}{2^{22}}$. We first combine \Cref{lem:moving mass,lem:sharpening} into a decoy pair at every scale $\Delta$ where there is enough probability mass to move, namely, any $\Delta\le \frac{(R^\ast)^2}{40}$ such that $m(\Delta)\ge \frac{20\Delta}{R^\ast}$. Let $G_\pm$ be given by \Cref{lem:moving mass}, and let $G_\pm'$ be the results of \Cref{lem:sharpening} with $K=\frac 5{R^\ast}$ applied to them, which is applicable as $\lvert R_{G_\pm}-R\rvert\le \frac{5\Delta}{R^\ast}$. Then every price has gap at least $\Delta$ under $G_+'$ or under $G_-'$, as this holds for $G_\pm$ (\Cref{lem:moving mass}) and sharpening decreases no gap (\Cref{lem:sharpening} 1). Also $N_{G_\pm'}\le \frac{65K}{(R^\ast)^2}\bar N=\frac{325}{(R^\ast)^3}\bar N$ (\Cref{lem:sharpening} 2). Moreover, by the triangle inequality for $H$ and $H^2\le \frac 12\chi^2$, and then by the $\chi^2$ bounds in \Cref{lem:moving mass,lem:sharpening}, we have
\begin{equation}\label{eq:moving and sharpening}
H^2(F,G_\pm')\le 2H^2(F,G_\pm)+\chi^2(G_\pm'\|G_\pm)\le \frac{76\Delta^2}{(R^\ast)^3m(\Delta)}+\frac{510\Delta}{(R^\ast)^4\bar N(\Delta)}.
\end{equation}
So $G_+'$ and $G_-'$ indeed form a $\Delta$-decoy pair (\Cref{def:decoy pair}) for every horizon $T'$ for which the right-hand side is at most $\frac 1{64T'}$. It only remains to pick an appropriate $\Delta$. Consider a $\Delta^\ast\in (0,1]$ with $\min\{T\Delta^\ast,N(\Delta^\ast)\}\ge V/2$. Except for the case where $V$ is a constant (Case 2 below), we apply the above construction to this $\Delta^\ast$.

\textit{Case 1: Large $N(\Delta^\ast)\ge \frac 1{c_0}$.} We use a $\Delta^\ast$-decoy pair for a shorter horizon $T'\le T$ (i.e., indistinguishable during the first $T'$ rounds, which already forces regret $\Omega(\Delta^\ast T')$). Let
\begin{equation*}
T':=\min\left\{T,\left\lfloor \frac{c_0N(\Delta^\ast)}{\Delta^\ast}\right\rfloor\right\},\qquad\text{where }\frac{c_0N(\Delta^\ast)}{\Delta^\ast}\ge c_0N(\Delta^\ast)\ge 1.
\end{equation*}
\begin{itemize}
\item \emph{The pair.} We move mass and then sharpen at $\Delta^\ast$, where $\Delta^\ast\le 2c_0\le \frac{(R^\ast)^2}{40}$ as $N(\Delta^\ast)\le \frac 2{\Delta^\ast}$. There is enough probability mass to move, as far apart near-optimal prices have large probability mass in between (as discussed in \Cref{sec:constructions}): Writing $\ell:=\ell_{\Delta^\ast}$ and $u:=u_{\Delta^\ast}$,
\begin{equation*}
m(\Delta^\ast)\overset{(a)}{\ge} \ell\, m(\Delta^\ast)\overset{(b)}{=}(u-\ell)\demand_F(u)-\left(R(u)-R(\ell)\right)\overset{(c)}{\ge} \frac{R^\ast}2\left(N(\Delta^\ast)-1\right)\Delta^\ast-\Delta^\ast\overset{(d)}{\ge} \frac{R^\ast}4\Delta^\ast N(\Delta^\ast),
\end{equation*}
where (a) uses $\ell\le 1$, (b) uses \Cref{eq:1-side Lip}, (c) uses $\demand_F(u)\ge R(u)\ge R^\ast-\Delta^\ast\ge \frac{R^\ast}2$, $u-\ell\ge (N(\Delta^\ast)-1)\Delta^\ast$ (\Cref{lem:covering}), and $R(u)-R(\ell)\le \Delta^\ast$, and (d) uses $N(\Delta^\ast)\ge \frac 1{c_0}\ge 2+\frac 4{R^\ast}$. In particular, $m(\Delta^\ast)\ge \frac{20\Delta^\ast}{R^\ast}$. The pair is close: As $\bar N\ge N$ and $\frac{\Delta^\ast}{N(\Delta^\ast)}\le \frac{c_0}{T'}$ by the choice of $T'$, the right-hand side of \Cref{eq:moving and sharpening} at $\Delta^\ast$ is at most
\begin{equation*}
\frac{304\Delta^\ast}{(R^\ast)^4N(\Delta^\ast)}+\frac{510\Delta^\ast}{(R^\ast)^4N(\Delta^\ast)}\le \frac{2^{10}c_0}{(R^\ast)^4T'}\le \frac 1{64T'}.
\end{equation*}
So moving mass and then sharpening gives a $\Delta^\ast$-decoy pair $G_+,G_-$ for the horizon $T'$.
\item \emph{The regret.} As $H^2(F,G_\pm)\le \frac 1{64T'}$, the first $T'-1$ values have total variation distance at most $\sqrt{2T'H^2(F,G_\pm)}\le \frac 14$ between $F$ and $G_\pm$, and hence at most $\frac 12$ between $G_+$ and $G_-$. So \Cref{lem:two-point} gives regret at least $\frac{\Delta^\ast T'}4$ on $G_+$ or on $G_-$, where, as $\lfloor x\rfloor\ge \frac x2$ for $x\ge 1$ and $c_0\le 1$,
\begin{equation*}
\Delta^\ast T'\ge \min\left\{T\Delta^\ast,\frac{c_0N(\Delta^\ast)}2\right\}\ge \frac{c_0}2\min\{T\Delta^\ast,N(\Delta^\ast)\}\ge \frac{c_0V}4.
\end{equation*}
\item \emph{The distance.} As $N(\Delta^\ast)\ge \frac V2$ and $N(\Delta^\ast)\le \frac 2{\Delta^\ast}$, we have $\frac{c_0N(\Delta^\ast)}{\Delta^\ast}\ge \frac{c_0}2N(\Delta^\ast)^2\ge \frac{c_0V^2}8$, so $T'\ge \min\{T,\frac{c_0V^2}{16}\}$, and hence $H^2(F,G_\pm)\le \frac 1{64T'}\le \max\{\frac 1{64T},\frac 1{4c_0V^2}\}$.
\end{itemize}

\textit{Case 2: Constant $V$}, namely $N(\Delta^\ast)<\frac 1{c_0}$ (which implies $V\le 2N(\Delta^\ast)<\frac 2{c_0}$ is at most a constant depending on $R^\ast$). We then take a tiny scale $\Delta:=\frac{4c_0}T=\frac{(R^\ast)^6}{2^{20}T}$ to force regret $\Omega(1)$. We construct the pair depending on whether $F$ has enough probability mass between its $\Delta$-optimal prices to move:
\begin{itemize}
\item \emph{Enough probability mass to move, i.e., $m(\Delta)\ge \frac{20\Delta}{R^\ast}$.} We move mass and then sharpen at $\Delta$, which is possible as $\Delta\le \frac{(R^\ast)^2}{40}$. As $\bar N\ge 1$, the right-hand side of \Cref{eq:moving and sharpening} is at most $\frac{4\Delta}{(R^\ast)^2}+\frac{510\Delta}{(R^\ast)^4}\le \frac{2^{10}\Delta}{(R^\ast)^4}\le \frac 1{64T}$, so this gives a $\Delta$-decoy pair.
\item \emph{Little probability mass to move, i.e., $m(\Delta)<\frac{20\Delta}{R^\ast}$.} Then we cannot move mass, but this also implies that under $F$, lowering the price from $p^\ast$ loses revenue quickly. So we keep $G_+:=F$, and let $G_-$ add a small mass just below $p^\ast$, which makes a price just below $p^\ast$ better than all prices above the added mass. Formally, let $\eta:=\frac{128\Delta}{(R^\ast)^3}=\frac{(R^\ast)^3}{2^{13}T}$, $x:=\frac{\eta p^\ast}2$, $y:=p^\ast-x$, and $\xi\in (0,\frac x2)$, and let
\begin{equation*}
G:=(1-\eta)F+\eta\,\text{Unif}[y,y+\xi],
\end{equation*}
so that $H^2(F,G)\le \eta$ by the convexity of $H^2$, and $\lvert R_G-R\rvert\le \eta$. Then:
\begin{enumerate}[label=\textup{(\alph*)}]
\item Under $G$, every price $p\ge y+\xi$ has gap at least $\Delta$, as the added buyers buy at $y$ but not at $p$. Indeed, $R(y)\ge R^\ast-x$ by \Cref{eq:1-side Lip}, so $R_G(y)\ge (1-\eta)(R^\ast-x)+\eta(p^\ast-x)=(1-\eta)R^\ast+\frac{\eta p^\ast}2$, whereas $R_G(p)=(1-\eta)R(p)\le (1-\eta)R^\ast$, and $\frac{\eta p^\ast}2\ge \frac{\eta R^\ast}2\ge \Delta$.
\item Under $F$, no price $p<y+\xi$ is $\Delta$-optimal, as $p$ is more than $\frac x2$ below $p^\ast$ and $F$ has little probability mass in between. Indeed, otherwise $\ell_\Delta\le p<p^\ast\le u_\Delta$, so by \Cref{eq:1-side Lip}, the gap of $p$ under $F$ would be
\begin{equation*}
(p^\ast-p)\demand_F(p^\ast)-p\intervalP_F([p,p^\ast))>\frac x2R^\ast-m(\Delta)>\frac{32\Delta}{R^\ast}-\frac{20\Delta}{R^\ast}>\Delta,
\end{equation*}
where we used $\demand_F(p^\ast)\ge R^\ast$, $p\intervalP_F([p,p^\ast))\le m(\Delta)$, and $\frac x2R^\ast\ge \frac{\eta(R^\ast)^2}4=\frac{32\Delta}{R^\ast}$.
\end{enumerate}
Adding mass may also lift a plateau of the revenue, so we sharpen $G_-$ by applying \Cref{lem:sharpening} to $G$ with $\eta$ in place of $\Delta$ and $K=1$. Then every price has gap at least $\Delta$ under $G_+=F$ by (b) or under $G_-$ by (a), as sharpening decreases no gap. Moreover, $N_{G_+}=N\le \bar N$, $N_{G_-}\le \frac{65}{(R^\ast)^2}\bar N\le \frac{325}{(R^\ast)^3}\bar N$, $H^2(F,G_+)=0$, and $H^2(F,G_-)\le 2H^2(F,G)+\chi^2(G_-\|G)\le 2\eta+\frac{102\eta}{(R^\ast)^3}\le \frac 1{64T}$. So $G_+,G_-$ indeed form a $\Delta$-decoy pair.
\end{itemize}

Therefore, in both cases, the regret on $G_+$ or on $G_-$ is at least $\min\{\frac{c_0}{16},\frac{c_0^2}2\}V=\frac{(R^\ast)^{12}}{2^{45}}V$, which proves claim (a). By \Cref{lem:envelope}, every decoy pair has $V_T(G_\pm)\le \frac{325}{(R^\ast)^3}V$, which gives claim (b). As $V^2\le 2T$, we have $\max\{\frac 1{64T},\frac 1{4c_0V^2}\}=\frac 1{4c_0V^2}=\frac{2^{20}}{(R^\ast)^6V^2}$, which proves claim (c).
\end{proof}

Together with \Cref{thm:instance upper} and \Cref{lem:V_T infinite}, this gives the near instance-optimality of \Cref{alg:size-2}:

\begin{corollary}[Restatement of \Cref{cor:instance optimality}; Near Instance-Optimality of \Cref{alg:size-2}]
Under the conditions of \Cref{thm:instance full},
\begin{enumerate}
\item \Cref{alg:size-2} has regret $\O(V_T(F)\log^5 T)$ on $F$ and $\O(R_F(p_F^\ast)^{-3}V_T(F)\log^5 T)$ on $G_+$ and on $G_-$.
\item Any algorithm, even with infinite menus, suffers $\Omega(R_F(p_F^\ast)^{12} V_T(F))$ regret on $G_+$ or on $G_-$.
\item If $F$ does not have interval dimension $0$, then $H^2(F,G_\pm)\to 0$ as $T\to \infty$ (see \Cref{lem:V_T infinite}).
\end{enumerate}
\end{corollary}
\begin{proof}
Let $G_\pm$ be given by \Cref{thm:instance full}. For the first claim, \Cref{thm:instance upper} gives regret $\O(V_T(F)\log^5T)$ on $F$ and $\O(V_T(G_\pm)\log^5T)$ on $G_\pm$, where $V_T(G_\pm)\le \frac{325}{R_F(p_F^\ast)^3}V_T(F)$ by \Cref{thm:instance full}(b). The second claim is \Cref{thm:instance full}(a). For the third claim, if $F$ does not have interval dimension $0$, then $V_T(F)\to\infty$ as $T\to\infty$ by \Cref{lem:V_T infinite}, so \Cref{thm:instance full}(c) gives $H^2(F,G_\pm)\to 0$.
\end{proof}

\subsection{\texorpdfstring{Proof of \Cref{thm:instance lower} (Near Instance-Optimality via Interval Dimensions)}{Proof: Near Instance-Optimality via Interval Dimensions}}\label{app:instance lower}

In the special case where $F$ has an interval dimension, we keep $F$ itself as one of the two instances. We make the proof sketch in \Cref{sec:constructions} precise. Besides explicit constants, the only new case is when the wide side is the lower one, $[\ell_\Delta,\ell_{\Delta'})$, where we move mass down instead of up (symmetric to the case we sketched in \Cref{sec:constructions}; Case 2 in the proof below).

\begin{lemma}[One-Sided Moving Mass]\label{lem:one-sided moving mass}
Let $0<\Delta'\le \Delta\le \frac{(R^\ast)^2}{16}$ be such that $\lambda:=\max\{u_\Delta-u_{\Delta'},\ell_{\Delta'}-\ell_\Delta\}\ge \frac{32\Delta}{(R^\ast)^2}$. Then there is a value distribution $G$ with $\chi^2(G\|F)\le \frac{96\Delta^2}{(R^\ast)^3\lambda}$ and $\lvert R_G(p)-R(p)\rvert\le \frac{4\Delta}{R^\ast}$ for all $p$, such that every price has gap more than $\Delta'$ under $F$ or at least $\Delta'$ under $G$.
\end{lemma}
\begin{proof}
As in \Cref{lem:moving mass}, $G$ moves mass $\theta$ via \Cref{eq:perturbation}, but only on the wide side: In Case 1, from $[b,u)$ up to $[u,1]$, which lifts $u$ above every price up to $b$; in Case 2, from $[a,1]$ down to $[\ell,a)$, which lowers every price from $a$ on below $\ell$. Write $\ell=\ell_\Delta$, $u=u_\Delta$, $a=\ell_{\Delta'}$, and $b=u_{\Delta'}$, so $\ell\le a\le p^\ast\le b\le u$, and every $p\notin[a,b]$ has gap more than $\Delta'$ under $F$. As $\Delta\le \frac{R^\ast}2$, each $p\in \{\ell,a,b,u\}$ has $R(p)=p\demand_F(p)\ge R^\ast-\Delta\ge \frac{R^\ast}2$, so $p\ge \frac{R^\ast}2$ and $\demand_F(p)\ge \frac{R^\ast}2$.

\textit{Case 1: $u-b=\lambda$.} Let $I:=[b,u)$, $\beta:=\demand_F(u)$, $h:=-\frac{\1_I}{\intervalP_F(I)}+\frac{\1_{[u,1]}}\beta$ (where $\1_I(v):=\1[v\in I]$), and $\theta:=\frac{\Delta+\Delta'}u\le \frac{4\Delta}{R^\ast}$. Then $\Psi=0$ on $[0,b]$, $\Psi(p)=\frac{\intervalP_F([b,p))}{\intervalP_F(I)}\in [0,1]$ on $[b,u]$, and $\Psi(p)=\frac{\demand_F(p)}\beta\in [0,1]$ on $[u,1]$. So every $p\le b$ has $R_G(p)=R(p)\le R^\ast$, whereas $R_G(u)=R(u)+\theta u\ge R^\ast+\Delta'$, and hence $p$ has gap at least $\Delta'$ under $G$. Applying \Cref{eq:1-side Lip} to $b<u$ gives $b\intervalP_F(I)=(u-b)\beta-(R(u)-R(b))\ge \frac{R^\ast}2\lambda-\Delta$, so $\intervalP_F(I)\ge \frac{R^\ast\lambda}4$ as $b\le 1$ and $\lambda\ge \frac{4\Delta}{R^\ast}$.

\textit{Case 2: $a-\ell=\lambda$.} Let $I:=[\ell,a)$, $\beta:=\demand_F(a)$, $h:=\frac{\1_I}{\intervalP_F(I)}-\frac{\1_{[a,1]}}\beta$, and $\theta:=\frac{\Delta+\Delta'}{R^\ast}\le \frac{2\Delta}{R^\ast}$. Then $\Psi=0$ on $[0,\ell]$, $\Psi(p)=-\frac{\intervalP_F([\ell,p))}{\intervalP_F(I)}\in [-1,0]$ on $[\ell,a]$, and $\Psi(p)=-\frac{\demand_F(p)}\beta\in [-1,0]$ on $[a,1]$. So $R_G(\ell)=R(\ell)\ge R^\ast-\Delta$, whereas every $p\ge a$ has $R_G(p)=R(p)(1-\frac\theta\beta)\le R^\ast(1-\theta)=R^\ast-\Delta-\Delta'$ as $\beta\le 1$, and hence $p$ has gap at least $\Delta'$ under $G$. Applying \Cref{eq:1-side Lip} to $\ell<a$ gives $\ell \intervalP_F(I)=(a-\ell)\beta-(R(a)-R(\ell))\ge \frac{R^\ast}2\lambda-\Delta$, so $\intervalP_F(I)\ge \frac{R^\ast\lambda}4$.

In both cases, $\theta\le \frac{4\Delta}{R^\ast}$, $\beta\ge \frac{R^\ast}2$, and $\intervalP_F(I)\ge \frac{R^\ast\lambda}4$, so $\frac\theta{\intervalP_F(I)}\le \frac{16\Delta}{(R^\ast)^2\lambda}\le \frac 12$ and $\frac\theta\beta\le \frac{8\Delta}{(R^\ast)^2}\le \frac 12$, i.e., $\lvert \theta h\rvert\le \frac 12$. Moreover, $\lvert R_G(p)-R(p)\rvert=\theta p\lvert \Psi(p)\rvert\le \theta\le \frac{4\Delta}{R^\ast}$, and as $\lambda\le 1$,
\begin{equation*}
\chi^2(G\|F)=\theta^2\left(\frac 1{\intervalP_F(I)}+\frac 1\beta\right)\le \frac{16\Delta^2}{(R^\ast)^2}\left(\frac 4{R^\ast\lambda}+\frac 2{R^\ast}\right)\le \frac{96\Delta^2}{(R^\ast)^3\lambda}.\qedhere
\end{equation*}
\end{proof}

\begin{theorem}[Restatement of \Cref{thm:instance lower}; Near Instance-Optimality via Interval Dimensions]
If $F$ has interval dimension $d\in (0,1)$ with multiplier $C$, there is a constant $c>0$ only depending on $R_F(p_F^\ast),C$, and $d$, such that: For any $T\ge \frac 1c$ satisfying $N_F(\Delta)\ge \frac{\Delta^{-d}}C$ with $\Delta=T^{-1/(d+1)}$, there is a value distribution $G$ that satisfies:
\begin{enumerate}[label=\textup{(\alph*)}]
\item Any algorithm, even with infinite menus, suffers regret at least $cT^{d/(d+1)}$ on $F$ or on $G$;
\item $G$ also has interval dimension $d$ with multiplier $C/c$; and
\item $H^2(F,G)\le \frac 1{cT}$.
\end{enumerate}
\end{theorem}
\begin{proof}
First, we claim that $R^\ast>0$. Indeed, as $R\ge 0$, every price is $\delta$-optimal for $\delta\ge R^\ast$, so $\lceil \frac 1\delta\rceil=N(\delta)\le C\delta^{-d}$ for all $\delta\in (0,1]$ with $\delta\ge R^\ast$, which forces $R^\ast\ge C^{-1/(1-d)}>0$ as $d<1$. Pick
\begin{equation*}
c:=\min\left\{\frac{(R^\ast)^4}{2^{20}C(4C^2)^{1/(1-d)}},\left(\frac{(R^\ast)^2}{128C}\right)^{(d+1)/d}\right\},
\end{equation*}
where the first term makes the regret and the distance large and small enough, respectively, and the second one makes $N(\Delta)$ large. Let $T\ge \frac 1c$ and $\Delta:=T^{-1/(d+1)}$ with $N(\Delta)\ge \frac{\Delta^{-d}}C$, and let $\Delta':=(4C^2)^{-1/(1-d)}\Delta\le \Delta$ be the smaller scale at which we separate $F$ from the decoy instance. As $C\ge N(1)=1$, we have $T\ge \frac{2^{20}}{(R^\ast)^4}\ge (\frac{32}{(R^\ast)^2})^{d+1}$, so $\Delta\le \frac{(R^\ast)^2}{32}$, and $N(\Delta)\ge \frac{T^{d/(d+1)}}C\ge \frac{128}{(R^\ast)^2}$.

\textit{Step 1: $\Delta'$-Optimal Prices Are Narrow.} As $F$ has interval dimension $d$ with multiplier $C$,
\begin{equation*}
u_{\Delta'}-\ell_{\Delta'}\le \Delta'N(\Delta')\le C\Delta'^{1-d}=\frac{\Delta^{1-d}}{4C}\le \frac 14\Delta N(\Delta).
\end{equation*}
As $u_\Delta-\ell_\Delta\ge \Delta(N(\Delta)-1)$ and $N(\Delta)\ge 4$, the width $\lambda:=\max\{u_\Delta-u_{\Delta'},\ell_{\Delta'}-\ell_\Delta\}$ of the wider side satisfies $\lambda\ge \frac 12(\Delta(N(\Delta)-1)-\frac 14\Delta N(\Delta))\ge \frac 14\Delta N(\Delta)$, which is at least $\frac{32\Delta}{(R^\ast)^2}$ and at least $\frac{\Delta^{1-d}}{4C}$.

\textit{Step 2: Decoy Instances.} Let $G_0$ be the value distribution in \Cref{lem:one-sided moving mass}, and let $G$ be the sharpened version from \Cref{lem:sharpening} with $K=\frac 4{R^\ast}$ applied to $G_0$, which is valid as $\Delta\le \frac{(R^\ast)^2}{32}=\frac{R^\ast}{8K}$. Then every price has gap at least $\Delta'$ under $F$ or under $G$, and $N_G\le \frac{260}{(R^\ast)^3}\bar N$ on $(0,1]$. Moreover, as $\bar N(\Delta)\ge N(\Delta)\ge \frac\lambda\Delta$,
\begin{equation*}
H^2(F,G)\le 2H^2(F,G_0)+\chi^2(G\|G_0)\le \frac{96\Delta^2}{(R^\ast)^3\lambda}+\frac{408\Delta}{(R^\ast)^4\bar N(\Delta)}\le \frac{504\Delta^2}{(R^\ast)^4\lambda}.
\end{equation*}

\textit{Step 3: Horizon.} Let $c_1:=\frac{(R^\ast)^4}{32256}$ and $T':=\min\{T,\lfloor \frac{c_1\lambda}{\Delta^2}\rfloor\}$, so that $H^2(F,G)\le \frac{\Delta^2}{64c_1\lambda}\le \frac 1{64T'}$. As $\frac\lambda{\Delta^2}\ge \frac{N(\Delta)}{4\Delta}\ge \frac{\Delta^{-(d+1)}}{4C}=\frac T{4C}$, and $\frac{c_1T}{4C}\ge \frac{2^{20}(4C^2)^{1/(1-d)}}{129024}\ge 1$ as $T\ge \frac 1c$, we have $T'\ge \frac{c_1T}{8C}$, using $\lfloor x\rfloor\ge \frac x2$ for $x\ge 1$.

\textit{Step 4: Regret.} The first $T'-1$ values have total variation distance at most $\sqrt{2T'H^2(F,G)}\le \frac 14$ between $F$ and $G$. So \Cref{lem:two-point} with $\Delta'$ in place of $\Delta$ shows that any algorithm suffers regret at least $\frac{\Delta'T'}4$ on $F$ or on $G$, where
\begin{equation*}
\frac{\Delta'T'}4\ge \frac{c_1\Delta'T}{32C}=\frac{(R^\ast)^4}{1032192\,C(4C^2)^{1/(1-d)}}T^{d/(d+1)}\ge cT^{d/(d+1)}.
\end{equation*}

\textit{Step 5: Distance and Interval Dimension.} We have $H^2(F,G)\le \frac 1{64T'}\le \frac C{8c_1T}=\frac{4032C}{(R^\ast)^4T}\le \frac 1{cT}$. Finally, $\bar N(\delta)\le C\delta^{-d}$ for all $\delta\in (0,1]$, as $\delta^{-d}$ is non-increasing and $N(\delta)=1\le C$ for $\delta\ge 1$. So $N_G(\delta)\le \frac{260C}{(R^\ast)^3}\delta^{-d}\le \frac Cc\delta^{-d}$ for all $\delta\in (0,1]$, i.e., $G$ has interval dimension $d$ with multiplier $\frac Cc$. Hence $G$ satisfies all three claims.
\end{proof}

\subsection{\texorpdfstring{Proof of \Cref{thm:lower MHR} (Near Instance-Optimality for MHR Distributions)}{Proof: Near Instance-Optimality for MHR Distributions}}\label{app:lower MHR}

As sketched in \Cref{sec:constructions}, the decoy instance $G$ multiplies the demand of $F$ by $e^{\phi}$, i.e., adds a ramp $\phi$ to the log-demand, which raises the revenue of $p^\ast+s$ above that of $p^\ast$ while keeping the hazard rate non-decreasing.

\begin{theorem}[Restatement of \Cref{thm:lower MHR}; Near Instance-Optimality for MHR Distributions]
Let $F$ be MHR. If on an open interval containing $p_F^\ast$, its hazard rate function $h(v)=\frac{f(v)}{\demand_F(v)}$ is differentiable with $h'$ bounded by some $[c,C]$ with $0<c<C$, then for every $T>0$, there is an MHR distribution $G$ with $H^2(F,G)\le \frac 1{64T}$, such that any algorithm, even with infinite menus, suffers regret $\mR_{\{F,G\}}^{(\infty)}(T)=\Omega(T^{1/3})$.
\end{theorem}
\begin{proof}
Write $\demand(p):=\Pr_{v\sim F}\{v\ge p\}$ and $R^\ast:=R(p^\ast)$, and take a closed interval $I\subseteq (0,1)$ with $p^\ast$ in its interior and $0<c\le C$ such that $c\le h'\le C$ on $I$. As $R'(p)=\demand(p)(1-ph(p))$ is continuous there and $p^\ast$ maximizes $R$, we have $p^\ast h(p^\ast)=1$. Let $C_1:=\frac 1{p^\ast}+2C$.

For $s>0$, let $L:=\frac{2C_1s}{cR^\ast}$, $a:=p^\ast-L$, $b:=p^\ast+s$, and $\phi(p):=\frac c2(\min\{p,b\}-a)_+^2$, which is $0$ below $a$, grows quadratically on $[a,b]$, and is constant above $b$. Let $G$ be the value distribution with $\demand_G(p)=\demand(p)e^{\phi(p)}$, so that $R_G=Re^\phi$. Let $s_0>0$ be such that every $s\le s_0$ satisfies
\begin{equation*}
\text{(i) }[a,b]\subseteq I,\qquad\text{(ii) }L\le \frac 1{2Cp^\ast},\qquad\text{(iii) }L+s\le 1,\qquad\text{(iv) }\frac c2(L+s)^2\le \frac 12,\qquad\text{(v) }s^2\le 2e^2R^\ast,
\end{equation*}
where (i) is used for the MHR property, (ii)--(iv) for the gaps and the distance, and (v) for the gaps under $F$. Fix any $s\le s_0$.
\begin{itemize}
\item \emph{$G$ is MHR.} Its hazard rate is $h-\phi'$, which equals $h$ outside $[a,b]$, is non-decreasing on $[a,b]$ as $\phi''=c\le h'$, jumps up at $b$, and is at least $h(a)\ge 0$.
\item \emph{Gaps under $F$.} By \Cref{lem:MHR tail}, the gap of $p^\ast+u$ under $F$ is at least $R^\ast(1-(1+x)e^{-x})$ with $x:=\frac u{p^\ast}$. Since $1-(1+x)e^{-x}=\int_0^xye^{-y}\,dy$ is at least $\frac{x^2}{2e}$ for $x\le 1$ and at least $1-\frac 2e>\frac 14$ for $x\ge 1$, and $\frac{R^\ast}{(p^\ast)^2}=\frac{\demand(p^\ast)}{p^\ast}\ge e^{-1}$, this gap is at least $\min\{\frac{u^2}{2e^2},\frac{R^\ast}4\}$. So by (v), every $p\ge p^\ast+\frac s2$ has gap at least $\frac{s^2}{8e^2}$ under $F$.
\item \emph{Gaps under $G$.} Every $p<p^\ast+\frac s2$ has gap at least $\frac{s^2}4$ under $G$. Indeed, for $x\in [0,s]$, we have $(p^\ast+x)h(p^\ast+x)-1\le (p^\ast+x)(h(p^\ast)+Cx)-1\le C_1x$, so $R'(p^\ast+x)\ge -C_1x$ and $R(b)\ge R^\ast-\frac{C_1s^2}2$. As $\phi$ is non-decreasing and $\phi(b)-\phi(p^\ast+\frac s2)\ge \frac{cLs}2=\frac{C_1s^2}{R^\ast}$, where $\frac{C_1s^2}{R^\ast}\le \phi(b)\le \frac 12$, we have
\begin{equation*}
\frac{R_G(b)-R_G(p)}{e^{\phi(p^\ast+s/2)}}\ge \left(R^\ast-\frac{C_1s^2}2\right)\left(1+\frac{C_1s^2}{R^\ast}\right)-R^\ast=\frac{C_1s^2}2\left(1-\frac{C_1s^2}{R^\ast}\right)\ge \frac{C_1s^2}4\ge \frac{s^2}4.
\end{equation*}
\item \emph{Distance.} $\chi^2(G\|F)\le \frac{702C_1^4s^3}{c(R^\ast)^3}$. Indeed, $\frac{dG}{dF}$ is $1$ below $a$, $e^{\phi(b)}$ above $b$, and $(1-\frac{\phi'}h)e^\phi$ on $[a,b]$, where $h\ge h(p^\ast)-CL\ge \frac 1{2p^\ast}$ and $f\le h\le C_1$. As $\phi\le \frac 12$ and $\phi'\le c(L+s)$, $\lvert \frac{dG}{dF}-1\rvert$ is at most $2\phi+\frac{2\phi'}h\le 5c(L+s)$ on $[a,b]$, and at most $2\phi(b)\le c(L+s)^2$ above $b$. So $\chi^2(G\|F)\le 25C_1c^2(L+s)^3+c^2(L+s)^4\le 26C_1c^2(L+s)^3$, where $L+s\le \frac{3C_1s}{cR^\ast}$ as $C_1\ge 2C\ge 2c$.
\end{itemize}

Therefore, every price has gap at least $\frac{s^2}{8e^2}$ under $F$ or under $G$. Take $s:=\min\{s_0,(\frac{c(R^\ast)^3}{22464C_1^4T})^{1/3}\}$, so that $H^2(F,G)\le \frac 12\chi^2(G\|F)\le \frac 1{64T}$. Then the first $T-1$ values have total variation distance at most $\sqrt{2TH^2(F,G)}\le \frac 12$ between $F$ and $G$, and \Cref{lem:two-point} with $\Delta=\frac{s^2}{8e^2}$ shows that $\mR_{\{F,G\}}^{(\infty)}(T)\ge \frac{s^2T}{32e^2}$. As $s=\Omega(T^{-1/3})$ with constants only depending on $F$, this gives the claimed $\Omega(T^{1/3})$ bound.
\end{proof}

\subsection{\texorpdfstring{Proof of \Cref{thm:fixed instance} (Anytime Lower Bounds)}{Proof: Anytime Lower Bounds}}\label{app:fixed instance}
\begin{definition}[Plateau]\label{def:plateau}
For a revenue function $R$ with maximum $L:=\max_{p\in [0,1]}R(p)$, a closed interval $P=[x-w,x+w]$ with $w>0$ is the \emph{plateau} of $R$, with center $x$, half-width $w$, and height $L$, if $R=L$ on $P$ and $R<L$ outside $P$. That is, the set of maximizers of $R$ is the interval $P$.
\end{definition}

We make the sketch in \Cref{sec:fixed instance} precise. Starting from a revenue function $R_0$ with a plateau $P_0$ around $\frac 12$, each scale $k=1,2,\ldots$ raises a part of the current plateau $P_{k-1}$ by $\Delta_k$, and the top of this part is the next plateau $P_k$. Odd scales are \emph{tilts}, which raise a random half of $P_{k-1}$ and fool infinite menus; even scales are \emph{needles}, which raise a random short piece of $P_{k-1}$ and fool posted prices. The heights shrink super-exponentially, so that the finer scales are invisible at the horizon of each scale.

\textit{Parameters.} Fix $d\in (0,1)$. Let $q:=\max\{5,\frac 2{1-d}\}$, let $\Delta_1>0$ be small enough, and let
\begin{equation*}
\Delta_{k+1}:=\Delta_k^q\quad\text{and}\quad w_k:=\Delta_{k+1}^{1-d}\qquad\text{for }k\ge 0.
\end{equation*}
Here $\Delta_k$ is the height of scale $k$, and $w_k$ is the half-width of the plateau $P_k$ (so $P_k$ has width $2w_k$). The exponent $1-d$ yields interval dimension $d$, and the large exponent $q$ makes the finer scales negligible. As $\Delta_1$ is small, we have $w_0\le 0.1$, and for all $k\ge 1$,
\begin{equation}\label{eq:small scales}
w_k\le \min\left\{\frac{w_{k-1}}4,\Delta_k\right\},\qquad \frac{10\Delta_k}{w_{k-1}}\le \frac 12,\qquad\text{and}\qquad \Delta_{>k}:=\sum_{m>k}\Delta_m\le 1.01\Delta_{k+1}.
\end{equation}
In words, the plateaus shrink by a factor of at least $4$, and each half-width $w_k$ is at most the height $\Delta_k$, which the construction below uses to fit the bumps into $P_{k-1}$; each height is small compared with the previous plateau, which keeps the tilts gentle (\Cref{lem:fixed scale distance}); and all finer scales together are about as high as the next one (\Cref{lem:fixed finer scales}).

\textit{Construction.} Let $x_0:=\frac 12$, $L_0:=\frac 14$, and $R_0(p):=\min\{p,\frac 14-\frac 14(\lvert p-x_0\rvert-w_0)_+\}$, so that $R_0=L_0$ on the plateau $P_0:=[x_0-w_0,x_0+w_0]$ and $R_0<L_0$ elsewhere. For $k\ge 1$, given $R_{k-1}$ with $R_{k-1}=L_{k-1}$ on the plateau $P_{k-1}:=[x_{k-1}-w_{k-1},x_{k-1}+w_{k-1}]$ and $R_{k-1}<L_{k-1}$ elsewhere, we draw an independent $\xi_k$, and let $R_k:=R_{k-1}+\Delta_k\phi_k$ and $L_k:=L_{k-1}+\Delta_k$, where the bump $\phi_k$ is as follows; scale $k$ is a tilt scale if $k$ is odd and a needle scale otherwise.
\begin{itemize}
\item \emph{Tilt.} $\xi_k$ is uniform over $\{-1,1\}$, and $\phi_k$ is $0$ outside the left (if $\xi_k=-1$) or the right (if $\xi_k=1$) half of $P_{k-1}$, rises linearly from both ends of this half, and equals $1$ on its middle part. Formally, $x_k:=x_{k-1}+\frac{\xi_kw_{k-1}}2$ and $\phi_k(p):=\min\{1,\frac{(w_{k-1}/2-\lvert p-x_k\rvert)_+}{w_{k-1}/2-w_k}\}$. The slopes of $\Delta_k\phi_k$ are $\pm\sigma_k$ with $\sigma_k:=\frac{\Delta_k}{w_{k-1}/2-w_k}\in [\frac{2\Delta_k}{w_{k-1}},\frac{4\Delta_k}{w_{k-1}}]$.
\item \emph{Needle.} $\xi_k$ is uniform over $\{1,2,\ldots,J_k\}$, where $J_k:=\lfloor \frac{w_{k-1}}{4\Delta_k}\rfloor\ge 8$, and $\phi_k$ is a narrow bump at the $\xi_k$-th of $J_k$ disjoint positions in $P_{k-1}$. Formally, $x_k:=x_{k-1}-w_{k-1}+4\Delta_k(2\xi_k-1)$ and $\phi_k(p):=\min\{1,\frac{(w_k+3\Delta_k-\lvert p-x_k\rvert)_+}{3\Delta_k}\}$, which is supported in $[x_k-4\Delta_k,x_k+4\Delta_k]\subseteq P_{k-1}$, where these supports are disjoint for different $\xi_k$. The slopes of $\Delta_k\phi_k$ are $\pm\sigma_k$ with $\sigma_k:=\frac 13$.
\end{itemize}
In both cases, $\phi_k=1$ exactly on $P_k:=[x_k-w_k,x_k+w_k]\subseteq P_{k-1}$, so $R_k=L_k$ on $P_k$ and $R_k<L_k$ elsewhere. Let $F_k$ be the value distribution with $\demand_{F_k}(p)=\frac{R_k(p)}p$, let $R:=\lim_kR_k$, let $F$ be the value distribution with $\demand_F(p)=\frac{R(p)}p$, which depends on $\xi:=(\xi_1,\xi_2,\ldots)$, and let $\mu$ be the distribution of $F$ when $\xi$ is drawn as above.

\textit{Horizons.} Scale $k$ forces regret at the horizon
\begin{equation*}
T_k:=\left\lfloor \frac{w_{k-1}}{2720\Delta_k^2}\right\rfloor\text{ on a tilt scale},\qquad T_k:=\left\lfloor \frac{J_k}{512\Delta_k^2}\right\rfloor\text{ on a needle scale},
\end{equation*}
up to which the two halves (with infinite menus) or the $J_k$ positions (with posted prices) cannot be told apart.

\begin{theorem}[Restatement of \Cref{thm:fixed instance}; Anytime Lower Bounds]
For every $d\in (0,1)$, there is a constant $c>0$ and a probability distribution $\mu$ over value distributions with interval dimension $d$ and multiplier $11$, such that for every algorithm, the following holds with probability $1$ over the randomized $F\sim \mu$:
\begin{enumerate}
\item Even with infinite menus, its regret on $F$ is at least $cT^{d/(d+1)}$ for infinitely many $T$.
\item When using posted prices, its regret on $F$ is at least $cT^{(d+1)/(d+2)}$ for infinitely many $T$.
\end{enumerate}
\end{theorem}
\begin{proof}
By \Cref{lem:fixed structure} below, $\mu$ is supported on value distributions with interval dimension $d$ and multiplier $11$. Fix an algorithm, and consider the first claim; the second claim is the same with needle scales in place of tilt scales. For a tilt scale $k$, let $B_k$ be the event that the regret on $F_k$ in $T_k$ rounds is at least $\frac{T_k\Delta_k}4$. It only depends on $\xi_1,\ldots,\xi_k$, and by \Cref{lem:fixed one scale}, its conditional probability given $\xi_1,\ldots,\xi_{k-1}$ is at least $\frac 12$. So for tilt scales $k_1<\cdots<k_n$, conditioning on $\xi_1,\ldots,\xi_{k_n-1}$ gives $\Pr\{B_{k_1}^c\cap\cdots\cap B_{k_n}^c\}\le \frac 12\Pr\{B_{k_1}^c\cap\cdots\cap B_{k_{n-1}}^c\}\le \cdots\le 2^{-n}$. Hence almost surely, $B_k$ holds for infinitely many tilt scales $k$, and for each of them, the regret on $F$ in $T_k$ rounds is at least $\frac{T_k\Delta_k}8$ by \Cref{lem:fixed finer scales}. Finally, $T_k\ge \frac{w_{k-1}}{5440\Delta_k^2}=\frac{\Delta_k^{-(1+d)}}{5440}$ on a tilt scale, so $T_k\Delta_k\ge 5440^{-1/(1+d)}T_k^{d/(d+1)}$; on a needle scale, $J_k\ge \frac{w_{k-1}}{8\Delta_k}$ and $T_k\ge \frac{J_k}{1024\Delta_k^2}\ge \frac{\Delta_k^{-(2+d)}}{8192}$, so $T_k\Delta_k\ge 8192^{-1/(2+d)}T_k^{(1+d)/(2+d)}$.
\end{proof}

We now prove the lemmas used above, in order of first use.

\begin{lemma}[Structure]\label{lem:fixed structure}
For every $\xi$, $F$ is a value distribution with revenue function $R$, whose maximum $L_\infty:=\lim_kL_k$ is attained only at $p^\ast=\lim_kx_k$. For $k\ge 1$ and $p\in P_{k-1}$, the gap $L_\infty-R(p)$ is $\Delta_k+\Delta_{>k}$ if $\phi_k(p)=0$, and $\sigma_k\operatorname{dist}(p,P_k)+\Delta_{>k}$ if $0<\phi_k(p)<1$; for $p\notin P_0$, it is at least $\frac 14(\lvert p-\frac 12\rvert-w_0)+\Delta_{>0}$. Moreover, $F$ has interval dimension $d$ with multiplier $11$.
\end{lemma}
\begin{proof}
All $\phi_k$ are supported in $P_0\subseteq [0.4,0.6]$, where $R_k\in [\frac 14,0.26]$ and hence $\frac{R_k(p)}p\in [0.41,0.66]$. On $P_0$, $R_k$ is $\frac 13$-Lipschitz, as the slopes of different scales lie on the disjoint sets $P_{k-1}\setminus P_k$ and $\sigma_k\le \frac 13$, so $(\frac{R_k}p)'=\frac 1p(R_k'-\frac{R_k}p)<0$ almost everywhere on $P_0$. Off $P_0$, $R_k=R_0$, and $\frac{R_0}p$ is non-increasing: It is $1$ where $R_0=p$; left of $P_0$ where $R_0<p$, its derivative is $-\frac{1+2w_0}{8p^2}$; right of $P_0$, its numerator decreases. So each $\frac{R_k}p$ is a continuous survival function, and so is its uniform limit $\frac{R}p$, as $0\le R_{k+1}-R_k\le \Delta_{k+1}$.

As the scales after $k$ are supported in $P_k$, we have $R=R_k=L_{k-1}+\Delta_k\phi_k$ on $P_{k-1}\setminus P_k$, where $\Delta_k(1-\phi_k(p))=\sigma_k\operatorname{dist}(p,P_k)$ if $\phi_k(p)>0$; this gives the gaps on $P_{k-1}$. Off $P_0$, $R=R_0\le \frac 14-\frac 14(\lvert p-\frac 12\rvert-w_0)$. Every $p\ne p^\ast$ lies in some $P_{k-1}\setminus P_k$ or off $P_0$, so its gap is positive.

For the interval dimension, let $\delta\in [\Delta_{k+1},\Delta_k)$ for some $k\ge 1$. By the gaps above, the prices in $P_{k-1}$ with $\phi_k=0$ and the prices off $P_{k-1}$ have gaps larger than $\delta$, and the $\delta$-optimal prices with $0<\phi_k<1$ lie within $\frac\delta{\sigma_k}$ of $P_k$. So $N_F(\delta)\le \frac{2w_k}\delta+\frac 2{\sigma_k}+1$, where $\frac{w_k}\delta\le \frac{\Delta_{k+1}^{1-d}}\delta\le \delta^{-d}$, and $\frac 2{\sigma_k}$ is $6$ on a needle scale and at most $\frac{w_{k-1}}{\Delta_k}=\Delta_k^{-d}\le \delta^{-d}$ on a tilt scale, so $N_F(\delta)\le 10\delta^{-d}$. If $\delta\ge \Delta_1$, then every $\delta$-optimal price is within $w_0+4\delta$ of $\frac 12$, so $N_F(\delta)\le \frac{2w_0}\delta+9\le 11\delta^{-d}$.
\end{proof}

\begin{lemma}[One Scale]\label{lem:fixed one scale}
Fix $k\ge 1$ and $\xi_1,\ldots,\xi_{k-1}$.
\begin{enumerate}
\item If $k$ is a tilt scale, then every algorithm, even with infinite menus and even if the seller observes $v_t$ in each round, has regret at least $\frac{T_k\Delta_k}4$ on $F_k$ in $T_k$ rounds for some $\xi_k\in \{-1,1\}$.
\item If $k$ is a needle scale, then every algorithm using posted prices has regret at least $\frac{T_k\Delta_k}2$ on $F_k$ in $T_k$ rounds for at least $\frac{J_k}2$ values of $\xi_k$.
\end{enumerate}
\end{lemma}
\begin{proof}
Write $G:=F_{k-1}$ and $w:=w_{k-1}$. Both claims compare $F_k$ with $G$, under which all prices in $P_{k-1}$ have the same revenue $L_{k-1}$: On a tilt scale, the two choices of $\xi_k$ form a two-point pair; on a needle scale, the $J_k$ choices form a needle in a haystack. For every $\xi_k$, the prices outside the support of $\phi_k$ have revenue at most $L_{k-1}$ under $F_k$, whereas $R_{F_k}(p_{F_k}^\ast)=L_{k-1}+\Delta_k$, so their gaps are at least $\Delta_k$.

For the first claim, let $F^\pm$ be $F_k$ with $\xi_k=\pm 1$. By \Cref{lem:fixed scale distance} below and the triangle inequality for $H$, $H^2(F^+,F^-)\le \frac{340\Delta_k^2}w$, so the first $T_k-1$ values have total variation distance at most $\sqrt{2T_kH^2(F^+,F^-)}\le \frac 12$ between $F^+$ and $F^-$. The supports of $\phi_k$ for $\xi_k=\pm 1$ only share $x_{k-1}$, where both vanish, so every price has gap at least $\Delta_k$ under $F^+$ or under $F^-$, and \Cref{lem:two-point} with the horizon $T_k$ gives the first claim.

For the second claim, let $F^{(j)}$ be $F_k$ with $\xi_k=j$, let $I_j$ be the support of $\phi_k$ in this case, and let $n_j$ be the number of rounds whose price lies in $I_j$. Under $G$ and $F^{(j)}$, posting $p$ observes Bernoulli variables with means $\frac{R_G(p)}p$ and $\frac{R_{F^{(j)}}(p)}p$, which coincide for $p\notin I_j$, and otherwise differ by at most $\frac{\Delta_k}p\le 2.5\Delta_k$ and lie in $[0.41,0.67]$, so their KL divergence is at most $\frac{(2.5\Delta_k)^2}{0.22}\le 32\Delta_k^2$. By the chain rule, the observations in $T_k$ rounds have KL divergence at most $32\Delta_k^2\E_G[n_j]$ between $G$ and $F^{(j)}$. As $\sum_j\E_G[n_j]\le T_k$, at least $\frac{J_k}2$ values of $j$ have $\E_G[n_j]\le \frac{2T_k}{J_k}$, for which this KL divergence is at most $\frac{64T_k\Delta_k^2}{J_k}\le \frac 18$, and the total variation distance is at most $\frac 14$ by Pinsker's inequality. Then $\E_{F^{(j)}}[n_j]\le \frac{2T_k}{J_k}+\frac{T_k}4\le \frac{T_k}2$, so the regret on $F^{(j)}$ is at least $\Delta_k\E_{F^{(j)}}[T_k-n_j]\ge \frac{T_k\Delta_k}2$.
\end{proof}

\begin{lemma}[Each Scale Is Close in Hellinger Distance]\label{lem:fixed scale distance}
For every $k\ge 1$ and every $\xi$, $H(F_k,F_{k-1})\le 2.2\sqrt{\Delta_k}$. Moreover, on a tilt scale $k$, $\lvert \frac{dF_k}{dF_{k-1}}-1\rvert\le \frac{10\Delta_k}{w_{k-1}}\le \frac 12$, and $H^2(F_k,F_{k-1})\le \frac{85\Delta_k^2}{w_{k-1}}$.
\end{lemma}
\begin{proof}
Write $G:=F_{k-1}$, $w:=w_{k-1}$, and $L:=L_{k-1}\in [\frac 14,0.26]$. The densities of $F_k$ and $G$ coincide outside the support of $\phi_k$, which lies in $P_{k-1}\subseteq [0.4,0.6]$. On this support, $G$ has density $\frac L{p^2}\le 1.7$, and $F_k$ has density $\frac{L+\Delta_k\phi_k}{p^2}-\frac{\Delta_k\phi_k'}p$.
\begin{itemize}
\item On a tilt scale, $\lvert \phi_k'\rvert\le \frac 4w$, so as $p\le 0.6$ and $w\le 0.1$,
\begin{equation*}
\left\lvert \frac{dF_k}{dG}-1\right\rvert\le \frac{\Delta_k}L+\frac{0.6\cdot 4\Delta_k}{wL}\le \frac{10\Delta_k}w\le \frac 12.
\end{equation*}
As the support of $\phi_k$ has length $w$ and hence $\intervalP_G$ at most $1.7w$, we get $H^2(F_k,G)\le \frac 12\chi^2(F_k\|G)\le \frac{85\Delta_k^2}w$. As $w=\Delta_k^{1-d}$ and $\Delta_k$ is small, this gives $H(F_k,G)\le \sqrt{85}\,\Delta_k^{(1+d)/2}\le 2.2\sqrt{\Delta_k}$.
\item On a needle scale, $\lvert \phi_k'\rvert\le \frac 1{3\Delta_k}$, so $\frac{dF_k}{dG}\in [0.19,1.81]$ on the support of $\phi_k$, which has length $8\Delta_k$ and hence $\intervalP_G$ at most $14\Delta_k$. So $H^2(F_k,G)\le \frac 12\cdot 0.81^2\cdot 14\Delta_k\le 4.6\Delta_k$, i.e., $H(F_k,G)\le 2.2\sqrt{\Delta_k}$.\qedhere
\end{itemize}
\end{proof}

\begin{lemma}[Finer Scales Do Not Affect the Revenue]\label{lem:fixed finer scales}
For every $\xi$ and $k\ge 1$, the regrets of an algorithm in $T_k$ rounds on $F$ and on $F_k$ differ by at most $\frac{T_k\Delta_k}8$ if $k$ is a tilt scale, and by at most $\frac{T_k\Delta_k}4$ if $k$ is a needle scale and the algorithm uses posted prices.
\end{lemma}
\begin{proof}
The finer scales change the revenue by at most $\Delta_{>k}$, and the value distribution by a squared Hellinger distance $\O(\Delta_{k+1})$, both negligible within $T_k$ rounds. Formally, by construction, $L_\infty-L_k=\Delta_{>k}$ and $0\le R-R_k\le \Delta_{>k}$.

On a tilt scale, the total revenue in $T_k$ rounds lies in $[0,T_k]$ and is a function of the values and the internal randomness, so the regrets differ by at most $T_k\Delta_{>k}+T_k\cdot \sqrt{2T_k}\,H(F,F_k)$. By \Cref{lem:fixed scale distance}, $H(F_m,F_{m-1})\le 2.2\sqrt{\Delta_m}$ for every $m$, so as $H$ is a metric and $F_m\to F$, we have $H(F,F_k)\le 2.3\sqrt{\Delta_{k+1}}$. So the regrets differ by at most $T_k(1.01\Delta_{k+1}+3.3\sqrt{T_k\Delta_{k+1}})\le \frac{T_k\Delta_k}8$, as $T_k\le \frac{\Delta_k^{-(1+d)}}{2720}$ and $\Delta_{k+1}=\Delta_k^q$ with $q>3+d$.

On a needle scale, the regret of a posted-price algorithm on a value distribution $G$ is $\E_G[\sum_t(R_G(p_G^\ast)-R_G(p_t))]$, so the regrets differ by at most $T_k\Delta_{>k}+T_k\cdot \mathrm{TV}$, where $\mathrm{TV}$ is the total variation distance between the observations under $F$ and under $F_k$. These Bernoulli variables have means that differ by at most $\frac{\Delta_{>k}}p\le 2.5\Delta_{>k}$, and lie in $[0.41,0.67]$ where they differ, so each has KL divergence at most $\frac{(2.5\Delta_{>k})^2}{0.22}\le 29\Delta_{>k}^2$, and by the chain rule and Pinsker's inequality, $\mathrm{TV}\le \sqrt{\frac{29T_k\Delta_{>k}^2}2}\le 3.9\sqrt{T_k}\Delta_{>k}$. So the regrets differ by at most $5T_k^{3/2}\Delta_{>k}\le \frac{T_k\Delta_k}4$, as $T_k\le \frac{\Delta_k^{-(2+d)}}{2048}$ and $q>\frac{4+d}2$.
\end{proof}

\subsection{Lower Bound for Lipschitz Experts}\label{app:experts}

\begin{definition}[Lipschitz Experts]\label{def:lipschitz experts}
In the \emph{Lipschitz experts} problem on $[0,1]$ \citep{kleinberg2019bandits}, in each round $t$, an algorithm picks an arm $x_t\in [0,1]$, then a payoff function $f_t:[0,1]\to [0,1]$ is drawn independently from an unknown distribution $P$ whose expected payoff $\mu(x):=\E_{f\sim P}[f(x)]$ is $1$-Lipschitz, and the algorithm receives $f_t(x_t)$ and observes $f_t$. Its regret is $T\max_x\mu(x)-\E[\sum_{t=1}^T\mu(x_t)]$.
\end{definition}
The separation from Lipschitz experts in \Cref{sec:lipschitz} uses the following lower bound, where the near-optimal arms concentrate as much as possible:

\begin{lemma}[Lipschitz Experts]\label{lem:experts lower bound}
For every $T\ge 1$, there are two payoff distributions $P_+,P_-$, under each of which the $\Delta$-optimal arms lie in an interval of length $4\Delta$ for every $\Delta>0$ (i.e., interval dimension 0), such that any Lipschitz expert algorithm has regret at least $\frac{\sqrt T}{64}$ under $P_+$ or under $P_-$.
\end{lemma}
\begin{proof}
Let $\epsilon:=\frac 1{8\sqrt T}$, $\mu_\pm(x):=\frac 12-\frac 12\lvert x-\frac 12\mp \epsilon\rvert$, which is 1-Lipschitz. Let $g:=\frac{\mu_++\mu_-}2\ge \frac 14$ and $h:=\frac{\mu_+-\mu_-}{4\epsilon}\in [-\frac 14,\frac 14]$. Let $P_+$ and $P_-$ draw every round $t$'s payoff function $f_t$ as follows:
\begin{equation*}
P_+:~f_t=\begin{cases}g+h&\text{w.p. }\frac{1+2\epsilon}2,\\ g-h&\text{w.p. }\frac{1-2\epsilon}2,\end{cases}\qquad\qquad P_-:~f_t=\begin{cases}g+h&\text{w.p. }\frac{1-2\epsilon}2,\\ g-h&\text{w.p. }\frac{1+2\epsilon}2,\end{cases}
\end{equation*}
so that their expected payoffs are $g\pm 2\epsilon h=\mu_\pm$. Under $P_\pm$, the $\Delta$-optimal arms form an interval of length at most $4\Delta$, and as the gaps of every arm under $P_+$ and under $P_-$ sum to at least $\epsilon$, every arm has gap at least $\frac\epsilon 2$ under one of them. Each payoff function only reveals which of $g\pm h$ is drawn, a coin whose squared Hellinger distance between $P_+$ and $P_-$ is $1-\sqrt{1-4\epsilon^2}\le 4\epsilon^2$, so the first $T-1$ payoff functions have total variation distance at most $\sqrt{8T\epsilon^2}\le \frac 12$. In each round $t$, the arm $x_t$ only depends on the first $t-1$ payoff functions and the internal randomness, so as every arm has gap at least $\frac\epsilon 2$ under $P_+$ or under $P_-$, the expected gaps of $x_t$ under $P_+$ and under $P_-$ sum to at least $\frac\epsilon 2(1-\frac 12)=\frac\epsilon 4$. Summing over $t$, one of the two regrets is at least $\frac{\epsilon T}8=\frac{\sqrt T}{64}$.
\end{proof}

\section{\texorpdfstring{Omitted Proofs in \Cref{sec:adv} (Adversarial Dynamic Pricing)}{Omitted Proofs: Adversarial Dynamic Pricing}}\label{app:adv}

We first verify that Hedge and Fixed-Share satisfy \Cref{def:online learning adv}, then prove the two lemmas on the estimates (\Cref{lem:adv estimator,lem:adv second moment}), and finally prove \Cref{thm:adv}. Throughout, $\E_t$ conditions on $v_t$ and the history before round $t$, so that it only averages over the draws $p_t\sim \bm w_t$ and $I_t\sim \bm \mu_t$. If $\mu_t(I)=0$, then $I_t\ne I$ almost surely, and we let $\hat \1_t(I):=0$ in \Cref{eq:importance weighting}.

\begin{definition}[Hedge and Fixed-Share]\label{def:hedge fixed share}
Let $\eta>0$ be the learning rate. Given the weights $\bm w_t\in \triangle[K]$ and the losses $\bm \ell_t$ in round $t$, let
\begin{equation*}
\tilde w_{t+1,i}:=\frac{w_{t,i}\exp(-\eta\ell_{t,i})}{\sum_{j\in [K]}w_{t,j}\exp(-\eta\ell_{t,j})},\quad \forall i\in [K].
\end{equation*}
Both algorithms start from $w_{1,i}=\frac 1K$ for all $i\in [K]$. \emph{Hedge} \citep{freund1997decision} sets $w_{t+1,i}=\tilde w_{t+1,i}$, i.e., $w_{t,i}\propto \exp(-\eta\sum_{\tau<t}\ell_{\tau,i})$. \emph{Fixed-Share} \citep{herbster1998tracking} mixes each update with the uniform distribution at rate $\frac 1T$, i.e., $w_{t+1,i}=(1-\frac 1T)\tilde w_{t+1,i}+\frac 1{KT}$.
\end{definition}

\begin{lemma}[Hedge and Fixed-Share Are Suitable]\label{lem:fixed share}
Let $\eta\le 1$. Hedge (\Cref{def:hedge fixed share}) satisfies \Cref{def:online learning adv} with the class $\mathcal C_0$ and $\textsc{Penalty}=\frac{\log K}\eta$. Fixed-Share satisfies it with the class $\mathcal C_{S-1}$ and $\textsc{Penalty}=\frac{S\log (KT)+1}\eta$ for every $S\ge 1$. Both have $\bm q_t\equiv 1$.
\end{lemma}
\begin{proof}
We bound the regret in each round by the change of $\log w_{t,i}$ for the comparator action $i$, and then sum these changes over the rounds.

\textit{One Round.} Fix $t$. As $\eta\le 1$ and $\ell_{t,j}\ge -1$, we have $\eta\ell_{t,j}\ge -1$, so $e^{-\eta\ell_{t,j}}\le 1-\eta\ell_{t,j}+\eta^2\ell_{t,j}^2$ by $e^{-x}\le 1-x+x^2$ for $x\ge -1$. Together with $\log y\le y-1$,
\begin{equation*}
\log \sum_{j\in [K]}w_{t,j}\exp(-\eta\ell_{t,j})\le -\eta\E_{j\sim \bm w_t}[\ell_{t,j}]+\eta^2\E_{j\sim \bm w_t}[\ell_{t,j}^2].
\end{equation*}
As $\log \frac{\tilde w_{t+1,i}}{w_{t,i}}=-\eta\ell_{t,i}-\log \sum_jw_{t,j}\exp(-\eta\ell_{t,j})$, for every $i\in [K]$,
\begin{equation}\label{eq:hedge one round}
\eta\left(\E_{j\sim \bm w_t}[\ell_{t,j}]-\ell_{t,i}\right)\le \log \frac{\tilde w_{t+1,i}}{w_{t,i}}+\eta^2\E_{j\sim \bm w_t}[\ell_{t,j}^2].
\end{equation}

\textit{Summing over Rounds.} Fix $(i_1^\ast,i_2^\ast,\ldots,i_T^\ast)\in \mathcal C_{S-1}$, and apply \Cref{eq:hedge one round} with $i=i_t^\ast$ in each round $t$. Regrouping the logarithms by rounds,
\begin{equation*}
\sum_{t=1}^T\log \frac{\tilde w_{t+1,i_t^\ast}}{w_{t,i_t^\ast}}=\log \frac 1{w_{1,i_1^\ast}}+\sum_{t=1}^{T-1}\log \frac{\tilde w_{t+1,i_t^\ast}}{w_{t+1,i_{t+1}^\ast}}+\log \tilde w_{T+1,i_T^\ast},
\end{equation*}
where $\log \frac 1{w_{1,i_1^\ast}}=\log K$ and $\log \tilde w_{T+1,i_T^\ast}\le 0$. After dividing by $\eta$, \Cref{def:online learning adv} holds with $\bm q_t\equiv 1$, as long as $\textsc{Penalty}$ bounds $\frac 1\eta$ times this sum, which we now do for each algorithm.

\textit{Hedge.} Every sequence in $\mathcal C_0$ has $i_{t+1}^\ast=i_t^\ast$, and $w_{t+1,i}=\tilde w_{t+1,i}$, so the middle sum is $0$, and $\textsc{Penalty}=\frac{\log K}\eta$.

\textit{Fixed-Share.} As $w_{t+1,i}\ge (1-\frac 1T)\tilde w_{t+1,i}$, each of the at most $T-1$ terms in the middle sum with $i_{t+1}^\ast=i_t^\ast$ is at most $\log \frac T{T-1}\le \frac 1{T-1}$, so these terms sum to at most $1$. As $w_{t+1,i}\ge \frac 1{KT}$ and $\tilde w_{t+1,i}\le 1$, each of the at most $S-1$ other terms is at most $\log (KT)$. Hence $\textsc{Penalty}=\frac{\log K+(S-1)\log (KT)+1}\eta\le \frac{S\log (KT)+1}\eta$.
\end{proof}

\begin{lemma}[Restatement of \Cref{lem:adv estimator}; Importance-Weighted Estimator]
For every round $t\in [T]$, for any price $p\in \candidateSet$ such that $w_t(p)>0$, we have $\E_t[\hat R_t(p)]=p\1[v_t\ge p]$. Meanwhile, $\hat R_t(p)\le 1$ is bounded from above.
\end{lemma}
\begin{proof}
By \Cref{lem:segment tree}, $\mI_\mT[0,p)$ partitions $[0,p)$ into tree intervals (and is empty if $p=0$), so $\1[v_t\ge p]=1-\sum_{I\in \mI_\mT[0,p)}\1_t(I)$. For every tree interval $I$ with $\mu_t(I)>0$, as $\Pr_t\{I_t=I\}=\mu_t(I)$, \Cref{eq:importance weighting} gives $\E_t[\hat \1_t(I)]=\1_t(I)$. If $w_t(p)>0$ and $p>0$, then every $I\in \mI_\mT[0,p)$ has $\mu_t(I)>0$ by \Cref{eq:exploration distribution}, as $p$ contributes $w_t(p)p^2>0$ to the sum defining $\mu_t(I)$. Hence, by \Cref{eq:income estimate},
\begin{equation*}
\E_t\left[\hat R_t(p)\right]=p\left(1-\sum_{I\in \mI_\mT[0,p)}\E_t\left[\hat \1_t(I)\right]\right)=p\left(1-\sum_{I\in \mI_\mT[0,p)}\1_t(I)\right)=p\1[v_t\ge p],
\end{equation*}
which also holds if $p=0$, as both sides are then $0$. Finally, as $\hat \1_t\ge 0$, we have $\hat R_t(p)\le p\le 1$ for every $p\in \candidateSet$.
\end{proof}

The proof of \Cref{thm:adv} also compares with prices that $\mA^{\text{OL}}$ may not play, for which the estimate is still optimistic in expectation:

\begin{lemma}[Optimistic Estimates for All Prices]\label{lem:adv estimator full}
For every round $t$ and every price $p\in \candidateSet$, we have $\E_t[\hat R_t(p)]\ge p\1[v_t\ge p]$.
\end{lemma}
\begin{proof}
For every tree interval $I$, as $\Pr_t\{I_t=I\}=\mu_t(I)$, \Cref{eq:importance weighting} gives $\E_t[\hat \1_t(I)]=\1_t(I)$ if $\mu_t(I)>0$, and $\hat \1_t(I)=0$ otherwise, so $\E_t[\hat \1_t(I)]\le \1_t(I)$. As $\mI_\mT[0,p)$ partitions $[0,p)$ by \Cref{lem:segment tree}, and $p\ge 0$,
\begin{equation*}
\E_t\left[\hat R_t(p)\right]=p\left(1-\sum_{I\in \mI_\mT[0,p)}\E_t\left[\hat \1_t(I)\right]\right)\ge p\left(1-\sum_{I\in \mI_\mT[0,p)}\1_t(I)\right)=p\1[v_t\ge p].\qedhere
\end{equation*}
\end{proof}

\begin{lemma}[Restatement of \Cref{lem:adv second moment}; Second Moment]
For every round $t$, we have
\begin{equation*}
\E_t\left[\sum_{p\in \candidateSet} w_t(p)\hat R_t^2(p)\right]\le 2+4(\log_2T+1)^2.
\end{equation*}
\end{lemma}
\begin{proof}
Fix a round $t$. As only $I_t$ can have $\hat \1_t(I)\ne 0$, we have $(\sum_{I\in \mI_\mT[0,p)}\hat \1_t(I))^2=\sum_{I\in \mI_\mT[0,p)}\hat \1_t^2(I)$, and $\E_t[\hat \1_t^2(I)]$ is $\frac{\1_t(I)}{\mu_t(I)}$ if $\mu_t(I)>0$ and $0$ otherwise. Let $\mT_t:=\{I\in \mT\mid \mu_t(I)>0\}$. Then
\begin{align*}
\E_t\left[\sum_{p\in \candidateSet}w_t(p)\hat R_t^2(p)\right]&\overset{(a)}{\le} 2\sum_{p\in \candidateSet}w_t(p)p^2+2\sum_{p\in \candidateSet}w_t(p)p^2\sum_{I\in \mI_\mT[0,p)\cap \mT_t}\frac{\1_t(I)}{\mu_t(I)}\\
&\overset{(b)}{=}2\sum_{p\in \candidateSet}w_t(p)p^2+2\sum_{I\in \mT_t}\frac{\1_t(I)}{\mu_t(I)}\sum_{p:\mI_\mT[0,p)\ni I}w_t(p)p^2\overset{(c)}{\le} 2+2Z_t\sum_{I\in \mT}\1_t(I),
\end{align*}
where (a) uses $\hat R_t^2(p)\le 2p^2+2p^2(\sum_{I\in \mI_\mT[0,p)}\hat \1_t(I))^2$; (b) swaps the summations; and (c) uses $\sum_pw_t(p)p^2\le 1$ and that, by \Cref{eq:exploration distribution}, the inner sum is $Z_t\mu_t(I)$ if $Z_t>0$ and is $0$ if $Z_t=0$. By \Cref{lem:segment tree} with $N=T$, the tree $\mT$ has at most $\log_2T+1$ depths, each $\mI_\mT[0,p)$ has at most two tree intervals of each depth, and $v_t$ lies in at most one tree interval of each depth. So $Z_t=\sum_pw_t(p)p^2\lvert \mI_\mT[0,p)\rvert\le 2(\log_2T+1)$ and $\sum_I\1_t(I)\le \log_2T+1$, which proves the lemma.
\end{proof}

\begin{theorem}[Restatement of \Cref{thm:adv}; Adversarial Dynamic Pricing]
For any comparator class $\mathcal C\subseteq \{0,\frac 1T,\ldots,1\}^T$, for any $\mA^{\text{OL}}$ satisfying \Cref{def:online learning adv} with $K=T+1$ and $\mathcal C$ translated to $[K]$, \Cref{alg:adv size-3} ensures
\begin{equation*}
\E\left[\sum_{t=1}^T p_t^\ast\1[v_t\ge p_t^\ast]\right]-\E\left[\sum_{t=1}^T\text{revenue}_t\right]\le \textsc{Penalty}(\mathcal C,T,T+1,\eta)+\O\left(\eta T\log^2 T+1\right),~\forall (p_1^\ast,\ldots,p_T^\ast)\in \mathcal C.
\end{equation*}
In particular, Hedge \citep{freund1997decision} with $\eta=(T\log T)^{-1/2}$ gives $\O\left(\sqrt{T\log^3 T}\right)$ regret against the best fixed price $\mathcal C_0$, and Fixed-Share \citep{herbster1998tracking} with $\eta=\sqrt{S/(T\log T)}$ gives $\O\left(\sqrt{ST\log^3 T}\right)$ against $\mathcal C_{S-1}$.
\end{theorem}
\begin{proof}
The first claim is \Cref{lem:adv exact} below, whose $\O(\cdot)$ term is $\eta T(2+4(\log_2T+1)^2)+2$. The rates compare with prices in $[0,1]$ rather than in $\candidateSet$, which \Cref{lem:adv rounding} below handles. As the rates are asymptotic, we assume $T\ge 3$; for Fixed-Share, we also assume $S\le T$, as otherwise $\sqrt{ST\log^3T}\ge T$ exceeds the trivial regret bound $T$. Then $\eta\le 1$ in both cases below, and $\log (T(T+1))\le 3\log T$.
\begin{itemize}
\item Hedge: By \Cref{lem:fixed share}, it satisfies \Cref{def:online learning adv} with the class $\mathcal C_0$ and $\textsc{Penalty}=\frac{\log (T+1)}\eta$. So for $\eta=(T\log T)^{-1/2}$, \Cref{lem:adv rounding} with $S=0$ bounds the regret against every fixed price in $[0,1]$ by
\begin{equation*}
\log (T+1)\sqrt{T\log T}+\sqrt{\frac T{\log T}}\left(2+4(\log_2T+1)^2\right)+3=\O\left(\sqrt{T\log^3T}\right).
\end{equation*}
\item Fixed-Share: By \Cref{lem:fixed share}, it satisfies \Cref{def:online learning adv} with the class $\mathcal C_{S-1}$ and $\textsc{Penalty}=\frac{S\log (T(T+1))+1}\eta$. So for $\eta=\sqrt{S/(T\log T)}$, \Cref{lem:adv rounding} with $S-1$ in place of $S$ bounds the regret against every sequence in $\mathcal C_{S-1}$ by
\begin{equation*}
(3S\log T+1)\sqrt{\frac{T\log T}S}+\sqrt{\frac{ST}{\log T}}\left(2+4(\log_2T+1)^2\right)+3=\O\left(\sqrt{ST\log^3T}\right).\qedhere
\end{equation*}
\end{itemize}
\end{proof}

\begin{lemma}[\Cref{thm:adv} with Explicit Constants]\label{lem:adv exact}
Let $\mathcal C\subseteq \candidateSet^T$, and let $\mA^{\text{OL}}$ satisfy \Cref{def:online learning adv} with $K=T+1$ and $\mathcal C$ translated to $[K]$. Then for every $(p_1^\ast,p_2^\ast,\ldots,p_T^\ast)\in \mathcal C$, \Cref{alg:adv size-3} ensures
\begin{equation*}
\E\left[\sum_{t=1}^T p_t^\ast\1[v_t\ge p_t^\ast]\right]-\E\left[\sum_{t=1}^T\text{revenue}_t\right]\le \textsc{Penalty}(\mathcal C,T,T+1,\eta)+\eta T\left(2+4(\log_2T+1)^2\right)+2.
\end{equation*}
\end{lemma}
\begin{proof}
We compare the revenue of the menus with that of the exploitation prices (Step 1), apply \Cref{def:online learning adv} to the estimated revenues on every sample path (Step 2), and take expectations using \Cref{lem:adv estimator,lem:adv estimator full,lem:adv second moment} (Step 3).

\textit{Step 1: Exploration Loses at Most $2\gamma$ Revenue per Round.} Fix a round $t$, and let $I_t=[a,b)$. The menu in round $t$ is built from the price distribution $\priceDist_t$ that puts mass $1-2\gamma$ on $p_t$ and mass $\gamma$ on each of $a$ and $b$. As all prices are in $[0,1]$,
\begin{equation*}
\text{revenue}_t=\E_{p\sim \priceDist_t}\left[p\1[v_t\ge p]\right]\ge (1-2\gamma)p_t\1[v_t\ge p_t]\ge p_t\1[v_t\ge p_t]-2\gamma.
\end{equation*}
As $p_t\sim \bm w_t$ under $\E_t$, we get $\E_t[\text{revenue}_t]\ge \sum_{p\in \candidateSet}w_t(p)p\1[v_t\ge p]-2\gamma$.

\textit{Step 2: Online Learning on Every Sample Path.} By \Cref{lem:adv estimator}, the loss $-\hat R_t(p)$ fed to $\mA^{\text{OL}}$ is at least $-1$. As \Cref{def:online learning adv} holds for every sequence of such losses, and $\bm q_t\le 1$, on every sample path,
\begin{equation*}
\sum_{t=1}^T\hat R_t(p_t^\ast)-\sum_{t=1}^T\sum_{p\in \candidateSet}w_t(p)\hat R_t(p)\le \textsc{Penalty}(\mathcal C,T,T+1,\eta)+\eta\sum_{t=1}^T\sum_{p\in \candidateSet}w_t(p)\hat R_t^2(p).
\end{equation*}

\textit{Step 3: Taking Expectations.} As $p_t^\ast$ is fixed in advance and $\bm w_t$ is determined by the history before round $t$, \Cref{lem:adv estimator full} gives $\E_t[\hat R_t(p_t^\ast)]\ge p_t^\ast\1[v_t\ge p_t^\ast]$, and \Cref{lem:adv estimator} gives $\E_t[\sum_{p\in \candidateSet}w_t(p)\hat R_t(p)]=\sum_{p\in \candidateSet}w_t(p)p\1[v_t\ge p]$, as only prices with $w_t(p)>0$ contribute. Taking expectations in Step 2 and bounding its last term by \Cref{lem:adv second moment},
\begin{equation*}
\E\left[\sum_{t=1}^T p_t^\ast\1[v_t\ge p_t^\ast]\right]-\E\left[\sum_{t=1}^T\sum_{p\in \candidateSet}w_t(p)p\1[v_t\ge p]\right]\le \textsc{Penalty}(\mathcal C,T,T+1,\eta)+\eta T\left(2+4(\log_2T+1)^2\right).
\end{equation*}
By Step 1, replacing the second expectation by $\E[\sum_t\text{revenue}_t]$ adds at most $2\gamma T=2$.
\end{proof}

The rates in \Cref{thm:adv} compare with prices in $[0,1]$ rather than in $\candidateSet$. Rounding them down to $\candidateSet$ adds no switches and loses at most $\frac 1T$ revenue per round:

\begin{lemma}[Rounding Comparators to the Grid]\label{lem:adv rounding}
Let $S\ge 0$, and let $\mA^{\text{OL}}$ satisfy \Cref{def:online learning adv} with $K=T+1$ and the class $\mathcal C_S$ translated to $[K]$. Then for every $(p_1^\ast,p_2^\ast,\ldots,p_T^\ast)\in \mathcal C_S$, i.e., every sequence of prices in $[0,1]$ that switches at most $S$ times, \Cref{alg:adv size-3} ensures
\begin{equation*}
\E\left[\sum_{t=1}^T p_t^\ast\1[v_t\ge p_t^\ast]\right]-\E\left[\sum_{t=1}^T\text{revenue}_t\right]\le \textsc{Penalty}(\mathcal C_S,T,T+1,\eta)+\eta T\left(2+4(\log_2T+1)^2\right)+3.
\end{equation*}
\end{lemma}
\begin{proof}
Let $\bar p_t:=\frac{\lfloor Tp_t^\ast\rfloor}T\in \candidateSet$. If $p_t^\ast=p_{t+1}^\ast$, then $\bar p_t=\bar p_{t+1}$, so $(\bar p_1,\bar p_2,\ldots,\bar p_T)$ also switches at most $S$ times, and \Cref{lem:adv exact} applies to it. It remains to show that rounding loses at most $\frac 1T$ in each round $t$:
\begin{equation*}
\bar p_t\1[v_t\ge \bar p_t]\overset{(a)}{\ge} \bar p_t\1[v_t\ge p_t^\ast]\overset{(b)}{\ge} \left(p_t^\ast-\frac 1T\right)\1[v_t\ge p_t^\ast]\ge p_t^\ast\1[v_t\ge p_t^\ast]-\frac 1T,
\end{equation*}
where (a) uses $\bar p_t\le p_t^\ast$ and $\bar p_t\ge 0$, and (b) uses $p_t^\ast-\bar p_t<\frac 1T$.
\end{proof}

\end{document}